\documentclass[11pt,a4paper]{article}
\usepackage{amscd,amsmath,amssymb,amsfonts,xspace,mathrsfs,mathtools,amsthm}
\usepackage{jheppub}
\renewcommand*{\backref}[1]{}
\renewcommand*{\backrefalt}[4]{{%
		\ifcase #1 %Not cited.%
		\or [Cited: p.~#2.]%
		\else [Cited: p. #2.]%
		\fi%
	}}
\usepackage{xr-hyper}
\usepackage[dvipsnames]{xcolor}
\usepackage{bm}
\usepackage{bbm}
\usepackage{graphicx}
\usepackage{dsfont}
\usepackage{tocloft}%toc on one page
\usepackage[color,matrix,arrow]{xy}
\usepackage{stmaryrd}
\SetSymbolFont{stmry}{bold}{U}{stmry}{m}{n}
\usepackage[shortlabels]{enumitem}
\usepackage{array,amsmath,booktabs}
\usepackage{slashed}
\usepackage{longtable}
\usepackage{tensor}
\usepackage{caption}
\usepackage{subcaption}
\usepackage{enumitem}
\usepackage{lineno}
\usepackage{diagbox}
\usepackage{pdflscape}
\usepackage{silence}
\usepackage{booktabs, cellspace, hhline}
\newcommand{\sss}[1]{\medskip \noindent  \textbf{#1}}
\usepackage{booktabs, cellspace, hhline}
\newcommand{\RN}[1]{%
  \textup{\uppercase\expandafter{\romannumeral#1}}%
}

\newtheorem{conjecture}{Conjecture}
\newtheorem{theorem}{Theorem}

\makeatletter
\def\th@plain{%
  \thm@notefont{}%
  \itshape
  \addtolength\thm@preskip\parskip
  \addtolength\thm@postskip\parskip
}
\makeatother
\newcommand{\Td}{{\rm Td}}

\newcommand{\cc}{\mathsf C}
\newcommand{\Eul}{{\rm Eul}}
\newcommand{\tra}{\mathsf{T}} 
\newcommand{\ch}{{\rm ch}}
\newcommand{\p}{\partial}
\newcommand{\R}{\mathbb{R}}
\newcommand{\Z}{\mathbb{Z}}
\newcommand{\mr}[1]{\mathrm{#1}}
\newcommand{\mf}[1]{\mathfrak{#1}}
\newcommand{\tex}[1]{\text{#1}}
\newcommand{\s}[1]{\mathcal{#1}}
\newcommand{\ov}[1]{\overline{#1}}

\newcommand{\coker}{{\rm coker}} 
\newcommand{\eg}{\textit{e.g.}}

\newcommand{\SW}{{\rm SW}}

\newcommand{\vd}{\text{vd}\,}

\newcommand{\tauD}{\tau_{\rm D}} 
\newcommand{\qD}{q_{\rm D}} 
\newcommand{\PSU}{{\rm PSU}} 
\newcommand{\SU}{{\rm SU}}

\renewcommand{\t}{\widetilde }

\newcommand{\smfr}[2]{\hbox{$\frac{#1}{#2}$}}

\newcommand{\diag}{\text{diag}} 
\newcommand{\ind}{{\rm ind}\,} 
\newcommand{\Ind}{{\rm Ind}\,}

\newcommand{\spinc}{\text{Spin}^c} 
\newcommand{\rk}{\text{rk}} 

\newcommand{\Disc}{{\rm Disc}}

\newcommand{\slz}{\text{SL}(2,\mathbb Z)}  
 
\newcommand{\Sp}{{\rm Sp}} 
\newcommand{\tq}{\mathtt{q}} 
\newcommand{\nstar}{\CN=2^*}

\newcommand{\jt}{\vartheta}

\newcommand{\chih}{\chi_{\rm h}}

\newcommand{\sw}{\mathcal{SW}}

\newcommand{\bea}{\begin{equation} \begin{aligned}}
\newcommand{\eea}{\end{aligned} \end{equation}}
\newcommand{\beasmall}{\begin{equation}\footnotesize\begin{aligned}}
\newcommand{\eeasmall}{\end{aligned}\footnotesize\end{equation}}
\newcommand{\be}{\begin{equation}} 
\newcommand{\ee}{\end{equation}} 
\newcommand{\bes}{\begin{equation*}}
\newcommand{\ees}{\end{equation*}}

\newcommand{\ip}{\text{Im}}
\newcommand{\im}{i}
\newcommand{\bk}{\bm{k}}
\newcommand{\bz}{\bm{z}}

\newcommand{\cross}{\mathcal C}
\newcommand{\tn}{\mathtt n}
\newcommand{\tA}{\mathtt A}
\newcommand{\tJ}{\mathtt J}
\newcommand{\tr}{\mathtt r}
\newcommand{\ts}{\mathtt s}
\newcommand{\tC}{\mathtt C}
\newcommand{\tD}{\mathtt D}
\newcommand{\tG}{\mathtt G}
\newcommand{\tH}{\mathtt H}
\newcommand{\tu}{\mathtt u}
\newcommand{\tU}{\mathtt U}
\newcommand{\tm}{\mathtt m}
\newcommand{\tx}{\mathtt x}
\newcommand{\tv}{\mathtt v}
\newcommand{\point}{{\mathrm{pt}}}
\newcommand{\mud}{\mu_{\mathrm D}}
\newcommand{\segre}{\mathscr S}
\newcommand{\segregen}{\mathcal G}
\newcommand{\universal}{\mathscr U}
\newcommand{\decoup}{\mathscr D}
\newcommand{\Mi}{\CM_k^{\rm i}}

\usepackage{environ}

\NewEnviron{eqsp}{%
\bea
\BODY
\eea
}

\newcommand{\CC}{\mathcal{C}}  
\newcommand{\cD}{\mathcal{D}}  
  
\newcommand{\CF}{\mathcal{F}} 
\newcommand{\CG}{\mathcal{G}} 

\newcommand{\CI}{\mathcal{I}}
\newcommand{\CJ}{\mathcal{J}}

\newcommand{\CL}{\mathcal{L}} 
\newcommand{\CM}{\mathcal{M}}  
\newcommand{\CN}{\mathcal{N}}
\newcommand{\CO}{\mathcal{O}} 

\newcommand{\CQ}{\mathcal{Q}}
\newcommand{\CR}{\mathcal{R}}
\newcommand{\CS}{\mathcal{S}}

\newcommand{\CV}{\mathcal{V}}

\newcommand{\CZ}{\mathcal{Z}}

\newcommand{\BR}{\mathbb{R}}

\newcommand{\BC}{\mathbb{C}}

\newcommand{\BZ}{\mathbb{Z}}
\newcommand{\BQ}{\mathbb{Q}}

\newcommand{\bfa}{{\boldsymbol a}}

\newcommand{\bfc}{{\boldsymbol c}}

\newcommand{\bfm}{{\boldsymbol m}} 
\newcommand{\bfn}{{\boldsymbol n}}

\newcommand{\bfk}{{\boldsymbol k}}

\newcommand{\bfmu}{{\mu}}

\newcommand{\Tr}{\text{Tr}}

\title{\texorpdfstring{$\boldsymbol{\mathrm{SU}(N)}$}{SU(N)} QCD, four-manifolds and Segre numbers}

\abstract{
We study correlation functions of topologically twisted $\mathcal N=2$ supersymmetric QCD with gauge group $\mathrm{SU}(N)$ on a compact four-manifold $X$. We derive the Coulomb branch integral for all $N$, including new couplings to background fluxes for the flavor symmetry. For $X$ of simple type with $b_2^+>1$, the correlators become sums over a finite number of strong-coupling singularities on the Coulomb branch, and can be expressed using a number of Coulomb branch couplings. The sum contains the $\mathcal N=1$ vacua, which become the supersymmetric vacua when the $\mathcal N=2$ theory is broken to $\mathcal N=1$ by an adjoint mass deformation, and whose contribution we attribute to the instanton branch.
We focus on asymptotically free $\mathrm{SU}(N)$ QCD with $N_f<2N$ equal mass hypermultiplets, and conjecture that its instanton contribution to topological correlators agrees with the generating function of virtual Segre numbers of rank $N$, which are intersection numbers on the moduli space of semistable sheaves on an algebraic surface. We present a precise dictionary for all $N$ between universal series for Segre numbers and Coulomb branch couplings at the $\mathcal N=1$ vacua. The calculation of these couplings requires new higher-rank techniques, involving Siegel modular forms, an exact $\mathbb Z_N$ symmetry, charge conjugation symmetry, factorisation of the Seiberg--Witten curve, and the blowup formula. For pure $\mathrm{SU}(N)$, our result agrees with earlier work by Mariño and Moore, and the twisted correlators give the $\mathrm{SU}(N)$ Donaldson invariants of $X$. For $\mathrm{SU}(3)$ QCD, we determine closed-form expressions for all 13 Coulomb branch couplings for each $N_f=1,\dots, 5$, and find precise agreement with results of Göttsche and Kool for virtual rank 3 Segre numbers.
}  

\author[a]{Elias Furrer,} \author[a,b]{Ranveer Kumar Singh} 
\affiliation[a]{New High Energy Theory Center and Department of Physics and Astronomy, Rutgers University, 126 Frelinghuysen Rd., Piscataway, NJ 08855, USA}
\affiliation[b]{Mani L. Bhaumik Institute for Theoretical Physics, Department of Physics and
Astronomy, University of California, Los Angeles, CA 90095, USA}

\preprint{}
\begin{document}
\maketitle
\baselineskip=18pt

\section{Introduction and summary}

\subsection*{Topological twisting and 4-manifold invariants}
One of the most important examples of the fruitful interaction between pure Mathematics and Physics is the application of supersymmetric quantum field theory (QFT) to 4-manifold topology. Path integrals in a general QFT are very difficult to define or compute but, as shown by Witten in~\cite{Witten:1988ze}, in topologically twisted 4d $\CN=2$ theories, the infinite-dimensional path integral for certain observables localizes to a finite-dimensional integral over certain moduli spaces. These correlation functions then lead to 4-manifold invariants. In~\cite{Witten:1988ze}, Witten gave a path integral representation for Donaldson invariants using the topologically twisted 4d $\CN=2$ super Yang--Mills (SYM) theory with gauge group $\SU(2)$. In this paper, we study the correlation functions of topologically twisted 4d $\CN=2$ super Yang--Mills theory with gauge group $\SU(N)$ and $N_f<2N$ flavors of fundamental hypermultiplets, henceforth called $\SU(N)$ SQCD, on a compact, oriented, Riemannian 4-manifold $X$ without boundary. 
\par 
Even though the path integral of the topological theory localizes to a finite-dimensional integral, its computation is not easy. It was proposed in~\cite{Witten:1994cg} that the path integral can be calculated using the low-energy effective theory. The low-energy effective theory of pure $\mr{SU(2)}$ SYM was understood by Seiberg and Witten in~\cite{Seiberg:1994rs,Seiberg:1994aj}. Witten then defined the Seiberg--Witten (SW) invariants using the Seiberg--Witten solution and proposed a relation between the Donaldson invariants and the Seiberg--Witten invariants based on physical insights from renormalization group flow~\cite{Witten:1994cg}. This proposal was proven by Moore--Witten in~\cite{Moore:1997pc}. Schematically, the relation reads
\begin{eqsp}\label{eq:ZD=Zu+ZSW}
    Z_{\mr{D}}=Z_{u}+Z_{\SW}~,
\end{eqsp}
where $Z_{\mr{D}}$ is the generating function of Donaldson invariants, $Z_u$, called the $u$-plane integral or Coulomb branch integral, is an integral over the regularized Coulomb branch of the theory (parametrized by $u\in\mathbb{C}$), and $Z_{\SW}$ is the Seiberg--Witten contribution from the singularities on the Coulomb branch. It turns out that $Z_{\SW}$ is considerably easier to evaluate compared to $Z_{\mr{D}}$. Moreover, for 4-manifolds $X$ with $b_2^+>1$, the $u$-plane integral vanishes and hence $Z_{\mr{D}}=Z_{\SW}$~\cite{Moore:1997pc}. These relations have been proven rigorously in the Mathematics literature, see for example~\cite{pidstrigach1995localisation,feehan2001pu,feehan2019so}.\par 
The next natural step is then to compute $Z_{\mr{D}}$ for other classes of 4d $\CN=2$ theories. Such calculations have been completed for many classes of Lagrangian 4d $\CN=2$ theories, see~\cite{Marino:1998bm,Hyun:1995hz,Manschot:2021qqe,Marino:1998uy}  for a sample of papers. Indeed, it can be shown that the path integral of any topologically twisted Lagrangian 4d $\CN=2$ theory localizes to a finite-dimensional integral over the moduli space of solutions to nonabelian monopole equations~\cite{labastida1995,Laba05,Labastida:1995zj,Moore:2024vsd,Hyun:1995mb}. For rank-one gauge groups, the low-energy effective theory is known in most cases, and hence one can again relate $Z_{\mr{D}}$ and $Z_{\SW}$ using the $u$-plane integral~\cite{Moore:1997pc,Aspman:2023ate,Aspman:2022sfj}.
\par 
The low energy effective theory has been determined for many higher-rank 4d $\CN=2$ theories.\footnote{See~\cite{Seiberg:1994aj,Seiberg:1994rs,Ne,Huang:2009md,Klemm:1994qs,Klemm:1994qj,Klemm:1995wp,Danielsson:1995is,Argyres:1995xn,Masuda:1996xj,Masuda:1997nv,Minahan:1995er,Minahan:1996ws,Argyres:1998bn,Argyres:1995jj,Galakhov:2013oja,Ferrari:1996sv,Argyres:1994xh,DHoker:2022loi,DHoker:2024vii,DHoker:2020qlp,Bonelli:2017ptp,DHoker:2024vii,Aharony:1996ks,AlvarezGaume:1997fg,AlvarezGaume:1997ek,Russo:2019ipg} for a sampling of results.} The Coulomb branch of a theory with gauge group of rank $r$ is an $r$-dimensional complex manifold (in fact a special K\"{a}hler manifold) with a complicated singularity structure. In particular, there are higher codimension subspaces of the Coulomb branch over which the SW curve degenerates and extra BPS states become massless. As a result, the expression for $Z_{\SW}$ for higher-rank gauge theories is considerably more complicated. The first attempt to compute $Z_{\mr{D}}$ for higher-rank gauge groups using the Coulomb branch integral appeared in~\cite{Marino:1998bm}, where the authors calculated the SW contribution for the pure $\SU(N)$ SYM theory. 
For 4-manifolds with $b_2^+>1$, this provided a complete expression for higher-rank Donaldson invariants, which were later defined in~\cite{Kronheimer2005}. 
\par 
An explicit computation of the Coulomb branch integral for higher rank presents new challenges. We list a selection of them:
\begin{enumerate}
    \item A central feature in all previous studies of rank-one theories is modularity: The simple monodromy groups on the Coulomb branch often translate to (algebraic functions of) modular forms~\cite{Nahm:1996di,Moore:1997pc,Malmendier:2008db,Korpas:2017qdo,Korpas:2019cwg,Korpas:2019ava,Aspman:2021vhs,Aspman:2021evt,Manschot:2021qqe,Kim:2025fpz,Magureanu:2022qym}. Even in cases where modularity is broken by mass parameters, expansions exist where modularity is restored, such as for small or large masses.     
    For higher rank, the modularity is elusive; even for pure $\SU(3)$ SYM it is not completely understood. As a consequence, we do not have a tool to study Coulomb branch couplings for higher-rank QCD as large or small mass series with modular coefficients.
    \item There are simply many more couplings for higher rank, which require new techniques for their calculation.
    \item Relatedly, for rank larger than one, there is a discrepancy between the number of Coulomb branch coordinates and elements of the coupling matrix: for $\SU(N)$ theories there are $N-1$ coordinates, while the coupling matrix has $\frac12 N(N-1)$ a priori distinct elements. The dependence of all couplings on the Coulomb branch coordinates is difficult to determine explicitly even for pure $\SU(3)$~\cite{Klemm:1995wp,Aspman:2020lmf}.
\end{enumerate}
Another approach to computing partition functions of topologically twisted Lagrangian theories is available for toric manifolds, where one can patch up the local contributions~\cite{Bonelli:2020xps,Closset:2022vjj,Bershtein:2015xfa,Khlaif:2026rnb,Dennemann:2026idg}. Since toric manifolds are locally $\mathbb{C}^2$, one can determine the local contributions using Nekrasov's instanton counting~\cite{Nekrasov:2003vi,Nekrasov:2012xe,Ne,NekOk,Lossev:1997bz}. This setup is particularly well understood in the algebro-geometric perspective~\cite{Nakajima:2003pg,Nakajima:2003uh,Nakajima:2005fg,Gottsche:2006tn,Gottsche:2006}

\subsection*{Topological correlators in \texorpdfstring{$\SU(N)$}{SU(N)} SQCD}
In this paper, we study the Coulomb branch integral and Seiberg--Witten contributions of the $\SU(N)$ SQCD with $N_f<2N$ hypermultiplets. The rank-one case, i.e. $N=2$, was extensively studied in~\cite{Malmendier:2008db,Aspman:2023ate,Aspman:2022sfj,Furrer:2026byd}. In particular, this article is a direct generalization of the results of~\cite{Furrer:2026byd} to higher rank. As was noticed in~\cite{Labastida:1997rg,Manschot:2021qqe,Aspman:2023ate,Aspman:2022sfj} and fully generalized in~\cite{Moore:2024vsd}, twisting the theory requires us to choose extra topological data on the 4-manifold. Following~\cite{Moore:2024vsd}, we describe the topological twisting of the $\SU(N)$ SQCD and discuss the precise cohomological condition required on the 4-manifold $X$. The twisted partition function $Z_{\mr{D}}(\bm{m})$ depends on the masses $\bm{m}=(m_1,\dots,m_{N_f})$ and other background fields (such as ``generalized spin-c structure''~\cite{Moore:2024vsd} and 't Hooft flux). The analogue of~\eqref{eq:ZD=Zu+ZSW} is then 
\begin{eqsp}
    Z_{\mr{D}}(\bm{m})=Z_{\mr{Coulomb}}(\bm{m})+Z_{\SW}(\bm{m})~,
\end{eqsp}
where $Z_{\mr{Coulomb}}$ is the Coulomb branch integral. Following~\cite{Marino:1998bm}, we derive the expression for $Z_{\mr{Coulomb}}(\bm{m})$. The singularity structure of the Coulomb branch and SW curves for $\SU(N)$ SQCD has been discussed in various references, see for example~\cite{Klemm:1994qs,Klemm:1995wp,Douglas:1995nw,Argyres:1994xh,Argyres:1995wt,Witten:1997sc}. The Coulomb branch has a singular locus which admits a stratification into loci of various codimensions. The SW contribution $Z_{\SW}(\bm{m})$ then naturally decomposes into a sum over contributions from the various strata. These contributions can be determined recursively by using the wall-crossing behaviour of $Z_{\mr{Coulomb}}$~\cite{Moore:1997pc,Marino:1998bm}. For 4-manifolds $X$ of SW simple-type with $b_2^+>1$, which we consider in the following, the only nonvanishing SW contribution comes from the top codimension singular points, which consist of superconformal fixed points, quark singularities, and the $\CN=1$ vacua. In the simplest case of pure $\SU(3)$, the superconformal points are the original Argyres--Douglas vacua~\cite{Argyres:1995jj}. It is believed (although unproven) that the contribution from SCFT points vanishes for the four-manifolds $X$ we consider. We are particularly interested in the contribution $Z_{\SW}^{\CN=1}(\bm{m})$ from the $\CN=1$ points on the Coulomb branch, which are the supersymmetric vacua of the $\CN=1$ gauge theory after the standard adjoint superpotential deformation of the $\CN=2$ vector multiplet.
The reason is the following: the UV partition function $Z_{\mr{D}}(\bm{m})$ is given by an integral over the moduli space $\CM$ of non-Abelian $\SU(N)$ multi-monopole equations~\cite{Labastida:1995zj}. After a localization with respect to the $\mr{U(1)}^{N_f}$ symmetry, $Z_{\mr{D}}(\bm{m})$ decomposes into a contribution $Z_{\mr{D}}^{\rm i}(\bm{m})$ from the instanton branch and a contribution $Z_{\mr{D}}^{\rm r}(\bm{m})$ from the reducible branch of $\CM$~\cite{LoNeSha}. Since there is no reducible branch for $N_f=0$ and the decoupling limit removes the quark singularities from the Coulomb branch, with the $\CN=1$ vacua giving the full answer for the $N_f=0$ theory, it is expected that~\cite{Furrer:2026byd}
\begin{eqsp}\label{eq:Z_Di=ZSWN=1}
    Z_{\mr{D}}^{\rm i}(\bm{m})=Z_{\SW}^{\CN=1}(\bm{m})~.
\end{eqsp}
We also denote the contribution from $\CN=1$ singularities by $Z_{\SW}^{\rm i}\coloneqq Z_{\SW}^{\CN=1}(\bm{m})$.
We derive the explicit form of $Z_{\SW}^{\rm i}$ for $\SU(N)$ QCD with arbitrary $N$, and express the result in terms of a set of universal functions (i.e. independent of the 4-manifold $X$). These universal functions are given by combinations of Seiberg--Witten couplings, such as couplings to the curvature of $X$ and background fluxes, evaluated at the $\CN=1$ vacua. We develop a variety of tools to compute these universal functions. In the process, we clarify several aspects of higher-rank Seiberg--Witten theory. 
\par
For comparison with results in the mathematical literature, we focus on the equal mass configuration $\bfm=(m,\dots, m)$ of $Z_{\SW}^{\rm i}(\bm{m})$. In the large $m$ limit, the $\CN=1$ singularities of the $\SU(N)$ SQCD flow to the $\CN=1$ vacua of the pure $\SU(N)$ theory (see section~\ref{sec:N=1_vacua}). Thus, the contribution $Z_{\SW}^{\rm i}(m)$ agrees with the full partition function of pure $\SU(N)$ theory for $m\to\infty$. More precisely, the contribution $Z_{\SW}^{\rm i}(m)$ is the product of the pure $\SU(N)$ partition function times a set of universal functions which are infinite series in $1/m$. 
\subsection*{Relation to virtual Segre numbers}
Another central aspect of this paper is the comparison with the algebraic geometry results. By a localization argument, the coefficients of the large equal mass expansion of $Z_{\SW}^{\rm i}(m)$ can be interpreted as intersection numbers of the instanton moduli space. Invoking the Donaldson--Uhlenbeck--Yau theorem~\cite{Donaldson1985,Uhlenbeck1986} allows us to make contact with the results in algebraic geometry~\cite{Lehn:1999,gottsche2022,Gottsche:2019meh,Marian2024,MARIAN2019,Marian2021,Marian:2015,Gottsche:1996,Gottsche:2006,Gottsche:2006tn,Mochizuki2009,Gttsche2024,Oberdieck_2022}. Indeed, L. Göttsche and M. Kool computed in~\cite{Gottsche:2020ass} certain intersection numbers, called \textit{virtual Segre numbers}, on the moduli space of semistable sheaves on algebraic surfaces. In more detail, let $S$ be an algebraic surface with $b_2^+(S)>1$, $x\in H_{\bullet}(S)$ a homology class, $\alpha$ a $K$-theory class on $S$, and $M$ the moduli space of semistable rank $N$ sheaves on $S$. Virtual $\SU(N)$ Segre numbers are defined by 
\begin{eqsp}
    \segre_{\alpha}\coloneqq \int_{M}c(\alpha_M)\mud(x)~,
\end{eqsp}
where $\alpha_M$ is the tautological class on the moduli space associated to $\alpha$, $c(\alpha_M)$ denotes the total Chern class and $\mud$ is the Donaldson map. Segre numbers generalize Donaldson invariants in the sense that for the trivial K-theory class $\alpha$, Segre numbers reduce to the $\SU(N)$ Donaldson invariants~\cite{Kronheimer2005}.  Let us write $\CG_{\segre}$ for the generating function of Segre numbers:
\begin{eqsp}
    \CG_{\segre}=\sum_{c_2\in \Z}z^{\frac{1}{2}\vd M}\segre_{\alpha}~,
\end{eqsp}
where the sum is over the second Chern class $c_2$ of sheaves in $M$, $z$ is a formal variable and $\vd M$ denotes the virtual dimension of $M$. 
There exists a universal series $\universal_{\alpha}$ valued in $\mathbb{C}[[z^{\frac{1}{2}}]]$, which is a certain product of $2N+7+\frac{N(N-1)}{2}$ universal functions $V_s,W_s,\dots$,
\begin{eqsp}
    \universal_{\alpha}=V_s^{c_2(\alpha)}W_s^{c_1(\alpha)^2}\times\dots~.
\end{eqsp}
From this, let $\CG_{\universal}$ be a certain projection (such as the even/odd part) of $\universal_\alpha$, depending on the details of $M$ and $S$. Göttsche--Kool conjectured~\cite{Gottsche:2020ass} 
that the generating function $\CG_{\segre}$ of Segre numbers is given by this projection of the universal series,
\begin{eqsp}
   \CG_{\segre}=\CG_{\universal}~.
\end{eqsp}
\par 
To compare the universal series $\universal_{\alpha}$ with the SW contribution of twisted $\SU(N)$ SQCD, we identify the $K$-theory class $\alpha$ with the background flavor bundle and the Chern classes $c_1(\alpha),c_2(\alpha)$ with the background fluxes. There are $N$ $\CN=1$ vacua, whose contribution we denote by $Z_{\SW,n}$, where $n=0,\dots, N-1$.
With the identification $z=-\frac{\Lambda_0^2}{4^{\frac1N}m^2}$, where $\Lambda_0$ is the scale of the $N_f=0$ theory, and $s=N_f$, we propose that 
\begin{eqsp}
   Z_{\SW,0}=\frac{1}{N}\universal_\alpha~. 
\end{eqsp}
The sum over all $\CN=1$ points then implements the projection to $\CG_{\universal}$ alluded to above, and we get 
\begin{eqsp}\label{eq:ZN=1SW=Gu}
    Z_{\SW}^{\rm i}=\CG_{\universal}~.
\end{eqsp}
We support the proposal by explicitly computing the couplings in SW theory and matching them with the universal series in~\cite{Gottsche:2020ass} for $N=3$. For $N=2$, it agrees with the earlier result in~\cite{Furrer:2026byd}.
%\par 
We notice that there are fewer physical couplings in SW theory compared to the number of universal functions. The precise dictionary then indicates that the universal functions in~\cite{Gottsche:2020ass} must satisfy nontrivial constraints, which we checked to hold again for $N=3$.
\subsection*{Summary of results}
We now briefly summarize the main results of this paper.
\begin{enumerate}
    \item We derive an expression for the Coulomb branch integral $Z_{\mr{Coulomb}}$ for $\SU(N)$ SQCD with background fluxes, see~\eqref{eq:uplintlagside}. Then, using the wall-crossing behaviour of $Z_{\mr{Coulomb}}$, we determine the SW contributions at the $\CN=1$ points, see~\eqref{eq:SW_cont_gen}. This generalizes the results of~\cite{Moore:1997pc,Marino:1998bm,Aspman:2023ate,Aspman:2022sfj}. Finally, we derive the expression for contact terms in terms of Riemann theta functions, see~\eqref{eq:HNf_gen} and~\eqref{eq:T_a,b_theta}. 
    \item We discuss various characterizations of the $\CN=1$ vacua (see section~\ref{sec:N=1_vacua}). Using the decoupling limit, we fix the coordinates of the $\CN=1$ points and using the factorization of the SW curve, we compute the period matrix at strong coupling. We clarify several features of $\SU(N)$ SQCD along the way, including the residual action of charge conjugation symmetry and $R$-symmetry, normalisations of periods, and new identities among the discriminant $\Delta_{N_f}$, the period matrix $A_{ab}=\frac{\partial a^a}{\partial u_b}$, and Siegel modular forms. For $\SU(3)$ with $N_f\leq 5$ flavors for instance, we find the identity  
    \bea\label{delta det = chi10 intro}
    \Delta_{N_f} \left(\det A\right)^{10}=c\, \chi_{10}~,
    \eea
    where $\chi_{10}$ is Igusa's cusp form and $c$ a constant. 
    Using the blow-up formula, we moreover determine the contact terms and other background couplings directly from the SW curve. We provide a complete analysis for all  $\SU(3)$ CB couplings in Section~\ref{sec:SU(3) calculation}.  
    \item We present a precise dictionary between the couplings in SW theory and universal functions in algebraic geometry such that~\eqref{eq:ZN=1SW=Gu} holds. This is summarized in Conjecture~\ref{conj:DZSW0=Ua}. We explicitly compute all couplings in SW theory as $z$-series for $\SU(3)$ and find perfect agreement with~\cite{Gottsche:2020ass}. For each $N_f=1,\dots, 5$, this amounts to calculating 13 physical couplings and matching with 16 universal functions. 
    We also clarify the physical origin of various symmetries of the universal functions observed in~\cite{Gottsche:2020ass}, and speculate on the Galois group of the cyclotomic field in which the universal series take values.
\end{enumerate}
\subsection*{Future directions}
\begin{enumerate}
\item It would be extremely interesting to compute the partition function for the superconformal case $N_f=2N$. The main obstacle is a better understanding of the singularity structure of the Coulomb branch. Moreover, the infrared theory retains the dependence on the ultraviolet coupling $\tau_{\rm uv}$, such that strong coupling expansions are expected to become modular forms in $\tau_{\rm uv}$. The SW curve for this case was obtained in~\cite{Argyres:1995wt}. This case is interesting for many reasons: via Argyres--Seiberg duality~\cite{Argyres:2007cn,Gaiotto:2009we}, a strong coupling limit of these theories is related to a class of non-Lagrangian theories called the $R_{0,N}$-theories~\cite{Chacaltana:2010ks,Gaiotto:2009we,Argyres:2007cn}. The $R_{0,3}$ theory is the famous $E_6$ Minahan--Nemeschansky theory with $E_6$ flavor symmetry. This would give us a way of computing the twisted partition function of a non-Lagrangian theory otherwise inaccessible via path integrals. The large $N$ limit of this theory is related to Supergravity on AdS via the AdS/CFT correspondence~\cite{Maldacena:1997re,Gaiotto:2009gz}. The structure of the twisted partition function might shed some light on a subsector of the observables in Supergravity. Large $N$ limits of twisted theories have been discussed in~\cite{Marino:1998bm,Papadodimas:2005wm,Gaiotto:2009gz,Labastida:1999ij,BenettiGenolini:2017zmu}.

The superconformal theories are on the boundary of the window of QCD with well-understood Coulomb branch geometries, as $0\leq N_f\leq 2N$ for theories with non-negative one-loop beta function coefficient $b_0=2N-N_f$. While the algebro-geometric framework is well-defined for any rank $s=\rk(\alpha)\in\mathbb Z$, the infrared calculation restricts~$s$ to the physical window. 

\item The higher-rank techniques we develop motivate the study of more theories, such as the $\SU(N)$ $\nstar$ theory with a massive adjoint hypermultiplet. In analogy with~\cite{Manschot:2021qqe}, it would be interesting to make connections with $\SU(N)$ Vafa--Witten invariants, for which many results are available~\cite{Gottsche:2021dye,Manschot:2017xcr,Labastida:1999ij,Tanaka:2017bcw,Gottsche:2018meg,Alexandrov:2020dyy}. In a different direction, it would be interesting to explore the compactification of higher-rank 5d theories on $X\times S^1$, and higher-rank class $\CS$ theories~\cite{Gaiotto:2009hg,Manschot:2021qqe,Moore:2024vsd,Kim:2025fpz,Gottsche:2006}.

\item Our analysis may shed some light on the explicit evaluation of Coulomb branch integrals for higher-rank theories, such as for $\mathbb P^2$. While it remains a central computational tool for rank one theories, our analysis together with~\cite{Aspman:2022sfj,Aspman:2023ate} suggests that modularity of the Coulomb branch measure is not required to perform the integral.

\item We only determine the instanton component of the correlators, which we attribute to the $\CN=1$ vacua. For QCD, there is certainly a contribution from the reducible branch, which in the IR corresponds to the quark singularities. To obtain the full correlation function, we need to consider generic masses rather than equal masses, since the contribution from the reducible branch is certainly singular in the equal mass limit. It would be very interesting to include these contributions, which would give a higher-rank analogue of~\cite{FMM:2026}.
\item Finally, there is a well-known correspondence between Segre numbers and Verlinde numbers. The Verlinde numbers are K-theoretic intersection numbers of the Chern character of the determinant line bundle. It would be interesting to extend our dictionary between physical parameters and the Chern data of Segre numbers to the Verlinde setup, which might allow to establish the physical realisation leading to the Segre--Verlinde correspondence and to explain its relation to strange duality~\cite{lepotier,johnson2020universal,Gttsche2018,Gttsche2019,Gottsche:2020ass,Marian2021,Gttsche2024,Marian2024}.
\end{enumerate}
\subsection*{Outline of the paper}
The rest of the paper is organised as follows. In section~\ref{sec:toptwist} we discuss the topological twisting of $\SU(N)$ SQCD and explain the instanton contribution to the correlation functions in section~\ref{sec:moduli_sp_corr}. In section~\ref{sec:SW geometry} we discuss the low-energy effective theory of $\SU(N)$ SQCD. We discuss the characterization of $\CN=1$ points and the calculation of the period matrix at the $\CN=1$ points in section~\ref{sec:N=1_vacua}. In section~\ref{sec:IR theory} we derive the expression of the Coulomb branch integral. Using the wall-crossing argument, we compute the SW contribution at the $\CN=1$ points in section~\ref{sec:SW_cont}. In section~\ref{sec:blowup} we use the blowup formula to compute the contact terms in terms of Riemann theta functions. In section~\ref{sec:top_cor_seg_num}, we discuss the relation to Segre numbers and present a precise dictionary between the universal functions of~\cite{Gottsche:2020ass} and the couplings in the SW theory. In section~\ref{sec:SU(3) calculation}, we discuss the computation of the universal series from SW theory for $\SU(3)$. We discuss new techniques to compute these couplings at the $\CN=1$ points. In appendix~\ref{app:modular forms}, we discuss some preliminaries on modular forms and hyperelliptic curves useful for the computation of couplings in SW theory. In appendix~\ref{app:corrections} we list the universal functions from~\cite{Gottsche:2020ass} and list some corrections to the universal functions. Finally, in appendix~\ref{app:lie_alg} we review some aspects of Lie algebras used in the discussion of Coulomb branch integrals. 

\sss{Acknowledgements.}
We would like to thank Cyril Closset, Thomas Dumitrescu, Osama Khlaif, Martijn Kool, Greg Moore, Ideal Majtara, Jan Manschot, and Vivek Saxena for discussions, correspondence and feedback on a draft of the paper. R.K.S. would like to thank Vivek Saxena and Runkai Tao for discussions on $u$-plane integrals over the years. The work of E.F. and R.K.S. is supported by the US Department of Energy under grant DE-SC0010008. The work of R.K.S. is partially supported by the Mani L. Bhaumik Institute for Theoretical Physics.
$\,$\\$\,$\\
\centerline{\emph{No part of this paper was written by AI.}}
\newpage
\section{Topological twisting of the UV theory}\label{sec:toptwist}
In this section, we discuss the topological twisting of $\SU(N)$ SQCD with $N_f$ flavors.\footnote{See~\cite{Hyun:1995mb,Moore:1997pc,Moore:2024vsd} for previous analyses.}
We first set up our notation for compact 4-manifolds.
\subsection{Compact 4-manifolds}
Let $X$ be a smooth, compact, oriented, Riemannian 4-manifold without boundary. We will denote by $b_i=b_i(X)$ the Betti numbers of $X$ and by $\chi$ the Euler characteristic of $X$.   Let $L$ be the image of $H^2(X,\Z)/\mathrm{Tors}\big(H^2(X,\Z)\big)$ in $H^2(X,\R)$. This is a lattice in a real vector space, so we can divide its elements without ambiguity. The intersection form on $H^2(X, \mathbb{Z})$ provides a natural non-degenerate bilinear form $B\colon(L \otimes \mathbb{R}) \times(L \otimes \mathbb{R}) \longrightarrow \mathbb{R}$ given by 
\begin{equation}\label{eq:int_pair_L}
\bk_1\bk_2\coloneqq B\left(\boldsymbol{k}_1, \boldsymbol{k}_2\right)\coloneqq \int_{X} \boldsymbol{k}_1 \wedge \boldsymbol{k}_2 ~.
\end{equation}
We will also use the notation $\bk^2\coloneqq Q(\bfk)\coloneqq B(\bk,\bk)$.
The restriction of $B$ to $L \times L$ is an integer-valued bilinear form. 
Let $H^{2, \pm}(X, \mathbb{R})$ denote the positive (resp. negative) definite subspace of $H^2(X, \mathbb{R})$ under the Hodge star:
\begin{equation}
H^{2, \pm}(X, \mathbb{R})\coloneqq \{\boldsymbol{k}\in H^{2}(X, \mathbb{R}):*\bfk =\pm \bfk\}~.    
\end{equation}
Set $b_2^{ \pm}\coloneqq b_2^{ \pm}(X)=\operatorname{dim}\left(H^{2, \pm}(X, \mathbb{R})\right)$.  The combination $\sigma=\coloneqq b_2^{+}-b_2^{-}$ is called the signature of $X$. In this paper, we will take $X$ to be simply connected, so that $b_1=0$, and that $X$ admits an almost-complex structure. 
This means that the tangent bundle $TX$ of $X$ is a complex vector bundle with first Chern class $c_1(TX)$ satisfying $c^2_1(TX)=2\chi+3\sigma$~\cite[Existence Theorem §10.1]{scorpan:2005}.
Moreover, this implies that $b_2^+$ is odd.
The cotangent bundle $T^*X$ of $X$ is also a complex vector bundle with the canonical bundle $K_X\coloneqq \mr{det}(T^*X)$ determined by the first Chern class 
\begin{eqsp}
    K\coloneqq c_1(K_X)=-c_1(TX)~.
\end{eqsp}
$K$ is called the canonical class and satisfies 
\begin{eqsp}
    K^2=2\chi+3\sigma~.
\end{eqsp}
A class $\bm{c}\in L$ is called a \textit{characteristic vector} if for all $\bk\in L$, we have 
\begin{eqsp}\label{characteristic vector}
    \bm{k}^2+\bm{c} \bk\in 2\Z~.
\end{eqsp}
Characteristic vectors satisfy $\bm{c}^2=\sigma\bmod 8$. 
Any integral lift of the second Stiefel--Whitney class $w_2(X)\in H^2(X,\Z_2)$ is a characteristic vector. Moreover, the canonical class $K$ is also an integral lift of $w_2(X)$ and hence a characteristic vector. We will often use the \textit{holomorphic Euler characteristic} $\chih$ defined by 
\begin{eqsp}
    \chih\coloneqq \frac{1}{4}(\chi+\sigma)=\frac{b_2^++1}{2}~.
\end{eqsp}
The name is justified by the fact that if $X$ is a complex manifold, then $\chih=1-h^{0,1}+h^{0,2}$, 
where $h^{i,j}$ are the Hodge numbers of $X$. It is clear that $\chih\in \Z_{\geq 1}$ for almost complex manifolds. \par
We will also restrict ourselves to $4$-manifolds with and $b_2^+\geq 1$.  For $b_2^+=1$, let $J\in H^{2, +}(X, \mathbb{R})$ be the unique generator of $H^{2, +}(X, \mathbb{R})$ satisfying 
\begin{equation}
    \star J=J,\quad Q(J)=\int_{X}J\wedge J=1,\quad dJ=0.
\end{equation}
Using $J$, we can project any vector $\bk\in L$ to its positive and negative parts as follows:
\begin{eqsp}
    \bk_+\coloneqq B(\bk,J)J~,\quad \bk_-\coloneqq \bk-\bk_+~.
\end{eqsp}
The canonical vector $K$ for $b_2^+=1$ satisfies $K^2=\sigma+8$.
\subsection{The structure groups and cohomological conditions}\label{sec:coh_cond_twist}
The twisting of the UV theory requires introduction of extra topological data on the 4-manifold $X$. The first step is to define the untwisted theory on a general non-spin manifold $X$. Let us begin by briefly reviewing the untwisted theory. The $\mathcal{N}=2$ SU($N$) gauge theory with  $N_f$ flavors is described in terms of $\mathcal{N}=1$ superfields by a chiral field strength multiplet $W_{~n}^m$ and a chiral multiplet $\Phi_{~n}^m$ both in the adjoint of the gauge group SU($N$), and $N_f$ hypermultiplets represented by $\mathcal{N}=1$ chiral multiplets $Q_m^i$ in the $\mathbf{N}$ (fundamental), and $\widetilde{Q}_i^m$ in the $\overline{\mathbf{N}}$ (antifundamental) representations of the gauge group. Flavor and color indices are respectively $i, j, k=1, \ldots, N_f$ and $m, n=1, \ldots, N$. 
One can write down the Lagrangian for this theory using  $\mathcal{N}=1$ gauge-invariant kinetic terms. The complexified gauge coupling constant is given by 
\begin{equation}
\tau\coloneqq \frac{\Theta}{2\pi}+\im\frac{4 \pi} {g^2} ~, 
\end{equation}
where $g$ is the gauge coupling. 
The superpotential is given by 
\begin{equation}
\mathcal{W}=\sqrt{2} \widetilde{Q}_i^m \Phi^{n}{}_{m} Q_n^i+\sqrt{2} M_{~j}^i \widetilde{Q}_i^m Q_m^j ~.
\end{equation}
The hypermultiplet mass matrix satisfies $\left[M, M^{\dagger}\right]=0$ and hence can be diagonalized by a flavor rotation to $M=\operatorname{diag}\left(m_1, \ldots, m_{N_f}\right)$. There is a $\mr{U}(1)^{N_f}$ flavor symmetry which gets enhanced to $\mr{U}(N_f)$ for massless hypermultiplets. 

\sss{Physical theory.} 
We first consider the physical (untwisted) theory with generic hypermultiplet masses on an arbitrary compact oriented Riemannian 4-manifold.\footnote{It is usually believed that the untwisted theory can only be put on a spin manifold because of the presence of fermions. As it was argued in~\cite{Moore:2024vsd}, this is not the correct viewpoint. We can study the untwisted theory on a non-spin 4-manifold by introducing background connections for a discrete subgroup of the structure group $G^{\mr{phys}}$ of the theory. This is equivalent to the existence of a principal $G^{\mr{phys}}$-bundle. See~\cite[Remark 2.3]{Moore:2024vsd} for the precise statement.} The structure group of the physical theory is 
\begin{eqsp}
    G^{\mr{phys}}\coloneqq \widetilde{G}^{\mr{phys}}/C^{\mr{phys}}=\frac{\mr{Spin(4)}\times \mr{SU(2)_R}\times \mr{U(1)}^{N_f}\times \SU(N)}{C^{\mr{phys}}}~,
\end{eqsp}
where $C^{\mr{phys}}$ is a subgroup of the center $\Z_2\times \Z_2\times\Z_2\times\mr{U(1)}^{N_f}\times \Z_N$ satisfying some conditions, see~\cite{Moore:2024vsd} for details. We will choose $C^{\mr{phys}}$ to be the \textit{maximal admissible subgroup} $C^{\mr{phys}}_{\mr{max}}$ given by 
\begin{eqsp}
C^{\mr{phys}}_{\mr{max}}\coloneqq \{(\zeta,\zeta,\zeta,(\mu_1,\dots,\mu_{N_f}),\omega):\zeta\mu_j\omega=1~\forall~j\}~.    
\end{eqsp}
Then we introduce a physical bundle $P^{\mr{phys}}\to X$ and as shown in~\cite{Moore:2024vsd}, the existence of $P^{\mr{phys}}$ imposes cohomological conditions on $X$. To make this precise, let $p_i,i=1,2,3,4$ be the projection onto each factor of $\widetilde{G}^{\mr{phys}}$. Then we have transfer of structure groups 
\begin{align}
\label{eq:p-star-bundles}
\begin{split}  
\mr{Fr}(X) & \coloneqq  (p_1)_*(P^{\mr{phys}})  ~,\\ 
P^{\mr{R}} & \coloneqq  (p_2)_*(P^{\mr{phys}}) ~,\\ 
P^{\mr{f}} & \coloneqq  (p_3)_*(P^{\mr{phys}}) ~,\\
P^{\mr{gauge}} & \coloneqq  (p_4)_*(P^{\mr{phys}}) ~,
\end{split}
\end{align}
where $\mr{Fr}(X)$ is the frame bundle of $X$ with structure group $\mr{SO(4)}$, $P^{\mr{R}}$ is the R-symmetry bundle with structure group $\mr{SO(3)_R}$, $P^{\mr{f}}$ is a $\mr{U(1)}^{N_f}/p_3(C^{\mr{phys}})$-bundle. We can write $P^{\mr{f}}$ as the fiber product 
\begin{eqsp}
P^{\mr{f}}=P^{\mr{f};1}\times_XP^{\mr{f};2}\times_X\dots\times_XP^{\mr{f};N_f}~,
\end{eqsp}
where each $P^{\mr{f},i}$ is a $\mr{U(1)}/\Z_{\mr{lcm}(2,N)}$-bundle since 
\begin{eqsp}
    p_3(C^{\mr{phys}}_{\mr{max}})\cong \Z_{\mr{lcm}(2,N)}\times \Z_{\mr{lcm}(2,N)}\times\dots\times\Z_{\mr{lcm}(2,N)}~.
\end{eqsp}
The gauge bundle $P^{\mr{gauge}}$ is an $\SU(N)/\Z_N$ bundle since $p_4(C_{\mr{max}}^{\mr{phys}})=\Z_N$. Let $w_2(X)\in H^2(X,\Z_2)$ be the second Stiefel--Whitney class of $X$, $\bm{c}_j\in H^2(X,\Z)$ with $j=1,\dots,N_f$ be the first Chern class of the $\mr{U(1)}$-bundle obtained by reduction of structure group of $P^{\mr{f};j}$ from $\mr{U(1)}/\Z_{\mr{lcm}(2,N)}$ to $\mr{U(1)}$ along the isomorphism $\mr{U(1)}/\Z_{\mr{lcm}(2,N)}\to \mr{U(1)}$, and $\mu(b)\in H^2(X,\Z_N)$ be the obstruction to lifting the structure group of $P^{\mr{gauge}}$ from $\SU(N)/\Z_N$ to $\SU(N)$. 
% Let $\mathcal{Z}$ be the group generated by $\Z_N$ and $-1$. Clearly,
% \begin{eqsp}
%     \mathcal{Z}\cong \Z_{\mr{lcm}(2,N)}~.
% \end{eqsp}
We have canonical homomorphisms $\rho_s:\Z_2\to \Z_{\mr{lcm}(2,N)}$ and $\rho^{\mr{gauge}}:\Z_N\to \Z_{\mr{lcm}(2,N)}$. 
Let $r_{n}:\Z\to\Z_{n}$ be the reduction modulo $n$ map. Then the cohomological condition can be written as~\cite{Moore:2024vsd} 
\begin{eqsp}\label{eq:coh_cond}
    \rho_s(w_2(X))+\rho^{\mr{gauge}}(\mu(b))+r_{\mr{lcm}(2,N)}(\bm{c}_j)=0\in H^2(X,\mathcal{Z})~,\quad j=1,\dots,N_f~.
\end{eqsp}
It is useful to write this condition in the integer cohomology lattice $L$ introduced in section~\ref{sec:toptwist}. To do this, we choose transition functions for the bundles in~\eqref{eq:p-star-bundles} on double overlaps $U_{pq}=U_p\cap U_q$ of a good cover $\{U_p\}$ of $X$:
\begin{equation}
\begin{aligned}
t_{pq}&: U_{pq} \longrightarrow \mathrm{SO}(4) \,, \\
t_{pq}^{\mathrm{R}} &: U_{pq} \longrightarrow \mathrm{SO}(3)_{\mathrm{R}} \,, \\
t_{pq}^{\mathrm{f};i} &: U_{pq} \longrightarrow \mathrm{U(1)}/\Z_{\mr{lcm}(2,N)}\;,i=1,\dots,N_f\,, \\
t_{pq}^{g} &: U_{pq} \longrightarrow \mathrm{SU}(N)/\Z_N \,,
\end{aligned}
\end{equation}
and their local lifts 
\begin{equation}
\begin{aligned}
(\widetilde{t}_{pq}^{+}, \widetilde{t}_{pq}^{-}) &: U_{pq} \longrightarrow \mathrm{Spin}(4) \cong \mathrm{SU}(2) \times \mathrm{SU}(2) \,, \\
\widetilde{t}_{pq}^{\mathrm{R}} &: U_{pq} \longrightarrow \mathrm{SU}(2)_{\mathrm{R}} \,, \\
\widetilde{t}_{pq}^{\mathrm{f};i} &: U_{pq} \longrightarrow \mathbb{R} \,, \\
\widetilde{t}_{pq}^{g} &: U_{pq} \longrightarrow \SU(N) \,.
\end{aligned}
\end{equation}
Note that the $\widetilde{t}^{\mr{f};i}_{pq}$ satisfies $t^{\mr{f};i}_{pq}=\exp(\frac{2\pi i\widetilde{t}^{\mr{f};i}_{pq}}{\mr{lcm}(2,N)})$.
On triple overlaps $U_{pqr}=U_p\cap U_q\cap U_r$, we have 
\begin{equation}
\begin{aligned}
\zeta_{pqr}^{\pm} &\coloneqq  \widetilde{t}_{pq}^{\pm} \widetilde{t}_{qr}^{\pm} \widetilde{t}_{rp}^{\pm}:U_{pqr}\longrightarrow \Z_2\,, \\
\zeta_{pqr}^{\mathrm{R}} &\coloneqq  \widetilde{t}_{pq}^{\mathrm{R}} \widetilde{t}_{qr}^{\mathrm{R}} \widetilde{t}_{rp}^{\mathrm{R}}:U_{pqr}\longrightarrow \Z_2 \,, \\
\zeta_{pqr}^{\mathrm{f};i} &\coloneqq  \exp\left(2\pi i\frac{\widetilde{t}_{pq}^{\mathrm{f}}+ \widetilde{t}_{qr}^{\mathrm{f}}+ \widetilde{t}_{rp}^{\mathrm{f}}}{\mr{lcm}(2,N)}\right):U_{pqr}\longrightarrow \Z_{\mr{lcm}(2,N)}\,, \\
\zeta_{pqr}^{g} &\coloneqq  \widetilde{t}_{pq}^{g} \widetilde{t}_{qr}^{g} \widetilde{t}_{rp}^{g}:U_{pqr}\longrightarrow \Z_N \,.
\end{aligned}
\end{equation}
These cocycles represent the cohomology classes 
\begin{eqsp}\label{eq:coh_class_cocycle}
    w_2(X)&=\left[\zeta_{pqr}^{\pm}\right]\in H^2(X,\Z_2)~,\quad \bm{c}_j=\left[\zeta_{pqr}^{\mathrm{f};j}\right]\bmod \mr{lcm}(2,N)\in H^2(X,\Z_{N})~,\\\mu(b)&=\left[\zeta_{pqr}^{g}\right]\in H^2(X,\Z_N)~.
\end{eqsp}
As discussed in~\cite{Moore:2024vsd}, for the bundle $P^{\mathrm{phys}}\to X$ to exist, we must have 
\begin{eqsp}
    \zeta_{pqr}^-=\zeta_{pqr}^+=\zeta_{pqr}^{\mathrm{R}}~,
\end{eqsp}
and 
\begin{eqsp}\label{eq:coh_cond_cocycle}
    \zeta_{pqr}^{+}\zeta_{pqr}^{g}\zeta_{pqr}^{\mathrm{f};i}=1\in\mr{U(1)}~.
\end{eqsp}
A choice of lift of the cohomology classes~\eqref{eq:coh_class_cocycle} corresponds to choosing cocycles 
\begin{eqsp}
    \eta_{pqr}^{s}&:U_{pqr}\longrightarrow \Z~,\\
    \eta_{pqr}^{\mathrm{f};j}&:U_{pqr}\longrightarrow \Z~,\\
    \eta_{pqr}^{g}&:U_{pqr}\longrightarrow \Z~,
\end{eqsp}
satisfying 
\begin{eqsp}
   \zeta_{pqr}^{\pm}=\exp\left(\frac{2\pi \im \eta_{pqr}^{s}}{2} \right)~,\\
   \zeta_{pqr}^{\mathrm{f};j}=\exp\left(\frac{2\pi \im \eta_{pqr}^{\mathrm{f};j}}{\mr{lcm}(2,N)} \right)~,\\
   \zeta_{pqr}^{g}=\exp\left(\frac{2\pi \im \eta_{pqr}^{g}}{N} \right)~.
\end{eqsp}
The cocycles $\eta_{pqr}$ then represent lifts of the classes $w_2(X),\bm{c}_j\bmod \mr{lcm}(2,N), \mu(b)$. 
The condition~\eqref{eq:coh_cond_cocycle} then implies 
\begin{eqsp}\label{eq:coh_cond_kj}
\frac{1}{\mr{lcm}(2,N)}\bm{c}_j+\frac{1}{2}c_1(\mf{s})+\frac{1}{N}c_1(E)=0\mod L~,    
\end{eqsp}
where $L$ is the lattice introduced above \eqref{eq:int_pair_L} and we used the fact that any integer lift of $w_2(X)$ represents a Spin${}^c$-structure and an integer lift of $\mu(b)$ is the first Chern class $c_1(E)$ of the associated bundle to $P^{\mathrm{gauge}}$ with the fundamental representation of SU$(N)$. Note that $E$ is only a $\PSU(N)=\SU(N)/\Z_N$-vector bundle if $\mu(b)\neq 0$ and the first term $\bm{c}_j/\mr{lcm}(2,N)$ is to be thought of as the first Chern class of the $\mr{U(1)}/\Z_{\mr{lcm}(2,N)}$-bundle $P^{\mr{f};j}$. A slightly less precise, but equivalent, way of stating the condition~\eqref{eq:coh_cond_kj} is as follows: let $\CL_j$ be the line bundle associated to the $\mr{U(1)}/\Z_{\mr{lcm}(2,N)}$-bundle $P^{\mr{f};j}$ and $S^{\pm}$ be the chiral spinor bundles. Then the tensor product of bundles 
\begin{eqsp}
    S^{\pm}\otimes E\otimes \CL_j~,
\end{eqsp}
exists as a $(\mr{Spin}(4)\times\mr{U(1)} \times \SU(N) )/C^{\mr{tw}}$-bundle, where $C^{\mr{tw}}$ is given by \eqref{eq:Ctw_def} below, even though the bundles may not exist individually (unless $X$ is spin, in which case $S^\pm$ exist and $\mu(b)=0$, in which case $E$ exists as an $\SU(N)$-bundle).

\sss{Twisted theory.} To topologically twist the theory, we choose the structure group of the twisted theory to be  
\begin{eqsp}
    G^{\mr{tw}}\coloneqq \widetilde{G}^{\mr{tw}}/C^{\mr{tw}}\coloneqq \frac{\mr{Spin(4)}\times\mr{U(1)}^{N_f}\times \SU(N)}{C^{\mr{tw}}}~,
\end{eqsp}
where 
\begin{eqsp}\label{eq:Ctw_def}
C^{\mr{tw}}=\{(\zeta,\zeta,(\mu_1,\dots,\mu_{N_f}),\omega)\in\Z_2\times\Z_2\times \mr{U(1)}^{N_f}\times\Z_N: \zeta\mu_j\omega=1~\forall~j\}~.    
\end{eqsp}
We have a twisting group homomorphism 
\begin{eqsp}
   \varphi\colon \widetilde{G}^{\mr{tw}}&\longrightarrow\widetilde{G}^{\mr{phys}} \\
   ((u_1,u_2),\mu,g)&\longmapsto ((u_1,u_2),u_2,\mu,g)~.
\end{eqsp}
$\varphi$ descends to a group homomorphism 
$\varphi^{\mr{tw}}\colon G^{\mr{tw}}\to G^{\mr{phys}}$. The twisted theory is defined by specifying a bundle with connection $(P^{\mr{tw}}\to X,\nabla^{\mr{tw}})$ with structure group $G^{\mr{tw}}$ and then using the transfer of structure group to write down the twisted action, see~\cite{Moore:2024vsd} for details. The essential point is that the cohomological condition~\eqref{eq:coh_cond} is enough for the existence of the twisted bundle $P^{\mr{tw}}$. It is important to note that there are four components of $\nabla^{\mr{tw}}$, out of which only one is dynamical and the remaining are background. We reproduce the table~\ref{tab:twconnGXRF} of connections from~\cite{Moore:2024vsd} for completeness. 
\begin{table}[!t]
    \centering
\begin{tabular}{|c|c|c|c|}
\hline Connection & Bundle & Local chart component & Dynamical/Background \\
\hline
$\left(p_1\right)_*\left(\varphi_*^{\mr{t w}}\left(\nabla^{\mr{t w}}\right)\right)$ & $\mr{Fr}(X)$ & $\omega_\mu$ & Background \\
$\left(p_2\right)_*\left(\varphi_*^{\mr{t w}}\left(\nabla^{\mr{t w}}\right)\right)$ & $P^{\mr{R}}$ & $\left(\omega_{+}\right)_\mu$ & Background \\
$\left(p_{3, j}\right)_*\left(\varphi_*^{\mr{t w}}\left(\nabla^{\mr{t w}}\right)\right)$ & $P^{\mr{f},j}$ & $\mathcal{A}_\mu^j$ & Background \\$\left(p_4\right)_*\left(\varphi_*^{\mr{t w}}\left(\nabla^{\mr{t w}}\right)\right)$ & $P^{\mr {gauge }}$ & $A_\mu$ & Dynamical \\
\hline
\end{tabular}
    \caption{Various dynamical and background connections in the physical theory obtained by push-forward of $\nabla^{\mr{tw}}$. Here, $\omega_{+}$ denotes the self-dual part of the Levi--Civita connection and $P^{\mr{f},j}$ is the $\mr{U(1)}$-bundle obtained from the $\mr{U(1)}^{N_f}$-bundle $P^{\mr{f}}$ by projecting onto the $j$-th factor.}
    \label{tab:twconnGXRF}
\end{table}

\subsection{Multi-monopole moduli space and correlation functions}\label{sec:moduli_sp_corr}
\sss{Bundles and characteristic classes.}
To describe the moduli problem relevant to our setup, we simplify notation and consider the case $G=\PSU(N)$ and denote by $P\to X$ a principal $G$-bundle\footnote{$P\to X$ will be the gauge bundle $P^{\mr{gauge}}$ from section~\ref{sec:coh_cond_twist} above.}. We let  $E$ be the rank $N$ vector bundle associated with $P$. 
In general, we have the Chern character
\bea\label{chern character}
\ch(E)=\rk E+c_1(E)+\tfrac12 c_1(E)^2-c_2(E)+\dots. 
\eea
It is related to the curvature $F$ of the connection on $E$ as 
\bea
\ch(E)={\rm Tr}\exp\left(\tfrac{i}{2\pi}F\right)~.
\eea
The instanton number is defined as
\bea
k(P)=-\frac{1}{16\pi^2N}\int_X {\rm Tr}(F^2)~,
\eea
where the trace is in the $\dim\SU(N)=N^2-1$ dimensional, adjoint representation of $\SU(N)$. It evaluates to
\bea\label{instanton number}
k(P)=c_2(E)-\frac{N-1}{2N}c_1(E)^2~.
\eea
Let $f_2(P)\in H^2(X,\Z_N)$ be the obstruction to lifting the PSU$(N)$-bundle $P$ to an $\SU(N)$-bundle. We can identify $f_2(P)$ with the characteristic class $\mu(b)$ of the background gerbe $b$ in the physical theory. Using the splitting principle, we can show that
\begin{eqsp}
    c_1(E)=\ov{f}_2(P)\bmod N~,
\end{eqsp}
where $\ov{f}_2(P)\in H^2(X,\Z)$ is an integral lift of $f_2(P)$ (see~\eg~\cite{Vafa:1994tf}). The instanton number thus satisfies 
\begin{equation}
    k(P)=\ov{f}_2(P)^2\left(\frac{1}{2N}-\frac{1}{2}\right)\bmod 1~.
\end{equation}
This relation is invariant under $\ov{f}_2(P)\to \ov{f}_2(P)+N\eta$ for $\eta\in H^2(X,\Z)$, which means that it is independent of the choice of the integral lift of $f_2(P)$.

\sss{Dirac operators and index bundles.}
For each hypermultiplet labelled by $j$, we moreover require a $\spinc$ Dirac operator coupled to the complex associated with the bundle $E$:
\bea
\slashed D_j\colon \Gamma(W_j^+\otimes E)\longrightarrow \Gamma(W_j^-\otimes E)~,
\eea
where $W_j^\pm$ are chiral $\spinc$ bundles.
Recall that $c_1(W_j^+)=w_2(X)\mod 2$. The first Chern class of the determinant line bundle is the characteristic class of the $\spinc$ structure:
\bea
c_1(\mathfrak s_j)=c_1(\det(W_j^\pm))~.
\eea
The Atiyah--Singer index theorem~\cite{Atiyah:1970ws} expresses the index as 
\bea
\ind \slashed D_j=\int_X \ch(E)e^{\frac12c_1(\mathfrak s_j)}\widehat A(X)~,
\eea
where $\widehat A(X)=1-\frac{1}{24}p_1(X)$. Combining~\eqref{chern character} and~\eqref{instanton number} with $\int_X p_1(X)=3\sigma$, this gives
\bea\label{ind D}
\ind \slashed D_j=-k-\frac N8\sigma+\frac N2\bfk_j^2~,
\eea
where 
\bea\label{bfk def}
\bfk_j:= -\frac12 c_1(\mathfrak s_j)-\frac 1Nc_1(E)~.
\eea
The Dirac index is an integer: inserting~\eqref{instanton number}, we obtain
\bea
\ind\slashed D_j=-c_2(E)+N\frac{c_1(\mathfrak s_j)^2-\sigma}{8}+\frac12\left(c_1(E)^2+c_1(\mathfrak s_j)c_1(E)\right)
\eea
Since $c_1(\mathfrak s_j)$ is a characteristic element, we have $c_1(\mathfrak s_j)^2=\sigma\mod 8$, such that together with the property~\eqref{characteristic vector}, all three terms are integers.
This fits in nicely with the cohomological condition \eqref{eq:coh_cond_kj} once we realize that the hypermultiplets are sections of the flavor bundle $\CL_j$ whose Chern class is $\bm{c}_j/\mr{lcm}(2,N))\equiv \bk_j\bmod L$. 
In the topological twist, we can choose a different chiral spin bundle $W^\pm_j$ and associated $\spinc$ structure $\mathfrak s_j$ for each hypermultiplet, $j=1,\dots, N_f$. We have
\bea\label{bfk_j definition}
\bfk_j=-\frac{1}{2N}c_1(W^+_j\otimes E)~,
\eea
which gives~\eqref{bfk def} since $W^+$ has rank 2, while $E$ has rank $N$.
Indeed, the cohomological condition~\eqref{eq:coh_cond_kj} gives $\bfc_j/\mr{lcm}(2,N)\equiv \bfk_j\mod L$.

To each Dirac operator $\slashed D_j$ we introduce the index bundles 
\bea\label{index bundle}
W^j&=\Ind\slashed D_j\,\,=[\ker \slashed D_j]\,\, - [\coker \slashed D_j]~, \\
V^j&=\Ind\slashed D^{\dagger}_j=[\ker \slashed D^{\dagger}_j] - [\coker \slashed D^{\dagger}_j]~.
\eea
They can be defined for fixed instanton charge $k=k(P)$, which we highlight in the notation $V_k^j$ and $W_k^j$. They are virtual bundles over the instanton moduli space, $V_k^j,W_k^j\in K(\Mi)$. Note further that $\ker \slashed D_j^\dagger\cong \coker \slashed D_j$, and thus as virtual bundles, $V^j=-W^j$. The virtual ranks are formal differences of the complex dimensions of kernel and cokernel, and $\rk\,  W^j=\ind \slashed D^j$. From~\eqref{ind D} it is clear that for large enough instanton numbers $k$, the rank of $W_k^j$ becomes negative, while the rank of $V_k^j$ becomes positive.

\sss{The moduli spaces.}
Let $\CM_{v(E)}^{\rm i}$ be the $\SU(N)$ instanton moduli space, where we fix the topological type / Chern data 
$v(E)=(\rk (E), c_1(E), c_2(E))$. Recall that $\rk (E)=N$, and $k$ depends on $c_1(E)$ and $c_2(E)$ through~\eqref{instanton number}. 
It has complex dimension~\cite{Kronheimer2005}
\bea
\vd \CM_{v(E)}^{\rm i}=2Nk-(N^2-1)\chih~.
\eea
Due to~\eqref{instanton number}, $2Nk$ is an integer, and for our choice of four-manifolds $\chih\in\mathbb Z_{>1}$, thus the moduli space has integral complex dimension (or even real dimension).
The $\CQ$-fixed equations of the $\SU(N)$ QCD with $N_f\leq 2N$ hypermultiplets read~\cite{Labastida:1995zj}
\bea
\label{Qfixedeqs_voa}
0&=\left(F^{a}_{\dot \alpha \dot \beta}\right)^++\frac{i}{2} \sum_{j=1}^{N_f}\bar M^j_{(\dot \alpha} T^a M^j_{\dot \beta)}~,\\
0&=\slashed DM^j~, \, j=1,\dots, N_f~,
\eea 
where $M^j\in\Gamma(W_j^+\otimes E)$ are the monopole fields. We denote the moduli space of solutions to the non-Abelian monopole equations modulo gauge transformations by $\CM^{\CQ,N_f}_{v(E)}$. Its virtual dimension is that of the instanton moduli space plus the sum of the numerical indices of the Dirac operators~\cite{Hyun:1995hz,Manschot:2021qqe},
\bea
\vd \CM^{\CQ,N_f}_{v(E)}=\vd \CM_{v(E)}^{\rm i}+\sum_{j=1}^{N_f}\ind \slashed D^j~,
\eea
which gives
\bea\label{vd MQ}
\vd \CM^{\CQ,N_f}_{v(E)}=(2N-N_f)k-(N^2-1)\chih-\frac{NN_f}{8}\sigma+\frac N2\sum_{j=1}^{N_f}\bfk_j^2~.
\eea
The dependence on the instanton number $k$ is through the one-loop beta function coefficient $2N-N_f$, as expected. For $N_f=1$ and in the spin case $w_2(X)=c_1(E)=\bfk_j=0$, this agrees with~\cite[(2.15)]{Labastida:1995zj}.

\sss{Correlation functions.}
The correlation functions on $X$ in the $\SU(N)$ theory with $N_f$ massive fundamental hypermultiplets reduce to a finite-dimensional integral over $\CM^{\CQ,N_f}_{v(E)}$.
Localisation to the fixed point locus in $\CM^{\CQ,N_f}_{v(E)}$ with respect to the $\mathbb T={\rm U}(1)^{N_f}$ action gives a sum of contributions from the instanton branch $\CM^{\rm i}_{v(E)}$ and the reducible branch $\CM^{\rm r}_{v(E)}$.\footnote{See~\cite{Mathai:1986tc,Atiyah:1990tm, Vafa:1994tf,Cordes:1994fc, LoNeSha,Lossev:1997bz, Manschot:2021qqe,Aspman:2022sfj,Aspman:2023ate,FMM:2026} for related work.}

Let $\CO$ be a $\CQ$-closed observable in the UV theory, and denote by 
\bea
\langle \CO\rangle=\langle \CO\rangle^{\rm i}+\langle \CO\rangle^{\rm r}
\eea
its correlation function on $X$, after localisation to the instanton and reducible branch. Our interest is in the contribution $\langle\CO\rangle^{\rm i}$
from the instanton branch.
A careful localisation argument~\cite{FMM:2026} shows that 
\bea  
\langle \CO\rangle^{\rm i} = \sum_k \Lambda_{N_f}^{\vd\CM_{v(E)}^{\CQ,N_f}}
\int_{\CM_{v(E)}^{\rm i}}\!\!\!\mud(\CO)
\prod_{j=1}^{N_f}
\Eul_{\mathbb T}(V^j_k;m_j)~,
\eea
where $\mud(\CO)$ is the associated Donaldson class of $\CO$. The sum runs over instanton numbers $k$~\eqref{instanton number}, that is, $c_2(E)$ with fixed $c_1(E)$. The characteristic class obtained by localisation is the equivariant Euler class of the adjoint Dirac index bundles $V_k^j$. 

To make the mass dependence explicit, we use that the mass $m_j$ of the $j$-th hypermultiplet is the equivariant weight of the torus action on the adjoint index bundle $V_k^j$ for the $j$-th Dirac operator. This gives:
\bea\label{Eul Vk}
\Eul_{\mathbb T}(V_k^j;m_j)=m_j^{\rk V_k^j}\sum_{\ell\geq 0}\frac{c_\ell(V_k^j)}{m_j^\ell}~.
\eea
As a result, the instanton contribution reads:
\bea
\label{O instanton}
\langle \CO\rangle^{\rm i}= \sum_k \Lambda_{N_f}^{\vd\CM_{v(E)}^{\CQ,N_f}}
\int_{\CM_{v(E)}^{\rm i}}\mud(\CO)
\prod_{j=1}^{N_f}
m_j^{\rk V_k^j}\sum_{\ell\geq 0}\frac{c_\ell(V_k^j)}{m_j^\ell}~.
\eea
This extends the earlier result~\cite[Equation (5.13)]{LoNeSha}.
To motivate the relation with Segre numbers, we need to identify the correct decoupling limit.

\sss{Decoupling limit.}
Increasing the mass of a hypermultiplet to infinity decouples it from the theory. For $N_f$ distinct masses, this decoupling limit is a double scaling limit, where $m_{N_f}\to\infty$, $\Lambda_{N_f}\to 0$, holding $\Lambda_{N_f-1}^{2N-N_f+1}=m_{N_f} \Lambda_{N_f}^{2N-N_f}$ finite.
Based on the $u$-plane integral and the instanton component of correlation functions, in~\cite{Marino:1998uy,Aspman:2022sfj,Aspman:2023ate,Furrer:2026byd} it is argued that the decoupling limit reads
\bea\label{decoupling formula}
\lim_{m_{N_f}\to\infty}\decoup_{N_f} \langle \CO\rangle_{N_f}^{\rm i}=\langle \CO\rangle_{N_f-1}^{\rm i}~,
\eea
where we determine the decoupling factor
\bea\label{decoupling factor}
\decoup_{N_f}&=\left(\frac{\Lambda_{N_f-1}}{\Lambda_{N_f}}\right)^{\vd \CM_k^{\CQ,{N_f}}+(2N-N_f)\rk W_k^{N_f}}\\
&= \left(\frac{\Lambda_{N_f-1}}{\Lambda_{N_f}}\right)^{(N^2+1)\chih-\frac{N^2}{4}c_1^2+(N^2-\frac12NN_f)\bfk_{N_f}^2+\frac N2(\bfk^2)}~,
\eea
for $\SU(N)$ SQCD, 
where $(\bfk^2)\coloneqq\bfk_1^2+\dots+ \bfk_{N_f}^2$.

For the situation of interest, we consider $N_f$  hypermultiplets with equal mass $m$, and want to decouple them simultaneously. This can be achieved by multiplying all decoupling factors for each of the $N_f$ hypermultiplets, and using the double scaling relation for each `intermediate' decoupling to obtain $\Lambda_{N_f}^{2N-N_f}=\Lambda_0^{2N}/m^{N_f}$ for the limit from $N_f$ to $0$ hypermultiplets. This gives the correlation function in the $N_f=0$ theory,
\begin{equation}
    \lim_{m\to\infty}\decoup\, 
    \langle \CO\rangle_{N_f}^{\rm i} =\langle \CO\rangle_{0}^{\rm i}~,
\end{equation}
with decoupling factor $\decoup=\prod_{j=1}^{N_f}\decoup_j$.
For $\SU(N)$ QCD with $N_f$ flavors, it reads\footnote{Proof: Write $\decoup=\prod_{j=1}^{N_f}\decoup_j$ as before, where the $\chih$ and $c_1^2$ terms are telescopic products, which derives the same terms in the final answer $\decoup$. The dependence on the background fluxes is more involved. We insert the relation~\eqref{decoupling limit SU(N)} everywhere, such that $\decoup$ depends on all the masses $m_i$, $i=1,\dots, N_f$ explicitly. 
To simplify the expression, we set $m_i=m$, such that the exponents become uniform. Inserting $m^{N_f}=\Lambda_0^{2N}/\Lambda_{N_f}^{2N-N_f}$ for the equal mass decoupling gives the answer.
}
\bea\label{eq:decouping_factor}
\decoup=\left(\frac{\Lambda_{0}}{\Lambda_{N_f}}\right)^{(N^2+1)\chih-\frac{N^2}4c_1^2+\frac{N^2}{N_f}(\bfk^2)}~.
\eea
This generalises~\cite[(2.23)]{Furrer:2026byd} from $N=2$ to all $N$.
\paragraph{Large mass expansion.} To see how the instanton component simplifies in the equal mass decoupling limit, we focus on the partition function $Z^{\mathrm{i}}_{\rm D}(\bm{m})=\langle \mathds{1}\rangle^{\mathrm{i}}$, where  $\bm{m}=(m,\dots,m)$. Including the  decoupling factor~\eqref{eq:decouping_factor}, it admits a nice large mass expansion:
\begin{eqsp}
\decoup Z^{\mathrm{i}}_{\rm D}(m)=\sum_{k}\left(\frac{\Lambda_0}{m}\right)^{\vd\CM_{v(E)}^{\mathrm{i}}}\sum_{\sum_jl_j=\vd\CM_{v(E)}^{\mathrm{i}}}\int_{\CM_{v(E)}^{\mathrm{i}}}\prod_{j=1}^{N_f}c_{l_j}(V_k^j)~,   
\end{eqsp}
The selection rule $\sum_jl_j=\vd\CM_{v(E)}^{\mathrm{i}}$ guarantees that we integrate top forms on $\CM_{v(E)}^{\mathrm{i}}$. Introducing the total vector bundle 
\begin{eqsp}\label{CV def}
    \CV_k\coloneqq \bigoplus_{j=1}^{N_f}V_k^j~,
\end{eqsp}
we have $c(\CV_k)=\prod_j c(V_k^j)$. We further define the instanton counting parameter 
\bea
q=\frac{\Lambda_0^2}{m^2}~,
\eea
such that the instanton contribution becomes
\begin{eqsp}\label{eq:large_mass_exp_UV}
 \decoup Z^{\mathrm{i}}_{\rm D}(\bm{m})=\sum_{k}q^{\frac12\vd\CM_{v(E)}^{\mathrm{i}}} \int_{\CM_{v(E)}^{\mathrm{i}}}c(\CV_k)~. 
\end{eqsp}
In section~\ref{sec:top_cor_seg_num}, we will match this series with the generating function of virtual Segre numbers, which are functions of $z$, where the counting parameter $q=-4^{\frac1N}z$ receives a different normalisation.\footnote{In principle, the equivariant weights $\nu_j$ for the adjoint index bundle $V_k^j$ need not  literally be the physical hypermultiplet mass $m_j$, but may differ by a convention-dependent numerical factor $\nu_j=\alpha_N m_j$. If that is the case, the instanton counting parameter matches perfectly with the IR calculation if $\alpha_N=\pm i4^{\frac{1}{2N}}$. 
This discrepancy is already observed for $N=2$ in~\cite{Furrer:2026byd,FMM:2026}.}

\section{Seiberg--Witten geometry of SU(\texorpdfstring{$N$}{N}) QCD}
\label{sec:SW geometry}
In this section, we introduce the Seiberg--Witten curves and important Coulomb branch functions, discuss the global symmetries and the $\CN=1$ vacua. 
\subsection{Curve and periods}
\sss{Coulomb branch.}
The vacuum moduli space of $\CN=2$ $\SU(N)$ gauge theory with $N_f\leq 2N$ fundamental hypermultiplets consists of several branches, characterised by the expectation values of the vector multiplet and hypermultiplet scalars. 
The Coulomb branch of the theory is parametrized by the vevs of the vector multiplet scalars. At generic points on the Coulomb branch, the SU($N$) gauge symmetry is broken to $\mathrm{U}(1)^{N-1}$ by the scalars, assuming the vev is in the Cartan subalgebra of $\mf{su}(N)$. More precisely, the adjoint-valued scalar $\phi$  has a vev 
\begin{equation}
    \langle \phi\rangle =\text{diag}(\phi_1,\dots,\phi_N) ~.
\end{equation}
Since $\langle\phi\rangle$ is traceless, we have $\sum_{m=1}^N\phi_m=0$.
The gauge-invariant coordinates on the Coulomb branch moduli space are given by the invariant polynomials of $u_a$ defined by the expression
\begin{equation}\label{char poly}
    \text{det}(x-\langle \phi\rangle)=\prod_{m=1}^{N}(x-\phi_m)=x^N-\sum_{a=1}^{N-1}u_ax^{N-a-1} ~.
\end{equation} 
The low-energy effective $\mathrm{U}(1)^{N-1}$ gauge theory at generic points on the Coulomb branch is characterised completely by a holomorphic function $\s{F}$ called the prepotential. The Seiberg--Witten solution gives the expression for the prepotential in terms of an auxiliary curve.

\sss{Higgs and Mixed Branches.}
The Higgs branches are parametrised by hypermultiplet expectation values satisfying the $D$- and $F$-flatness conditions, while the adjoint scalar is constrained so that part of the gauge symmetry remains unbroken. More generally, there exist mixed branches, on which both Coulomb and Higgs-branch moduli are simultaneously non-zero. These branches intersect the Coulomb branch at isolated singular loci, referred to as their \emph{roots}, where additional massless states appear.

For $N_f<N$, the Higgs branches are all non-baryonic and are labelled by an integer $r$ from 1 to $\min(\lfloor\frac{N_f}{2}\rfloor,N-2)$,
corresponding to vacua in which $r$ quark flavors acquire expectation values~\cite{Argyres:1996eh}. At the root of the $r$-th branch, the low-energy theory contains an unbroken $\SU(r)$ gauge symmetry together with Abelian factors and massless monopoles and quarks. When $N_f\geq N$, an additional Higgs branch, the \emph{baryonic branch}, appears. Its root intersects the Coulomb branch at a distinguished singular point, the \emph{baryonic root}, where the low-energy theory is described by a magnetic gauge group $\SU(N_f-N)$ with massless magnetic quarks~\cite{Argyres:1996eh}. In particular, the case $N_f=N$ marks the transition between theories without a baryonic branch and those admitting a non-Abelian magnetic description: the baryonic branch first appears, but its magnetic gauge group degenerates, leaving a purely Abelian low-energy theory. We find that much of the analysis simplifies greatly for $N_f=N$, with many a priori infinite series collapsing to a simple term or polynomial.

\sss{SW curves and discriminant.}
Introduce the following notations 
\allowdisplaybreaks
\begin{subequations} \label{SU(N) curve}
    \begin{align}
    P_{N_f}(X) & \coloneqq X^N-\sum_{a=1}^{N-1}u_aX^{N-a-1} ~,\label{eq:P(x)Nf<=2N}\\B(X) & \coloneqq  \Lambda_{N_f}^{2 N-N_f}\prod_{j=1}^{N_f}\left(X+m_j\right) ~,\label{eq:B(x)Nf<2N}\\t_p(\vec{m})&\coloneqq \sum_{j_1<\cdots<j_p} m_{j_1} \cdots m_{j_p}~,\\ T(X)&\coloneqq \left\{\begin{aligned}&\sum_{p=0}^{N_f-N} t_p(\vec{m}) X^{N_f-N-p}~,\quad N_f-N\geq 0~,\\&0,\quad N_f-N< 0~,\end{aligned}\right.\\A(X)&\coloneqq P(X)+\frac{1}{4}\Lambda_{N_f}^{2 N-N_f} T(X)\label{eq:A(x)Nf<2N}~.
\end{align}  
\end{subequations}
The Seiberg--Witten  curve $\Sigma$ for $\mathrm{SU}(N),N_f< 2N$ gauge theory is given by~\cite{DHoker:1996pva,Hanany:1995na}\footnote{See also~\cite{EGUCHI1996430,DHoker:1999yni,Argyres:1995wt,Abolhasani:1996np,Minahan:1995er,Minahan:1996ws,Klemm:1995wp}.}
\begin{equation}\label{y^2 curve}
 \Sigma\colon\quad   Y^2=A(X)^2-B(X)~.
\end{equation}
We define the physical discriminant uniformly as
\bea\label{SU(N) discriminant}
\Delta_{N_f}=\frac{1}{2^{2N}\Lambda_{N_f}^{N(2N-N_f)}}\Disc_X(Y^2)~.
\eea
In this way, the discriminant has mass dimension $[\Delta_{N_f}]=N(2N+N_f-2)$.
The decoupling limit for generic masses is $m_{N_f}\to\infty$, $\Lambda_{N_f}\to 0$ with 
\bea\label{decoupling limit SU(N)}
\Lambda_{N_f-1}^{2N-N_f+1}=m_{N_f} \Lambda_{N_f}^{2N-N_f}~.
\eea
Then the decoupling formula for the discriminant reads
\bea
\lim_{m_{N_f}\to \infty}\frac{1}{m_{N_f}^{N}}\Delta_{N_f}=\Delta_{N_f-1}~,
\eea
which justifies the normalisation~\eqref{SU(N) discriminant}.
Setting the masses $m_i=m$ equal, we can simultaneously decouple all $N_f$ hypermultiplets, using the relation
\bea\label{decoupling SU(N)}
\Lambda_{N_f}^{2N-N_f}=\frac{\Lambda_0^{2N}}{m^{N_f}}~.
\eea

\sss{Period matrix and discriminant identity.}
The SW differential is for generic masses given by~\cite{DHoker:1996pva}  
\begin{equation}\label{SW differential}
    \lambda=\frac{1}{2\pi i}\left(A^{\prime}-\frac{1}{2}(A-Y) \frac{B^{\prime}}{B}\right) \frac{XdX}{Y}~,
\end{equation}
with $A$, $B$ and $Y$ as in the curve above. In fact, using $(A-Y)(A+Y)=B$, one can write this as a logarithmic derivative, 
\bea
\lambda=\frac{X}{2\pi i}\, \mathrm d \log(A+Y)~.
\eea
The periods $a^a$ and $a^a_{\rm D}$ are obtained by contour integrals of the Seiberg--Witten differential,
\bea\label{a and aD}
a^a=\oint_{\alpha_a}\lambda~, \qquad
a^a_{\rm D}=\oint_{\beta_a}\lambda~,
\eea
where the cycles $\alpha_a,\beta_a\in H_1(\Sigma,\mathbb Z)$ are a symplectic basis of the integral homology on the curve $\Sigma$.
For an electromagnetic charge $\gamma=(\bfn_m,\bfn_e)\in H_1(\Sigma,\mathbb Z)$, the Seiberg--Witten central charge is 
\bea
Z_\gamma=\oint_\gamma \lambda=\bfn_m\cdot \bfa_{\rm D}+ \bfn_e\cdot \bfa~,
\eea
where we collect $\bfa=(a^1,\dots, a^{N-1})$, and similarly  $\bfa_{\rm D}=(a^1_{\rm D},\dots, a^{N-1}_{\rm D})$.

We also require the period matrix for the calculation of topological correlators. Following~\cite{Klemm:1995wp}, for $\SU(N)$ we define the $g=N-1$ holomorphic differentials (abelian differentials of the first kind), 
\bea\label{omega_j differentials}
\omega_{a}=\frac{X^{N-a-1}\mathrm dX}{2\pi i \, Y}~.
\eea
One can show that these differentials are obtained from the SW differential $\lambda$ by differentiation, and we get~\cite{Argyres:1995wt}
\bea
\omega_a=\frac{\partial \lambda}{\partial u_a}\in H^{1,0}(\Sigma)~,
\eea
where the equality is only modulo exact differentials.
For this, we need to choose a sheet of the curve $\Sigma$~\eqref{y^2 curve}.

From these differentials $\omega_a$, one constructs a set of periods $\int_{\gamma_m}\omega_a$. Here, $\gamma_m$ with $m=1,\dots, 2g$ is any basis of $H_1(\Sigma_g,\mathbb Z)$. The periods combine to a $(g,2g)$-dimensional period matrix $\Pi_{am}=\int_{\gamma_m}\omega_a$.
We choose a symplectic homology basis, denoting $\alpha_a=\gamma_a$ and $\beta_a=\gamma_{g+a}$ for $a=1,\dots, g$, with intersection pairing 
\bea
\alpha_a\cap \beta_b=\delta_{ab}~, \quad  \alpha_a\cap\alpha_b=\beta_a\cap\beta_b=0~.
\eea
These cycles have been obtained carefully in previous work~\cite{Klemm:1994qs,Klemm:1995wp,Douglas:1995nw}.
Writing the periods as $\Pi\eqqcolon (\widetilde A,B)$, the period matrix is
\bea\label{tau matrix}
\tau=\widetilde A^{-1}B~, \qquad \tau_{ab}=\frac{\partial a_{\mr{D},a}}{\partial a^b}~.
\eea
By Riemann's second relation, ${\rm Im}\tau>0$ is a positive definite matrix. We require in particular the periods
\bea\label{A tilde matrix}
\widetilde A_{ab}&=\frac{\partial a^b}{\partial u_a}=\oint_{\alpha_b}\omega_a~, \\
B_{ab}&=\frac{\partial a_{\mr{D},b}}{\partial u_a}=\oint_{\beta_b}\omega_a~.
\eea
We find it more convenient to work instead with the transposes $A=\widetilde A^\tra$ and $\widetilde B=B^\tra$, such that
\bea\label{A_ab}
A_{ab}&=\frac{\partial a^a}{\partial u_b}=\oint_{\alpha_a}\omega_b=\frac{1}{2\pi i}\oint_{\alpha_a}\frac{X^{N-b-1}\mathrm dX}{Y}~, \\
\widetilde B_{ab}&=\frac{\partial a_{\mr{D},a}}{\partial u_b}=\oint_{\beta_a}\omega_b=\frac{1}{2\pi i}\oint_{\beta_a}\frac{X^{N-b-1}\mathrm dX}{Y}~. 
\eea
The matrix $A$ is invertible away from the singular points, we denote its inverse by
\bea\label{J matrix}
J_{ab}=(A^{-1})_{ab}=\frac{\p u_a}{\p a^b}~.
\eea

The determinant $\det A$ of the period matrix is related to the discriminant $\Delta_{N_f}$~\eqref{SU(N) discriminant} in an interesting way. As a consequence of Theorem~\ref{thm:discriminant} in Appendix~\ref{subsec:hyperelliptic-discriminants},  there exists a constant $c_N$ that depends only on $\Lambda_{N_f}$, and a Siegel modular form $\varphi_{N-1}$ of genus $N-1$, such that 
\bea
\Delta_{N_f}^{\binom{2N-2}{N}}\left(\det A\right)^{4\binom{2N-1}{N}}=c_N\varphi_{N-1}(\Omega)~,
\eea
where the exponents are binomial coefficients. This identity is in part enabled by the fact that both sides vanish at strong coupling singularities, which are at the boundary of the Siegel upper-half plane, and the period determinant has the appropriate modular weight to make the product a modular form. For rank 1 theories, this becomes the well-known identity $\Delta \left(\frac{da}{du}\right)^{12}=c_2\, \eta^{24}(\tau)$, which has been useful in various contexts~\cite{Aspman:2021vhs,Manschot:2021qqe,Furrer:2026byd}. See Appendix~\ref{subsec:hyperelliptic-discriminants} for a discussion of this identity.

\subsection{Global symmetries}
The theory of interest has an intricate set of global symmetries, all of which are relevant to our calculation.

\sss{The global symmetries of $\SU(N)$ QCD.} To start, consider pure $\SU(N)$ $\CN=2$ supersymmetric Yang--Mills theory (SYM) on $\mathbb R^4$, without hypermultiplets. Its global symmetries are the usual $R$-symmetry, an ordinary global 0-form symmetry, and a higher-form symmetry. 
The classical $R$-symmetry is $\SU(2)_{\rm R}\times {\rm U}(1)_r$. The $\SU(2)_{\rm R}$ subgroup remains an exact symmetry quantum mechanically, and is used in the topological twist to redefine the fields. 
The ${\rm U}(1)_r$ subgroup, on the other hand, is broken in the microscopic theory by the ABJ anomaly to a $\mathbb Z_{4N}$ subgroup; and its anomaly becomes the ghost number anomaly in the twisted theory. On the Coulomb branch, the $\mathbb Z_{4N}$ group acts effectively only as a subgroup $\mathbb Z_N^{(r)}$~\cite{Argyres:1994xh}.

Since there is no matter in pure SYM, there is no continuous flavor symmetry. However, there is another global symmetry which acts on the fundamental Wilson lines, it is the global centre 1-form symmetry (see~\cite{Aharony:2013hda,Gaiotto:2014kfa}). It is isomorphic to $\mathbb Z_N^{(1)}$, being the centre of the Lie group $\SU(N)$, or equivalently, the group of outer automorphisms of the affine algebra $\widehat{\mathfrak{su}}(N)$. Even though they are isomorphic as groups, the one-form symmetry $\mathbb Z_N^{(1)}$ should not be confused with the discrete $R$-symmetry $\mathbb Z_N^{(r)}$, as the $R$-symmetry acts on local Coulomb branch operators, while the centre symmetry acts on Wilson lines and trivially on local operators. Yet, the two symmetries have a mixed 't Hooft anomaly (see e.g.~\cite{Cordova:2018acb,Closset:2023pmc}), which breaks one symmetry if the other is gauged.\footnote{The $\CN=1$ points we discuss in detail below can be viewed as different SPT phases for the centre one-form symmetry~\cite{Gaiotto:2014kfa}.}
Finally, the pure $\SU(N)$ has a charge conjugation symmetry $\mathsf C$, which is a global $\mathbb Z_2$ 0-form symmetry $\mathbb Z_2^{(0)}$, for $N\geq 3$. It is generated by outer automorphisms of the finite Dynkin diagram for $\mathfrak{su}(N)$, as we describe below. Charge conjugation also acts on the centre symmetry by inversion, e.g. $\mathsf C\colon q\mapsto -q$ for some charge $q\in\mathbb Z_N^{(1)}$ (see~\eg~\cite{Delmastro:2020dkz}).
Thus, the global quantum symmetry schematically contains 
\bea\label{global symmetries}
\mr{SU(2)_R}\times \mathbb Z_{4N}^{(r)}
\times \mathbb Z_N^{(1)}
\rtimes \mathbb Z_2^{\mathsf C}~.
\eea
By the CPT theorem, the $\SU(N)$ theory in addition has time reversal and parity symmetries. However, they do not act on gauge-invariant local operators, and we do not study them further.\footnote{See~\cite{DHoker:2024vii} for a more detailed account of these symmetries.}

Adding massive hypermultiplets modifies this symmetry group significantly. While the $\SU(2)_R$ symmetry is still preserved, the other three symmetries are broken. First, the residual action of the ${\rm U}(1)_r$ symmetry on the Coulomb branch is explicitly broken by the masses~\cite{Seiberg:1994aj}. The generic masses of the hypermultiplets furthermore break the charge conjugation symmetry of the vector multiplet. Similarly, the one-form symmetry is broken since fundamental Wilson lines can end on quarks. On the other hand, SQCD exhibits a global flavor symmetry ${\rm U}(N_f)\simeq [\SU(N_f)\times {\rm U}(1)_B]/\mathbb Z_{N_f}$, where $\mr{U(1)}_B$ is the baryon-number symmetry~\cite{Argyres:1996eh}. Generic masses break this symmetry to its maximal torus ${\rm U}(1)^{N_f}$. Since the flavor symmetry does not act on the Seiberg--Witten curve, we do not discuss it further and refer to the classical literature~\cite{Hanany:1995na,Argyres:1995wt,Argyres:1996eh,Seiberg:1994aj} for a detailed analysis.

Instead, we consider the special case of $\SU(N)$ QCD with $N_f$ hypermultiplets of equal and large mass $m$.
Since in the limit $m\to\infty$, the hypermultiplets decouple and we recover pure $\SU(N)$ SYM, the global symmetries~\eqref{global symmetries} are expected to emerge asymptotically. Indeed, as will be discussed in the remainder of this Section, the residual $\mathbb Z_N$ R-symmetry action and the charge conjugation symmetry are particularly important physical inputs in the calculation.

\sss{Charge conjugation symmetry.}
The pure $\SU(N)$ theory has a charge conjugation symmetry $\mathsf C$ for $N\geq 3$, which originates from the outer automorphism group
\bea
{\rm Out}(\SU(N))\cong \mathbb Z_2
\eea
of the $A_{N-1}$ Dynkin diagram:
\bea
\includegraphics[width=0.45\textwidth]{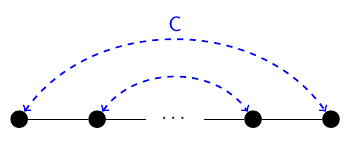}
\eea 
It exchanges the fundamental weights, and so acts on the simple roots by a Dynkin reversal 
\bea
\mathsf C\colon \alpha_b\mapsto \alpha_{N-b}~.
\eea
In the Lie algebra, it can be represented by the map $\phi\mapsto -\phi^\tra$. This applies to the adjoint scalar $\phi$ in the $\CN=2$ vector multiplet, from which we determine the characteristic polynomial $\det(x-\phi)$~\eqref{char poly}. Since the Casimirs $u_a$ are traces of $\phi^{a+1}$, charge conjugation acts on the Coulomb branch as
\bea
\cc\colon u_a\mapsto (-1)^{a+1}u_a~.
\eea
To see how this symmetry persists in the quantum theory, we check its effect on the Seiberg--Witten curve. For generic masses, one can check that the full $\SU(N)$ QCD curve~\eqref{SU(N) curve} has a $\mathbb Z_2$ symmetry
\bea\label{Cc curve}
 X\mapsto -X~, \quad Y\mapsto (-1)^N Y~, \quad u_a\mapsto (-1)^{a+1}u_a~, \quad m_j\mapsto -m_j~,
\eea
together with a more peculiar transformation $\Lambda_{N_f}^{2N-N_f}
\mapsto
(-1)^{N_f}\Lambda_{N_f}^{2N-N_f}$.\footnote{This symmetry is studied for pure $\SU(N)$ in~\cite{Arias-Tamargo:2023duo,DHoker:2024vii}.} In the decoupling limit~\eqref{decoupling SU(N)} for equal masses, $\Lambda_{N_f}^{2N-N_f}$ becomes $\Lambda_0^{2N}/m^{N_f}$. We thus see that this phase for $\Lambda_{N_f}$ is induced simply by $m\mapsto -m$ in the decoupling limit. As a consequence, for large and equal masses, the four transformations~\eqref{Cc curve} become an exact symmetry. Moreover, one can show that the Seiberg--Witten differential~\eqref{SW differential} receives a sign,
\bea
\cc\colon\quad \lambda\mapsto -\lambda~.
\eea
From~\eqref{A_ab} it is clear that the conjugation then maps the periods $A_{ab}=\frac{\p a^a}{\p u_b}$  and $\tilde B_{ab}=\frac{\partial a_{\mr{D},a}}{\partial u_b}$ as 
\bea\label{C on Aab}
\cc\colon\quad  
\begin{aligned} A_{ab}&\mapsto (-1)^b A_{(N-a)b}~, \\ B_{ab}&\mapsto (-1)^b B_{(N-a)b}
\end{aligned}
\eea
since the cycles $\alpha_a\mapsto \alpha_{N-a}$ are mapped according to Dynkin reversal. The action on the period matrix is not a simple matrix transformation. If we define $R_{ab}=\delta_{b,N-a}$ and $D_{ab}=(-1)^b\delta_{ab}$, then the map reads $A\mapsto RAD$.  To determine the action on the inverse $J=A^{-1}$~\eqref{J matrix}, we observe that $R^2=D^2=1$ are involutions, such that $J\mapsto (RAD)^{-1}=DJR$, or
\bea\label{cc duda}
\cc\colon \quad J_{ab}\mapsto (-1)^a J_{a(N-b)}~.
\eea
Since on the period vectors $\bfa$ and $\bfa_{\rm D}$~\eqref{a and aD}, conjugation acts as $\bfa \mapsto -R\bfa$ and $\bfa_{\rm D}\mapsto -R\bfa_{\rm D}$, one can show that
\bea\label{C on tau}
\cc\colon \quad \tau_{ab}\mapsto \tau_{N-a,N-b}~.
\eea
A special situation occurs for even $N$, when the middle Dynkin node $a=\frac12 N$ is self-conjugate. This leads to special reality conditions.

\subsection{\texorpdfstring{$\CN=1$}{N=1} vacua}\label{sec:N=1_vacua}

\sss{Singular loci and $\CN=1$ vacua.}
Whenever the discriminant $\Delta$ vanishes, the Seiberg--Witten curve becomes singular. Physically, a one-cycle $\gamma$ on the curve has a vanishing period $Z_\gamma$, and so the corresponding BPS particle whose charge is associated with $\gamma$ becomes massless. Thus the discriminant locus $\Delta=0$ is the locus of massless charged states. For pure $\SU(2)$ SYM, these are the singular points $u=\pm \Lambda_0^2$, where a monopole and a dyon become massless, respectively.

For higher-rank theories, the discriminant locus
\bea\label{singular locus}
\cD=\{u\colon \Delta_{N_f}(u)=0\}
\eea
has a richer structure. The generic elements of this divisor are smooth complex hypersurfaces, i.e. codimension-one strata $\cD^{(1)}$. Physically, they correspond to points on the Coulomb branch where a single BPS hypermultiplet becomes massless. There are also higher-codimension $\ell$ strata, whose smooth components we denote by $\cD^{(\ell)}$, where several singularity conditions are satisfied simultaneously. The corresponding massless BPS states can be mutually local or mutually non-local. For instance, pure $\SU(3)$ SYM contains two points on the Coulomb branch where mutually non-local BPS states become massless---these are the first examples of Argyres--Douglas superconformal theories~\cite{Argyres:1995jj}. In general, the singular locus admits the stratification
\bea\label{stratification}
\cD=\bigcup_{\ell=1}^{r}\cD^{(\ell)}~,
\eea
where the codimension $\ell$ ranges from $1$ to the rank $r$ of the gauge group.
Of particular interest is the top-codimension locus $\cD^{(r)}$, which under favourable conditions is the only locus that contributes to twisted partition functions.
It may contain isolated singularities of different types, including multicritical superconformal points such as higher-rank analogues of the Argyres--Douglas theories~\cite{Argyres:1995jj,Argyres:1995xn,EGUCHI1996430}.

A physical interpretation of some of the top-codimension singularities is obtained by softly breaking the $\CN=2$ theory to $\CN=1$ with an adjoint mass deformation $W=\mu\,\Tr\Phi^2$.
As shown by Seiberg and Witten for gauge group $\SU(2)$~\cite{Seiberg:1994rs}, supersymmetric vacua of the deformed theory occur precisely at singular points on the Coulomb branch where massless BPS states satisfy the low-energy F-term equations. For a gauge group $G$ of rank $r$, the expected generalisation is that this requires $r$ mutually local massless hypermultiplets, and hence the $\CN=1$ vacua are localised at the maximal, codimension-$r$ degenerations of the Seiberg--Witten curve. For pure $\CN=2$ SYM with gauge group $G$, these consist of $h^\vee(G)$ isolated points, in agreement with the $h^\vee(G)$ supersymmetric vacua of pure $\CN=1$ SYM and the supersymmetric index~\cite{Witten:1994cg,Marino:1998bm,Witten:2000nv}.\footnote{$h^\vee(G)$ is the dual Coxeter number of $G$.}
Specialising to pure $G=\SU(N)$, we have $r=N-1$ and $h^\vee=N$, so that $\cD^{(N-1)}$ contains the $N$ $\CN=1$ vacua. We denote them by $\vec \tu_0^{(n)}$, or in components, $\tu_{a;0}^{(n)}$, with $n=0,\dots, N-1$, where the subscript denotes $N_f=0$.

In the present paper, we consider $\SU(N)$ SQCD with $N_f=0,\dots,2N-1$ fundamental hypermultiplets. In this case, the top-codimension locus $\cD^{(N-1)}$ contains, in addition to the $\CN=1$ vacua and multicritical superconformal points, also further isolated singularities, including isolated singularities associated with the collision of quark singularities. The top codimension then decomposes as 
\bea\label{top codimension}
\cD^{(N-1)}
=
\cD_{\CN=1}
\cup
\cD_{\rm SCFT}
\cup
\cD_{\rm coll}~,
\eea
whose elements we label by $\vec{\tu}^{(n)}$, $\vec{\tu}^{(\alpha)}_{\rm SCFT}$, and $\vec{\tu}^{(\beta)}_{\rm coll}$, respectively. At each of the $N$ $\CN=1$ points, precisely $N-1$ mutually local hypermultiplets become massless.

\sss{The $\CN=1$ vacua of pure $\SU(N)$.}
Due to the complexity of the curves~\eqref{y^2 curve} for $\SU(N)$ QCD with generic masses, it is difficult to write down explicit expressions for the Casimirs $\tu_a$ at the $\CN=1$ points. Rather, we will define them through the decoupling limit, for which we require the description in the pure $\SU(N)$ theory.

For $N_f=0$, the curve becomes
\bea\label{pure SU(N) curve}
Y^2=P_0(X)^2-\Lambda_0^{2N}~,
\eea
with $P_0(X)$ as in~\eqref{eq:P(x)Nf<=2N}. The $\CN=1$ vacua are the points $\vec \tu$ where the $\alpha$-cycles simultaneously degenerate, and $N-1$ mutually local magnetic monopoles become massless. It is found in~\cite{Douglas:1995nw} that this requires~\eqref{pure SU(N) curve} to have $N-1$ double zeros and two single zeros, which is elegantly solved by going to Chebyshev polynomials,\footnote{See also~\cite{Edelstein:2000jc,Laba05,Marino:1998bm}.}
\bea\label{Chebyshev}
P_0(X)=2q^NT_N\left(\frac{X}{2q}\right)~,
\eea
where $T_N$ is the $N$-th Chebyshev polynomials of the first kind, defined as $T_n(\cos\theta)=\cos(n\theta)$, and $q\coloneqq 2^{-\frac1N}\Lambda_0$.\footnote{See e.g.~\cite{Abramovitz:1964} for important identities.} To see this, we insert into the curve, use the Chebyshev identity $T_N(z)^2-1=(z^2-1)U_{N-1}(z)^2$ to the Chebyshev polynomials $U_N$ of the second kind, defined by $U_{n}(\cos \theta )\sin \theta ={\sin }{\big (}(n+1)\theta {\big )}$, such that the curve at the $\CN=1$ points becomes
\bea\label{SU(N) chebyshev}
Y^2= q^{2N-2}(X^2-4q^2)
U_{N-1}\left(\frac{X}{2q}\right)^2.
\eea
The zeros of the $U$ functions are $X_k=2q\cos\frac{k\pi }{N}$ with $k=1,\dots, N-1$, which occur with multiplicity two. Since the discriminant $\Disc_X(Y^2)=0$ vanishes when the polynomial has a repeated root, the moduli corresponding to~\eqref{Chebyshev} are singular points of maximal degeneracy, with two single zeros $X=\pm 2q$ and $N-1$ double zeros $X_k$. 

Using the Taylor expansion of $T_N$, we can expand~\eqref{Chebyshev} and compare with~\eqref{eq:P(x)Nf<=2N} to give an explicit form for the moduli $\tu_{a,0}^{(0)}$ for arbitrary $N$. To find all $N$ $\CN=1$ vacua $\tu_{a,0}^{(n)}$ with $n=0,\dots, N-1$, we use that they are arranged in a $\mathbb Z_N$ symmetric fashion. Since $\tu_a$ has dimension $a+1$, we have then $\tu_{a,0}^{(n)}=\varepsilon_N^{n(a+1)}\tu_{a,0}^{(0)}$, with $\varepsilon_N\coloneqq e^{2\pi i/N}$. This gives
\bea\label{u N=1 SU(N)}
\tu_{a;0}^{(n)}=\begin{cases}(-1)^{\frac{a+3}{2}}\varepsilon_N^{n(a+1)}
\frac{N}{N-\frac{a+1}{2}}
\binom{N-\frac{a+1}{2}}{\frac{a+1}{2}}
2^{-\frac{a+1}{N}}
\Lambda_0^{a+1}~, \quad &a \text{ odd}, \\
0~, \quad &a \text{ even}.
\end{cases}
\eea
The first Casimir $a=1$ is of particular significance below. We have for all $N$:
\bea\label{u1;0}
\tu_{1;0}^{(0)}=2^{-\frac 2N}N\Lambda_0^2~.
\eea
We print the whole vector $\vec \tu_{0}^{(0)}$ with $n=0$ for small values of $N$:
\bea
N=2\colon\qquad &\vec \tu_{0}^{(0)}=(\Lambda_0^2)~, \\
N=3\colon\qquad &\vec \tu_{0}^{(0)}=( 2^{-\frac23}3\Lambda_0^2,\, 0)~, \\
N=4\colon\qquad &\vec \tu_{0}^{(0)}=(2^{\frac32}\Lambda_0^2,\, 0,\, -\Lambda_0^4)~, \\
N=5\colon\qquad &\vec \tu_{0}^{(0)}=(2^{-\frac25}5\Lambda_0^2,\, 0,\, 2^{-\frac45}5\Lambda_0^4)~, \\
N=6\colon\qquad &\vec \tu_{0}^{(0)}=(2^{\frac23}3\Lambda_0^2,\, 0,\, -2^{-\frac23}9\Lambda_0^4,\, 0, \, \Lambda_0^6)~, \\
\eea
and so forth.

\sss{The $\CN=1$ vacua of $\SU(N)$ QCD.}
To define the $\CN=1$ vacua of $\SU(N)$ QCD with arbitrary masses, we use the fact that the low-energy theory has a well-defined decoupling limit: when $m_i\to\infty$, in the scaling limit~\eqref{decoupling limit SU(N)} the curve for $\SU(N)$ with $N_f$ hypers flows to the one for $N_f=0$ hypers. Correspondingly, the finitely many top codimension singularities of the QCD theory move either to the infinite region in moduli space, or approach the singularities of the pure $\SU(N)$ theory. We \emph{define} the $\CN=1$ vacua $\vec \tu_{N_f}^{(n)}$ of the $\SU(N)$ theory with $N_f$ hypers through this decoupling limit:
\bea\label{N=1 sing definition}
\lim_{m_i\to\infty}\vec \tu_{N_f}^{(n)}=\vec \tu_{0}^{(n)}~.
\eea
This fixes in addition the labelling $(n)$ through~\eqref{u N=1 SU(N)}.

It should be noted that there are several ways to characterise the $\CN=1$ vacua in a given theory, which can be useful in different contexts.
One characterisation motivated by the result~\eqref{SU(N) chebyshev} is the fact that the curve factorises at these maximally singular points, 
\bea\label{curve factorised_sec3}
Y^2=H_{N_f}(X)^2F_{N_f}(X)~.
\eea
where $H_{N_f}$ is of degree $N-1$, and $F_{N_f}$ is quadratic.
While for pure $\SU(N)$ this factorisation is conveniently described using Chebyshev polynomials, for more general theories, the characterisation of $\CN=1$ vacua through factorisation requires more advanced techniques such as elliptic functions and matrix models~\cite{Cachazo:2002zk, Cachazo:2001jy, Janik:2005sk, Ahn:2004ym, Janik:2007es,Laba05}. For our specific calculation of $\SU(3)$ QCD in section~\ref{sec:SU(3) calculation}, we will make heavy use of this non-trivial factorisation, and we demonstrate it as well in section~\ref{sec:beyond SU(3)} for $\SU(4)$.

Finally, the $\CN=1$ vacua can also be obtained as the non-degenerate critical points of the discriminant. This was pointed out at least for $N=3$ in~\cite{Argyres:1994xh}:
\bea\label{sing points pure SU(N)}
\Delta=0~, \quad \frac{\p \Delta}{\partial u_a}=0~, \quad \det_{ab} \frac{\p^2\Delta}{\partial{u_a}\partial_{u_b}}\neq 0~.
\eea
It was argued that the second equation gives the top codimension singularities, which are the three $\CN=1$ points together with two Argyres--Douglas (AD) points. The third condition then eliminates the AD points. We do not adopt this definition, or use these equations for the calculation, since they lead to a system of coupled polynomial equations of large degree which is difficult to solve in practice. In section~\ref{sec:SU(3) calculation}, we develop an alternative strategy to obtain the $\CN=1$ moduli for generic theories, by expressing them as roots of a univariate rather than a multivariate polynomial. Similarly, it is clear that the factorised form~\eqref{curve factorised_sec3} implies the first two equations in~\eqref{sing points pure SU(N)}, while the third equation will be modified for arbitrary $N$.

\sss{$\mathbb Z_N$ symmetry of the $\CN=1$ vacua.}
For equal masses, there is a simplifying feature that the $\CN=1$ vacua come in an orbit under the discrete R-symmetry. In our case this residual symmetry is a $\mathbb Z_N$ symmetry, and we make the action precise in this section.

For this, consider the $\SU(N)$ curve~\eqref{SU(N) curve}, and set all $m_i=m$. 
As we will discuss below, an interesting scaling limit is
\bea\label{eq:XY_xy_ua_def}
X=mx~, \qquad Y=m^N y~, \qquad u_a=m^{a+1}T_a~.
\eea
The rescaling of the coordinates $X$ and $Y$ is certainly well-defined because $m\neq 0$, since we are interested in the large $m$ limit. The Casimirs $u_a$ can always be written like this, since $u_a$ has dimension $a+1$, and $T_a$ is thus dimensionless.
A short calculation gives
\bea
P(X)&=m^N p(x)~, \\
T(X)&=m^{N_f-N}t(x)~,
\eea
where we define the polynomials
\bea
p(x)&=x^N-\sum_{a=1}^{N-1}T_a x^{N-a-1}~,\\
t(x)&=\sum_{p=0}^{N_f-N}\binom{N_f}{p}x^{N_f-N-p}~,
\eea
such that
\bea
A(X)&=P(X)+\tfrac14\Lambda_{N_f}^{2N-N_f}T(X) \\
&=m^Np(x)+\tfrac14\Lambda_{N_f}^{2N-N_f}m^{N_f-N}t(x)~.
\eea
We further simplify 
\bea
\frac14\Lambda_{N_f}^{2N-N_f}m^{N_f-N}=\frac14 \frac{\Lambda_0^{2N}}{m^N}=m^N(-z)^N~,
\eea
where in the first equality we use the decoupling formula~\eqref{decoupling SU(N)}, and in the second one we substitute
\bea\label{z parameter}
z = -\frac{\Lambda_0^2}{4^{\frac1N}m^2}~,
\eea
which we will identify later with the instanton counting parameter. The same can be done for the remaining component $B(X)$ of the curve, and we obtain: 
\bea
A(X)&=m^N(p(x)+(-z)^Nt(x))~, \\
B(X)&=4(-z)^Nm^{2N}(x+1)^{N_f}~. 
\eea
Putting everything together, the curve $Y^2=A(X)^2-B(X)$ becomes
\bea\label{dimensionless curve N=1}
y^2=\left(p(x)+(-z)^Nt(x)\right)^2-4(-z)^N(x+1)^{N_f}~.
\eea
In the decoupling limit, $z$ is the only dimensionless variable related to $m$. 
From $u_a=m^{a+1}T_a$, we know that $T_a$ depends on $m$ as well, but consequently it can only do so through $z$, hence $T_a=T_a(z)$. 

Since the coefficients of the dimensionless Seiberg--Witten curve depend on $z$ only through $z^N$, 
it has a $\mathbb Z_N$ symmetry:\footnote{The minus sign is irrelevant for this analysis; it is fixed by conventions below and allows us to obtain $z$-series in many cases with rational rather than complex coefficients.}
\bea\label{Z_N action on z}
\mathbb Z_N\colon\quad z\longmapsto \varepsilon_N z~,
\eea
with $\varepsilon_N=e^{2\pi i/N}$, but only if it acts simultaneously as 
\bea\label{Z_N action on T}
T_a(z)\longmapsto T_a(\varepsilon_Nz)~.
\eea
We interpret this $\mathbb Z_N$ symmetry below as the exact symmetry permuting the $N$ $\CN=1$ vacua, and prove the scaling $u_a=m^{a+1}T_a(z)$ by finding explicit closed-form functions $T_a(z)$.

\sss{Period matrix at $\CN=1$ points.}
The factorised form $Y^2=H(X)^2F(X)$~\eqref{curve factorised_sec3} of the curve makes it possible to evaluate the period matrix $A_{ab}$ at the $\CN=1$ points. For this, we suppress the labels $N_f$ and write the polynomials as
\bea\label{H and F}
H(X)&=\prod_{a=1}^{N-1}(X-\ts_a)~, \\
F(X)&=(X-\tr_1)(X-\tr_2)~.
\eea
The roots $\ts_a$ and $\tr_a$ depend on the parameters such as the mass $m$ and the scale $\Lambda_{N_f}$.
The period matrix $A_{ab}$~\eqref{A_ab} is defined as the contour integral of the 1-form $\omega_b$ along the 1-cycles $\alpha_a$. We claim that at the $\CN=1$ points, the 1-cycles $\alpha_a$ are small contours around the roots $\ts_a$, which are clockwise for $a$ odd and counter-clockwise for $a$ even. Therefore, the period matrix at the $\CN=1$ point becomes
\bea\label{tA ab}
\tA_{ab}&=\frac{(-1)^{a+1}}{2\pi i}\oint_{\ts_a}\frac{X^{N-b-1}\mathrm dX}{Y}=
\mathop{\mathrm{Res}}\limits_{X=\ts_a}\frac{(-1)^{a+1}X^{N-b-1}}{H(X)\sqrt{F(X)}} \\
&=\frac{(-1)^{a+1}\ts_a^{N-b-1}}{H'(\ts_a)\sqrt{F(\ts_a)}}=\frac{(-1)^{a+1}\ts_a^{N-b-1}}{\prod_{c\neq a}(\ts_a-\ts_c)\sqrt{(\ts_a-\tr_1)(\ts_a-\tr_2)}}~.
\eea
Thus the strong coupling period matrix is fully determined once we know the roots~$\ts_a$ and~$\tr_i$.

\sss{Complex conjugation as charge conjugation.}
To determine how the charge conjugation acts on this matrix, it is useful to make a certain choice of reality condition. In the following sections, we are interested in a map $\varkappa\colon i\mapsto -i$, which is charge conjugation on the below $z$-series as long as $z$ is considered real. Since $z = -\frac{\Lambda_0^2}{4^{\frac1N}m^2}$~\eqref{z parameter}, we have 
% that $z^{\frac12}\sim \frac{\Lambda}{m}$ if $N$ is even, and 
that $z^{\frac12}\sim i \frac{\Lambda_0}{m}$. Thus $z^{\frac12}$ is real if $\bar m=-m$. That is, we momentarily choose $m\in i\mathbb R$.
This choice is \emph{not} a constraint on the low-energy theory, it is merely a necessary condition to make charge conjugation precise. We also see in~\eqref{Cc curve} that $\mathsf C\colon m\mapsto -m$, which is the action of complex conjugation if $m\in i\mathbb R$.

Also from~\eqref{eq:XY_xy_ua_def} we see that since $T_a$ is a function of $z$ (which we consider real), we get 
\begin{eqsp}
    \bar X=-X~,\quad \bar Y=(-1)^NY~,\quad \bar u_a=(-1)^{a+1}u_a~.
\end{eqsp}
Thus complex conjugation also acts on $X,Y,u_a$ in the same way as charge conjugation, see~\eqref{Cc curve}.

To make the action of complex conjugation on the curve precise, let $\sigma=(-1)^N$. From the curve $Y^2=P(X)$ specialised to the $\CN=1$ vacua, e.g.~\eqref{dimensionless curve N=1} but with $(X,Y)$ instead of $(x,y)$, one can show that $\overline{P(X)}=P(- X)=(-1)^NP(X)$, such that the curve admits the anti-holomorphic involution $(X,Y)\mapsto (- \bar X,\bar Y)$. 
In terms of the factorisation~\eqref{curve factorised_sec3}, the reality property becomes
$\overline{H(X)^2F(X)}=H(- X)^2F(- X)$. This implies that the roots $\ts_a$ and $\tr_j$ of $H$ and $F$ come in pairs of (anti)-complex conjugates, where we choose the labelling 
\bea
\bar\ts_a&=-\ts_{N-a}~, \\
\bar\tr_1&=-\tr_{2}~.
\eea
Then from the result~\eqref{tA ab} one can show that
\bea\label{conj A}
\bar \tA_{ab} = (-1)^b \tA_{(N-a)b}~.
\eea
Comparing with~\eqref{C on Aab}, we find that charge conjugation at the $\CN=1$ points simply becomes complex conjugation. The action is trivially inherited to the inverse $J=A^{-1}$, which becomes $\tJ_{ab}$ at strong coupling, and we get the analogue of~\eqref{cc duda}:
\bea\label{J CC}
\bar \tJ_{ab} = (-1)^a \tJ_{a(N-b)}~.
\eea

The dual $B$-periods~\eqref{A_ab} are more difficult to evaluate at the $\CN=1$ points. The corresponding cycles do not shrink, and the dual periods generally contain logarithmic divergences at the discriminant locus. The diagonal couplings $\tau_{aa}$ tend to zero, while the off-diagonals $\tau_{a\neq b}$ have finite limits. However, a similar argument shows that $\mathtt {\tilde B}_{ab} = (-1)^b \mathtt {\tilde B}_{(N-a)b}$ for the $\tilde B$ matrix at strong coupling, which reflects also~\eqref{C on Aab}. As a consequence, the coupling matrix $\tau_{ab}$ at strong coupling, satisfies
\bea\label{tau CC}
\bar \tau_{ab}=\tau_{(N-a)(N-b)}~,
\eea
which again agrees with the action of the charge conjugation~\eqref{C on tau}. We have thus shown that if we consider $m\in i\mathbb R$, then charge conjugation acts on strong coupling expansions as complex conjugation.

\section{Twisted low energy effective theory}
\label{sec:IR theory}
In this section, we write down the expression for the Coulomb branch integral and derive the Seiberg--Witten contributions at the strong coupling singularities. 
\subsection{Coulomb branch integral and SW contribution}\label{sec:Coul_int}
\sss{Contributions to partition function.}
A fundamental result in physical mathematics~\cite{Bah:2022wot} is the relation between Donaldson and Seiberg--Witten invariants~\cite{Moore:1997pc}. Schematically, the relation takes the following form:
\begin{eqsp}
Z_{\mr{D}}=Z_{\mr{Coulomb}}+Z_{\mr{SW}}~,
\end{eqsp}
where $Z_{\mr{D}}$ is the twisted UV partition function, $Z_{\mr{Coulomb}}$ is the integral over the regularised Coulomb branch, and $Z_{\mr{SW}}$ is the contribution from the singularities on the Coulomb branch. This reflects the global structure of the Coulomb branch:
\bea
\CM_{\rm Coulomb}=\mathbb C^r\setminus \cD~,
\eea
where $\cD$ is the singular locus~\eqref{singular locus} and $r$ is the rank of the Coulomb branch. More precisely, due to the stratification~\eqref{stratification} we can write
\bea
Z_\SW=\sum_\ell Z_\SW^{(\ell)}~.
\eea
The contributions $Z_\SW^{(\ell)}$ from a stratum $\cD^{(\ell)}$ are obtained as follows. The contribution $Z_\SW^{(1)}$ can be computed by wall-crossing from $Z_{\mr{Coulomb}}$. Its wall-crossing, in turn, determines $Z_\SW^{(2)}$, and the higher-codimension contributions follow  recursively~\cite{Marino:1998bm}. If $b_2^+>1$ and $X$ is of Seiberg--Witten simple type, all contributions vanish except the one from the top-codimension stratum $Z_\SW^{(r)}$. In this case, using~\eqref{top codimension} we can write for $\SU(N)$ SQCD:
\bea\label{ZSW decomp}
Z_\SW=Z_\SW^{\CN=1}+Z_\SW^{\rm SCFT}+Z_\SW^{\rm coll}~.
\eea
It is expected albeit not proven in~\cite{Marino:1998bm} that $Z_\SW^{\rm SCFT}=0$. We assume this in the following as well.\footnote{It would be good to prove this rigorously, which we hope to do in future work.}
In analogy with previous work on rank-one theories~\cite{Manschot:2021qqe,Furrer:2026byd,Aspman:2022sfj,Aspman:2023ate,Kim:2025fpz}, we notice that the decomposition~\eqref{ZSW decomp} is compatible with the decomposition of the UV partition function into the instanton component and the reducible component, after localisation with respect to the ${\rm U}(1)^{N_f}$ flavor symmetry acting on the non-Abelian multi-monopole moduli space $\CM_{v(E)}^{\CQ,N_f}$~\eqref{Qfixedeqs_voa}. In view of this agreement, we denote
\bea
Z_\SW^{\rm i}\coloneqq Z_\SW^{\CN=1}~,
\eea
which we expect in particular to agree with the instanton component~\eqref{O instanton} after localisation. We label the $h^\vee(\SU(N))=N$ points by $\vec{\tu}^{(n)}_{N_f}$, where $n=0,\dots,N-1$. The objective of this paper is to compute the sum
\bea
Z_\SW^{\rm i}=\sum_{n=0}^{N-1}Z_{\SW,n}~.
\eea

\sss{Generalities on low energy effective theory.}
Consider now a general low energy effective theory for a gauge group with rank $r>1$ whose effective theory has gauge group $\mr{U(1)}^r$. From the general discussion of~\cite{Marino:1998bm,Manschot:2021qqe,Aspman:2022sfj}, for four-manifolds with $b_2^{+}=1$, the Coulomb branch integral reduces to an integral over zero modes. 
The bosonic zero modes of the theory are the scalars $a^I$ and field strengths $F^I$ with (anti)-self-dual projections $F_{ \pm}^I$. For $b_1=0$, the fermionic zero modes are the 0-forms $\eta^I$ and self-dual 2-forms $\chi^I$. We also have self-dual 2-form auxiliary fields $D^{I}$. The gauge coupling matrix is given by
\begin{equation}
\tau_{I J}=\frac{\partial^2 \s{F}}{\partial a^I \partial a^J} ~,
\end{equation}
where $\s{F}$ is the prepotential of the theory. 
Denote $y_{I J}=\mr{Im}\left(\tau_{I J}\right)$ and
\begin{equation}
\s{F}_{I J K}=\frac{\partial \s{\tau}_{J K}}{\partial a^I}=\frac{\partial^3\CF}{\partial a^I\partial a^J\partial a^K} ~.
\end{equation}
The zero mode Lagrangian for the effective theory is then given by~\cite{Marino:1998bm}
\begin{align}\label{eq:genlaglowen}
\mathcal{L}&=  \frac{\im}{16 \pi}\left(\ov{\tau}_{I J} F_{+}^I \wedge F_{+}^J+\tau_{I J} F_{-}^I \wedge F_{-}^J\right)-\frac{1}{8 \pi} y_{I J} D^I \wedge D^J \\
&\quad +\frac{i \sqrt{2}}{16 \pi} \overline{\mathcal{F}}_{I J K} \eta^I \chi^J \wedge\left(D+F_{+}\right)^K ~.
\end{align}
The action of the scalar supercharge on the zero modes is given by
\begin{equation}
\begin{split}
&\left[\mathcal{Q}, A^I\right]=\psi^I=0~, \hspace{0.55cm}\left[\mathcal{Q}, \psi^I\right]=4 \sqrt{2} d a^I~, \\
&\left[\mathcal{Q}, a^I\right]=0~,\quad\quad\quad\quad  \left[\mathcal{Q}, \ov{a}^I\right]=\sqrt{2} i \eta^I~,\\
&\left[\mathcal{Q}, \eta^I\right]=0~,\quad\quad\quad\quad \left[\mathcal{Q}, \chi^I\right]=i\left(F_{+}^I-D_{+}^I\right)~, \\
&\left[\mathcal{Q}, D^I\right]=\left(d \psi^I\right)_{+}=0 ~.
\end{split}
\end{equation}
The expression for $\mathcal{Q}$ in terms of field variables is given by
\begin{equation}\label{eq:supcharloeener}
\mathcal{Q}=\sqrt{2} i \eta^I \frac{\partial}{\partial \ov{a}^I}+i\left(F_{+}^I-D_{+}^I\right) \frac{\partial}{\partial \chi^I} ~.
\end{equation}

The Lagrangian $\mathcal{L}$ for zero modes can be written as a topological, holomorphic term plus a $\mathcal{Q}$-exact term,
\begin{equation}
\mathcal{L}=\frac{i}{16 \pi} \tau_{I J} F^I \wedge F^J+\{\mathcal{Q}, W\} ~,
\end{equation}
with
\begin{equation}
W=-\frac{i}{8 \pi} y_{I J} \chi^I\left(F_{+}^J+D^J\right) ~.
\end{equation}
From this it is clear that $[\mathcal{Q}, \mathcal{L}]=0$.

\subsection{The Coulomb branch integral}
As mentioned above, the twisted UV theory has background gauge fields. The IR theory at generic points on the Coulomb branch is determined by a $\mr{U(1)}^{N-1}$ abelian gauge theory with the background gauge fields for the $\mr{U(1)}^{N_f}$ flavor symmetry and background metric on $X$. Thus, the effective theory must be modelled, as in~\cite{Aspman:2022sfj,Manschot:2021qqe}, by a $\mr{U(1)}^{N-1+N_f}$ abelian gauge theory with the vectormultiplet associated with the $\mr{U}(1)^{N_f}$ factor ``frozen'' as described below. The Lagrangian of this theory is as in~\eqref{eq:genlaglowen} with $r=N+N_f-1$. We will identify the index $I=1, \ldots, N-1$ with the unbroken $\mr{U}(1)^{N-1}$ of $\operatorname{SU}(N) / \mathbb{Z}_N$ and $I=N, \ldots, N_f+N-1$ with the $\mr{U}(1)^{N_f}$ factor of the gauge group. 
\\\\
\textbf{Notation.} We will use the convention that lower case indices $a, b, c, \ldots$ run over $1, \ldots, N-1$, and $u, v, w, \ldots$ run over $N, \ldots, N+N_f-1$ and upper case indices $I, J, K, \ldots$ run from $1, \ldots, N+N_f-1$. We will also use the index $i,j,k,...$ running over $1,2,...,N_f$ labelling the background fields. In particular, we have the relation $u=i+N-1$. 
\\\\
The masses $m_1, \ldots, m_{N_f}$ of the hypermultiplets are the VEVs of the scalar fields of the $\mr{U}(1)$ vector multiplets:
\begin{equation}
\left\langle\phi^u\right\rangle=a^u=m_{u-N+1}, \quad u=N, \ldots, N_f+N-1.
\end{equation}
The field strength $F^u$ provides the coupling of the theory to the background connections $\mathcal{A}_\mu^{u-N+1}$ defined in Table~\ref{tab:twconnGXRF}. We set 
\begin{equation}
\left[F^{u}\right]=4 \pi \bm{k}_{u-N+1} ~,\quad u=N,N+1,\dots,N-1+N_f~,
\end{equation}
where $\bk_j$ are the background fluxes satisfying~\eqref{bfk def}. Note that
\begin{equation}\label{eq:kmexp}
\bm{k}_j\in\begin{cases}
\frac{L}{\mr{lcm}(2, N)}, & \mu(b)\neq 0 \\
 \frac{L}{2}, & \mu(b)=0~.
\end{cases}
\end{equation}
As in~\cite{Manschot:2021qqe}, we set the fermionic fields to zero:
\begin{equation}
\eta^I=\chi^I=0, \quad I=N, \ldots, N+N_f-1 \text {. }
\end{equation}
From the SUSY transformation, we see that the SUSY transformation of $\ov{a}^I$ vanishes for $I=N, \ldots, N+N_f-1$. The vanishing of the SUSY variation of $\chi^I$ implies
\begin{equation}\label{eq:Du=F+v}
D^I=F_{+}^I, \quad I=N \ldots, N+N_f-1 .
\end{equation}
The scalar supercharge~\eqref{eq:supcharloeener} then takes the form
\begin{equation}
\mathcal{Q}=\sum_{I=1}^{N-1}\left(\sqrt{2} \im \eta^I \frac{\partial}{\partial \ov{a}^{I}}+\im\left(F_{+}^I-D_{+}^I\right) \frac{\partial}{\partial \chi^I}\right) \text {. }
\end{equation}

The zero-mode Lagrangian is given by
\begin{equation}\label{eq:lagQW}
\mathcal{L}=\frac{\im}{16 \pi} \tau_{IJ} F^I \wedge F^J+\{\mathcal{Q}, W\} ~,
\end{equation}
where now $W$ is given by
\begin{equation}
W=-\frac{\im}{8 \pi} \sum_{I=1}^{N-1}\sum_{J=1}^{N+N_f-1} y_{I J} \chi^I\wedge\left(F_{+}+D\right)^J
\end{equation}
and in the first term in~\eqref{eq:lagQW}, the sum is over $I, J=1, \ldots, N+N_f-1$.

We can now expand the zero-mode Lagrangian as
\begin{equation}
\begin{split}
\mathcal{L}&=  \frac{\im}{16 \pi} \tau_{I J} F^I \wedge F^J-\frac{\im}{8 \pi}\left(\sqrt{2} \im \eta^a \frac{\partial}{\partial \ov{a}^a}+\im\left(F_{+}^a-D_{+}^a\right) \frac{\partial}{\partial \chi^a}\right)\left(y_{b J} \chi^b\wedge\left(F_{+}^J+D^J\right)\right) \\
&= \frac{\im}{16 \pi} \tau_{I J} F^I \wedge F^J+\frac{\sqrt{2}}{8 \pi} \eta^a \frac{\partial}{\partial \ov{a}^a}\left(\frac{\tau_{b J}-\ov{\tau}_{b J}}{2 \im}\right) \chi^b \wedge\left(F_{+}+D\right)^J \\
& \quad +\frac{1}{8 \pi}\left(F_{+}^a-D_{+}^a\right) \wedge\left(F_{+}^J+D^J\right) y_{b J} \delta_a^b \\
&=  \frac{\im}{16 \pi} \tau_{I J} F^I \wedge F^J+\frac{\im \sqrt{2}}{16 \pi} \s{\overline{F}}_{a b J} \eta^a \chi^b \wedge\left(F_{+}+D\right)^J +\frac{1}{8 \pi} y_{a b}\left(F_{+}^a-D_{+}^a\right) \wedge\left(F_{+}^b+D^b\right)\\
& \quad +\frac{1}{8 \pi} y_{a u}\left(F_{+}^a-D_{+}^a\right) \wedge\left(F_{+}^u+D^u\right) ~.
\end{split}
\end{equation}
Substituting $D^u=F_{+}^u$ from~\eqref{eq:Du=F+v}, we obtain 
\begin{align}
\mathcal{L} &=\frac{\im}{16 \pi} \tau_{I J} F^I \wedge F^J+\frac{\im \sqrt{2}}{16 \pi} \overline{\s{F}}_{a b c} \eta^a \chi^b \wedge\left(F_{+}+D\right)^c+\frac{\im \sqrt{2}}{8 \pi} \overline{\s{F}}_{a b u} \eta^a \chi^b \wedge F_{+}^u \nonumber \\
&\quad  +\frac{1}{8 \pi} y_{a b} F_{+}^a \wedge F_{+}^b-\frac{1}{8 \pi} y_{a b} D^a \wedge D^b+\frac{1}{4 \pi} y_{a u}\left(F_{+}^a-D_{+}^a\right) \wedge F_{+}^u ~.
\end{align}
We now want to integrate out $D^a$. To do this, we need to perform a Gaussian path integral
\begin{equation}
\int[\s{D} D] \exp \left(-\left[\int_{X}-\frac{1}{8 \pi} y_{a b} D^a \wedge D^b-\frac{y_{a u}}{4 \pi} D^a \wedge F_{+}^u+\frac{\im \sqrt{2}}{16 \pi} \overline{\s{F}}_{a b c} \eta^a \chi^b \wedge D^c\right]\right) ~.
\end{equation}
This can be evaluated using 
\begin{equation}
\begin{split}
\int[D \vec{\varphi}] \exp &\left(-\int d^d x\left[\frac{1}{2} \vec{\varphi}(x)^{T} K \vec{\varphi}(x)+\vec{J}(x)^{T} \vec{\varphi}(x)\right]\right) \\
&=\exp \left(\int d^d x\left[\frac{1}{2} \vec{J}(x)^{T} K^{-1} \vec{J}(x)\right]\right)(2 \pi)^{n / 2}(\operatorname{det} K)^{-1 / 2} ~,
\end{split}
\end{equation}
where $\vec{\varphi}(x) \in \mathbb{R}^n$ and $K \in \mathrm{GL}(n, \mathbb{R})$.
We first analytically continue $D \rightarrow \im D$ so that the integral takes the form
\begin{equation}
\begin{aligned}
\im \int[\s{D} D] &\exp \left(-\left[\int_{X} \frac{1}{2} \frac{y_{a b}}{4 \pi} D^a \wedge D^b-\frac{\im y_{a u}}{4 \pi} D^a \wedge F_{+}^u-\frac{\sqrt{2}}{16 \pi} \overline{\s{F}}_{a b c} \eta^a \chi^b \wedge D^c\right]\right) \\
& =\im \int[\s{D} D] \exp \left(-\left[\int_{X} \frac{1}{2} \frac{y_{a b}}{4 \pi} D^a \wedge D^b-\left(\frac{\im y_{a u}}{4 \pi} F_{+}^u+\frac{\sqrt{2}}{16 \pi} \overline{\s{F}}_{bca} \eta^b \chi^c\right) \wedge D^a\right]\right) \\
& =\exp \left(\int _ { X } \left[-\frac{y_{a u}\left(y^{-1}\right)_{a b} y_{b v}}{8 \pi} F_{+}^u \wedge F_{+}^v+\frac{\sqrt{2}\im}{32 \pi} y_{a u}\left(y^{-1}\right)_{a b} \overline{\s{F}}_{b c d} \eta^c \chi^d \wedge F_{+}^u\right.\right.  \\&\left.\left.+\frac{\sqrt{2} \im}{32 \pi} y_{b u}\left(y^{-1}\right)_{a b} \overline{\s{F}}_{a c d} \eta^c \chi^d \wedge F_{+}^u +\frac{1}{64 \pi} \overline{\s{F}}_{c d a}\left(y^{-1}\right)_{a b} \overline{\s{F}}_{b e f} \eta^c \chi^d \wedge\eta^e \chi^f\right]\right) \frac{(8 \pi^2)^{\frac{N-1}{2}} \im }{\sqrt{\mr{det} y_{a b}}} \\
& =\exp \left(-\int_{X}\left[\frac{y_{a u}\left(y^{-1}\right)_{a b} y_{b v}}{8 \pi} F_{+}^u \wedge F_{+}^v-\frac{\sqrt{2}\im}{16 \pi} y_{a u}\left(y^{-1}\right)_{a b} \overline{\s{F}}_{b c d} \eta^c \chi^d \wedge F_{+}^u\right.\right. \\
& \left.\left.-\frac{1}{64 \pi} \overline{\s{F}}_{c d a}\left(y^{-1}\right)_{a b} \overline{\s{F}}_{b e f} \eta^c \chi^d\wedge \eta^e \chi^f\right]\right) \frac{(8 \pi^2)^{\frac{N-1}{2}} \im}{\sqrt{\mr{det} y_{a b}}},
\end{aligned}
\end{equation}
where we used the fact that $\overline{\s{F}}_{a b c}, y_{a b},\left(y^{-1}\right)_{a b}$ are symmetric.
Thus the effective action is given by
\begin{equation}\label{eq:twilowenerlag}
\begin{split}
\mathcal{L}&=\frac{\im}{16 \pi} \tau_{I J} F^I \wedge F^J+\frac{\im \sqrt{2}}{16 \pi} \overline{\s{F}}_{a b c} \eta^a \chi^b \wedge F_{+}^c+\frac{\im \sqrt{2}}{8 \pi} \overline{\s{F}}_{a b u} \eta^a \chi^b \wedge F_{+}^u +\frac{1}{8 \pi} y_{a b} F_{+}^a \wedge F_{+}^b\\
&+\frac{1}{4 \pi} y_{a u} F_{+}^a \wedge F_{+}^u+\frac{y_{a u}\left(y^{-1}\right)_{a b} y_{b v}}{8 \pi} F_{+}^u \wedge F_{+}^v -\frac{\sqrt{2}\im}{16 \pi} y_{a u}\left(y^{-1}\right)_{a b} \overline{\s{F}}_{b c d} \eta^c \chi^d \wedge F_{+}^u\\
& -\frac{1}{64 \pi} \overline{\s{F}}_{c d a}\left(y^{-1}\right)_{a b} \overline{\s{F}}_{b e f} \eta^c \chi^d\wedge \eta^e \chi^f +\log \frac{\sqrt{\mr{det} y_{a b}}}{(8 \pi^2)^{\frac{N-1}{2}} \im} \\
& =\frac{\im}{16 \pi} \tau_{I J} F^I \wedge F^J+\frac{1}{8 \pi} y_{a b} F_{+}^a \wedge F_{+}^b+\frac{1}{4 \pi} y_{a u} F_{+}^a \wedge F_{+}^u +\frac{y_{a u}\left(y^{-1}\right)_{a b} y_{b v}}{8 \pi} F_{+}^u \wedge F_{+}^v\\&+\frac{\im \sqrt{2}}{16 \pi} \overline{\s{F}}_{a b c} \eta^a \chi^b \wedge F_{+}^c+\frac{\im \sqrt{2}}{8 \pi} \overline{\s{F}}_{a b u} \eta^a \chi^b \wedge F_{+}^u -\frac{\sqrt{2}\im}{16 \pi} y_{a u}\left(y^{-1}\right)_{a b} \overline{\s{F}}_{b c d} \eta^c \chi^d \wedge F_{+}^u\\
& -\frac{1}{64 \pi} \overline{\s{F}}_{c d a}\left(y^{-1}\right)_{a b} \overline{\s{F}}_{b e f} \eta^c \chi^d\wedge \eta^e \chi^f +\log \frac{\sqrt{\mr{det} y_{a b}}}{(8 \pi^2)^{\frac{N-1}{2}}\im} \\
& =\frac{\im}{16 \pi} \tau_{I J} F^I \wedge F^J+\frac{1}{8 \pi} y_{a b} F_{+}^a \wedge F_{+}^b+\frac{1}{4 \pi} y_{a u} F_{+}^a \wedge F_{+}^u+\frac{y_{a u}\left(y^{-1}\right)_{a b} y_{b v}}{8 \pi} F_{+}^u \wedge F_{+}^v \\
& +\frac{\im \sqrt{2}}{8 \pi} \overline{\s{F}}_{a b u} \eta^a \chi^b \wedge F_{+}^u+\frac{\im \sqrt{2}}{16 \pi} \overline{\s{F}}_{a b c} \eta^a \chi^b \wedge\left(F_{+}^c-y_{d u}\left(y^{-1}\right)_{c d} F_{+}^u\right) \\
& -\frac{1}{64 \pi} \overline{\s{F}}_{c d a}\left(y^{-1}\right)_{a b} \overline{\s{F}}_{b e f} \eta^c \chi^d\wedge \eta^e \chi^f+\log \frac{\sqrt{\mr{det} y_{a b}}}{(8 \pi^2)^{\frac{N-1}{2}} \im} ~.
\end{split}
\end{equation}
We now want to integrate over fermion zero modes $\eta_0^c, \chi_0^d$. For $b_2^+=1$, let $J$ be the generator of the selfdual 2-forms of $H^2(X, \mathbb{R})$ normalised such that 
\begin{equation}
    \int_{X} J \wedge J=1 ~.
\end{equation}
In the Lagrangian we can replace $\eta^a$ by its zero mode $\eta_0^a$, and $\chi^c$ by
\begin{equation}
\chi^c=\chi_0^c J ~.
\end{equation}
The path integral we must perform now is
\begin{equation}
\int \prod_{a=1}^{N-1} d \eta_0^a d \chi_0^a \exp \left(-\int_{X} \mathcal{L}\right) ~.
\end{equation}
We want to evaluate the path integral
\begin{equation}
\begin{split}\label{eq:ferm_0-mode_int}
I \equiv \int \prod_{a=1}^{N-1} d \eta_0^a d \chi_0^a &\exp \left\{-\left[\frac{\im \sqrt{2}}{8 \pi} \overline{\s{F}}_{a b u} \eta_0^a \chi_0^b \int_{X} J \wedge F_{+}^u-\frac{1}{64 \pi} \overline{\s{F}}_{c d a}\left(y^{-1}\right)_{a b} \overline{\s{F}}_{b e f} \eta_0^c \chi_0^d \eta_0^e \chi_0^f\right.\right.\\
& \hspace{3cm}\left.\left.+\frac{\im \sqrt{2}}{16 \pi} \overline{\s{F}}_{a b c} \eta_0^a \chi_0^b \int_{X} J \wedge\left(F_{+}^c-y_{d u}\left(y^{-1}\right)_{c d} F_{+}^u\right)\right]\right\}~.
\end{split}
\end{equation}
For $N=3$, we can explicitly perform the integral. We get 
\begin{equation}
\begin{split}
    I=&-\frac{1}{\pi^2 2^7}\left(\overline{\mathcal{F}}_{11 a} \overline{\mathcal{F}}_{22 b}-\overline{\mathcal{F}}_{12 a} \overline{\mathcal{F}}_{12 b}\right)\bigg\{B( J ,\left.F_{+}^a-y_{c u}\left(y^{-1}\right)_{a c} F_{+}^u\right)B(J,\left.F_{+}^b-y_{d v}\left(y^{-1}\right)_{b d} F_{+}^v\right)\\&- 4\pi(y^{-1})_{ab} \bigg\}-\frac{1}{\pi^2 2^5}\left(\overline{\mathcal{F}}_{11 u} \overline{\mathcal{F}}_{22 v}-\overline{\mathcal{F}}_{12 u} \overline{\mathcal{F}}_{12 v}\right)B(J, F_{+}^u)B(J ,F_{+}^v)~.
\end{split}
\end{equation}
For $N \geq 4$, this integral is more complicated, but is of the form 
\begin{equation}
    I\sim\text{det}_{ab}(\overline{\mathcal{F}}_{abc}F_+^c)+\text{det}_{ab}(\overline{\mathcal{F}}_{abu}F_+^u)+\dots
\end{equation}
where $\dots$ must be considered as contact terms. We proceed with the above integral~\eqref{eq:ferm_0-mode_int}.
\subsubsection{Observables and contact terms}
In the low-energy path integral on the Coulomb branch, we can include observables. There are two basic observables which we will consider---the point observable and the surface observable. The point observable, which we denote by $U$, is any general gauge-invariant function on the Lie algebra $\mf{su}(N)$. We will restrict to the linear combination of the Casimirs of $\mf{su}(N)$~\cite{Marino:1998bm}: 
\begin{eqsp}
    U\coloneqq \sum_{a=1}^{N-1}p^a\frac{\mr{Tr}\,\langle\phi\rangle^{a+1}}{\Lambda_{N_f}^{a+1}}~,
\end{eqsp}
where $p^a$ are formal fugacities and $\langle\phi\rangle$ denotes the VEV of the adjoint scalar field of the UV theory valued in the Cartan subalgebra of $\mf{su}(N)$. The surface observables are obtained by applying the one-form operator implementing the descent equations to any general gauge-invariant function $V$ on $\mf{su}(N)$. We will consider $V=u_a$, then the corresponding surface observable is given by 
\begin{eqsp}\label{surface observable}
    \tilde{I}^a(S)\coloneqq \frac{\im}{\sqrt{2}\pi}\int_{S}\left(\frac{1}{32}\frac{\partial^2u_a}{\partial a^b\partial a^c}\psi^b\wedge\psi^c-\frac{\sqrt{2}}{4}\frac{\partial u_a}{\partial a^b}(F_-+D)^b\right)~,
\end{eqsp}
where $S\in H_2(X,\Z)$ is the fugacity for the surface observable. 
For $b_2^+=1$, we see that the surface observable takes the form
\begin{eqsp}
\tilde{I}^a(S)=-\frac{\im}{4\pi}\frac{\partial u_a}{\partial a^b}\int_{S}(F_-+D)^b~.   
\end{eqsp}
We can then consider the formal sum of surface observables
\begin{eqsp}
    \tilde{I}(S)\coloneqq \sum_{a=1}^{N-1}\frac{f_a}{\Lambda_{N_f}^{a}}\tilde{I}^a(S)~.
\end{eqsp}
We will consider the generating function $    \left\langle e^{U+\tilde{I}(S)}\right\rangle$ of the correlation functions of these observables. 
It is useful to introduce the notation 
\begin{eqsp}\label{eq:def_V_a}
    V_a\coloneqq \sum_{b=1}^{N-1}\frac{f_b}{\Lambda_{N_f}^b}\frac{\partial u_b}{\partial a^a}~.
\end{eqsp}
When we insert $\tilde{I}(S)$ in the path integral and integrate out $D$, it is equivalent to shifting 
\begin{eqsp}
    \im y_{aj}F^j_+\to \im y_{aj}F^j_++V_aS_+~, 
\end{eqsp}
in~\eqref{eq:twilowenerlag}. This results in terms of the 
form 
\begin{eqsp}\label{eq:cont_terms_couplings}
    \exp\left(\frac{(y^{-1})_{ab}}{8\pi} S_+^2 V_aV_b\right)~,\quad \exp\left(\frac{(y^{-1})_{ab}y_{aj}}{4\pi}\int_X F^j_+\wedge S_+\right)~.
\end{eqsp}
These couplings must be supplemented by contact terms in the low-energy path integral. The contact term for the first term in~\eqref{eq:cont_terms_couplings} takes the form~\cite{Laba05,Mari_o_1999}:
\begin{eqsp}
  \exp\left(\sum_{a,b}f_af_bT_{a,b}(\vec{u})\right)~.  
\end{eqsp}
The function $T_{a,b}(\vec{u})$ can be computed using the blow-up formula~\cite{Mari_o_1999}, which we discuss in Section~\ref{sec:blowup} below. In the case when $f^a=0$ for $a\neq 1$, or equivalently, $\tilde{I}(S)=\tilde{I}^1(S)$, the contact term $G_{N_f}\coloneqq T_{1,1}$ can be written in terms of the prepotential.
Let $\tau_{\text{uv}}$ be the UV coupling for $N_f=2N$ and for $N_f<2N$ be defined by $\Lambda_{N_f}^{2N-N_f}=e^{\im\pi\tau_{\text{uv}}}$. Then the contact term for the first coupling is given by~\cite{Marino:1998bm}
\begin{eqsp}\label{eq:contact_G}
    G_{N_f}\coloneqq T_{1,1}=\frac{4}{\pi \im}\frac{\partial^2\CF}{\partial \tau_{\text{uv}}^2}=\frac{1}{\pi\im}\frac{\partial u_1}{\partial \tau_{\text{uv}}}~,
\end{eqsp}
where we used Matone's relation~\cite{Matone:1995rx}
\begin{eqsp}
\frac{\partial \CF}{\partial \tau_{\text{uv}}}  =\frac{1}{4}u_1~.  
\end{eqsp}
For $N_f<2N$, we can equivalently write 
\begin{eqsp}
    G_{N_f}=\frac{4\pi \im}{(2N-N_f)^2\Lambda_{N_f}^2}(\Lambda_{N_f}\partial_{\Lambda_{N_f}})^2\CF~.
\end{eqsp}
The contact term for the second coupling in~\eqref{eq:cont_terms_couplings} is given by~\cite{Manschot:2021qqe,Furrer:2026byd}
\begin{eqsp}
H_{N_f,j}=\frac{4\sqrt{2}\pi}{2N-N_f}\frac{\partial^2\CF}{\partial\Lambda_{N_f}\partial m_j}~,\quad j=1,\dots,N_f~.    
\end{eqsp}
\subsubsection{Sum over fluxes}\label{sec:sum_flux}
The Coulomb branch integral of the $\mr{S U}(N)$ theory with $N_f$ fundamental flavors involves a photon partition function, which is a sum over the lattice of fluxes $F^a=4 \pi \lambda^a$. Also, we have  $F^j=4 \pi \bm{k}_j$, where $\bm{k}_j$ is given by~\eqref{bfk def}. Let us first describe the lattice $\Gamma$ we need to sum over. The photon partition function for the $\mr{U}(1)^{N-1}$ low energy effective theory involves a sum over electric line bundles with a fixed choice of magnetic flux ('t Hooft flux) characterising the UV $\SU(N) / \mathbb{Z}_N$ bundle. In general, the magnetic flux is a cohomology class in $H^2(X,\Lambda_{\text{w}}/\Lambda_{\text{r}})$ where $\Lambda_{\text{w}}$ and $\Lambda_{\text{r}}$ are the weight and root lattices of the gauge group. For $\SU(N)$ we have $\Lambda_{\text{w}}/\Lambda_{\text{r}}\cong\Z_N$ so that we can consider $\mu(\mf{b})\in H^2(X,\Lambda_{\text{w}}/\Lambda_{\text{r}})$, where we use the symbol $\mu(\mf{b})$ for the 't Hooft flux to differentiate it from its representative $\mu(b)$ in the group $H^2(X,\Z_N)$. Now, one can choose a lift $\ov{\mu}(\mf{b})\in H^2(X,\Lambda_{\text{w}})$. This means that there exists classes $\pi^a \in H^2(X,\Z)$ such that
\begin{equation}\label{eq:intliftf2ma}
\ov{\mu}(\mf{b})=\sum_{a=1}^{N-1} \pi^a \vec{m}_a\in H^2(X,\Lambda_{\text{w}}) ,    
\end{equation}
where $\vec{m}_a\in\Lambda_{\text{w}}$ is a set of minimal weights of $\SU(N)$ which are in one to one correspondence with the cosets $\Lambda_{\text{w}}/\Lambda_{\text{r}}$ (see Appendix~\ref{app:lie_alg} for relevant definitions). One way to relate $\ov{\mu}(\mf{b})$ and $\mu(b)$ is to argue as in~\cite{Vafa:1994tf}. Let $E$ be the vector bundle associated to the $\SU(N)/\Z_N$-bundle $P^{\mr{gauge}}\to X$ with the fundamental representation of $\SU(N)$. We can use the splitting principle to write the vector bundle $E$ with 't Hooft flux $\mu(b)$ as
\begin{equation}\label{eq:Esplitprin}
    E=\bigoplus_{i=1}^{N}\s{L}^{a_i}~,
\end{equation}
where $\s{L}$ is a line bundle with first Chern class 
\begin{equation}
    c_1(\s{L})\equiv \mu(b)\bmod N~,
\end{equation}
and one can take, for example
\begin{equation}\label{eq:vecaweight}
    a_i=\frac{1}{N},\quad 1\leq i\leq N-1,\quad a_N=-\frac{N-1}{N}~.
\end{equation}
Then the direct sum of line bundles in~\eqref{eq:Esplitprin} is a well-defined $\SU(N)/\Z_N$-vector bundle with 't Hooft flux $\mu(b)$~\cite{Vafa:1994tf}. The instanton number of this bundle is given by 
\begin{equation}
    k(P^{\mr{gauge}})=\ov{\mu}(b)^2\left(\frac{1}{2N}-\frac{1}{2}\right)\bmod 1~,
\end{equation}
where $\ov{\mu}(b)\in H^2(X,\Z)$ is an integral lift of $\mu(b)\in H^2(X,\Z_N)$. This expression is the instanton number of any $\SU(N)/\Z_N$-vector bundle with 't Hooft flux $\mu(b)$. 
Moreover, the vector $\vec{a}=(a_1,\dots,a_{N-1},a_N)$ with 
\begin{equation}
    a_i\in \Z/N,\quad a_i-a_j\in\Z,\quad \sum_{i=1}^Na_i=0~,
\end{equation} 
represents a weight vector $\vec{a}\in \Lambda_{\rm w}$. One can replace $\vec{a}$ by a general weight vector $\vec{w}\in\Lambda_{\rm w}$, in which case the 't Hooft flux of the vector bundle~\eqref{eq:Esplitprin} is $[\vec{w}]\cdot c_1(\s{L})$, where $[\vec{w}]$ is the coset of $\vec{w}$ in $\Lambda_{\rm w}/\Lambda_{\rm r}$. This gives the representation of the 't Hooft flux $\mu(b)$ as an element $\mu(\mf{b})\in H^2(X,\Lambda_{\rm w}/\Lambda_{\rm r})$. A choice of lift of $\mu(\mf{b})$ to $\ov{\mu}(\mf{b})\in H^2(X,\Lambda_{\rm w})$ is simply $\vec{w}\cdot c_1(\s{L})$. The instanton number of the bundle is given by 
\begin{equation}
    k(P^{\mr{gauge}})=-\frac{\langle\vec{w},\vec{w}\rangle}{2}c_1(\s{L})^2\bmod 1~.
\end{equation}
Since the instanton number is independent of the choice of integral lift $\ov{\mu}(b)$, we will choose any such lift and take $c_1(\s{L})=\ov{\mu}(b)$. Thus, the representation of the 't Hooft flux in $H^2(X,\Lambda_{\rm w}/\Lambda_{\rm r})$ is given by 
\begin{equation}\label{eq:mff2Elift}
 \mu(\mf{b})=[\vec{w}]\cdot\ov{\mu}(b),\quad  \ov{\mu}(\mf{b})= \vec{w}\cdot\ov{\mu}(b)~. 
\end{equation}
Expanding the coset $[\vec{w}]$ in terms of the minimal weights $\vec{m}_a$, we can find $\pi^a$ in~\eqref{eq:intliftf2ma} in terms of $\ov{\mu}(b)$. Here we will simply use the expression~\eqref{eq:mff2Elift}.  
From~\cite{Marino:1998bm}, 
the sum over fluxes is over
\begin{equation}
\vec{\lambda}=\vec{\lambda}_{\mathbb{Z}}+\ov{\mu}(\mf{b})~,
\end{equation}
where
\begin{equation}
\vec{\lambda}_{\mathbb{Z}}=\sum_{a=1}^{N-1} \lambda_{\mathbb{Z}}^a \vec{\alpha}_a, 
\end{equation} 
where $\vec{\alpha}_a$ is a set of simple roots, $\lambda_{\mathbb{Z}}^a\in H^2(X, \mathbb{Z})$. This can be interpreted as a sum over all possible lifts of $\mu(\mf{b})\in H^2(X,\Lambda_{\text{w}}/\Lambda_{\text{r}})$. Let us expand the weight vector $\vec{w}$ in terms of the simple roots~\cite{humphreys1972introduction}:
\begin{equation}
    \vec{w}=w^a\vec{\alpha}_a,\quad w^a\in\Z/N~.
\end{equation}
Thus we can write
\begin{equation}
\vec{\lambda}=\sum_{a=1}^{N-1} \lambda^a \vec{\alpha}_a ~,
\end{equation}
where
\begin{equation}\label{eq:lambamabsum}
\lambda^a=\lambda_{\mathbb{Z}}^a+w^a\ov{\mu}(b) \in \frac{1}{N}L ~,
\end{equation}
where $L$ is the lattice introduced in section~\ref{sec:toptwist}.
For example, for the weight vector $\vec{a}$ in~\eqref{eq:vecaweight} we have 
\begin{equation}
    \vec{a}=\sum_{a=1}^{N-1}\frac{a}{N}\vec{\alpha}_a~,
\end{equation}
and the sum runs over 
\begin{equation}\label{eq:special_lambdaa}
    \lambda^a=\lambda^a_\Z+\frac{a}{N}\ov{\mu}(b)~.
\end{equation}
So the lattice over which the sum is indexed is given by $\Gamma+\vec{\beta}$ where $\Gamma=L^{N-1}$, and $\vec{\beta}$ is a vector with components $\beta^a=w^a\ov{\mu}(b)$. 
\par
To include the observables, we introduce the notation
\begin{eqsp}
    F^a_S&\coloneqq \im y_{aj}F^j_++V_aS_+~,
    \\
    \bm{z}^a_S&\coloneqq \im y_{aj}\bm{k}_{j+}+\frac{1}{4\pi}V_aS_+~.
\end{eqsp}
The sum over fluxes can then be written as
\begin{equation}
\begin{aligned}
\Psi&=\frac{(4 \pi)^{N-1} i}{\left(\text {det } y_{a b}\right)^{1 / 2}}\sum_{\lambda \in \Gamma+\vec{\beta}} \exp \left[-\frac{i}{16 \pi} \int_{X} \tau_{a b} F_{-}^a \wedge F_{-}^b-\frac{i}{16 \pi} \int_{X} \ov{\tau}_{a b} F_{+}^a \wedge F_{+}^b-\frac{i}{8 \pi} \int_{X} \tau_{a j} F_{-}^a \wedge F_{-}^j\right. \\
&\left.\hspace{3cm}-\frac{i}{8 \pi} \int_{X} \ov{\tau}_{a j} F_{+}^a \wedge F_{+}^j+\frac{\left(y^{-1}\right)_{a b} }{8 \pi} \int_{X} F_{S}^a \wedge F_{S}^b-\frac{\im}{4\pi\Lambda_{N_f}}\frac{\partial u_1}{\partial a^a}\int_X S\wedge F^a_-\right] \\
& \times\int \prod_{a=1}^{N-1} d \eta_0^a d \chi_0^a \exp \left[-\frac{i \sqrt{2}}{8 \pi} \overline{\s{F}}_{a b j} \eta_0^a \chi_0^b B\left(F_{+}^j, J\right)+\frac{1}{64 \pi} \overline{\s{F}}_{c d a}\left(y^{-1}\right)_{a b} \overline{\s{F}}_{b e f} \eta_0^c \chi_0^d \eta_0^e \chi_0^f\right. \\
& \hspace{7.75cm}\left.-\frac{i \sqrt{2}}{16 \pi} \overline{\s{F}}_{a b c} \eta_0^a \chi_0^b B\left(F_{+}^c+\im  \left(y^{-1}\right)_{c d} F_{S}^d, J\right)\right]\\
& =\frac{(4 \pi)^{N-1} i}{\left(\operatorname{det} y_{a b}\right)^{1 / 2}} \sum_{\lambda \in \Gamma+ \vec{\beta}} \exp \left[- \pi i \tau_{a b} B\left(\lambda_{-}^a, \lambda_{-}^b\right)- \pi i \ov{\tau}_{a b} B\left(\lambda_{+}^a, \lambda_{+}^b\right)-2 \pi i \tau_{a j} B\left(\lambda_{-}^a, \bm{k}_j\right)\right. \\
& \hspace{3cm}\left.-2 \pi i \ov{\tau}_{a j} B\left(\lambda_{+}^a, \bm{k}_j\right)+2 \pi \left(y^{-1}\right)_{a b} B\left(\bm{z}_{S}^a, \bm{z}_{S}^b\right)-\frac{\im}{\Lambda_{N_f}}\frac{\partial u_1}{\partial a^a}B(S,\lambda^a_-)\right] \\
& \times \int \prod_{a=1}^{N-1} d \eta_0^a d \chi_0^a \exp \left[-\frac{i}{\sqrt{2}} \overline{\s{F}}_{a b j} \eta_0^a \chi_0^b B\left(\bm{k}_j, J\right)\right. -\frac{i}{2\sqrt{2}} \overline{\s{F}}_{a b c} \eta_0^a \chi_0^b B\left(\lambda_{+}^c+\im \left(y^{-1}\right)_{c d} \bm{z}^d_S, J\right)\\
&\hspace{9.5cm}\left. +\frac{1}{64 \pi} \overline{\s{F}}_{c d a}\left(y^{-1}\right)_{a b} \overline{\s{F}}_{b e f} \eta_0^c \chi_0^d \eta_0^e \chi_0^f\right]
\end{aligned}
\end{equation}
where we used the fact that $\omega_{ \pm} \wedge \eta_{\mp}=0$. Here, each $\bm{k}_u$ for every $u=N,\dots,N+N_f-1$ is required for consistency of the topologically twisted theory as described in Section~\ref{sec:toptwist}. There is a global phase factor of
\begin{equation}
\exp \left[-i \pi B\left(\left(\vec{\lambda}-\vec{\lambda}_0\right) \cdot \vec{\rho}, w_2(X)\right)\right]
\end{equation}
which can be derived using Witten's analysis in~\cite{Witten:1995gf}. Here $\vec{\lambda}_0\in\Gamma+\vec{\beta}$ is a fixed vector and $\vec{\rho}$ is the sum of all fundamental weights of $\SU(N)$. Thus, this global phase can also be written as 
\begin{equation}\label{eq:globphasesum}
\exp \left[-i \pi \sum_{a=1}^{N-1}B\left(\lambda_\Z^a-\lambda_{\Z0}^a, w_2(X)\right)\right].    
\end{equation}
Thus the sum over fluxes takes the form
\begin{equation}\label{eq:Psisumfinal}
\begin{aligned}
\Psi=\frac{(4 \pi)^{N-1} i}{\left(\operatorname{det} y_{a b}\right)^{1 / 2}} &\sum_{\lambda \in \Gamma+\vec{\beta}} \exp \left[- \pi i \tau_{a b} B\left(\lambda_{-}^a, \lambda_{-}^b\right)- \pi i \ov{\tau}_{a b} B\left(\lambda_{+}^a, \lambda_{+}^b\right)-2 \pi i \tau_{a j} B\left(\lambda_{-}^a, \bm{k}_j\right)\right. \\
&-2 \pi i \ov{\tau}_{a j} B\left(\lambda_{+}^a, \bm{k}_j\right)+2 \pi \left(y^{-1}\right)_{a b} B\left(\bm{z}_{S}^a, \bm{z}_{S}^b\right)-\frac{\im}{\Lambda_{N_f}}\frac{\partial u_1}{\partial a^a}B(S,\lambda^a_-)\\&-i \pi B\left(\left(\lambda-\lambda_0\right) \cdot \rho, w_2(X)\right)\big{]}\times \int \prod_{a=1}^{N-1} d \eta_0^a d \chi_0^a \exp \left[-\frac{i}{\sqrt{2}} \overline{\s{F}}_{a b j} \eta_0^a \chi_0^b B\left(\bm{k}_j, J\right)\right.\\&  -\frac{i}{2\sqrt{2}} \overline{\s{F}}_{a b c} \eta_0^a 
\chi_0^b B\left(\lambda_{+}^c+\im \left(y^{-1}\right)_{c d} \bm{z}^d_S, J\right) \left.+\frac{1}{64 \pi} \overline{\s{F}}_{c d a}\left(y^{-1}\right)_{a b} \overline{\s{F}}_{b e f} \eta_0^c \chi_0^d \eta_0^e \chi_0^f\right] \text {. } \\
&
\end{aligned}
\end{equation}
We want to write it in terms of the theta function as in~\cite{Marino:1998bm}. 
Introduce the theta function
\begin{equation}
\begin{aligned}
\Theta_{\Gamma}\left(\tau_{a b}, \alpha_a, \beta^a, \xi_a\right)&=  \exp \left[- \pi i B\left(\alpha_a, \beta^a\right)+\frac{\pi}{2}\left(y^{-1}\right)^{a b}\left\{ B(\xi_{a,+},\xi_{b,+})-B(\xi_{a,-}, \xi_{b,-})\right\}\right] \\
& \times \sum_{\lambda \in \Gamma}\exp\left[- \pi i \ov{\tau}_{a b} B\left(\hat{\lambda}_{+}^a, \hat{\lambda}_{+}^b\right)- \pi i \tau_{a b} B\left(\hat{\lambda}_{-}^a, \hat{\lambda}_{-}^b\right)\right. -2 \pi i B\left(\hat{\lambda}^a, \xi_a\right)\\&\left.\hspace{8cm}+2 \pi iB\left(\hat{\lambda}^a, \alpha_a\right)\right] ,
\end{aligned}
\end{equation}
where $\hat{\lambda}^a=\lambda^a+\beta^a$.
Then taking 
\begin{equation}
\begin{aligned}
\alpha_a & =\frac{1}{2} w_2(X), \quad a=1, \ldots, N-1 \\
\xi_a & =\frac{\sqrt{2}}{8 \pi} \overline{\s{F}}_{a b c} \eta_0^b \chi_0^c J+\tau_{aj}(\bm{k}_j)_{-}+\ov{\tau}_{aj}(\bm{k}_j)_++\frac{1}{2\pi}V_aS_-~, \\
\beta^a & =w^a\ov{\mu}(b)~,
\end{aligned}
\end{equation}
and performing a tedious but straightforward computation, we see that the sum over fluxes can be written as
\begin{equation}\label{eq:Psithetafinal}
\begin{aligned}
\Psi&=\frac{(4 \pi)^{N-1} i}{\left(\mathrm{det}\, y_{a b}\right)^{1 / 2}}\exp\left[2\pi(y^{-1})_{ab}B(\widetilde{\bm{z}}^a_S,\widetilde{\bm{z}}^b_S-\ov{\tau}_{bj}(\bm{k}_j)_+)+i\pi B\left(\alpha_a, \beta^a\right)\right]\\&\times \int \prod_{a=1}^{N-1} d \eta_0^a d \chi_0^a \,\Theta_{\Gamma}\left(\tau_{a b}, \alpha_a, \beta^a, \xi_a\right) \exp \left[\frac{1}{2\sqrt{2}}(y^{-1})_{ab}\ov{\CF}_{acd}\eta_0^c\chi_0^d B\left(J,\widetilde{\bm{z}}^b_S-\frac{3}{2}\ov{\tau}_{bj}\bm{k}_j\right) \right.
\\&\left.\hspace{8cm}-\frac{i}{\sqrt{2}} \overline{\s{F}}_{a b j} \eta_0^a \chi_0^b B\left(\bm{k}_j, J\right)\right],
\end{aligned}
\end{equation}
where 
\begin{eqsp}
    \widetilde{\bm{z}}^a_S\coloneqq \frac{1}{2}\tau_{aj}\bm{k}_j+\frac{1}{4\pi}V_aS~.
\end{eqsp}
The theta function satisfies the following transformation property~\cite{Marino:1998bm,Moore:1997pc}:
\begin{eqsp}\label{eq:theta_tra1}
    \Theta_{\Gamma}&\left(-(\tau^{-1})_{ab},\alpha_a,\beta^a,-(\ov{\tau}^{-1})_{ab}\xi_{b,+}-(\tau^{-1})_{ab}\xi_{a,-}\right)\\&=\frac{1}{\sqrt{|\Gamma^\vee/\Gamma|}}\left(\text{det}\,i\ov{\tau}_{ab}\right)^{b_+/2}\left(\text{det}\,-i\tau_{ab}\right)^{b_-/2}\Theta_{\Gamma^\vee}(\tau_{ab},\beta^a,-\alpha_a,\xi_a)~,
\end{eqsp}
where $\Gamma^\vee$ is the dual lattice of $\Gamma$. If there is a characteristic vector~\eqref{characteristic vector} $w_2$ such that 
\begin{eqsp}
    B(\lambda^a,w_2)=B(\lambda^a,\lambda^a)\bmod 2~,
\end{eqsp}
then for an $(N-1)\times (N-1)$ symmetric matrix   $\theta_{ab}$ with integer entries, we have the transformation law
\begin{eqsp}\label{eq:theta_tra2}
\Theta_{\Gamma}\left(\tau_{ab}+\theta_{ab}, \alpha_a, \beta^a, \xi_a\right)=\exp \left[\frac{\pi i}{2} \sum_a B\left(w_2, \theta_{aa} \beta^a\right)\right] \Theta_{\Gamma}\left(\tau_{ab}, \alpha_a-\frac{1}{2} \theta_{aa} w_2-\theta_{ab} \beta^b, \beta^a, \xi_a\right) ~.
\end{eqsp}
Finally, we have the transformation
\begin{eqsp}\label{eq:theta_tra3}
    \Theta_{\Gamma}\left((A \tau A^t)_{ab}, \alpha_a, \beta^a, \xi^a\right)=\Theta_{\Gamma}\left(\tau_{ab}, (A^{-1} \alpha)_a, (A^t \beta)^a, (A^{-1} \xi)_a\right)~,\quad A\in \text{GL}(N-1,\Z)~.
\end{eqsp}
The transformations~\eqref{eq:theta_tra1}--\eqref{eq:theta_tra3} generate the group $\mathrm{Sp}(2N-2,\Z)$, which is the duality group of the low energy effective theory on the Coulomb branch. 
\subsubsection{Final expression for the Coulomb branch integral}
To write down the full expression for the Coulomb branch integral, we need to find the gravitational couplings. Let us start with the low energy effective theory. 
The gravitational coupling is given by~\cite{Moore:1997pc,Marino:1998bm}
\begin{equation}
A(\vec{u})^\chi B(\vec{u})^\sigma
\end{equation}
where
\begin{equation}\label{gravitational couplings}
\begin{aligned}
& A=\alpha_{N_f}\left(\operatorname{det} \frac{\partial u_a}{\partial a^b}\right)^{\frac12}~, \\
& B=\beta_{N_f} \Delta_{N_f}^{\frac18}~,
\end{aligned}
\end{equation}
and $\Delta_{N_f}$ is the physical discriminant of the SW curve. 

Since the couplings $A$ and $B$ are contained in the low-energy effective action as $\chi\log A+\sigma\log B$, both $A$ and $B$ are necessarily dimensionless. As the discriminant $\Delta_{N_f}$~\eqref{SU(N) discriminant} and the (inverse) period matrix have mass dimension
\bea
\left[\Lambda_{N_f}\right]=N(2N+N_f-2)~,
\eea
\bea
\left[\det \frac{\partial u_a}{\partial a^b}\right]=\tfrac12N(N-1)~,
\eea
we can thus write the gravitational couplings as
\bea\label{alpha beta SU(3)} 
\alpha_{N_f}&=\Lambda_{N_f}^{-\frac14N(N-1)}\alpha_{N_f,0}~, \\
\beta_{N_f}&=\Lambda_{N_f}^{-\frac18N(2N+N_f-2)}\beta_{N_f,0}~,
\eea
where $\alpha_{N_f,0}$ and $\beta_{N_f,0}$ are dimensionless complex numbers. These numbers may not depend on $N_f$ at all, or only through a phase, and will be fixed later. See also~\cite{Marino:1998bm,Manschot:2019pog,John:2022yql}.

As explained above, we will take the masses of the flavors to be the VEVs of the background $\mr{U}(1)$ vector multiplets which are ``frozen'' as described above. We thus have the following identification of couplings
\begin{equation}\label{eq:tauabuvcalFder}
\begin{aligned}
& \tau_{a b}=\frac{\partial^2 \s{F}}{\partial a^a \partial a^b}, \\
& \tau_{a j}=\frac{\partial^2 \s{F}}{\partial a^a \partial m_j}, \\
& \tau_{ij}=\frac{\partial^2 \s{F}}{\partial m_i \partial m_j}~.
\end{aligned}
\end{equation}
We further introduce
\begin{equation}\label{Cij def}
C_{ij}\coloneqq \exp \left(-\pi \im \tau_{ij}\right) \text {. }
\end{equation}
Then we have 
\begin{eqsp}\label{eq:uplintlagside}
\left\langle e^{U+\tilde{I}(S)}\right\rangle_{\mr{Coulomb}}=K\int_{\s{M}_{\rm Coulomb}}[d a d \overline{a}] &\exp\left(U+S^2\sum_{a,b=1}^{N-1}f_af_bT_{a,b}(\vec{u})+\sum_{j=1}^{N_f}H_{N_f,j}B(S,\bm{k}_j)\right)\\&\times A(\vec{u})^\chi B(\vec{u})^\sigma \prod_{i,j=1}^{N_f}C_{ij}^{B\left(\bm{k}_i, \bm{k}_j\right)} \, \Psi~,
\end{eqsp}
where $K$ is a normalization factor to be fixed later and $\s{M}_{\rm Coulomb}$ denotes the regularized Coulomb branch of the theory.
\subsection{Wall-crossing and Seiberg--Witten contributions}\label{sec:SW_cont}
In this section, we would like to determine the SW contribution from the highest codimension singular subloci of the Coulomb branch. More specifically, we will be interested in the contributions from the $\CN=1$ singular points. 
\par
Recall from section~\ref{sec:Coul_int} that the SW contribution $Z_{\SW}$ to the twisted partition function can be written as a sum over the SW contribution $Z_{\SW}^{(\ell)}$ from codimension $\ell$ singular subloci of the Coulomb branch. For the rank 1 case, it was argued in~\cite{Moore:1997pc} that the SW contributions from the codimension 1 singular subloci, which are a finite set of points, can be obtained from the wall-crossing of the $u$-plane integral. This was generalized to higher rank in~\cite{Marino:1998bm}. The wall-crossing of the higher rank Coulomb branch integral determines the SW contribution $Z_{\SW}^{(1)}$ from codimension 1 singular loci. Once $Z_{\SW}^{(1)}$ is determined, we can determine $Z_{\SW}^{(2)}$ from the wall-crossing of $Z_{\SW}^{(1)}$. We continue this process recursively to determine the highest codimension $r=N-1$ SW contribution $Z_{\SW}^{(r)}$. It is clear that this computation cannot be performed for arbitrary $r$. We follow the idea of~\cite{Marino:1998bm} by first determining the SW contributions for $N=3$ and then use the final form of the expression to generalize it to arbitrary $N$.    
\subsubsection{Wall-crossing at monopole locus}
We will closely follow the derivation of wall-crossing in~\cite{Marino:1998bm}. We first analyse the wall-crossing of~\eqref{eq:uplintlagside} at the monopole locus $a^2=0$. We will restrict to the SU(3) case first and then generalize the final result to SU$(N)$. Using the transformation~\eqref{eq:theta_tra1}, we see that the shift in the lattice vector $\lambda^2$ can be taken to be $\frac{1}{2}w_2(X)$, i.e., $\lambda^2$ parametrizes Spin${}^{c}$-structures on $X$. Once we make such a symplectic transformation, we are still allowed to perform another symplectic transformation which fixes $a^2$ but changes $a^1$. As in the pure $\text{SU}(3)$ case, the singular behaviour of coupling near $a^2=0$ is
\begin{equation}
\tau_{22}=\frac{1}{2 \pi i} \log a^2+\ldots, \quad \ov{\CF}_{222}=-\frac{1}{2 \pi i} \frac{1}{\ov{a}^2}+\cdots,
\end{equation}
and all other couplings are regular. We have
\begin{equation}
\operatorname{det} y_{a b}=y_{22}\left(y_{11}+O\left(1 / y_{22}\right)\right),
\end{equation}
and
\begin{equation}
y^{-1}_{ab}=\left(\begin{array}{ll}
y_{11}^{-1}+O\left(1 / y_{22}\right) & O\left(1 / y_{22}\right) \\
O\left(1 / y_{22}\right) & O\left(1 / y_{22}\right)
\end{array}\right) .
\end{equation}
The possible discontinuities of the Coulomb branch integral come from terms involving $\frac{1}{\sqrt{y_{22}} a^2}$ and, as we will see below, they occur when
\begin{equation}\label{eq:l2_wall_cross_cond}
\left(\lambda^2\right)^2<0, \quad \lambda_{+}^2=0~ .
\end{equation}
These are the standard conditions for wall-crossing of  $\operatorname{SW}\left(\lambda^2\right)$~\cite{Witten:1994cg}. Let us now analyse the integrand of~\eqref{eq:uplintlagside} near $a^2= 0$.
The fermionic zero-mode integral is given by
\begin{equation}\label{eq:ferm_zero_int}
\begin{aligned}
& \int \prod_{a=1}^{N-1} d \eta_0^a d \chi_0^a \exp \left[-\frac{i}{\sqrt{2}} \overline{\mathcal{F}}_{a b j} \eta_0^a \chi_0^b B\left(\boldsymbol{k}_j, J\right)-\frac{i}{2 \sqrt{2}} \overline{\mathcal{F}}_{a b c} \eta_0^a \chi_0^b B\left(\lambda_{+}^c+i\left(y^{-1}\right)_{c d} \bm{z}_S^d, J\right)\right. \\
& \left.\quad+\frac{1}{64 \pi} \overline{\mathcal{F}}_{c d a}\left(y^{-1}\right)_{a b} \overline{\mathcal{F}}_{b e f} \eta_0^c \chi_0^d \eta_0^e \chi_0^f\right] \\
& =-\frac{1}{\pi^2 2^7}\left(\ov{\CF}_{11 a} \ov{\CF}_{22 b}-\ov{\CF}_{12 a} \ov{\CF}_{12 b}\right)\left\{(4 \pi)^2 B\left(J, \lambda_{+}^a+i\left(y^{-1}\right)_{a c} \bm{z}_S^c\right) B\left(J, \lambda_{+}^b+i\left(y^{-1}\right)_{b d} \bm{z}_S^d\right)\right.\\&\hspace{12cm}-4 \pi\left(y^{-1}\right)_{a b}\Big{\}} \\
& -\frac{1}{\pi^2 2^5}(4 \pi)^2\left(\ov{\CF}_{11 i} \ov{\CF}_{22 j}-\ov{\CF}_{12 i} \ov{\CF}_{12 j}\right) B\left(J, (\bk_i)_{+}\right) B\left(J, (\bk_j)_{+}\right) .
\end{aligned}
\end{equation}
Therefore, in the $y_{22} \rightarrow 0$ limit, the fermionic zero-mode integral in~\eqref{eq:ferm_zero_int} takes the form
\begin{equation}\label{eq:ferm_0int_a20}
\begin{aligned}
& -\frac{1}{\pi^2 2^7} \ov{\CF}_{111} \ov{\CF}_{222}(4 \pi) B\left(\lambda^2, J\right)\left[4 \pi B\left(J, \lambda_{+}^1+i\left(\left.y_{11}\right|_{a^2=0}\right)^{-1} \bz_S^1\right)\right] \\
& =-\frac{1}{32 \pi} \ov{\CF}_{111} \ov{\CF}_{222} B\left(\lambda^2, J\right)\left[4 \pi B\left(J, \lambda^1\right)+4 \pi i\left(\left.y_{11}\right|_{a^2=0}\right)^{-1} B\left(J, \bz_S^1\right)\right]~ .
\end{aligned}
\end{equation}
Let us define
\begin{equation}
\ov{\tau}_{11}^{(0)}\coloneqq \ov{\tau}_{11}\left(a^2, a^2=0\right), \quad\left(y_{11}\right)_{(0)}=\frac{1}{2 i}\left(\tau_{11}-\ov{\tau}_{11}^{(0)}\right) ~.
\end{equation}
Then we can write~\eqref{eq:ferm_0int_a20} as
\begin{equation}
-\frac{1}{32 \pi} \ov{\CF}_{111} \ov{\CF}_{222} B\left(\lambda^2, J\right)\left[4 \pi B\left(J, \lambda^1\right)+4 \pi i\left(y_{11}\right)_{(0)}^{-1} B\left(J, \bz_S^1\right)\right]~.
\end{equation}
So, for a fixed $\lambda^2$ satisfying~\eqref{eq:l2_wall_cross_cond}, the integrand of~\eqref{eq:uplintlagside} in the limit $a^2 \rightarrow 0$ takes the form 
\begin{equation}
\begin{aligned}
& \exp \left(U +S^2 \sum_{a,b=1}^{N-1}f_af_bT_{a,b}(\vec{u})+\sum_{j=1}^{N_f}H_{N_f, j} B\left(S, \bk_j\right)-\im V_2 B\left(S, \lambda_{-}^2\right)\right) A^{\chi} B^\sigma \prod_{i, j} C_{i j}^{B\left(\bk_i, \bk_j\right)} \\
& \times\left(-\frac{1}{32 \pi}\right)\left(-\frac{1}{2 \pi i \ov{a}^2}\right) \frac{B\left(\lambda^2, J\right)}{\sqrt{y_{22}}} \exp \left[-\pi i \tau_{22}\left(\lambda_{-}^2\right)^2-\pi i \ov{\tau}_{22} B\left(\lambda_{+}^2\right)^2\right] \\
& \times \sum_{\lambda^1 \in L+\gamma} \exp \left[-2 \pi i \tau_{12} B\left(\lambda_{-}^1, \lambda_{-}^2\right)-2 \pi i \tau_{a j} B\left(\lambda^a_{-}, \bk_j\right)\right] \Psi\left(\lambda^1\right),
\end{aligned}
\end{equation}
where $e^{i \phi\left(\lambda^2\right)}$ is a global phase which depends on $\lambda^2$ and is obtained through the appropriate symplectic transformation to the monopole locus,
\begin{equation}
\begin{aligned}
\Psi\left(\lambda^1\right) & =\frac{(4 \pi)^2 i}{\sqrt{\left(y_{11}\right)_{(0)}}} \frac{\partial \ov{\tau}_{11}^{(0)}}{\partial \ov{a}^1} \exp \left[2 \pi\left(y_{11}\right)_{(0)}^{-1} B\left(\bz_S^1, \bz_S^1\right)\right] \\
& \times \exp \left[-\pi i \tau_{11}\left(\lambda_{-}^1\right)^2-i \pi \ov{\tau}_{11}^{(0)}\left(\lambda_{+}^1\right)^2-2 \pi i \tau_{a i} B\left(\lambda_{-}^a, k^i\right)-\im V_1 B\left(S, \lambda_{-}^1\right)-i \pi B\left(\lambda^1, \alpha^1\right)\right] \\
& \times\left[4 \pi B\left(J, \lambda^1\right)+4 \pi i\left(y_{11}\right)_{(0)}^{-1} B\left(J, \bz_S^1\right)\right]~,
\end{aligned}
\end{equation}
and $\gamma$ is the shift in $\lambda^1$ depending on the choice of symplectic basis. Note that we have only retained terms with $\lambda_{+}^2$ dependence, which are responsible for wall-crossing. For example, we removed $\exp \left[-2 \pi i \ov{\tau}_{2 j} B\left(\lambda^2_+,\bk_j\right)\right]$. We first evaluate the $a^2$-integral in the vicinity of the monopole locus. To do this, let us write the integrand as
\begin{equation}
e^{i \phi\left(\lambda^2\right)} \left(-\frac{1}{32 \pi}\right)\left(-\frac{1}{2 \pi i \ov{a}^2}\right) \frac{B\left(\lambda^2, J\right)}{\sqrt{y_{22}}} \exp \left[-\pi i \tau_{22}\left(\lambda_{-}^2\right)^2-\pi i \ov{\tau}_{22} B\left(\lambda_{+}^2\right)^2\right] \sum_{\lambda^1 \in L+\gamma} F_{\lambda^1}(a, \ov{a})
\end{equation}
where 
\begin{equation}
\begin{aligned}
F_{\lambda^1}(a, \ov{a})&=\exp \left(U +S^2 \sum_{a,b=1}^{N-1}f_af_bT_{a,b}(\vec{u})+\sum_{j=1}^{N_f}H_{N_f, j} B\left(S, \bk_j\right)-\im V_2 B\left(S, \lambda_{-}^2\right)\right) \\
& \times A^{\chi} B^\sigma \prod_{i, j} C_{i j}^{B\left(\bk_i, \bk_j\right)} \exp \left[-2 \pi i \tau_{12} B\left(\lambda_{-}^1, \lambda_{-}^2\right)-2 \pi i \tau_{a j} B\left(\lambda_{-}^a, \bk_j\right)\right] \Psi\left(\lambda^1\right) .
\end{aligned}
\end{equation}
Let $q_{I J}\coloneqq e^{2 \pi i \tau_{I J}}$. Using holomorphicity of the couplings, we can expand
\begin{equation}
F_{\lambda^1}(a, \ov{a})=\sum_{d \in \mathbb{Z}} C_{\lambda^1}\left(d, a^1, \ov{a}^{1}\right) q_{22}^d .
\end{equation}
The relevant integral is
\begin{eqsp}
\CI(J)\coloneqq \int d a^2 d \ov{a}^2\left(-\frac{1}{32 \pi}\right)\left(-\frac{1}{2 \pi i \ov{a}^2}\right) \frac{B\left(\lambda^2, J\right)}{\sqrt{y_{22}}} &\exp \left[-\pi i \tau_{22}\left(\lambda_-^2\right)^2-\pi i \ov{\tau}_{22} B\left(\lambda_{+}^2\right)^2\right]\\&\times C_{\lambda^1}\left(d, a^1, \ov{a}^2\right) q_{22}^d .
\end{eqsp}
We have
\begin{equation}
\tau_{2 2}=\frac{1}{2 \pi i} \log a^2+\ldots \implies \frac{\partial \tau_{22}}{\partial a^2}=2 \pi i q_{22}~.
\end{equation}
Writing $\tau_{22}=x+i y$ and using $d \tau_{22} \wedge d \ov{\tau}_{22}=2 i d y\wedge d x$, we get
\begin{equation}\label{eq:CI_wall_cross_cal}
\begin{aligned}
\CI(J) & =-\frac{2 i}{32 \pi} 2 \pi i C_{\lambda^1}\left(d, a^{1}, \ov{a}^{1}\right) \int d y d x\, \frac{1}{\sqrt{y}} e^{\left.-i \pi x\left(\lambda_{-}^2\right)^2+\left(\lambda_{+}^2\right)^2-d+1\right)} e^{-\pi y\left(\left(\lambda_{+}^2\right)^2-\left(\lambda_{-}^2\right)^2+d+1\right)} B\left(\lambda_{+}^2, J\right) \\
& =\frac{1}{8} C_{\lambda^1}\left(d, a^{1}, \ov{a}^{1}\right) \delta_{d,\frac{(\lambda^2)^2}{2}-1} \int_0^{\infty} d y\, e^{-2 \pi y\left(\lambda_{+}^2\right)^2} B\left(\lambda_{+}^2, J\right) \\
& =\frac{1}{8} C_{\lambda^1}\left(\frac{(\lambda^2)^2}{2}-1, a^2, \ov{a}^2\right) \frac{\sqrt{\pi}}{\sqrt{2 \pi\left(\lambda_{+}^2\right)^2}} B\left(\lambda_{+}^2, J\right) \\
& =\frac{1}{8 \sqrt{2}} C_{\lambda^1}\left(\frac{(\lambda^2)^2}{2}-1, a_{,}^1 \ov{a}^2\right) \frac{B\left(\lambda_{+}^2, J\right)}{\left|B\left(\lambda_{+}^2, J\right)\right|}~ .
\end{aligned}
\end{equation}
So the discontinuity is $\CI(J)$ at $B\left(\lambda_{+}^2, J\right)=0$ is
\begin{equation}
\begin{aligned}
\CI\left(J^{+}\right)-\CI\left(J^{-}\right) & =\frac{\sqrt{2}}{8} C_{\lambda^1}\left(\frac{(\lambda^2)^2}{2}-1, a^1, \ov{a}^1\right) \\
& =\frac{\sqrt{2}}{8} \underset{a^2=0}{\operatorname{Res}} \, q_{22}^{-\frac{(\lambda^2)^2}{2}} F_{\lambda^1}(a, \ov{a})~.
\end{aligned}
\end{equation}
We thus get
\begin{equation}\label{eq:WC1}
\begin{aligned}
\text{WC}\left(\lambda^2\right)=\frac{\sqrt{2}}{8} e^{i \phi\left(\lambda^2\right)} &\int_{\mathcal{D}} d a^2 d \ov{a}^{2} \underset{a^2=0}{\operatorname{Res}}\left[A^{\chi} B^\sigma \prod_{i,j} C_{ij}^{B\left(\bk_i, \bk_j\right)} \sum_{\lambda^1 \in L+\gamma} q_{22}^{-\left(\lambda^2\right)^2 / 2} q_{12}^{-B\left(\lambda^1, \lambda^2\right)} \Psi\left(\lambda^1\right)\right. \\
& \left.\exp \left(U +S^2 \sum_{a,b=1}^{N-1}f_af_bT_{a,b}(\vec{u})+\sum_{j=1}^{N_f}H_{N_f, j} B\left(S, \bk_j\right)-\im V_2 B\left(S, \lambda^2\right)\right)\right]~,
\end{aligned}
\end{equation}
where we replaced $\lambda_{-}^2$ by $\lambda^2$ since $\lambda_{+}^2=0$.
\subsubsection{Wall-crossing of wall-crossing}
We now compute the ``wall-crossing of wall-crossing'' at the $\mathcal{N} = 1$ points. As in~\cite{Marino:1998bm}, near the $\mathcal{N}=1$ points, we have $a^1 \rightarrow 0$. The shift $\gamma$ for $\lambda^1$ is again $\alpha^1=\frac{1}{2} w_2(X)$ and hence $\lambda^1$ also parametrizes spin${}^c$ structures. For $a^1 \rightarrow 0$, $\tau_{11}$ has the behaviour
\begin{equation}
\tau_{11}=\frac{1}{2 \pi i} \log a^1+\ldots,\quad  \frac{\partial \ov{\tau}_{11}^{(0)}}{\partial \ov{a}^2}=-\frac{1}{2 \pi i} \frac{1}{\ov{a}^1}+\dots,
\end{equation}
and hence there is wall-crossing for $\text{WC}\left(\lambda^2\right)$. The wall-crossing of $\text{WC}\left(\lambda^2\right)$ is again for
\begin{equation}\label{eq:l1_wall_cross_cond}
\left(\lambda^2\right)^2<0, \quad \lambda_{+}^1=0 ~.
\end{equation}
The calculation of $\text{WC}\left(\text{WC}\left(\lambda^2\right)\right)\left(\lambda^1\right)$ is similar to the calculation of $\text{WC}\left(\lambda^2\right)$.
For a fixed $\lambda^1$ satisfying~\eqref{eq:l1_wall_cross_cond}, we want to find wall-crossing of the integral
\begin{eqsp}
\frac{\sqrt{2}}{8} e^{i \phi\left(\lambda^{1}, \lambda^2\right)} e^{2 \pi i B\left(\lambda^a, \lambda_0^a\right)}(4 \pi)^3 i \int_{\mathcal{D}} d a^1 d \ov{a}^2\frac{B\left(J, \lambda^1\right)}{\sqrt{\left(y_{11}\right)_{(0)}}} \frac{\partial \ov{\tau}_{11}^{(0)}}{\partial \ov{a}^1} \exp \left[-\pi i \tau_{11}^{(0)}\left(\lambda^1_-\right)^2-i \pi \ov{\tau}_{11}^{(0)}\left(\lambda^1_+\right)^2\right]  \\
\times \underset{a^2=0}{\operatorname{Res}}\left[\exp \left(U +S^2 \sum_{a,b=1}^{N-1}f_af_bT_{a,b}(\vec{u})+\sum_{j=1}^{N_f}H_{N_f, j} B\left(S, \bk_j\right)-\im \sum_{a=1}^{N-1}V_a B\left(S, \lambda^a\right)\right)\right. \\
\left.A^{\chi} B^\sigma \prod_{i, j} C_{i j}^{B\left(\bk_i, \bk_j\right)} q_{22}^{-\left(\lambda^2\right)^2 / 2} q_{12}^{-B\left(\lambda^1, \lambda^2\right)} \prod_{a, i} q_{a i}^{-B\left(\lambda^a_-, \bk_i\right)}\right],
\end{eqsp}
where $e^{i \phi\left(\lambda^1, \lambda^2\right)}$ is a phase obtained from the appropriate symplectic transformation and $e^{2 \pi i B\left(\lambda^a, \lambda_0^a\right)}$ generalizes a similar phase in~\cite{Moore:1997pc,Marino:1998bm}.
We have kept the terms in $\Psi\left(\lambda^1\right)$ which survive the limit $a^2 \rightarrow 0$, in particular we have put $\tau_{11}^{(0)}=\tau_{11}\left(a^1=a^2=0\right)$. At the end we will sum over all the $\mathcal{N}=1$ points. Let us again expand 
\begin{eqsp}
G(a, \ov{a})= \, & \underset{a^2=0}{\operatorname{Res}}\left[\exp \left(U +S^2 \sum_{a,b=1}^{N-1}f_af_bT_{a,b}(\vec{u})+\sum_{j=1}^{N_f}H_{N_f, j} B\left(S, \bk_j\right)-\im \sum_{a=1}^{N-1}V_a B\left(S, \lambda^a\right)\right)\right. \\
& \left.A^{\chi} B^\sigma \prod_{i, j} C_{i j}^{B\left(\bk_i, \bk_j\right)} q_{22}^{-\left(\lambda^2\right)^2 / 2} q_{12}^{-B\left(\lambda^1, \lambda^2\right)} \prod_{a, i} q_{a i}^{-B\left(\lambda^a_-, \bk_i\right)}\right]\\
= \, & \sum_{d \in \mathbb{Z}} D(d) q_{11}^d .
\end{eqsp}
Using the same computation as in~\eqref{eq:CI_wall_cross_cal}, we obtain
\begin{equation}
\begin{aligned}
\CJ(J): & =\int_{\mathcal{D}} d a^1 d \ov{a}^1 \frac{1}{\sqrt{y_{11}}} \frac{-1}{2 \pi i a^1} \exp \left[-\pi i \tau_{11}\left(\lambda_{-}^1\right)^2-i \pi \ov{\tau}_{11}\left(\lambda_{+}^1\right)^2\right] B\left(J, \lambda^1\right) q_{11}^d \\
& =-32 \pi D\left(\frac{\left(\lambda^1\right)^2}{2} -1\right) \frac{B\left(\lambda_{+}^1, J\right)}{\left|B\left(\lambda_{+}^1, J\right)\right|}~.
\end{aligned}
\end{equation}
So the discontinuity is $\CJ(J)$ at $B\left(\lambda_{+}^1, J\right)=0$ is
\begin{equation}
\begin{aligned}
\CJ\left(J^{+}\right)-\CJ\left(J^{-}\right) & =-32 \pi D\left(\frac{\left(\lambda^1\right)^2}{2} -1\right) \\
& =-32 \pi \underset{a^1=0}{\operatorname{Res}} q_{11}^{-\left(\lambda^1\right)^2 / 2} G(a, \ov{a})~.
\end{aligned}
\end{equation}
We thus have 
\begin{equation}\label{eq:WC2}
\begin{aligned}
& \text{WC}\left(\text{WC}\left(\lambda^2\right)\right)\left(\lambda^1\right) \\
& =-\sqrt{2} i(4 \pi)^4 e^{i \phi\left(\lambda^2, \lambda^2\right)} e^{2 \pi i B\left(\lambda^a, \lambda_0^a\right)} \\
&\quad \times \underset{a^1, a^2=0}{\operatorname{Res}}\left[A^{\chi} B^\sigma \prod_{i, j} C_{i j}^{B\left(\bk_i, \bk_j\right)} \prod_{a, b} q_{a b}^{-\frac{1}{2} B\left(\lambda^a, \lambda^b\right)} \prod_{a, i} q_{a i}^{-B\left(\lambda^a, \bk_i\right)}\right] \\
&\quad \times \left. \exp \left(U +S^2 \sum_{a,b=1}^{N-1}f_af_bT_{a,b}(\vec{u})+\sum_{j=1}^{N_f}H_{N_f, j} B\left(S, \bk_j\right)-\im \sum_{a=1}^{N-1}V_a B\left(S, \lambda^a\right)\right)\right]
\end{aligned}
\end{equation}
We now compute the SW contribution by comparing to the wall-crossing formula derived above.
For $\SU(3)$, there are three $\mathcal{N}=1$ points. On general grounds~\cite{Marino:1998bm,Moore:1997pc}, we have
\begin{equation}
\begin{aligned}
Z_{\text{D}} & =Z_{\text {Coulomb}}+\sum_n \int_{\mathcal{D}_n^{(1)}} d a_n^1 d \ov{a}_n^1 \sum_{\lambda^1, \lambda^2} \int_{\CM_{\text{SW}}\left(\lambda^2\right)} \mu_{\lambda^1, \lambda^2}^n\left(a_n^1, \ov{a}_n^1, a^2\right) \\
& +\sum_{n=1}^3 \sum_{\lambda^1, \lambda^2} \int_{\CM_{\text{SW}}\left(\lambda^1\right) \times \CM_{\text{SW}}\left(\lambda^2\right)} \Phi_n\left(a^1, a^2\right)~.
\end{aligned}
\end{equation}
The functions $\mu_{\lambda^1, \lambda^2}^n\left(a_n^1, \ov{a}_n^1, a^2\right)$ and $\Phi_n\left(a^1, a^2\right)$ can be obtained from~\eqref{eq:WC1} and~\eqref{eq:WC2} respectively as in~\cite{Moore:1997pc,Marino:1998bm}.
For $b_2^{+}>1$, the contributions only from isolated singularities are nonvanishing. As argued in section~\ref{sec:Coul_int}, in the large equal mass limit, only the $\mathcal{N}=1$ singularities contribute. So we will only be interested in $\Phi_{n}\left(a^1, a^2\right)$. As discussed in~\cite{Marino:1998bm}, comparing with~\eqref{eq:WC2} the SW contribution from an $\CN=1$ singularity has the form 
\begin{equation}
\begin{aligned}
\Phi_n\left(a^1, a^2\right)= & K_n e^{i \phi_n\left(\lambda^1, \lambda^2\right)} e^{2 \pi i B\left(\lambda^a, \lambda_0^a\right)} \widetilde{A}^{\chi / 4} \widetilde{B}^{\sigma / 8} \prod_{i, j} C_{i j}^{B\left(\bk_i, \bk_j\right)} \prod_{a, b} \widetilde{q}_{a b}^{\frac{1}{2} B\left(\lambda^a, \lambda^b\right)} \prod_{a, i} q_{a i}^{-B\left(\lambda^a, \bk_i\right)} \\
& \times \exp \left(U +S^2 \sum_{a,b=1}^{N-1}f_af_bT_{a,b}(\vec{u})+\sum_{j=1}^{N_f}H_{N_f, j} B\left(S, \bk_j\right)-\im \sum_{a=1}^{2}V_a B\left(S, \lambda^a\right)\right),
\end{aligned}
\end{equation}
where $K_n$ is a constant which will be fixed by comparing to the conjecture of~\cite{Gottsche:2020ass} and
\begin{equation}
\begin{aligned}
& \widetilde{q}_{a a}=\frac{a^a}{q_{a a}}, \quad a=1,2, \\
& \widetilde{q}_{a b}=q_{a b}^{-1}, \quad a, b=1,2,  a \neq b, \\
& \widetilde{A}=\alpha_{N_f}^4\left(\operatorname{det} \frac{\partial u_a}{\partial a^b}\right)^2, \\
& \widetilde{B}=\beta_{N_f}^8 \frac{\Delta_{N_f}}{a^1 a^2} .
\end{aligned}
\end{equation}
The extra factors of $a^a$ come from the coupling
\begin{equation}
\sum_{a=1}^{N-1} \int_X c(\vec{u}) F^a \wedge F^a
\end{equation}
in the Lagrangian at the SW points, see~\cite{Moore:1997pc} for details. Let us rewrite $\Phi_n(a^1,a^2)$ as 
\begin{eqsp}
\Phi_n\left(a^1, a^2\right)&=  K_n e^{i \phi_n\left(\lambda^1, \lambda^2\right)} e^{2 \pi i B\left(\lambda^a, \lambda_0^a\right)}\prod_{a=1}^2(a^a)^{d_{\lambda^a}/2-d_{\lambda^a}/2}\widetilde{q}_{aa}^{(\lambda^a)^2/2}
\\
&\times
\widetilde{A}^{\chi / 4} \widetilde{B}^{\sigma / 8} \prod_{i, j} C_{i j}^{B\left(\bk_i, \bk_j\right)} \prod_{1\leq a< b\leq 2} q_{a b}^{-B\left(\lambda^a, \lambda^b\right)} \prod_{a, i} q_{a i}^{-B\left(\lambda^a, \bk_i\right)} \\
& \times \exp \left(U +S^2 \sum_{a,b=1}^{N-1}f_af_bT_{a,b}(\vec{u})+\sum_{j=1}^{N_f}H_{N_f, j} B\left(S, \bk_j\right)-\im \sum_{a=1}^{2}V_a B\left(S, \lambda^a\right)\right),    
\end{eqsp}
where 
\begin{equation}
d_{\lambda^a}=\left(\lambda^a\right)^2-\frac{2 \chi+3 \sigma}{4}~.
\end{equation}
Using the fact that
\begin{equation}
\int_{\CM_{\text{SW}}\left(\lambda^a\right)}\left(a^a\right)^{d_{\lambda^a} / 2}=\operatorname{SW}\left(\lambda^a\right),
\end{equation}
we can write the SW contribution at the singularity $n$ for the $\text{SU}(3)$ SQCD as 
\begin{equation}
\begin{aligned}
& \left\langle e^{U +\tilde{I}(S)}\right\rangle_{\lambda^1,\lambda^{2} }^{(n)} \\
& =K_n \alpha_{N_f}^{\chi} \beta_{N_f}^\sigma e^{i \phi_n(\lambda_1^1,  \lambda^{2})} e^{2 \pi i B\left(\lambda^a, \lambda_0^a\right)}\left(\prod_{a=1}^{2} \text{SW}\left(\lambda^a\right)\right) \underset{{a^1,a^{2}=0}}{\operatorname{Res}}\left[\prod_{a=1}^{2}\left(a^a\right)^{-\frac{d_{\lambda^a}}{2}-1}\widetilde{q}_{a a}^{\left(\lambda^a\right)^2 / 2}\right. \\
&\times  \prod_{1 \leq a<b \leq 2} q_{a b}^{-B\left(\lambda^a, \lambda^b\right)}\left(\frac{\Delta_{N_f}}{\prod_{a=1}^{2} a^a}\right)^{\sigma/8} \prod_{i, j=1}^{N_f} C_{i j}^{B\left(\bk_i, \bk_j\right)} \prod_{a=1}^{2} \prod_{i=1}^{N_f} q_{a i}^{-B\left(\lambda^a, \bk_i\right)} \\
&\times \left.\left(\operatorname{det} \frac{\partial u_a}{\partial a^b}\right)^{\chi / 2} \exp \left(U +S^2 \sum_{a,b=1}^{2}f_af_bT_{a,b}(\vec{u})+\sum_{j=1}^{N_f}H_{N_f, j} B\left(S, \bk_j\right)-\im \sum_{a=1}^{2}V_a B\left(S, \lambda^a\right)\right)\right]~.
\end{aligned}
\end{equation}
This formula is amenable to generalization to $\SU(N)$, we simply replace the range of $a,b$ from $1,2$ to $1,2,\dots,N-1$. We get 
\begin{equation}
\begin{aligned}
& \left\langle e^{U +\tilde{I}(S)}\right\rangle_{\lambda^1,\ldots, \lambda^{N-1} }^{(n)} \\
& =K_n \alpha_{N_f}^{\chi} \beta_{N_f}^\sigma e^{i \phi_n(\lambda_1^1, \ldots, \lambda^{N-1})} e^{2 \pi i B\left(\lambda^a, \lambda_0^a\right)}\left(\prod_{a=1}^{N-1} \text{SW}\left(\lambda^a\right)\right) \underset{a^1,\dots,a^{N-1}=0}{\operatorname{Res}}\left[\prod_{a=1}^{N-1}\left(a^a\right)^{-\frac{d_{\lambda^a}}{2}-1}\widetilde{q}_{a a}^{\left(\lambda^a\right)^2 / 2}\right. \\
&\times  \prod_{1 \leq a<b \leq N-1} q_{a b}^{-B\left(\lambda^a, \lambda^b\right)}\left(\frac{\Delta_{N_f}}{\prod_{a=1}^{N-1} a^a}\right)^{\sigma/8} \prod_{i, j=1}^{N_f} C_{i j}^{B\left(\bk_i, \bk_j\right)} \prod_{a=1}^{N-1} \prod_{i=1}^{N_f} q_{a i}^{-B\left(\lambda^a, \bk_i\right)} \\
&\times \left.\left(\operatorname{det} \frac{\partial u_a}{\partial a^b}\right)^{\chi / 2} \exp \left(U +S^2 \sum_{a,b=1}^{N-1}f_af_bT_{a,b}(\vec{u})+\sum_{j=1}^{N_f}H_{N_f, j} B\left(S, \bk_j\right)-\im \sum_{a=1}^{N-1}V_a B\left(S, \lambda^a\right)\right)\right]~.
\end{aligned}
\end{equation}
This generalises the pure $\SU(N)$ result~\cite[(9.6)]{Marino:1998bm} to include matter hypermultiplets and the couplings to background fluxes. 

The residue can be evaluated for 4-manifolds $X$ of Seiberg--Witten simple type, where $\SW(\lambda)$ is only nonzero if the dimension $d_\lambda$ of the SW moduli space vanishes, 
\begin{eqsp}\label{eq:sw_simple_type}
    d_\lambda=(\lambda)^2-\frac{2\chi+3\sigma}{4}=0~.
\end{eqsp}
In this case, since 
\begin{eqsp}
    \widetilde{q}_{a a}^{\left(\lambda^a\right)^2 / 2}\prod_{1 \leq a<b \leq N-1} q_{a b}^{-B\left(\lambda^a, \lambda^b\right)}\left(\frac{\Delta_{N_f}}{\prod_{a=1}^{N-1} a^a}\right)^{\chi / 2}~
\end{eqsp}
is regular at the $\CN=1$ points, the residue can be evaluated to 
\begin{equation}\label{eq:SW_cont_gen}
\begin{aligned}
& \left\langle e^{U +\tilde{I}(S) }\right\rangle_{\lambda^1,\ldots, \lambda^{N-1} }^{(n)} \\
& =K_n \alpha^{\chi} \beta^\sigma e^{i \phi_n(\lambda^1, \ldots, \lambda^{N-1})} e^{2 \pi i B\left(\lambda^a, \lambda_0^a\right)}\left[\left(\prod_{a=1}^{N-1} \text{SW}\left(\lambda^a\right)\widetilde{q}_{a a}^{\left(\lambda^a\right)^2 / 2}\right) \right.\\
&\times  \prod_{1 \leq a<b \leq N-1} q_{a b}^{-B\left(\lambda^a, \lambda^b\right)}\left(\frac{\Delta_{N_f}}{\prod_{a=1}^{N-1} a^a}\right)^{\sigma / 8} \prod_{i, j=1}^{N_f} C_{i j}^{B\left(\bk_i, \bk_j\right)} \prod_{a=1}^{N-1} \prod_{i=1}^{N_f} q_{a i}^{-B\left(\lambda^a, \bk_i\right)} \\
&\times \left.\left(\operatorname{det} \frac{\partial u_a}{\partial a^b}\right)^{\chi / 2} \exp \left(U +S^2 \sum_{a,b=1}^{N-1}f_af_bT_{a,b}(\vec{u})+\sum_{j=1}^{N_f}H_{N_f, j} B\left(S, \bk_j\right)-\im \sum_{a=1}^{N-1}V_a B\left(S, \lambda^a\right)\right)\right]^{(n)}~,
\end{aligned}
\end{equation}
where $n=0,\dots,N-1$ labels the $\CN=1$ points and the superscript $[\cdots]^{(n)}$ indicates that the expression inside the square brackets is to be evaluated at the $\CN=1$ point $n$. This result generalizes the pure $\SU(N)$ formula~\cite[(9.17)]{Marino:1998bm} for simple type to twisted QCD. 
\par It will be useful to write the answer for the case when 
\begin{eqsp}
    U=pu_1/\Lambda^2_{N_f}~,\quad \tilde{I}(S)=\frac{1}{\Lambda_{N_f}}\tilde{I}^1(S)~.
\end{eqsp}
In terms of the contact term~\eqref{eq:contact_G}, we get 
\begin{equation}\label{eq:SW_cont_GK_comp}
\begin{aligned}
& \left\langle e^{pu_1/\Lambda^2_{N_f} +\tilde{I}^1(S) / \Lambda_{N_f}}\right\rangle_{\lambda^1,\ldots, \lambda^{N-1} }^{(n)} \\
& =K_n \alpha^{\chi} \beta^\sigma e^{i \phi_n(\lambda^1, \ldots, \lambda^{N-1})} e^{2 \pi i B\left(\lambda^a, \lambda_0^a\right)}\left[\left(\prod_{a=1}^{N-1} \text{SW}\left(\lambda^a\right)\widetilde{q}_{a a}^{\left(\lambda^a\right)^2 / 2}\right) \right.\\
&\times  \prod_{1 \leq a<b \leq N-1} q_{a b}^{-B\left(\lambda^a, \lambda^b\right)}\left(\frac{\Delta_{N_f}}{\prod_{a=1}^{N-1} a^a}\right)^{\sigma / 8} \prod_{i, j=1}^{N_f} C_{i j}^{B\left(\bk_i, \bk_j\right)} \prod_{a=1}^{N-1} \prod_{i=1}^{N_f} q_{a i}^{-B\left(\lambda^a, \bk_i\right)} \\
&\times \left.\left(\operatorname{det} \frac{\partial u_a}{\partial a^b}\right)^{\chi / 2} \exp \left(pu_1/\Lambda^2_{N_f} +S^2 G_{N_f}+\sum_{j=1}^{N_f}H_{N_f, j} B\left(S, \bk_j\right)-\frac{if_1}{\Lambda_{N_f}} \frac{\partial u_1}{\partial a^a} B\left(S, \lambda^a\right)\right)\right]^{(n)}~.
\end{aligned}
\end{equation}

\sss{Consistency check: SU(2).}
Let us check that this expression returns the correct answer for the rank 1 case. The couplings to the $\spinc$ structures and observables rather trivially reduce to the known quantities for $\SU(2)$. The main check is to compute the couplings to $\chi$ and $\sigma$, or equivalently $\chih$ and $K^2$ (see section~\ref{sec:toptwist} for notations). In the $N=2$ case, there are only two independent couplings $\tilde\tm_{N_f}$ and $\tilde\tx_{N_f}$, which are defined as~\cite{Moore:1997pc,Furrer:2026byd}
\bea\label{su(2) couplings}
u_{\rm D}(\tauD)&=\tu_{N_f}+\tilde\tm_{N_f} \qD+\CO(\qD^2)~, \\
\tilde\tx_{N_f}&= (\tfrac{da}{du})_{\rm D}(\tauD)\Big|_{\qD^0}~,
\eea
in a local $\qD$ series around the cusp $\tau_{\rm D}=i\infty$. For this, we define the couplings more appropriate to $\SU(N)$~\eqref{eq:SW_couplings}:
\bea\label{SU(2) from SU(N)}
\tm_{N_f}=  \left.\frac{\Delta_{N_f}}{a} \right|_{\tu_{N_f}}~, \qquad
\tn_{N_f}=  \left.\frac{\Delta_{N_f}}{\qD} \right|_{\tu_{N_f}}~, \qquad
\tx_{N_f}=  \left.\frac{du}{da} \right|_{\tu_{N_f}}~,
\eea
through which we will express the full $\SU(N)$, and determine these for $N=2$.
Clearly, $\tilde \tx_{N_f}=\tx_{N_f}^{-1}$.
To obtain the other two couplings, we make use of the identity
\bea
\Delta_{N_f}\left(\frac{da}{du}\right)^{12}=c_{N_f}\eta^{24}~,
\eea
which, according to Theorem~\ref{thm:discriminant} in appendix~\ref{subsec:hyperelliptic-discriminants}, exists for any rank 1 model. Here, $\eta$ is the Dedekind eta function~\eqref{dedekind eta}, and $c_{N_f}$ is a normalisation factor that has been explicitly determined in~\cite{Aspman:2021vhs}. Taking the modular $S$-transformation of this identity and expanding it to first order in $q_{\rm D}$, using $\Delta_{N_f}=(u-\tu_{N_f})\Delta'(\tu_{N_f})+\dots$, we find $\Delta'(\tu_{N_f})=c_{N_f}\tilde \tm_{N_f}^{-1}\tilde\tx_{N_f}^{-12}$.

The first two $\SU(2)$ quantities in~\eqref{SU(2) from SU(N)} are obtained as first-order degenerations of these functions, where both numerator and denominator vanish at the singular point. We evaluate them using l'Hôpital's rule, such that
\bea
\tm_{N_f}&=  \left.\frac{\partial\Delta_{N_f}}{\partial a} \right|_{\tu_{N_f}}=\left.\frac{\partial u}{\partial a}\Delta'(\tu_{N_f}) \right|_{\tu_{N_f}}=c_{N_f}\tilde \tm_{N_f}^{-1}\tilde\tx_{N_f}^{-13}~, \\
\tn_{N_f}&=  \left.\frac{\Delta_{N_f}}{\qD} \right|_{\tu_{N_f}}=\left.\frac{\partial u}{\partial q_{\rm D}}\Delta'(\tu_{N_f}) \right|_{\tu_{N_f}}=c_{N_f}\tilde\tx_{N_f}^{-12}~,
\eea
where we have used the definition~\eqref{su(2) couplings} of $\tilde\tm_{N_f}$. Then the couplings $f_{\chih}$ and $f_{K^2}$ to $\chih$ and $K^2$ become
\bea
f_{\chih}&=\Lambda_{N_f}^{-1}\alpha_{N_f}^{12}\beta_{N_f}^{-8}\tx_{N_f}^6\tm_{N_f}^{-1} =\Lambda_{N_f}^{-1}\alpha_{N_f}^{12}\beta_{N_f}^{-8}
c_{N_f}^{-1}\tilde\tm_{N_f}\tilde\tx_{N_f}^{7}~, \\
f_{K^2}&=\alpha_{N_f}^{-1}\beta_{N_f}\tx_{N_f}^{-\frac12}\tn_{N_f}^{\frac18}=\alpha_{N_f}^{-1}\beta_{N_f}c_{N_f}^{\frac18}\tilde\tx_{N_f}^{-1}~.
\eea
These agree precisely with the contribution found for $\SU(2)$~\cite[(2.36)]{Furrer:2026byd}, accounting for a slight redefinition of the gravitational couplings $\alpha_{N_f}$ and $\beta_{N_f}$ due to a change in conventions. Moreover, it was checked in \emph{loc.cit.} that it agrees perfectly with Moore--Witten~\cite[(11.28)]{Moore:1997pc}.

\subsection{The blowup formula and contact terms}\label{sec:blowup}
In this section, we use the blowup formula to derive physical couplings.\footnote{See also~\cite{Edelstein:1999xk,Edelstein:2000wg,Edelstein:2000aj,Edelstein:2000jc,Edelstein:1998sp,Edelstein:1999fz,Edelstein:1999tb,Fintushel1996, Gttsche2024,Nakajima:2003pg,Nakajima:2005fg,Nakajima2011,Kim:2019uqw,Bonelli:2024wha,Bonelli:2025juz,Bonelli:2025bmt} for a selection of important results on the blowup equation.}

\sss{The blowup function in the UV.}
We will denote the blowup of $X$ by $\widetilde{X}$. Let $C^{\prime}$ be the exceptional divisor of the blowup satisfying $C^{\prime 2}=-1$. It is orthogonal to all curves pulled back from $X$ by the blowdown map $\phi: \widetilde{X} \rightarrow X$. The topological quantities of $X$ and $\widetilde{X}$ are related as
\begin{equation}
\begin{aligned}
b_2(\widetilde{X}) & =b_2(X)+1, \\
\chi(\widetilde{X}) & =\chi(X)+1, \\
\sigma(\widetilde{X}) & =\sigma(X)-1, \\
K_{\widetilde{X}} & =\phi^* K_X+C^{\prime} .
\end{aligned}
\end{equation}
We denote the Poincaré dual of $C^{\prime}$ also by $C^{\prime}$. In the chamber $C_{+}^{\prime}=0$, we can choose a basis of 
\begin{eqsp}
H^2(\widetilde{X})=H^2(X)\oplus C'\mathbb{R}~,    
\end{eqsp}
such that
\begin{eqsp}
\widetilde{\lambda}_{\mathbb{Z}+}^a&=\left(\lambda_{\mathbb{Z}+}^a, 0\right), \quad \widetilde{\lambda}_{\mathbb{Z}-}^a=\left(\lambda_{\mathbb{Z}-}^a, n^a\right), \\
\widetilde{\beta}^a&=\left(\beta^a, \beta^{\prime a}\right), \quad \widetilde{\bk}_j=\left(\bk_j, k^{\prime}_j\right)~,
\end{eqsp}
where 
\begin{eqsp}
n^a \in \mathbb{Z}, \beta^{\prime a} \in \mathbb{Z} / N,\quad  k_j^{\prime} \in \begin{cases}
\mathbb{Z} / \operatorname{lcm}(2, N),&\beta^{\prime a}\neq 0,
\\
\Z/2,&\beta^{\prime a}=0.
\end{cases}
\end{eqsp}
Here, $\beta^{\prime a}$ depends on the gerbe connection of the gauge bundle $\widetilde{P}^{\mr{gauge}}\rightarrow \widetilde{X}$, which satisfies
\begin{equation}
\int_X c_2\left(P^{\mr{gauge}}\right)=\int_{\widetilde{X}} c_2(\widetilde{P}^{\mr{gauge}})~ .
\end{equation}
The blowup formula provides non-trivial results particularly when the virtual dimensions~\eqref{vd MQ} of the nonabelian monopole moduli space $\CM^{\CQ,N_f}_{v(E)}$ for $X$ corresponding to the vector bundle $E$ associated to $P^{\mr{gauge}}\to X$ and $\CM^{\CQ,N_f}_{v(\widetilde{E})}$  for $\widetilde{X}$ corresponding to the vector bundle $\widetilde{E}$ associated to $\widetilde{P}^{\mr{gauge}}\to \widetilde{X}$ are equal.
The difference of the virtual dimensions  is given by
\begin{eqsp}\label{eq:diff_vd_mon}
    \vd \CM^{\CQ,N_f}_{v(E)}-\vd \CM^{\CQ,N_f}_{v(\widetilde{E})}=-(2N-N_f)\sum_{a=1}^{N-1}(\beta^{\prime a})^2-\frac{NN_f}{8}+\frac{N}{2}\sum_{j=1}^{N_f}k_j^{\prime 2}~.
\end{eqsp}
The simplest choice of $\beta^{\prime a},k_j'$ such that the difference vanishes\footnote{For $\beta'^a$ non-zero, the solutions for the $k_j'$ are more complicated for $N\geq 3$. For $N=2$, a simple choice for  $\beta'^a\neq 0$ was used in~\cite{Furrer:2026byd}.}
\begin{eqsp}\label{eq:choice_ba_kj_blowup}
 \beta'^a&=0~,\quad  a=1,\dots, N-1~,
 \\
 k_j'&=\pm\frac12~,\quad j=1,\dots, N_f~.
\end{eqsp}
We can write~\cite{Marino:1998bm} 
\begin{eqsp}
    \beta^{\prime a}=m^{a}_be^b~,\quad e^b\in\Z~,
\end{eqsp}
where $\vec{m}^a$ are the minimal weights of $\mf{su}(N)$.
We also deform the surface $S$ to 
\begin{eqsp}
S^{\prime}=S+t C^{\prime},\quad t \in \mathbb{R}~.    
\end{eqsp}
Then the blowup formula relates the correlation functions on $X$ and $\widetilde{X}$ and schematically has the form 
\begin{eqsp}
    \left\langle e^{U+\tilde{I}(S')}\right\rangle_{\widetilde{X},\{\widetilde{\beta}^{a}\},\{\widetilde{\bk}_{j}\}}=\left\langle e^{U+\tilde{I}(S)}\CR\left(\bm{\tau}, t, \{\beta^{\prime a}\},\left\{k_j^{\prime}\right\}\right)\right\rangle_{X,\{\beta^{a}\},\{\bk^{j}\}}~,
\end{eqsp}
where $\CR(\bm{\tau}, t, \{\beta^{\prime a}\},\{k_j^{\prime}\})$ is the blowup function or the blowup factor. 

\sss{The blowup function in the IR.}
The blowup function can be determined in the infrared using the Coulomb branch integral~\cite{Moore:1997pc}. We begin by noting that in the chamber $C_{+}^{\prime}=0$, the Siegel-Narain theta function factorises as
\begin{eqsp}\label{Psi blowup}
\widetilde{\Psi} & =\sum_{\vec{n} \in \mathbb{Z}^{N-1}} \exp \left[\pi i \tau_{a b}\left(n^a+\beta^{\prime a}\right)\left(n^b+\beta^{\prime b}\right)+2 \pi i \tau_{a j}\left(n^a+\beta^{\prime a}\right) k^{\prime j}\right.\\&\left.\hspace{2.5cm}+ \frac{\im t_a}{\Lambda_{N_f}^a}\frac{\partial u_a}{\partial a^b} \left(n^b+\beta^{\prime b}\right)-i \pi \sum_{a=1}^{N-1}\left(n^a+\beta^{\prime a}\right)\right] \Psi~,
\end{eqsp}
where we introduced the notation 
\begin{eqsp}
    t_a\coloneqq tf_a~.
\end{eqsp}
We can write the relation between $\widetilde{\Psi}$ and $\Psi$ in terms of a Riemann theta function $\Theta$:
for $\vec{\gamma}, \vec{\delta} \in \mathbb{R}^{N-1}$ and $\vec{z} \in \mathbb{C}^{N-1}$, define
\begin{equation}\label{eq:def_Riemann_theta}
\Theta[\vec{\gamma}, \vec{\delta}](\vec{z} \mid \bm{\tau})\coloneqq \sum_{\vec{n} \in \mathbb{Z}^{N-1}} \exp \left[i \pi \tau_{a b}\left(n^a+\delta^a\right)\left(n^b+\delta^b\right)+2 \pi i\left(n^a+\delta^a\right)\left(z^a+\gamma^a\right)\right]~.
\end{equation}
Then taking
\begin{equation}
\begin{aligned}
& \delta^a=\beta^{\prime a}, \quad \vec{\gamma}=\vec{\Delta}=-\frac{1}{2}(1,1, \ldots, 1), \\
& \zeta^a=\frac{1}{2 \pi}\sum_{b=1}^{N-1} \frac{t_b}{\Lambda_{N_f}^b}\frac{\partial u_b}{\partial a^a}+\tau_{a j} \bk^{\prime }_j .
\end{aligned}
\end{equation}
we can write~\eqref{Psi blowup} as
\begin{eqsp}
  \widetilde{\Psi}=  \Theta\left[\vec{\Delta}, \vec{\beta}^{\prime}\right](\vec{\zeta} \mid \bm{\tau}) \Psi~.
\end{eqsp}
Consequently, the blowup factor is given by
\begin{equation}\label{eq:blow_func_IR}
\CR\left(\bm{\tau}, t, \{\beta^{\prime a}\},\left\{k_j^{\prime}\right\}\right)=A B^{-1} \prod_{i,j=1}^{N_f} C_{ij}^{-k^{\prime}_i k^{\prime}_j} \exp \left(-\sum_{a,b=1}^{N-1}t_at_b T_{a,b}(\vec{u})-t k^{\prime}_j H_{N_f, j}\right) \Theta\left[\vec{\Delta}, \vec{\beta}^{\prime}\right](\vec{\zeta} \mid \bm{\tau}) .
\end{equation}
This generalises the $\SU(2)$ result~\cite[(2.52)]{Furrer:2026byd} to all $N$.
We rewrite this equation as 
\begin{eqsp}\label{eq:ABR=theta}
    \log\Theta[\vec{\Delta},\vec{\beta}'](\vec{\zeta}|\bm{\tau})-\log\left(A^{-1} B \prod_{i,j=1}^{N_f} C_{ij}^{k^{\prime}_i k^{\prime}_j}\right)&-\sum_{a,b=1}^{N-1}t_at_b T_{a,b}(\vec{u})-t k^{\prime}_j H_{N_f, j}\\&-\log\CR\left(\bm{\tau}, t, \{\beta^{\prime a}\},\left\{k_j^{\prime}\right\}\right)=0~.
\end{eqsp}
Let us substitute the Taylor series in $t$ of the theta function and the blowup function:
\begin{eqsp}
    \log\Theta[\vec{\Delta},\vec{\beta}'](\vec{\zeta}|\bm{\tau})&=\sum_{k\geq 0}\frac{t^k}{k!(2\pi)^k}\sum_{\substack{a_1,\dots,a_k\\b_1,\dots,b_k}}\left(\prod_{i=1}^{k}\frac{f_{a_i}}{\Lambda_{N_f}^{a_i}}\frac{\partial u_{a_i}}{\partial a^{b_i}}\right)\Theta_{\log}^{(b_1,\dots,b_k)} [\vec{\Delta},\vec{\beta}'](\vec{\zeta}|\bm{\tau})~,
    \\
\log\CR\left(\bm{\tau}, t, \{\beta^{\prime a}\},\left\{k_j^{\prime}\right\}\right)&=\sum_{k\geq 0}\frac{t^k}{k!}\CR^{\log}_{k}\left(\bm{\tau}, 0, \{\beta^{\prime a}\},\left\{k_j^{\prime}\right\}\right)~,
\end{eqsp}
where we introduced logarithmic derivatives:
\bea\label{logarithmic theta derivative}
    \Theta^{(a_1,\ldots,a_k)}_{\log}[\vec\Delta,0](\vec z\mid\bm\tau)\coloneqq \frac{\partial}{\partial z^{a_1}}\cdots \frac{\partial}{\partial z^{a_k}}\log\Theta[\vec\Delta,0](\vec z\mid\bm\tau)~,
\eea
and the coefficients $\CR^{\log}_{k}$ can be determined from the derivatives of the blowup function. We now put $\beta^{\prime a}=0$ in accordance with~\eqref{eq:choice_ba_kj_blowup}.
Comparing the coefficient of $t^0$ on both sides of~\eqref{eq:ABR=theta} gives 
\begin{eqsp}\label{eq:ABC=theta_kj}
    A^{-1}B\prod_{i,j=1}^{N_f}C_{ij}^{k_i'k_j'}=\Theta[\vec{\Delta},0](\tau_{aj}k_j'|\bm{\tau})~.
\end{eqsp}
Choosing $k_j'=\frac{1}{2}$ gives
\begin{equation}\label{eq:ABCjk=theta}
A^{-1} B \prod_{i,j=1}^{N_f} C_{ij}^{\frac{1}{4}}=\Theta[\vec{\Delta}, 0]\left(\left.\left\{\frac{\vec{\tau}}{2} \right\} \right\rvert\, \bm{\tau}\right)~,
\end{equation}
where $\vec{\tau}=(\tau^1,\dots,\tau^{N-1})$ with
\begin{eqsp}
    \tau^a\coloneqq \sum_{j=1}^{N_f}\tau_{aj}~.
\end{eqsp}
Comparing the $t^1$ coefficient of both sides of~\eqref{eq:ABR=theta} with $k_j'=\frac{1}{2}$, we get
\begin{equation}
\sum_{j=1}^{N_f} H_{N_f, j}=\frac{1}{\pi} \sum_{a,b=1}^{N-1}\frac{f_b}{\Lambda_{N_f}^b} \frac{\partial u_b}{\partial a^a}  \Theta^{(a)}_{\log}[\vec{\Delta}, 0]\left(\left.\left\{\frac{\vec{\tau}}{2}\right\} \right\rvert\, \bm{\tau}\right)+\CR^{\log}_1(\bm{\tau},0,\{0\},\{1/2\})~.
\end{equation}
\sss{Contact terms from blowup.}
As a consequence of a theorem by Nakajima and Yoshioka~\cite{Nakajima2011}, our blowup function~\eqref{eq:blow_func_IR} is expected to have a gap:
\bea\label{eq:blowup_exp_t3}
\CR\left(\bm{\tau}, t,\{0\},\left\{\tfrac{1}{2}\right\}\right)=1+\CO(t^n)~ , \quad 
    n=\max(N+1,2N-N_f)~.
\eea
This implies that
\begin{eqsp}\label{eq:R_klog=0}
\CR^{\log}_k(\bm{\tau},0,\{0\},\{1/2\})=0~,\quad k\leq n-1~.
\end{eqsp}
In particular, this gives us
\begin{equation}\label{eq:HNf_gen}
\sum_{j=1}^{N_f} H_{N_f, j}=\frac{1}{\pi} \sum_{a,b=1}^{N-1}\frac{f_b}{\Lambda_{N_f}^b} \frac{\partial u_b}{\partial a^a}  \Theta^{(a)}_{\log}[\vec{\Delta}, 0]\left(\left.\left\{\frac{\vec{\tau}}{2}\right\} \right\rvert\, \bm{\tau}\right)~.
\end{equation}
For the case when the surface observable is $\tilde{I}(S)=f_1\tilde{I}^1(S)/\Lambda_{N_f}$, we have
\begin{eqsp}\label{eq:HNfj=theta}
 \sum_{j=1}^{N_f} H_{N_f, j}=\frac{f_1}{\pi\Lambda_{N_f}} \sum_{a=1}^{N-1} \frac{\partial u_1}{\partial a^a}  \Theta^{(a)}_{\log}[\vec{\Delta}, 0]\left(\left.\left\{\frac{\vec{\tau}}{2}\right\} \right\rvert\, \bm{\tau}\right)~.   
\end{eqsp}
Comparing the coefficient of $t^2$ with $k_j'=\frac{1}{2}$ we get 
\begin{eqsp}\label{eq:Tabsum_ab}
    \sum_{a,b=1}^{N-1}f_af_bT_{a,b}(\vec{u})&=\frac{1}{8\pi^2}\sum_{a,b,c,d=1}^{N-1}\frac{f_af_b}{\Lambda_{N_f}^{a+b}}\frac{\partial u_a}{\partial a^c}\frac{\partial u_b}{\partial a^d}\Theta^{(c, d)}_{\log}[\vec{\Delta}, 0]\left(\left.\left\{\frac{\vec{\tau}}{2}\right\} \right\rvert\, \bm{\tau}\right)+\frac{1}{2}\CR^{\log}_2(\bm{\tau},0,\{0\},\{1/2\})~.
\end{eqsp}
Using~\eqref{eq:R_klog=0} for $k=2$ and 
comparing the coefficients of $f_a,f_b$ in~\eqref{eq:Tabsum_ab} gives 
\begin{eqsp}
\label{eq:T_a,b_theta}
    T_{a,b}(\vec{u})&=\frac{1}{8\pi^2\Lambda_{N_f}^{a+b}}\sum_{c,d=1}^{N-1}\frac{\partial u_a}{\partial a^c}\frac{\partial u_b}{\partial a^d}\Theta^{(c, d)}_{\log}[\vec{\Delta}, 0]\left(\left.\left\{\frac{\vec{\tau}}{2}\right\} \right\rvert\, \bm{\tau}\right)
\end{eqsp}
Using the identity 
\begin{eqsp} \Theta^{(a,b)}[\vec{\gamma},\vec{\delta} ](\vec{z} \mid \bm{\tau})=(4\pi i)\partial_{\tau_{ab}}\log\Theta[\vec{\gamma},\vec{\delta} ](\vec{z} \mid \bm{\tau}) ~,  
\end{eqsp}
we see that~\eqref{eq:T_a,b_theta} agrees with~\cite[Equation 10.98]{Laba05} once we restore the factor of the scale $\Lambda_{N_f}$.\footnote{see also~\cite{Edelstein:1998sp,Edelstein:1999fz,Edelstein:1999tb,Edelstein:1999xk,Edelstein:2000aj,Edelstein:2000jc,Edelstein:2000wg}}  
In the special case when $\tilde{I}(S)=\tilde{I}^1(S)/\Lambda_{N_f}$, we have
\begin{eqsp}
 T_{1,1}(\vec{u})=G_{N_f} & =\frac{1}{2}\left(\frac{1}{2 \pi \Lambda_{N_f}}\right)^2 \frac{\partial u_1}{\partial a^a} \frac{\partial u_1}{\partial a^b} \Theta^{(a, b)}_{\log}[\vec{\Delta}, 0]\left(\left.\left\{\frac{\vec{\tau}}{2}\right\} \right\rvert\, \bm{\tau}\right)~,
\end{eqsp}
or equivalently
\begin{eqsp}\label{eq:GNf=theta}
  8\left(\pi \Lambda_{N_f}\right)^2 G_{N_f}=\frac{\partial u_1}{\partial a^a} \frac{\partial u_1}{\partial a^b} \Theta^{(a, b)}_{\log}[\vec{\Delta}, 0]\left(\left.\left\{\frac{\vec{\tau}}{2}\right\} \right\rvert\, \bm{\tau}\right)~.
\end{eqsp}
Equations~\eqref{eq:HNfj=theta} and~\eqref{eq:GNf=theta} generalise the previous $\SU(2)$ result~\cite[(2.58)]{Furrer:2026byd} to all $N$.
We note that 
\eqref{eq:ABR=theta} along with~\eqref{eq:R_klog=0} for $3\leq k\leq n-1$ gives us theta function identities:
\begin{eqsp}\label{eq:theta_id_blowup}
\sum_{\substack{a_1,\dots,a_k\\b_1,\dots,b_k}}\left(\prod_{i=1}^{k}\frac{f_{a_i}}{\Lambda_{N_f}^{a_i}}\frac{\partial u_{a_i}}{\partial a^{b_i}}\right)\Theta_{\log}^{(b_1,\dots,b_k)} [\vec{\Delta},\vec{\beta}']\left(\left.\left\{\frac{\vec{\tau}}{2}\right\} \right\rvert\, \bm{\tau}\right)=0~,\quad 3\leq k\leq n-1~. 
\end{eqsp}
While the relations~\eqref{eq:HNfj=theta} and~\eqref{eq:GNf=theta} give us consistent results (see section~\ref{sec:SU(3) calculation} below), we have checked for $N=3,N_f\leq 5$ that the identities~\eqref{eq:theta_id_blowup} do \emph{not} hold at the $\CN=1$ singularities on the Coulomb branch unless $f_a=0$.

\sss{Contact terms for equal masses.}
Specializing to equal masses, we see that the couplings $C_{j k}$ become $N_f$ equal diagonal couplings $C_{j j}$ and $N_f\left(N_f-1\right)$ equal off-diagonal couplings $C_{j k}, j \neq k$. The contact terms $H_{N_f, j}$ are all equal, which we now denote by $H_{N_f}$, and the couplings $\tau_{a j}$ are also equal, which we denote by $v^a$. We thus have
\bea\label{eq:C11C12=theta}
 A^{-1} B C_{11}^{\frac{N_f}{4}} C_{12}^{\frac{N_f(N_f-1)}{4}}&=\Theta[\vec{\Delta}, 0]\left(\left.\left\{\frac{N_f}{2} v_a\right\} \right\rvert\, \bm{\tau}\right), \\
 H_{N_f}&=\frac{f_1}{\pi \Lambda_{N_f} N_f} \frac{\partial u_1}{\partial a^a} \Theta_{\log}^{(a)}[\vec{\Delta}, 0]\left(\left.\left\{\frac{N_f}{2} v_a\right\} \right\rvert\, \bm{\tau}\right), \\
 G_{N_f}&=\frac{1}{2}\left(\frac{1}{2 \pi \Lambda_{N_f}}\right)^2 \frac{\partial u_1}{\partial a^a} \frac{\partial u_1}{\partial a^b} \Theta^{(a, b)}_{\log}[\vec{\Delta}, 0]\left(\left.\left\{\frac{N_f}{2} v_a\right\} \right\rvert\, \bm{\tau}\right) ~ .
\eea
To solve for the coupling $C_{12}$, we will need one more equation involving $C_{11}$ and $C_{12}$. To get one more equation, we use ~\eqref{eq:ABC=theta_kj} with the choice $k^{\prime}_j=\frac{1}{2}$ for $1 \leq j \leq \ell, k^{\prime}_j=-\frac{1}{2}$ for $\ell+1 \leq j \leq N_f$. We obtain 
\begin{equation}\label{eq:C11C12l=theta}
A^{-1} B C_{11}^{\frac{N_f}{4}} C_{12}^{\frac{(2 \ell-N_f)^2-N_f}{4}}=\Theta[\vec{\Delta}, 0]\left(\left.\left\{\frac{N_f-2 \ell}{2} v_a\right\} \right\rvert\, \bm{\tau}\right)~.
\end{equation}
From~\eqref{eq:C11C12=theta} and~\eqref{eq:C11C12l=theta} we have
\begin{equation}
C_{12}^{\frac{N_f(N_f-1)}{4}-\frac{(2 \ell-N_f)^2-N_f}{4}}=\frac{\Theta[\vec{\Delta}, 0]\left(\left.\left\{\frac{N_f}{2} v_a\right\} \right\rvert\, \bm{\tau}\right)}{\Theta[\vec{\Delta}, 0]\left(\left.\left\{\frac{\left(N_f-2 \ell\right)}{2} v_a\right\} \right\rvert\, \bm{\tau}\right)}~,
\end{equation}
or equivalently
\begin{equation}\label{eq:C12=theta}
C_{12}=\frac{\Theta[\vec{\Delta}, 0]\left(\left.\left\{\frac{N_f}{2} v_a\right\} \right\rvert\, \bm{\tau}\right)^{\frac{1}{\ell(N_f-\ell)}}}{\Theta[\vec{\Delta}, 0]\left(\left.\left\{\frac{\left(N_f-2 \ell\right)}{2} v_a\right\} \right\rvert\, \bm{\tau}\right)^{\frac{1}{\ell(N_f-\ell)}}} .
\end{equation}
Note that for $N_f=1$, the coupling $C_{12}$ does not exist. We have
\begin{equation}\label{Nf=1 blowup 1st}
N_f=1: \quad A^{-1} B C_{11}^{\frac{1}{4}}=\Theta[\Delta, 0]\left(\left.\left\{\frac{1}{2} v_a\right\} \right\rvert\, \tau\right)
\end{equation}
To solve for $C_{11}$, we will need to determine $A^{-1}B$. We determine $A^{-1}B$ for $N=3$ in terms of the Igusa cusp form and the discriminant of the SW curve in~\eqref{delta det = chi10}. 
\section{Topological correlators and Segre numbers}
\label{sec:top_cor_seg_num}
With the topological correlators set up, we explain in this section the relation to the virtual Segre numbers. We conjecture a precise dictionary between universal functions and Coulomb branch couplings, the sum over the $\CN=1$ vacua and the generating function of Segre numbers, and expand on several properties of the universal functions using the physical symmetries. The bulk of the evidence for the conjectures~\ref{conj:DZSW0=Ua} and~\ref{conj:DZSWi=GU} will be demonstrated in Section~\ref{sec:SU(3) calculation}.
\subsection{Segre numbers and universal functions}
\label{sec: segre numbers}
In this section, we briefly review the definition of virtual Segre numbers and their conjectural generating function in terms of universal functions. We follow the previous work~\cite{Gottsche:2020ass,Gottsche:2019meh,Gttsche2024} by L. Göttsche and M. Kool for the definitions and notations.

\sss{Virtual Segre numbers.}
Throughout this section, we denote by  $S$ a smooth projective surface over $\mathbb C$. We will assume that $b_1(S)=0$ and the geometric genus $p_g(S)$ is positive. A necessary condition for the geometric genus to be non-zero is $b_2^+(S)>1$. 
% The geometric genus counts the number of independent holomorphic 2-forms on $S$, which is non-zero if  $b_2^+(S)>1$. 
% Let $K(S)$ be the  Grothendieck group of coherent sheaves on $S$.  is the free abelian group generated by isomorphism classes of coherent sheaves on $S$, with relations coming from short exact sequences.
One can impose stability conditions on sheaves on $S$  determined by a polarisation $H$ on $S$. The relevant stability condition is the Gieseker $H$-semi-stability. 

Fix an integer $\rho\in\mathbb Z_{>0}$ and integer classes $c_1\in H^2(S,\mathbb Z)$ and $c_2\in H^4(S,\mathbb Z)$. We then consider the moduli space of rank $\rho$ Gieseker $H$-semistable torsion-free sheaves on $S$ with Chern classes $c_1$, $c_2$. We will use the notation $M\coloneqq M_S^H(\rho,c_1,c_2)$ for this moduli space. We assume that $M$ does not contain strictly semistable sheaves with respect to $H$. 
A standard computation gives the (complex) virtual dimension of $M$: 
\begin{equation} \label{vdim M}
    \vd M=2\rho c_2-(\rho-1)c_1^2-(\rho^2-1)\chih~.
\end{equation}
The moduli space $M$ in physics literature is the 
moduli space of $\SU(N)$ instantons, where the dictionary reads
\bea
\rho=N~, \quad c_2=k+\frac{N-1}{2N}c_1^2~,
\eea
where $k=k(P)$ is the instanton number defined in~\eqref{instanton number}.
As in Donaldson theory, one defines Chern classes on $M$ using a universal sheaf $\mathbb E$ on $S\times M$. To a class $\alpha \in K(S)$, one can associate a tautological class $\alpha_M$ on the moduli space $M$. Its Chern character is
\bea \label{eq:chalphaM main}
{\rm ch}(\alpha_M)={\rm ch}(-\pi_{M!}(\pi^*_S\alpha\cdot \mathbb E \cdot {\rm det}(\mathbb E)^{-\frac 1\rho}))~,
\eea
where $\pi_S\colon S\times M\to S$ and $\pi_M\colon S\times M\to M$ are projections. The Chern character is valued in the Chow ring $A^*(M)_{\mathbb Q}$ with rational coefficients.

Let $c_i(\alpha_M)$ denote the Chern classes obtained from the Chern character $\ch(\alpha_M)$. We write the total Chern class as 
\begin{eqsp}
c(\alpha_M)=\sum_i c_i(\alpha_M)~.    
\end{eqsp}
Let $M^{[\rm vir]}\in H_{2\vd M}(M)$ denote the virtual fundamental class constructed by T. Mochizuki~\cite{Mochizuki2009}.  Virtual Segre numbers are then defined as the integrals of $c(\alpha_M)$ over $M^{[\rm vir]}$:
\begin{eqsp}
    \segre_\alpha\coloneqq \int_{M^{[\rm vir]}}c(\alpha_M)~.
\end{eqsp}
As in Donaldson theory, we can
include the point and surface classes. Let $\point\in H^4(S,\mathbb Z)$ denote the Poincaré dual of the point class, and $p'$ be a formal variable. To define the surface observable, we choose $L\in\mathrm{Pic}(S)$, and introduce the formal variable $x'$. Let $\mud:H^{\ell}(S,\Z)\to H_{4-\ell}(M,\mathbb{Q})$ be the Donaldson map. 
Then the virtual Segre number of $M$ with fixed $(\rho,c_1,c_2)$ is defined by
\begin{equation}\label{virtual Segre def}
    \segre_\alpha[p',x']\coloneqq\int_{M^{[\rm vir]}}c(\alpha_M)  \exp(\mud(\point)p'+\mud(L)x')~.
\end{equation}
We define the generating function of virtual Segre numbers by summing over $c_2$ in $\segre_\alpha[p',x']$ with fixed rank $\rho$ and first Chern class $c_1$: 
\bea\label{segre gen}
\segregen_\segre[p',x']=\sum_{c_2}z^{\frac{1}{2} \vd M}\segre_\alpha[p',x']~.
\eea
For $\alpha=0$, the virtual Segre numbers $\segre_{\alpha}[p',x']$ reduce to the ${\rm SU}(\rho)$ Donaldson invariants of $S$. In this case, the Göttsche--Kool conjecture below reduces to an algebraic version of the Mariño--Moore conjecture~\cite{Marino:1998bm}. 
Virtual Segre numbers also generalize ordinary Segre numbers: for $\rho=1$ and $c_1=0$, the moduli space $M=M_S^H(1,0,n)\cong S^{[n]}$ becomes the Hilbert scheme of $n$ points on $S$. The total Chern class is given by   $c(\alpha_M)=c(\alpha^{[n]})$ where $\alpha^{[n]}\in K(S^{[n]})$ is the tautological class on $S^{[n]}$. Then the virtual Segre numbers of $S$ reduce to the ordinary Segre numbers.\footnote{\label{footnote Segre} To be precise, ordinary \textit{Segre numbers} are the  coefficients of the generating series $\sum_{n=0}^\infty z^n \int_{S^{[n]}} c(\alpha^{[n]})\in \mathbb Q[[z]]$, 
where $c$ is the total Chern class. For $\alpha=-V$ , where $V$ is the class of a vector bundle, $c(\alpha^{[n]})=s(V^{[n]})$, where $s$ denotes the total Segre class. }

For a class $a\in H^2(S,\Z)$, let $\sw(a)$ denote the Seiberg--Witten invariants in the algebraic geometry convention, e.g. in Mochizuki~\cite{Mochizuki2009}. Recall that there are finitely many SW basic classes, i.e., classes $a\in H^2(S,\Z)$  such that $\sw(a)\neq 0$. SW basic classes satisfy $a^2=aK$, where $K$ is the canonical class of $S$. For a class $\lambda\in H^2(S,\Z)$, we will denote by $\SW(\lambda)$ the Seiberg--Witten invariant in the differential geometry and physics convention. The simple type condition reads $\lambda^2=\frac14K^2$. The two conventions are related by 
$a=\tfrac12K+\lambda$, such that 
(see \eg~\cite[Proposition 1.4.5]{Mochizuki2009})
\bea\label{SW shift}
\sw(a)= \SW(\lambda)~.
\eea
Set $\varepsilon_\rho=\exp(2\pi i /\rho)$. We now state the Göttsche--Kool conjecture~\cite[Conjecture 2.8]{Gottsche:2020ass} in full generality.
\begin{samepage}
\begin{conjecture}[Göttsche--Kool]\label{conj_GK}
Let $M$ be as above, and $s\in\mathbb Z$. There exist universal functions
\bea\label{Cz Cz12}
Q_s, R_s, T_s, V_s, W_s, X_s  &\in \mathbb C[[z]]~, \\
S_s, S_{j,s}, Y_s, Y_{j,s}, Z_s,  Z_{jk,s} &\in \mathbb C[[z^{\frac{1}{2}}]]~,
\eea
for all $1\leq j\leq k\leq \rho-1$, such that for any $\alpha\in K(S)$ with $\rk(\alpha)=s$ we have 
{\normalfont
\bea\label{GK conjecture}
\segre_\alpha[p',x']=\universal_\alpha[p',x'] \Big|_{z^{\frac12\vd M}}~,
\eea
}
where 
{\normalfont
\bea\label{universal series}
\universal_\alpha[p',x']& \coloneqq \rho^{2-\chih+K^2}V_s^{c_2(\alpha)}W_s^{c_1(\alpha)^2}X_s^{\chih} Y_s^{c_1(\alpha)K}Z_s^{K^2}e^{x'^2 Q_s + c_1(\alpha)x' R_s + x'K S_{s}  +  p'\, T_s} \\
&\quad \times \sum_{(a_1,\dots, a_{\rho-1})}\prod_{j=1}^{\rho-1}\varepsilon_\rho^{j a_j c_1}\sw(a_j) Y_{j,s}^{c_1(\alpha)a_j}e^{a_j x' S_{j,s}} \!\!\!\!\!\prod_{1\leq j\leq k\leq \rho-1}Z_{jk,s}^{a_ja_k}~,
\eea
}
and 
\bea\label{changeofvar}
z = t (1+(1-\tfrac{s}{\rho}) t)^{1-\frac{s}{\rho}}~.
\eea
Here, 
\bea\label{Vs Ws Xs Qs Rs Ts} 
V_s(t) &= (1+(1-\tfrac{s}{\rho})t)^{1-s} (1+(2-\tfrac{s}{\rho})t)^s (1+(1-\tfrac{s}{\rho})t)^{\rho-1} , \\ 
W_s(t) &= (1+(1-\tfrac{s}{\rho})t)^{\frac{1}{2}s-1} (1+(2-\tfrac{s}{\rho})t)^{\frac{1}{2}(1-s)} (1+(1-\tfrac{s}{\rho})t)^{\frac{1}{2} - \frac{1}{2} \rho}, \\
X_s(t) &=  (1+(1-\tfrac{s}{\rho})t)^{\frac{1}{2} s^2-s} (1+(2-\tfrac{s}{\rho})t)^{-\frac{1}{2}s^2+\frac{1}{2}} (1+(1-\tfrac{s}{\rho})(2-\tfrac{s}{\rho})t)^{-\frac{1}{2}} (1+(1-\tfrac{s}{\rho})t)^{-\frac{(\rho-1)^2}{2\rho} s}, \\
Q_s(t) &= \tfrac{1}{2}t(1+(1-\tfrac{s}{\rho}) t), \quad R_s(t) = t, \quad T_s(t) = \rho t (1+ \tfrac{1}{2}(1-\tfrac{s}{\rho})(2-\tfrac{s}{\rho})t),
\eea
and moreover $Y_s$, $Y_{j,s}$,  $Z_s$, $Z_{jk,s}$, $S_s$, $S_{j,s}$ are all algebraic functions.
\end{conjecture}
\end{samepage}
Let us introduce the generating function 
\begin{eqsp}\label{eq:uni_series}
    \CG_{\universal}[p',x']\coloneqq \left.\universal_{\alpha}[p',x']\right|_{z^{\left\{\frac{1}{2}\vd M|c_2\in\Z\right\}}}~,
\end{eqsp}
where the notation on the right hand side of~\eqref{eq:uni_series} means that we expand the universal series~\eqref{universal series} as a $z$-series and only retain terms of the form $z^{m}$ with $m=\frac{1}{2}\vd M$ for some value of $c_2\in\Z$. Then Conjecture~\ref{conj_GK} is equivalent to the statement that 
\begin{eqsp}\label{eq:GK_conj_gen_ser}
    \CG_{\universal}[p',x']=\CG_{\segre}[p',x']~,
\end{eqsp}
with $\CG_\segre$ defined in~\eqref{segre gen}.

\subsection{The dictionary}
In this section, we compare the virtual Segre series with the SQCD partition function. 

\sss{The K-theory class $\alpha$.} Let us unpack the formula~\eqref{universal series}.
To relate the Segre numbers to intersection numbers on the instanton moduli space, we have to identify the $K$-theory class $\alpha$ with the corresponding bundles over $X$. Since $\alpha$ is fixed in the computation and has rank $s=N_f$, it must be the bundle corresponding to the background gauge fields in the twisted theory. The background field in the twisted theory is the $\mr{U(1)}^{N_f}$-connection for the flavor group. This leads to the proposal
\bea\label{alpha}
\alpha\cong K_X^{\frac12}\otimes \bigoplus_{j=1}^{s} \CL_j~,
\eea
where $K_X$ is the canonical bundle of $X$, and $\oplus_{j}\CL_j$ is the rank $s$ bundle associated to the $\mr{U(1)}^{N_f}$-flavor bundle $P^{\mr{f}}$ defined in~\eqref{eq:p-star-bundles}. Note that $\alpha$ is in general not a well-defined bundle; nevertheless, we can work in terms of its Chern data. To determine the Chern classes, it is useful to write $\alpha$ as a sum over line bundles $\alpha=\oplus_{j=1}^s L_j$ where 
$L_j=K_X^{\frac12}\otimes \CL_j$, with $c_1(L_j)=\frac12 K+\bfk_j$. 
% since $2\bfk_j=c_1(W_j^+\otimes E)$. 
By the splitting principle, we have that
$c_1(\alpha)=\sum_j c_1(L_i)$ and $c_2(\alpha)=\sum_{i<j} c_1(L_i)c_1(L_j)$, such that we have the identifications
\begin{eqsp}\label{eq:c12alpha}
c_1(\alpha) &= \frac{s}{2}K + (\boldsymbol{k}) \;, \\
c_2(\alpha) &= \frac{s(s-1)}{8}K^2 + \frac{s-1}{2}K(\boldsymbol{k}) + \llbracket \boldsymbol{k}^2 \rrbracket \;. 
\end{eqsp}
These agree with the ones found for $\rho=N=2$ in~\cite{Furrer:2026byd}, and are in particular independent of the rank $\rho=N$.

\sss{The tautological bundle $\alpha_M$.}
Given the class $\alpha$ on the four-manifold, the tautological class $\alpha_M$~\eqref{eq:chalphaM main} is essentially the algebro-geometric analogue of the Dirac index bundle on the instanton moduli space. 
The universal sheaf $\mathbb E\to S\times M$ has the defining property that for a point $E$ in $M$, its restriction to the corresponding fibre is the sheaf represented by $E$. While a class $\alpha\in K(S)$ describes a fixed bundle on $S$, the class $\mathbb E$ then describes the family of gauge bundles parametrised by $M$.
The product $\pi_S^* \alpha\cdot \mathbb E\in K(S\times M)$ is therefore described the family of gauge bundles $E$ twisted by $\alpha$.
It is convenient to use the determinant-normalised universal sheaf
\bea\label{E0 sheaf}
\mathbb E_0\coloneqq\mathbb E\cdot\det(\mathbb E)^{-\frac1\rho}~.
\eea
Since $\det(\mathbb E_0)=\CO$ is the trivial sheaf and thus $c_1(\mathbb E_0)=0$, the determinant factor extracts the $\SU(\rho)$ part of the sheaf $\mathbb E$ by removing the trace or ${\rm U}(1)$ part of the ${\rm U}(\rho)$ sheaf, similar to how a ${\rm U}(N)$ vector bundle has structure group $\SU(N)$ if its first Chern class vanishes. When the root $\det(\mathbb E)^{-\frac1\rho}$ does not exist, the Chern character is defined by a formal application of the Grothendieck--Riemann--Roch formula (see also~\cite[Remark 2.5]{Gottsche:2020ass}). We thus wish to derive~\eqref{eq:chalphaM main}
\bea\label{eq:chalphaM e0}
\alpha_M=-\pi_{M!}(\pi_S^*\alpha\cdot \mathbb E_0)
\eea
from our choice of $\alpha$~\eqref{alpha}.

To determine $\alpha_M$ from $\alpha$, consider $N_f=1$ for simplicity. Since $\alpha=\CL_j\times K_S^{\frac12}$ is a line bundle with $c_1(K_S)=-c_1(S)$, we have $\ch(\alpha)=e^{c_1(\alpha)}=e^{\frac12(c(\mathfrak s)-c_1(S))}$. To compute the Chern character of the index bundle $W=\Ind\slashed D$, we use the family index theorem~\cite{Atiyah:1970ws},
\bea
\ch(\Ind\slashed D)=\pi_{M*}\left(\ch(\mathbb E_0)e^{\frac12c(\mathfrak s)}\widehat A(S)\right)~.
\eea
Rather than the index itself, we consider $V= \Ind \slashed D^\dagger= -\Ind \slashed D$, as in~\eqref{index bundle}. Using the fundamental relationship between the Todd class and the $\widehat A$ class, $\Td(S)=e^{\frac12c_1(S)}\widehat A(S)$, we express the Chern character as $
\ch(V)=-\pi_{M*}\left(\ch(\pi_S^*\alpha \cdot \mathbb E_0)\Td(S)\right)$.
Applying Grothendieck--Riemann--Roch (see e.g.~\cite{Borel1958}) to the projection $\pi_M\colon S\times M\to M$, we obtain
\bea
\ch(V)=\ch\left(-\pi_{M!}\left(\pi_S^*\alpha \cdot \mathbb E_0)\right)\right)=\ch(\alpha_M)~,
\eea
where in the last equation we use the definition~\eqref{eq:chalphaM e0}. That is, $V\cong \alpha_M$ in K-theory.  Generalising now to $N_f$ flavors, we have
\bea\label{alphaM V}
\alpha_M\cong \CV~,
\eea
where we recall $\CV\coloneqq \bigoplus_{j=1}^{N_f}V^j$ from~\eqref{CV def}. Indeed, the instanton contribution to QCD partition functions is in~\eqref{eq:large_mass_exp_UV} expressed as a generating function of intersection numbers of the total Chern class $c(\CV)$ on the instanton moduli space $\CM^{\rm i}$, motivating the equivalence with the definition~\eqref{virtual Segre def} of Segre numbers
in the algebraic setup.

\sss{The dictionary.}
We will consider $\mathrm{SU}(N)$ SQCD with equal and large masses. To compare with Conjecture~\ref{conj_GK}, we need to fix the normalization $K_n$ and the global phases $\phi_n$ in the SW contribution~\eqref{eq:SW_cont_gen}. We propose that $K_n=N$.  To fix the phase $\phi_n$, recall that $\lambda_0^a \in L + \beta^a$. We make the choice~\eqref{eq:special_lambdaa} of the shift of the lattice $L$, so that $\beta^a=\frac{a}{N}\ov{\mu}(b)$, and we get
\begin{equation}
e^{2\pi\im \lambda^a \lambda_0^a} = e^{\frac{2\pi\im}{N} (a\lambda^a c_1)} ,
\end{equation}
where $c_1$ is the first Chern class of the gauge bundle over $X$. Here $\ov{\mu}(b)$ is an integral lift of the 't Hooft flux $\mu(b)$. 
We label the $\CN=1$ singularities such that $n=0$ corresponds to the value $\vec{u}=\vec{\tu}^{(0)}_{N_f}$ to be matched with the Göttsche--Kool conjecture.
To be compatible with the Göttsche--Kool conjecture, we propose the following 
\begin{eqsp}
\phi_n(\lambda^1,\dots,\lambda^{N-1})&=\frac{\pi}{N}\sum_{a}aKc_1~.
\end{eqsp}
The phase in~\eqref{eq:SW_cont_gen} becomes 
\begin{equation}
e^{\im \phi_0(\lambda^1, \dots, \lambda^{N-1})} e^{\frac{2\pi\im}{N} \sum_{a}a\lambda^a c_1} = \varepsilon_N^{\sum_{a}a \frac{(K + 2\lambda^a)}{2} c_1} .
\end{equation}
We now turn to the couplings. For equal masses, the background couplings $C_{ij},q_{ai}, H_{N_f,i}$ are all equal for various values of $i,j$. For comparison with the Göttsche--Kool conjecture, we will also specialize the point observable to 
\begin{eqsp}
    U=pu_1/\Lambda_{N_f}^2~.
\end{eqsp}
Let us denote the parameters $u_a$ at an $\CN=1$ singularity by $\vec{\tu}_{N_f}^{(n)}=(\tu^{(n)}_{1;N_f},\dots,\tu^{(n)}_{N-1;N_f})$. We introduce the following notations:\footnote{We adapt to the mathematical notation and suppress the dependence on the rank $N$.}
\begin{subequations}\label{eq:SW_couplings}
\begin{align}
\tC_{N_f}&\coloneqq C_{ii}|_{\vec{\tu}^{(0)}_{N_f}}~,
\label{eq:CNf}
&\qquad
\hat{\tC}_{N_f}&\coloneqq C_{ij}|_{\vec{\tu}^{(0)}_{N_f}}~, \\
\tD_{a,N_f}&\coloneqq q_{ai}^{N_f}|_{\vec{\tu}^{(0)}_{N_f}}~, \label{tDa}
&\qquad
\tG_{N_f}&\coloneqq G_{N_f}|_{\vec{\tu}^{(0)}_{N_f}}~, \\
\tH_{N_f}&\coloneqq H_{N_f,i}|_{\vec{\tu}^{(0)}_{N_f}}~,
&\qquad
\tm_{N_f}&\coloneqq\left.\frac{\Delta_{N_f}}{\prod_{a=1}^{N-1}a^a}\right|_{\vec{\tu}^{(0)}_{N_f}}~, \\
\tn_{N_f}&\coloneqq\left.\frac{\Delta_{N_f}}{\prod_{a=1}^{N-1}q_{aa}}\right|_{\vec{\tu}^{(0)}_{N_f}}~,
&\qquad
\tq_{ab,N_f}&\coloneqq q_{ab}|_{\vec{\tu}^{(0)}_{N_f}}~, \\
\tx_{a,N_f}&\coloneqq\left.\frac{\p u_1}{\p a^a}\right|_{\vec{\tu}^{(0)}_{N_f}}~,
&\qquad
\tx_{N_f}&\coloneqq\left.\left(\det_{ab} \frac{\p u_a}{\p a^b}\right)\right|_{\vec{\tu}^{(0)}_{N_f}}~,
\label{eq:xNf}
\end{align}
\end{subequations}
which are independent of the flavor label $i=1,\dots, N_f$, while $j\neq i$.
Throughout this section, we will denote the surface fugacity $S$ by $x$ to avoid confusion with other quantities denoted by $S$. Let us denote the SW contribution at the $\CN=1$ singularity $(n)$ by 
\begin{eqsp}
    Z_{\SW,n}[p,x]\coloneqq\left\langle e^{pu_1/\Lambda^2_{N_f} +\tilde{I}^1(x) / \Lambda_{N_f}}\right\rangle^{(n)}~.
\end{eqsp}
Including the decoupling factor~\eqref{eq:decouping_factor}, we obtain from~\eqref{eq:SW_cont_GK_comp}:
\begin{eqsp}\label{eq:ZSW1_physics}
 \decoup Z_{\SW,0}[p,x]&=N \left( \left(\frac{\Lambda_{0}}{\Lambda_{N_f}}\right)^{(N^2+1)} \Lambda_{N_f}^{1-N}\alpha_{N_f}^{12} \beta_{N_f}^{-8} \tx_{N_f}^6 \tm_{N_f}^{-1} \right)^{\chih}  \hat{\tC}_{N_f}^{2 \, \llbracket \bk^2 \rrbracket} \left(\left(\frac{\Lambda_{0}}{\Lambda_{N_f}}\right)^{\frac{N^2}{N_f}} \tC_{N_f}\right)^{(\bk^2)} \\
&\times \left( \left(\frac{\Lambda_{0}}{\Lambda_{N_f}}\right)^{-\frac{N^2}{4}}\alpha_{N_f}^{-1} \beta_{N_f} \tx_{N_f}^{-\frac{1}{2}} \mathtt{n}_{N_f}^{\frac{1}{8}}  \right)^{K^2} e^{\frac{p}{\Lambda_{N_f}^2} \tu_{1;N_f}  + x^2 \tG_{N_f} + \tH_{N_f} x(\bk)}
\\&\times\sum_{\lambda^1, \dots, \lambda^{N-1}} \prod_{a=1}^{N-1} \varepsilon_N^{
a\left(\frac K2+\lambda^a\right)c_1}\SW(\lambda^a) \, \tD_{a,N_f}^{\lambda^a(\bk)} e^{-\frac{i}{\Lambda_{N_f}}\tx_{a,N_f}x\lambda^a} \!\!\!\!\!\!\!\prod_{1 \le a < b \le N-1} \tq_{ab,N_f}^{-\lambda^a \lambda^b}~.
\end{eqsp}
Using the identification~\eqref{eq:c12alpha} of Chern classes, and the change in convention~\eqref{SW shift} for the SW invariant, we find
\bea\label{universal SU(N)}
\universal_\alpha[p',x']
&=
\rho^2
\left(\frac{X_s}{\rho}\right)^{\chih}
\left(
\rho\,
V_s^{\frac{s(s-1)}{8}}
W_s^{\frac{s^2}{4}}
Y_s^{\frac{s}{2}}
Z_s
\prod_{j=1}^{\rho-1}Y_{j,s}^{\frac{s}{4}}
\prod_{j<k}Z_{jk,s}^{\frac14}
\prod_{j=1}^{\rho-1}Z_{jj,s}^{\frac12}
\right)^{K^2}\\
&\quad\times
\left(
V_s^{\frac{s-1}{2}}
W_s^s
Y_s
\prod_{j=1}^{\rho-1}Y_{j,s}^{\frac12}
\right)^{K(\boldsymbol{k})}
\left(V_sW_s^2\right)^{\llbracket \boldsymbol{k}^2\rrbracket}
W_s^{(\boldsymbol{k}^2)}
\\
&\quad\times
\exp\left(
p'T_s+
x'^2Q_s
+x'(\boldsymbol{k})R_s
+\frac 12x'K\left(
sR_s
+2S_s
+\textstyle\sum_{j}S_{j,s}
\right)
\right)\\
&\quad\times
\sum_{\lambda^1,\dots,\lambda^{\rho-1}}
\prod_{j=1}^{\rho-1}
\varepsilon_\rho^{
j\left(\frac K2+\lambda^j\right)c_1
}
\,
\mathrm{SW}(\lambda^j)
\,
Y_{j,s}^{\lambda^j(\boldsymbol{k})}
\,
\exp\left(\lambda^j x'S_{j,s}\right)\\
&\quad\times
\left(
Y_{j,s}^{\frac{s}{2}}
Z_{jj,s}
\prod_{i=1}^{j-1}Z_{ij,s}^{\frac12}
\prod_{k=j+1}^{\rho-1}Z_{jk,s}^{\frac12}
\right)^{\lambda^j K}
\prod_{1\leq j<k\leq \rho-1}
Z_{jk,s}^{\lambda^j\lambda^k}~,
\eea
Comparison of~\eqref{eq:ZSW1_physics} with the universal series~\eqref{universal SU(N)} gives the following conjecture.
\begin{samepage}
\begin{conjecture}\label{conj:DZSW0=Ua}
Let $X$ be a smooth, oriented, compact and simply-connected four-manifold of Seiberg–Witten simple type, with $b_2^+ > 1$. Let $Z_{\mathrm{SW},0}$ be the contribution of the $\CN=1$ singularity $\vec{u}=\vec{\tu}^{(0)}_{N_f}$ to the topological partition function of $\mathcal{N} = 2$ $\mathrm{SU}(N)$ SQCD with $1\leq N_f <2N$ hypermultiplets with large equal mass $m$ on $X$, and $\decoup$ the decoupling factor~\eqref{eq:decouping_factor}. Then, $\decoup Z_{\mathrm{SW},0}$~\eqref{eq:ZSW1_physics} agrees with the universal function $\universal_\alpha$~\eqref{universal series} with $\rho=N$, $c_1 = \ov{\mu}(b)$, $\mathrm{rk}(\alpha) = s = N_f$, and Chern classes~\eqref{eq:c12alpha}: 
\begin{equation}\label{eq:ZSW0=universal}
\decoup Z_{\mathrm{SW},0} [p, x] = \frac{1}{N}\mathscr{U}_\alpha [p', x'] .
\end{equation}
The functions agree in the decoupling limit $m^{N_f} \Lambda_{N_f}^{2N-N_f} = \Lambda_0^{2N}$ as series in 
\begin{equation}\label{z in terms of m}
z = -\frac{\Lambda_0^2}{4^{\frac1N}m^2}=-\frac{1}{4^{\frac1N}}\left(\frac{\Lambda_{N_f}}{m}\right)^{\frac{2N-N_f}{N}}~ ,
\end{equation}
with the change of variables 
\begin{eqsp}\label{change of variables}
    z = t\left(1 + \left(1 - \tfrac{s}{\rho}\right)t\right)^{1 - \frac{s}{\rho}}~.
\end{eqsp} 
The observables are related as $p' = -\left( \frac{m}{\Lambda_s} \right)^2 p$ and $x'=\im \frac{m}{\Lambda_s} x$, and the universal functions are given in terms of Coulomb branch functions as
\allowdisplaybreaks
\begin{subequations}\label{dictionary}
\begin{alignat}{2}
Q_s &= -\left( \frac{\Lambda_s}{m} \right)^2\tG_s \;, \qquad
&R_s &= -\im\left( \frac{\Lambda_s}{m} \right) \tH_s \;, \\
S_{a,s} &= -\frac{1}{m} \tx_{a,s} \;, 
&S_s &= \frac{\im s}{2} \left( \frac{\Lambda_s}{m} \right) \tH_s + \frac{1}{2 m} \sum_{a=1}^{\rho-1} \tx_{a,s} \;
\\ \label{Ts dictionary}
T_s&=-\frac{1}{m^2}\tu^{(0)}_{1;s}~,
&V_s &= \hat{\tC}_s^2 \tC_s^{-2} \left( \frac{\Lambda_0}{\Lambda_s} \right)^{-\frac{2\rho^2}{s}} \;, \\
W_s &= \left( \frac{\Lambda_0}{\Lambda_s} \right)^{\frac{\rho^2}{s}} \tC_s ~,
&X_s &= \rho \alpha_s^{12} \beta_s^{-8} \left( \frac{\Lambda_0}{\Lambda_s} \right)^{\rho^2+1}\Lambda_s^{1-\rho} \tx_s^6 \tm_s^{-1} \;, \\
Y_{a,s} &= \tD_{a,s} \;,
&Y_s &= \left( \frac{\Lambda_0}{\Lambda_s} \right)^{-\frac{\rho^2}{s}} \tC_s^{-1} \hat{\tC}_s^{-(s-1)} \prod_{a=1}^{\rho-1} \tD_{a,s}^{-\frac{1}{2}} \;, 
\\
Z_{ab,s} &= \tq_{ab,s}^{-1} \;, (a<b) 
&Z_{bb,s} &= 
%\tD_{b,s}^{-\frac{s}{2}} \prod_{a=1}^{b-1} \tq_{ab,s}^{\frac{1}{2}} \prod_{c=b+1}^{\rho-1} \tq_{bc,s}^{\frac{1}{2}}=
\tD_{b,s}^{-\frac{s}{2}} \prod_{b\neq a=1}^{\rho-1} \tq_{ab,s}^{\frac{1}{2}} \;, 
\end{alignat} \begin{equation}
    Z_s = \rho^{-1} \alpha_s^{-1} \beta_s \tx_s^{-\frac{1}{2}} \mathtt{n}_s^{\frac{1}{8}} \hat{\tC}_s^{\frac{s(s-1)}{4}} \tC_s^{\frac{s}{4}}  \prod_{a=1}^{\rho-1}\tD_{a,s}^{\frac{s}{4}}\prod_{a<b}\tq_{ab,s}^{-\frac{1}{4}} \;,
\end{equation}\end{subequations} \end{conjecture}
\end{samepage}
%\noindent with $1\leq a<b\leq \rho-1$.
To compare with the universal Segre series~\eqref{universal series}, we notice that we have $\rho+1$ constraints among the universal functions:
\begin{subequations}\label{constraints}
\begin{align}
    V_s^{s-1} W_s^{2s} Y_s^2 \prod_{j=1}^{\rho-1} Y_{j,s}&=1~,
    \\ \label{sRs +..}
    sR_s + 2S_s + \sum\limits_j S_{j,s}&\overset{*}{=}0~,
    \\
    Y_{j,s}^s Z_{jj,s}^2 \prod_{i=1}^{j-1} Z_{ij,s} \prod_{k=j+1}^{\rho-1} Z_{jk,s} &=1,\quad j=1,\dots,\rho-1~. \label{constraint 3}
\end{align}
\end{subequations}
The equation~\eqref{sRs +..} follows from the absence of the coupling $B(x,K)$ of the surface $x$ to the canonical class $K$ in the physical partition function. The $*$ highlights a subtlety, as we find it to be not true in general with the universal functions determined in~\cite{Gottsche:2020ass}. We comment further on this ambiguity in Appendix~\ref{app:corrections}.

\sss{Counting universal series.} Since we have related the universal series in the Göttsche--Kool conjecture to couplings in SW theory, we must make sure that the number of couplings is equal to the number of universal series minus the number of constraints, since it is not obvious from our dictionary~\eqref{dictionary}. The number of couplings in~\eqref{eq:SW_couplings} including the coordinate $\tu^{(0)}_{1;N_f}$ is
\begin{eqsp}
  \#\,\text{couplings} =  7+N+\frac12N(N-1)~.
\end{eqsp}
On the other hand, the number of universal series is
\bea
  \#\,\text{universal series} = 7+2N+\frac12N(N-1)~.
\eea
There are 
\bea
\#\,\text{constraints}=N+1~,
\eea
such that
\bea
 \#\,\text{universal series}-\#\,\text{constraints} =  \#\,\text{couplings}-1~.
\eea
Thus, for all $N$, we have an additional relation among the physical couplings we have not used. 
For instance, for $\SU(2)$, this gives 10 couplings and 12 universal series, with 3 constraints. However, we have the accidental case where $\det \frac{du_a}{da^b}=\frac{du}{da}$, such that there are only 9 a priori independent couplings, and the counting is $12-3=9$.
For $\SU(3)$, there are 16 universal series with 4 constraints, while we have 13 physical couplings. We find in Section~\ref{sec:SU(3) calculation} that $\tx_s$ is related to $\tn_s$ as in~\eqref{tn in terms of tx}, reducing the effective number of couplings to 12, such that we have $16=12-4$.

\subsection{Sum over \texorpdfstring{$\CN=1$}{Nf} vacua}
In this section, we sum over the SW contributions from $\CN=1$ points and argue that it agrees with the generating function of the virtual Segre numbers. Our physical motivation for this comparison parallels the motivation in~\cite{Furrer:2026byd}. 

\sss{Motivation.}
We stress again that once we make the identification of the Chern character as
\begin{eqsp}
 (\rho,c_1,c_2)=\left(N,\ov{f}_2(P),k+\frac{N-1}{2N}\ov{f}_2(P)^2\right)~,   
\end{eqsp}
then the virtual dimension~\eqref{vd MQ} of the SU($N$) (or PSU($N$)) instanton moduli space $\CM^{\CQ,0}_{v(E)}$ matches the virtual dimension~\eqref{vdim M} of $M$: 
\begin{eqsp}
\vd M=\vd\CM_{v(E)}^{\CQ,0}=\vd\CM_{v(E)}^{\mathrm{i}}~.
\end{eqsp}
In gauge theory, the path integral naturally sums over all isomorphism classes of $\mathrm{PSU}(N)$-bundles and hence the topological partition function includes a sum over $c_2$. As in~\cite{Furrer:2026byd}, if we identify $q=\left(\frac{\Lambda_0}{m}\right)^2$, then the instanton component of the topological UV partition function $Z^{\mathrm{i}}(\bm{m})$ with equal masses and with the decoupling factor included admits the $q$-expansion:
\begin{eqsp}
\decoup Z^{\mathrm{i}}(\bm{m})=\sum_{k}q^{\frac{1}{2}\vd\CM_{v(E)}^{\mathrm{i}}}  \int_{\vd\CM_{v(E)}^{\mathrm{i}}}c(\CV_k)~,     
\end{eqsp}
see~\eqref{eq:large_mass_exp_UV}. 
In Section~\ref{sec:Coul_int}, we argued that the Seiberg--Witten contribution from the $\CN=1$ points on the Coulomb branch with equal and large masses is identified with $Z^{\mathrm{i}}(\bm{m})$. Since we expect $\CV\cong\alpha_M$~\eqref{alphaM V} for our choice of class $\alpha$, whenever $X$ is algebraic we expect that the integral counts the Segre numbers. Up to a small renormalisation of the instanton counting parameter $q$ vs $z$, which we discussed in Section~\ref{sec:moduli_sp_corr}, this motivates the comparison:
\begin{eqsp}\label{eq:ZSW=GS}
    \decoup Z^{\mr{i}}_{\mathrm{SW}}=\CG_{\segre}~.
\end{eqsp}
To get $Z^{\mr{i}}_{\mathrm{SW}}$, we need to sum over the $N~\CN=1$ singularities. For the pure SU$(N)$ theory, the $\mr{U(1)_R}$ symmetry of the UV theory breaks to the discrete $\Z_{4N}$ subgroup and the further $\Z_N<\Z_{4N}$ subgroup permutes the $N~\CN=1$ points on the Coulomb branch. Since SQCD with large equal masses can be viewed as a deformation of the pure $\SU(N)$ theory, the $\CN=1$ points of QCD are small deformations of the ones of the pure theory, and so we expect that the $\Z_{4N}$ R-symmetry persists as an asymptotic symmetry and the $\Z_N$ subgroup still permutes the $\CN=1$ points. 

Indeed, as we showed in section~\ref{sec:SW geometry}, the $\CN=1$ QCD vacua in the large and equal mass limit still exhibit the full $\mathbb Z_N$ symmetry, generated by the $\mathbb Z_N$ action $z\mapsto \varepsilon_N^n z$, where $z\sim 1/m^2$~\eqref{z parameter} becomes the appropriate variable. This argument is based on the scaling limit of the curve, as all the Casimirs $u_a$ can be written as $m^{a+1}T_a(z)$, where $T_a(z)$ are well-defined $z$-series. We complete this argument by determining the universal functions $T_a$ for $N=3$ in section~\ref{sec:SU(3) calculation}.
Therefore, the singularities $\vec{\tu}^{(n)}_{N_f}$, considered as $z$-series, get permuted under the $\mathbb Z_N$ action as well.
Retaining the notation and identifications from Conjecture~\ref{conj:DZSW0=Ua}, we obtain the
\begin{samepage}
\begin{conjecture}\label{conj:DZSWi=GU}
The contribution $Z_\SW^{\rm i}$~\eqref{ZSWi} of the instanton component to the topological $\SU(N)$ SQCD partition function on $X$ agrees with the generating function $\CG_\universal$ of universal functions for virtual $\SU(N)$ Segre numbers~\eqref{eq:uni_series},
\bea
\decoup Z_\SW^{\rm i}[p,x]=\segregen_\universal[p',x']~.
\eea
\end{conjecture}
\end{samepage}
\noindent In the remainder of this subsection, we will provide evidence for this claim.

\sss{Sum over $\CN=1$ vacua.}
First, to deduce the SW contribution $Z_{\mr{SW},n}$ of an arbitrary $\CN=1$ point $\vec{\tu}^{(n)}_{N_f}$ from the ``starting point'' $\vec{\tu}^{(0)}_{N_f}$, which allows to compute the total partition function
\bea\label{ZSWi}
Z_\SW^{\rm i}=\sum_{n=0}^{N-1}Z_\SW^{(n)}~,
\eea
we simply sum over R-symmetry translates of $Z_{\mr{SW},0}$. For this, we will need the R-charge of $Z_{\mr{SW},0}$. Let $\mathsf R(\CO) $ denote the R-charge of an operator $\CO$. The R-charge of various ingredients of $Z_{\mr{SW},0}$ has been computed in~\cite{Marino:1998bm},
\begin{eqsp}
    \mathsf R( \tm_{N_f}^{\sigma / 8}) &=\frac{\sigma}{2}N(N-1)-\frac{\sigma}{4}(N-1)~,
    \\
    \mathsf R(\tx_{N_f}^{\chi / 2})&=\frac{\chi}{2}N(N-1)~,
\\
\mathsf R(\tm^{-1}_{N_f}\tn_{N_f})&=(N-1)\frac{2\chi+3\sigma}{4}~,
\end{eqsp}
where we used the SW simple type condition~\eqref{eq:sw_simple_type}. We get   
\begin{eqsp}
    Z_{\mr{SW}}^{\mr{i}}&=N\sum_{n=0}^{N-1}\varepsilon_N^{\frac{n}{2}[(N^2-1)\chih+(N-1)\ov{f}_2(P)^2]} \Bigg{[}\left( \Lambda_{N_f}^{1-N}\alpha_{N_f}^{12} \beta_{N_f}^{-8} \tx_{N_f}^6 \tm_{N_f}^{-1} \right)^{\chih}  \hat{\tC}_{N_f}^{2 \, \llbracket \bk^2 \rrbracket}  \tC_{N_f}^{(\bk^2)}\ \\
&\times \left( \alpha_{N_f}^{-1} \beta_{N_f} \tx_{N_f}^{-\frac{1}{2}} \mathtt{n}_{N_f}^{\frac{1}{8}}  \right)^{K^2} e^{\frac{p}{\Lambda_{N_f}^2} \tu_{1;N_f}  + x^2 \tG_{N_f} + \tH_{N_f} x(\bk)}
\\&\times\sum_{\lambda^1, \dots, \lambda^{N-1}} \left.\prod_{a=1}^{N-1} \varepsilon_N^{
a\left(\frac K2+\lambda^a\right)c_1}\SW(\lambda^a) \, \tD_{a,N_f}^{\lambda^a(\bk)} e^{-\frac{i}{\Lambda_{N_f}}\tx_{a,N_f}x\lambda^a}\!\!\!\!\!\!\! \prod_{1 \le a < b \le N-1} \tq_{ab,N_f}^{-\lambda^a \lambda^b}\right]^{(n)}
\end{eqsp}
where we included a global phase $e^{\pi i(N-1)\ov{f}_2(P)^2}$ which generalizes the $\SU(2)$ phase $e^{\pi i\ov{w}_2(E)^2}$ of~\cite{Moore:1997pc,Furrer:2026byd}. We will show below that this phase is crucial for $Z_{\SW}^{\mathrm{i}}$ to be real.
\par
For the moment, let us drop the dependence of the universal Segre series on $p',x'$ in favour of $z$ and write $\universal_\alpha(z)$. The $\Z_N$-action on the variable $z$ is given by $z\mapsto \varepsilon_N^nz$.  Let us introduce the notation
\begin{eqsp}\label{kappa}
    \kappa\coloneqq (N^2-1)\chih+(N-1)\ov{f}_2(P)^2~,
\end{eqsp}
with $\ov{f}_2(P)^2=c_1^2$.
Then we see that the  the $\Z_N$-translates of the universal Segre series $\universal_\alpha(z)$ via the $\Z_N$-action on $z$ along with the equality~\eqref{eq:ZSW0=universal} implies
\begin{eqsp}\label{DZSW sum over U_a}
    \decoup Z_{\mr{SW}}^{\mr{i}}=\frac{1}{N}\sum_{n=0}^{N-1}\varepsilon_N^{\frac{n\kappa}{2}}\universal_\alpha(\varepsilon_N^{n}z)~,
\end{eqsp}
Let us now prove that this is consistent with our proposal ~\eqref{eq:ZSW=GS} and the Göttsche--Kool conjecture~\eqref{eq:uni_series},\eqref{eq:GK_conj_gen_ser},  i.e.,
\begin{eqsp}\label{eq:projection_universal}
    \left.\universal_{\alpha}[p',x']\right|_{z^{\left\{\frac{1}{2}\vd M|c_2\in\Z\right\}}}=\frac{1}{\rho}\sum_{n=0}^{N-1}\varepsilon_{\rho}^{\frac{n\kappa}{2}}\universal_\alpha(\varepsilon_{\rho}^{n}z)~.
\end{eqsp}
First, $\kappa$ is simply the ``off-set'' of the virtual dimension of $M$: using~\eqref{vdim M}, we see that
\begin{eqsp}
   \frac{1}{2}\vd M=\rho c_2-\frac{\kappa}{2}~. 
\end{eqsp}
Since $c_2\in\Z$, we have 
\begin{eqsp}\label{eq:GU_series}
    \CG_{\universal}(z)\in z^{-\frac{\kappa}{2}}\mathbb{C}[[z^{\rho}]]~.
\end{eqsp}
Thus $\kappa\bmod 2\rho$ determines the powers of $z$ that appear in $\CG_{\universal}(z)$. More precisely, we have 
\begin{eqsp}
    \frac{1}{2}\vd M\in \frac{2\rho-\ell}{2}+\rho\Z~,\quad \kappa\equiv\ell\bmod 2\rho~.
\end{eqsp}
Next, we observe that using the constraints~\eqref{constraints}, we can simplify the universal Segre series to 
\begin{eqsp}\label{eq:universal_simplified}
   \universal_\alpha[p',x']
&=
\rho^2
\left(\frac{X_s}{\rho}\right)^{\chih}
\left(
\rho\,
Y_s^{\frac{s}{4}}
Z_s
\prod_{j=1}^{\rho-1}Z_{jj,s}^{\frac{1}{4}}
\right)^{K^2}
\left(V_sW_s^2\right)^{\llbracket \boldsymbol{k}^2\rrbracket}
W_s^{(\boldsymbol{k}^2)}
e^{
x'^2Q_s
+x'(\boldsymbol{k})R_s
+p'T_s}\\
&\quad\times
\sum_{\lambda^1,\dots,\lambda^{\rho-1}}
\prod_{j=1}^{\rho-1}
\varepsilon_\rho^{
j\left(\frac K2+\lambda^j\right)c_1
}
\,
\mathrm{SW}(\lambda^j)
\,
Y_{j,s}^{\lambda^j(\boldsymbol{k})}
\,
\exp\left(\lambda^j x'S_{j,s}\right)
\prod_{1\leq j<k\leq \rho-1}
Z_{jk,s}^{\lambda^j\lambda^k}~. 
\end{eqsp}
Since all universal series are elements of $\mathbb{C}[[z^{\frac{1}{2}}]]$, the universal Segre series also satisfies $\universal_{\alpha}(z)\in \mathbb{C}[[z^{\frac{1}{2}}]]$.
Since the universal Segre series is a formal 
$z^{\frac12}$-series, we rearrange the sum to write 
\begin{eqsp}\label{eq:U_alpha_decomp}
  \universal_{\alpha}(z)=\sum_{\substack{\ell=0}}^{2\rho-1}z^{\frac{2\rho-\ell}{2}}\universal_{\alpha,\ell}(z)~,  
\end{eqsp}
for some series $\universal_{\alpha,\ell}(z)\in\mathbb{C}[[z^{\rho}]]$. This is possible since any half-integer $h/2$ can be written as 
\begin{eqsp}
    \frac{h}{2}=m\rho+\frac{2\rho-\ell}{2}~,
\end{eqsp}
by choosing $\ell=-h\bmod 2\rho$ and for some $m\in\Z$. To perform this sum, we notice that not all integers $\ell$ occur. 
More precisely, we can compute the monodromy of $\universal_\alpha(z)$ around $z=0$, which determines the nature of the series depending on the parity of $\kappa$. We discuss this monodromy in the section below, and list the result here: $  \universal_\alpha(e^{2\pi i}z)=(-1)^{\kappa}\universal_\alpha(z)$~\eqref{upsilon universal}.  This implies that only the $\ell\equiv \kappa\bmod 2$ terms appear in~\eqref{eq:U_alpha_decomp}:
\begin{eqsp}\label{eq:U_alpha_decomp_l=k}
  \universal_{\alpha}(z)=\sum_{\substack{\ell=0\\\ell\equiv \kappa\bmod 2}}^{2\rho-1}z^{\frac{2\rho-\ell}{2}}\universal_{\alpha,\ell}(z)~.  
\end{eqsp}
Taking now the $\mathbb Z_r$ images gives 
\begin{eqsp}\label{eq:universal_ZN_action}
 \universal_{\alpha}(\varepsilon_\rho^nz)=\sum_{\substack{\ell=0\\\ell\equiv \kappa\bmod 2}}^{2\rho-1}\varepsilon_{\rho}^{\frac{n(2\rho-\ell)}{2}}z^{\frac{2\rho-\ell}{2}}\universal_{\alpha,\ell}(z)~.   
\end{eqsp}
Substituting this in the RHS of~\eqref{eq:projection_universal} gives 
\begin{eqsp}
    \frac{1}{\rho}\sum_{n=0}^{\rho-1}\varepsilon_{\rho}^{\frac{n\kappa}{2}}\sum_{\substack{\ell=0\\\ell\equiv \kappa\bmod 2}}^{2\rho-1}\varepsilon_{\rho}^{\frac{n(2\rho-\ell)}{2}}z^{\frac{2\rho-\ell}{2}}\universal_{\alpha,\ell}(z)=\frac{1}{\rho}\sum_{n=0}^{\rho-1}\sum_{\substack{\ell=0\\\ell\equiv \kappa\bmod 2}}^{2\rho-1}e^{\frac{2\pi in(\kappa-\ell)}{2\rho}}z^{\frac{2\rho-\ell}{2}}\universal_{\alpha,\ell}(z)~.
\end{eqsp}
Exchanging the sum using the geometric sum,
\begin{eqsp}\label{eq:roots_unity_sum}
    \sum_{n=0}^{\rho-1}e^{\frac{2\pi in(\kappa-\ell)}{2\rho}}=\begin{cases}
        \rho&\text{if}~~\frac{\kappa-\ell}{2}\equiv 0\bmod \rho,\\
        0&\text{otherwise},
    \end{cases}
\end{eqsp}
we conclude that 
\begin{eqsp}
    \frac{1}{\rho}\sum_{n=0}^{\rho-1}\varepsilon_{\rho}^{\frac{n\kappa}{2}}\universal_\alpha(\varepsilon_{\rho}^{n}z)\in z^{-\frac{\kappa}{2}}\mathbb{C}[[z^{\rho}]]~,
\end{eqsp}
which proves~\eqref{eq:projection_universal}.
We have thus shown Conjecture~\ref{conj:DZSWi=GU}, up to the assumption on the monodromy of the universal series $\universal_\alpha$. This monodromy property is a consequence of a rich system of symmetries of these universal series, which we explore in the next section.

\subsection{Monodromy, conjugation and Galois groups}\label{sec:monodromy}
The universal series $\universal_\alpha$ has several interesting symmetries, which we explore in this section. To define the relevant operators, let $L\in\mathbb C[[z^{\frac12}]]$ be a formal series. The operator $\upsilon$ probes the monodromy of $L$ at $z=0$, that is, it maps $\upsilon\colon z\mapsto e^{2\pi i}z$, or better, $z^{\frac12}\mapsto -z^{\frac12}$. The second operator $\varkappa$ acts as complex conjugation $\varkappa\colon i\mapsto -i$ on the coefficients of $L$. Writing  $L(z)=\sum_n a_n z^{\frac n2}$, one may express these operators as follows:\footnote{In~\cite{Gottsche:2020ass}, $\varkappa$ is denoted by $\tau$}
\bea\label{upsilon varkappa}
\upsilon L(z)&\coloneqq \sum_n (-1)^n a_n z^{\frac n2}~, \\
\varkappa L(z)&\coloneqq \sum_n \bar a_n z^{\frac n2}~.
\eea
Clearly, $\upsilon L$ and $\varkappa L\in \mathbb C[[z^{\frac12}]]$ are new formal series.
Let us discuss these operations first in turn, and finally combined.

\sss{Monodromy operator $\upsilon$.}
To determine the nature of the universal series $\universal_\alpha$~\eqref{eq:universal_simplified}, we calculate the monodromy phase of $\universal_\alpha(z)$ around $z=0$.
From~\eqref{Cz Cz12} is it known that $Q_s$, $R_s$, $T_s$, $V_s$, $W_s$, $X_s\in\mathbb C[[z]]$ are $z$-series, and thus have trivial monodromy. The functions based on the letters $S$, $Y$ and $Z$ have a priori non-trivial monodromy. We conjecture that they are
\allowdisplaybreaks
\begin{subequations}\label{eq:SYZ_monodromy}
\begin{align} 
\upsilon S_s&= S_s+\sum_{j=1}^{\rho-1} S_{j,s}~, \\
\upsilon S_{j,s}&=-S_{\rho-j,s}~, \\
\upsilon Y_{s}&=Y_s\prod_{j=1}^{\rho-1}Y_{j,s}~, \\
\upsilon Y_{j,s}&=Y_{\rho-j,s}^{-1}~, \\
\upsilon Z_{s}&=Z_s \prod_{j=1}^{\rho-1}Z_{jj,s} \prod_{j<k}Z_{jk,s}~, \\
\upsilon Z_{jj,s}&=Z_{(\rho-j)(\rho-j),s}^{-1}\prod_{i=1}^{\rho-j-1}Z_{i(\rho-j),s}^{-1}\prod_{k=\rho-j+1}^{\rho-1}Z_{(\rho-j)k,s}^{-1}~, \\
\upsilon Z_{ij,s}&=Z_{(\rho-j)(\rho-i),s}~, \quad i<j~.
\end{align}\end{subequations}
These monodromy relations hold for all cases checked in ranks $\rho=2,3,4$.

We furthermore notice that the first line of~\eqref{eq:universal_simplified} (all factors before the sum over basic classes) is a $z$-series with integer exponents if and only if the $K^2$ coupling is a $z$-series with integer exponents. Using the relations~\eqref{eq:SYZ_monodromy} together with the constraint~\eqref{constraint 3}, multiplied over all $j$, we obtain
\bea
\upsilon Y_s^{\frac{s}{4}}
Z_s
\prod_{j=1}^{\rho-1}Z_{jj,s}^{\frac{1}{4}}= Y_s^{\frac{s}{4}}
Z_s
\prod_{j=1}^{\rho-1}Z_{jj,s}^{\frac{1}{4}}~,
\eea
such that the first line of~\eqref{eq:universal_simplified} is invariant under $\upsilon$.
The only monodromy can therefore originate from the second line. 
Using~\eqref{eq:SYZ_monodromy}, we change variables $\lambda^j\mapsto -\lambda^{\rho-j}$, use the (anti)-symmetry~\cite{Witten:1994cg}
\bea\label{SW -}
\SW(-\lambda)=(-1)^{\chih}\SW(\lambda)~,
\eea
and the fact that $2\lambda^j$ is a characteristic vector~\eqref{characteristic vector}, to find:\footnote{The second equality uses the fact that $\rho^2$ has the same parity as $\rho$}
\bea\label{upsilon universal}
    \upsilon \universal_\alpha&=(-1)^{(\rho-1)(\chih+c_1^2)}\universal_\alpha \\
    &=(-1)^{\kappa}\universal_\alpha~,
\eea
with $\kappa$ as before~\eqref{kappa}. This generalises~\cite[(3.31)]{Furrer:2026byd} to all ranks $\rho$. This is the hypothesis for the putative proof of Conjecture~\ref{conj:DZSWi=GU}, which then boils down to proving~\eqref{eq:SYZ_monodromy}.

The monodromy operator descends from charge conjugation symmetry. Since we identify $z\sim 1/m^2$~\eqref{z in terms of m}, we have $z^{\frac12}\sim 1/m$ and thus $z^{\frac12}\mapsto -z^{\frac12}$ is induced by $m\to -m$. This is part of the charge conjugation $\mathsf C$ action~\eqref{Cc curve}.
It is important to distinguish it however from a Coulomb branch monodromy, or even a monodromy in complexified mass-parameter space: Since $m\mapsto e^{-\pi i}m$, this is not a monodromy at $m=\infty$, rather, it is a deck transformation or monodromy of the double cover parametrised by $z\sim 1/m^2$.

\sss{Complex conjugation $\varkappa$.} Since $\universal_\alpha$ counts intersection numbers on the instanton moduli space, its coefficients should be real. The conjecture~\ref{conj:DZSW0=Ua} then implies that the instanton component $\decoup Z^{\mathrm{i}}_{\SW}$ to the partition function is real as well. To show this, we will use the relation~\eqref{DZSW sum over U_a}. The universal series known in the literature suggest that all universal functions $Q_s$, $R_s$, $T_s$, $V_s$, $W_s$, $X_s$ have real coefficients, except $S_{j,s},Y_{j,s},Z_{jk,s}$, which satisfy the following relations under complex conjugation:
\bea\label{eq:conjugation_universal_func}
   \varkappa S_{j,s}&=S_{\rho-j,s}~, \\\varkappa Y_{j,s}&=Y_{\rho-j,s}~, \\\varkappa Z_{jk,s}&=Z_{(\rho-k)(\rho-j),s}~.
\eea
We again conjecture these to hold for all ranks $\rho$, and checked it in the examples $\rho=2,3,4$.

Using~\eqref{eq:conjugation_universal_func} and changing variables $j\to\rho-j$ in the product, swapping $\lambda^j\to\lambda^{\rho-j}$ in the sum in~\eqref{eq:universal_simplified}, and the fact that $K$ and $2\lambda^j$ are characteristic vectors, we find that the universal series $\universal_\alpha$ is real,  
\begin{eqsp}
    \varkappa\,\universal_{\alpha}(z)=e^{-2\pi i\rho c_1^2}\universal_{\alpha}(z)=\universal_{\alpha}(z)~.
\end{eqsp}
In particular, each of the series $\universal_{\alpha,\ell}\in\BC[[z^\rho]]$ is real, i.e., $\universal_{\alpha,\ell}\in\BR[[z^\rho]]$. Thus, we conclude that 
\begin{eqsp}\label{eq:conjugate_ZSWi}
     \varkappa\,  Z^{\mathrm{i}}_{\SW}&=\frac{1}{\rho}\sum_{n=0}^{\rho-1}\varepsilon_\rho^{-\frac{n\kappa}{2}}\sum_{\substack{\ell=0\\\ell\equiv\kappa\bmod 2}}^{2\rho-1}\varepsilon_\rho^{-\frac{n(2\rho-\ell)}{2}}z^{\frac{2\rho-\ell}{2}}\universal_{\alpha,\ell}(z)
    \\&=\frac{1}{\rho}\sum_{\substack{\ell=0\\\ell\equiv\kappa\bmod 2}}^{2\rho-1}\sum_{n=0}^{\rho-1}\varepsilon_\rho^{\frac{n(\ell-\kappa)}{2}}z^{\frac{2\rho-\ell}{2}}\universal_{\alpha,\ell}(z)~.
\end{eqsp}
From~\eqref{eq:roots_unity_sum} we see that $
    \sum_{n=0}^{\rho-1}\varepsilon_\rho^{\frac{n(\ell-\kappa)}{2}}=\sum_{n=0}^{\rho-1}\varepsilon_\rho^{\frac{n(\kappa-\ell)}{2}}$.
Then backtracking the steps in~\eqref{eq:conjugate_ZSWi}, we get 
\begin{eqsp}
  \varkappa\,Z^{\mathrm{i}}_{\SW}= Z^{\mathrm{i}}_{\SW}~. 
\end{eqsp}

As for the monodromy $\upsilon$, the complex conjugation $\varkappa$ is also a consequence of charge conjugation symmetry. 
Using the dictionary~\eqref{dictionary}, the first and the third equation in~\eqref{eq:conjugation_universal_func} are equivalent to the statements that $ \varkappa\,\tx_{a,s}=-\tx_{N-a,s}$ and $\varkappa\,\tau_{ab}=\tau_{(N-a)(N-b)}$. These hold due to the charge conjugation symmetry, as explained in section~\ref{sec:SW geometry}---see in particular~\eqref{J CC} and~\eqref{tau CC}. The second identity can be explained by a similar argument, relating the couplings $\tD_{a,s}$ to the SW curve using the blowup function, as explained in~\eqref{sec:blowup}.

\sss{Monodromy $\circ$ conjugation.}
The only step we are unable to prove in Conjecture~\ref{conj:DZSWi=GU} is the monodromy relations~\eqref{eq:SYZ_monodromy} of the functions $S$, $Y$ and $Z$. In fact, they give strong constraints for rank $\rho> 4$, where less is known about them~\cite{Gottsche:2020ass,Gttsche2024}.
Combining~\eqref{eq:SYZ_monodromy} with~\eqref{eq:conjugation_universal_func}, we find that the combination of monodromy and conjugation can be expressed in terms of an inversion, 
\bea\label{upsilon kappa}
\upsilon\varkappa\, S_{j,s}&=-S_{j,s}^{-1}~, \\
\upsilon\varkappa\, Y_{j,s}&=Y_{j,s}^{-1}~, \\
\upsilon\varkappa\, Z_{jj,s}&=Y_{j,s}^{s} Z_{jj,s}~.
\eea
To obtain the last identity, we use again the constraint~\eqref{constraint 3}. In fact, it is useful to express the combinations through auxiliary sets of functions~\cite{Gottsche:2020ass},
\bea
S_{J,s}=S_s+\sum_{j\in J}S_{j,s}~, \quad
Y_{J,s}\coloneqq Y_s\prod_{j\in J}Y_{j,s}, \quad Z_{J,s}\coloneqq Z_s\prod_{i\le j\in J}Z_{ij,s}~,
\eea
which are defined for all subsets $J\subset [\rho-1]\coloneqq \{1,\dots, \rho-1\}$. Then, combining~\eqref{eq:SYZ_monodromy} with~\eqref{eq:conjugation_universal_func}, one can show that
\bea
\upsilon\varkappa\, S_{J,s}=S_{J^c,s}~,\\
\upsilon\varkappa\, Y_{J,s}=Y_{J^c,s}~,\\
\upsilon\varkappa\, Z_{J,s}=Z_{J^c,s}~,\\
\eea
where $J^c=[\rho-1]\setminus J$ denotes the set complement. These identities are conjectured in~\cite[Remark 4.5] {Gottsche:2020ass}.
Since both conjugation and monodromy act on the indices as a Dynkin reversal, $j\mapsto \rho-j$, we understand all of these identities as a consequence of charge conjugation symmetry, as we explained in section~\ref{sec:SW geometry}. 
As a simple consequence,~\eqref{upsilon kappa} allows us to predict the power series
\bea
S_{j,s}&\in i\mathbb Q[[z]]+z^{\frac12}\mathbb Q[[z]]~, \\
\log Y_{j,s}&\in i\mathbb Q[[z]]+z^{\frac12}\mathbb Q[[z]]~,  \\
Y_{j,s}^{\frac s2}Z_{jj,s} &\in \mathbb Q[[z]]
+i z^{\frac12}\mathbb Q[[z]]~.
\eea
For the last identity, we find for $\rho=3$ that the complex part vanishes entirely.

\sss{Galois groups.}
It appears that even more can be said about the Segre series. Denoting any of the series $S_{J,s}$, $Y_{J,s}$ or $Z_{J,s}$ by $L_{J,s}$, they are elements of $\mathbb C[[z^{\frac12}]]$, while all other series (with single subscript $s$) are in $\mathbb Q[[z]]$.  It is observed in~\cite{Gottsche:2020ass} that the power series for rank $\rho$ lie in certain Galois extensions $K_\rho$ of the rationals $\mathbb Q$,
\bea
L_{J,s}\in K_\rho[[z^{\frac12}]]~.
\eea
This is expected from the fact that we have several actions on the contributions $Z_{\SW,0}$, or equivalently, on the universal series $\universal_\alpha$. For instance, charge conjugation $\cc$ is a non-trivial map, and thus should be reflected in a non-trivial fashion in the partition function: it acts in particular as the complex conjugation $\varkappa\colon i\mapsto -i$, which is respected by the universal functions as described above. 

At the same time, the $\mathbb Z_N$ symmetry of the $\CN=1$ vacua translates into the projection~\eqref{eq:projection_universal} of $\universal_\alpha$ to certain conditions. Since this projection involves the phase $\varepsilon_\rho$
of $z$, but the universal series are functions of $z^{\frac12}$, we expect the coefficients of $\universal_\alpha$ to contain such roots $\varepsilon_\rho^{\frac12}=\varepsilon_{2\rho}$ of unity. A natural proposal is thus to look for a field extension of $\mathbb Q$ by $i$ and $\varepsilon_{2\rho}$, giving the field extension
\bea\label{field extension}
K_\rho=\mathbb Q(i,\varepsilon_{2\rho})~.
\eea
Since $i=\varepsilon_4$, we may write this as $K_\rho=\mathbb Q(\varepsilon_{m})$ with $m={\rm lcm}(4,2\rho)$, therefore $K_\rho$ is simply a cyclotomic field. Then 
\bea
G_\rho={\rm Gal}(K_\rho,\mathbb Q)
\eea
is the cyclotomic Galois group, which encodes the automorphisms of $K_\rho$ leaving $\mathbb Q$ invariant. The order of the Galois group is given by Euler's totient, $|G_\rho|=\varphi(m)$.
Let us check this proposal for low rank.

The rank $\rho=2$ is the unique case where $\SU(\rho)$ does not have outer automorphisms, which means that charge conjugation acts trivially. Since charge conjugation acts as complex conjugation on the $z$-series, we therefore expect $\varkappa\colon i\mapsto -i$ to have trivial action. At the same time, $\varepsilon_{2\rho}=i$, but if the power series contained $i$ there would be a nontrivial action of $\varkappa$, and so all series are rational:
\bea\label{K2}
K_2=\mathbb Q~.
\eea
This holds for all $\SU(2)$ Segre series~\cite{Gottsche:2020ass,Furrer:2026byd}.
For $\rho=3$, we get $\mathbb Q(i,\varepsilon_6)=\mathbb Q(i,\sqrt3)$, such that
\bea\label{K3}
K_3=\mathbb Q(i,\sqrt3)
\eea
with Galois group $G_3\cong\mathbb Z_2\times \mathbb Z_2$, which is generated by $\varkappa$ and $\sigma\colon \sqrt3\mapsto -\sqrt3$. For $\rho=4$ we have instead $\mathbb Q(i,\varepsilon_8)=\mathbb Q(i,\sqrt2)$, such that
\bea\label{K4}
K_4=\mathbb Q(i,\sqrt2)~,
\eea
again with $G_4\cong\mathbb Z_2\times \mathbb Z_2$,  generated by $\varkappa$ and $\sigma\colon \sqrt2\mapsto -\sqrt2$. All these three~\eqref{K2}--\eqref{K4} agree precisely with the remarks in section 4.5 of~\cite{Gottsche:2020ass}, particularly Conjectures 4.2, 4.3 and 4.4.
For $\rho=5$, our proposal predicts 
\bea
K_5=\mathbb Q(i,\varepsilon_{10})
=\mathbb Q(i,\varepsilon_5)
=\mathbb Q(\varepsilon_{20})~.
\eea
This is a cyclotomic extension of degree $[K_5:\mathbb Q]=\varphi(20)=8$, 
with Galois group
\bea
G_5
\cong\mathbb Z_{20}^\times
\cong\mathbb Z_4\times\mathbb Z_2~.
\eea
In addition to the involution $\varkappa\colon i\mapsto-i$, it contains the order-four automorphism $\sigma\colon \varepsilon_5\mapsto\varepsilon_5^2$.
For $\rho=6$, we propose $K_6=\mathbb Q(i,\varepsilon_{12})=\mathbb Q(i,\sqrt3)$ and so $K_6\cong K_3$ with isomorphic Galois group, and so forth. 

Another piece of evidence for the proposal $K_\rho$~\eqref{field extension} comes from the all-$N$ result~\eqref{tq jk SU(N)} of the off-diagonal coupling $\tq_{jk;0}$ for $N_f=0$. Since $\tq_{jk;0}$ is the $m\to\infty$ limit of the series $\tq_{jk;N_f}$ with $N_f$ flavors, it is the first term in the $z$-series. After a short calculation, one finds that $\tq_{jk;0}\in\mathbb Q(\varepsilon_{2\rho})$, since the sine functions are the imaginary parts of $\varepsilon_{2\rho}$. Thus necessarily $\mathbb Q(\varepsilon_{2\rho})\subseteq K_\rho$ for all $\rho$. The cyclotomic field can thus be understood as a direct consequence of the Chebyshev geometry of the maximally degenerated SW curve.

\subsection{Pure \texorpdfstring{$\SU(N)$}{SU(N)}}
Conjecture~\ref{conj:DZSWi=GU} and the dictionary holds as well for $s=N_f=0$, which we discuss in this section.

\sss{Universal series for $s=0$.} Clearly, for $s=0$ the partition function must be independent of $m$. This is not completely obvious from the expression
\eqref{eq:ZSW1_physics} since the universal functions are a $z$-series and $z$ has explicit mass dependence. But it is not difficult to check that the $z$-dependence of $\decoup Z_{\mr{SW},1}$ is such that the final answer is independent of $z$. To see this, first note that there are no background fluxes and we can set $\bk_j=0$. The $\chih$ coupling is 
\begin{eqsp}
    \left( \Lambda_{N_f}^{1-N}\alpha_{0}^{12} \beta_{0}^{-8} \tx_{0}^6 \tm_{0}^{-1} \right)=X_0/\rho~,
\end{eqsp}
and from~\eqref{Vs Ws Xs Qs Rs Ts} the universal function $X_0$ is given by 
\begin{eqsp}
    X_0=1~.
\end{eqsp}
Thus, the $z$-dependence dropped out!
The $\tu_{1;0}$-term in the exponential is 
\begin{eqsp}
    \tu^{(0)}_{1;0}=-m^2T_0=-\rho m^2t(1+t)=-\rho m^2z=\frac{\rho}{4^{\frac{1}{\rho}}}\Lambda_0^2~.
\end{eqsp}
This agrees with~\eqref{u1;0} for all $\rho=N$.
Similarly, we have 
\begin{eqsp}
    \tG_0=-\left(\frac{m}{\Lambda_0}\right)^2Q_0=-\frac{1}{2}\left(\frac{m}{\Lambda_0}\right)^2t(1+t)=-\frac{1}{2}\left(\frac{m}{\Lambda_0}\right)^2z=\frac{1}{4^{\frac{1}{\rho}}2}~.
\end{eqsp}
For $\rho=2,3,4$, using the explicit form of the functions $S_0,S_{a,0},Z_{aa,0},Z_{ab,0}$ given in~\cite{Gottsche:2020ass} (see also~\cite[Conjecture 4.4]{Gttsche2024} for expression for general $\rho$), we find that $S_0$ and $S_{a,0}$ are either constants or proportional to $z^{\frac{1}{2}}$. In more detail, we have
\begin{eqsp}
    &\rho=2:\quad Z_0=1,\quad Z_{11,0}=1,\quad S_0=z^{\frac{1}{2}},\quad S_{1,0}=-2z^{\frac{1}{2}}~,
    \\
    &\rho=3:\quad Z_0=2,\quad Z_{11,0}=Z_{22,0}=Z_{12,0}=1,\quad  S_0=-\sqrt{3}z^{\frac{1}{2}},\quad S_{1,0}=S_{2,0}=\sqrt{3}z^{\frac{1}{2}}~,
    \\
    &\rho=4:\quad Z_0=2(2+\sqrt{2}),\quad Z_{11,0}=(2+\sqrt{2})^{-1},\quad Z_{22,0}=(2-\sqrt{2})^{2}/2,
    \\&\quad\quad\quad\quad\, Z_{33,0}=(2+\sqrt{2})^{-1},\quad Z_{12,0}=2(2-\sqrt{2})^{-2},\quad Z_{13,0}=(2+\sqrt{2})^{2},
    \\&\quad\quad\quad\quad\, Z_{23,0}=2(2-\sqrt{2})^{-2}~, \quad S_0=(1+\sqrt{2})z^{\frac{1}{2}},\quad S_{1,0}=-\sqrt{2}z^{\frac{1}{2}},\\&\quad\quad\quad\quad\, S_{2,0}=-2z^{\frac{1}{2}}~,\quad S_{3,0}=0~. 
\end{eqsp}
Note that the functions $S_{a,0}$ are proportional to $z^{\frac{1}{2}}$ and $Z_{ab,0}$ are constants. We expect that this behaviour is so for all $\rho$. 
The coupling to $K^2$ is evaluated to 
\begin{eqsp}\label{eq:K^2_coup_s=0}
   &\rho=2:\quad 2^{K^2}~,
   \\
   &\rho=3:\quad 6^{K^2}~,
   \\
   &\rho=4:\quad (4\sqrt{2}(2+\sqrt{2})^{\frac{3}{2}})^{K^2}~.
\end{eqsp}
Thus, the coupling to $K^2$ in~\eqref{eq:ZSW1_physics} is also independent of $m$. 
Finally, we see that 
\begin{eqsp}
    \tx_{a,0}=-mS_{a,0}~,
\end{eqsp}
is independent of $m$. 

\sss{Pure $\SU(N)$ partition function.} 
The SW contribution $Z_{\SW,0}$ then takes the form
\begin{eqsp}
    \decoup Z_{\SW,0}=N^{-\chih+1}b_N^{K^2}&\exp\left[\frac{1}{4^{\frac{1}{N}}2}\left(2pN\Lambda_0+x^2\right)\right]
    \\
&\times\sum_{\lambda^1,\dots,\lambda^{N-1}}\prod_{a=1}^{N-1}\varepsilon_N^{a(\frac{K}{2}+\lambda^a)c_1}\SW(\lambda^a)e^{-\frac{1}{2^{1/N}}c_{a,N}x\lambda^a}\prod_{1\leq a<b\leq N-1}Z_{ab,0}^{\lambda^a\lambda^b}~,
\end{eqsp}
where $b_N,c_{a,N}$ are constants independent of any parameters. For $N=2,3,4$, $b_N$ is given in~\eqref{eq:K^2_coup_s=0} and $c_{a,N}$ is given by 
\begin{eqsp}
    c_{a,N}=z^{-\frac12}S_{a,0}~.
\end{eqsp}
We can write down the full partition function for $s=0$ as follows. First, since $\decoup=1$, we have $Z_{\mr{SW},0}=\universal_{\alpha}/N$.
As we show below in~\eqref{DZSW sum over U_a}, the full instanton contribution, which indeed is the full partition function for $s=0$, is given by
\begin{eqsp} Z_{\mr{D}}=Z_{\SW}^{\mr{i}}=\sum_{n=0}^{N-1}Z_{\SW,n}=\frac{1}{N}\sum_{n=0}^{N-1}\varepsilon_N^{\frac{n\kappa}{2}}\universal_{\alpha}(\varepsilon_N^{n}z)~,
\end{eqsp}
with $\kappa$~\eqref{kappa} as before.
Since $\universal_{\alpha}$ is independent of $z$, we get 
\begin{eqsp}
   Z_{\mr{D}}=\frac{1}{N}\universal_{\alpha}\sum_{n=0}^{N-1} \varepsilon_N^{\frac{n\kappa}{2}}~.
\end{eqsp}
For $\kappa\in 2\Z+1$, we have 
\begin{eqsp}
    \sum_{n=0}^{N-1} \varepsilon_N^{\frac{n\kappa}{2}}=\frac{1-(-1)^{\kappa}}{1-\varepsilon_N^{\kappa/2}}=\frac{2}{1-\varepsilon_N^{\kappa/2}}=:\varphi_{\kappa,N}~.
\end{eqsp}
For $\kappa\in 2\Z$, we have 
\begin{eqsp}
   \sum_{n=0}^{N-1} \varepsilon_N^{\frac{n\kappa}{2}}=\begin{cases}
       0&\text{if}~\frac{\kappa}{2}\not\equiv 0\bmod N,
      \\
      N&\text{if}~\frac{\kappa}{2}\equiv 0\bmod N.
   \end{cases} 
\end{eqsp}
The full partition function is thus given by  
\begin{eqsp}
    Z_{\mr{D}}=\begin{cases}
       \frac{1}{N}\varphi_{\kappa,N}\universal_{\alpha}&\text{if}~\kappa\in 2\Z+1,
       \\
0&\text{if}~\kappa\in 2\Z,~\frac{\kappa}{2}\not\equiv 0\bmod N,
      \\
      \universal_{\alpha}&\text{if}~\kappa\in 2\Z,~\frac{\kappa}{2}\equiv 0\bmod N. 
    \end{cases}
\end{eqsp}

\sss{Selection rule.}
Let us prove the following selection rule for the partition function without point and surface observable insertions:
\begin{eqsp}\label{eq:u_alpha0Kodd}
    \universal_{\alpha}^{\mr{no\,obs}}=0\quad \text{for}\quad\kappa\in2\Z+1~.
\end{eqsp}
Let us write $\universal_{\alpha}^{\mr{no \,obs}}$ explicitly:
\begin{eqsp}
    \universal_{\alpha}^{\mr{no\,obs}}=\rho^{-\chih+2}b_\rho^{K^2}\sum_{\lambda^1,\dots,\lambda^{\rho-1}}\prod_{j=1}^{\rho-1}\varepsilon_\rho^{j(\frac{K}{2}+\lambda^j)c_1}\SW(\lambda^j)\prod_{1\leq j<k\leq \rho-1}Z_{jk,0}^{\lambda^j\lambda^k}~.
\end{eqsp}
Change variables in the product to $j\to \rho-j$, then changing the variables in the sum to $\lambda^{\rho-j}\to-\lambda^j$ and using~\eqref{SW -} together with the fact
\footnote{Recall that in general, $Z_{jk,s}$ satisfies $\upsilon Z_{jk,s}=Z_{(\rho-k)(\rho-j),s}$ where $\upsilon$ denotes the monodromy operation which takes $z\to e^{2\pi i}z$ (see~\eqref{eq:SYZ_monodromy}). For $s=0$, we see (for $\rho=2,3,4$ and conjecture for all $\rho$) that $Z_{jk,0}$ is $z$-independent and hence monodromy free.}
that $Z_{jk,0}=Z_{(\rho-k)(\rho-j),0}$,
we find
\begin{eqsp}
    \universal_{\alpha}^{\mr{no\,obs}}&=(-1)^{(\rho-1)(\chih+c_1^2)}\universal_{\alpha}^{\mr{no\,obs}}
    \\
    &=(-1)^{\kappa}\universal_{\alpha}^{\mr{no\,obs}}~.
\end{eqsp}
This immediately implies~\eqref{eq:u_alpha0Kodd}. It is also a consequence of the monodromy analysis~\eqref{upsilon universal}: If $\universal_\alpha$ does not depend on $z$, it is a constant $z$-series, and thus $\upsilon\universal_\alpha=\universal_\alpha$. Therefore, $\kappa$ odd implies $\universal_\alpha=0$. 

\sss{Example: K3.}
The partition function further simplifies for $X=K3$. For $K3$, $\chih=2$ is even and hence for zero 't Hooft flux $f_2(P)=0$, $\frac{\kappa}{2}\not\equiv 0\bmod N$. Thus
\begin{eqsp}
 Z_{\mr{D}}^{K3,f_2(P)=0}=0~.   
\end{eqsp}
To get a non-zero partition function, we need to turn on 't Hooft flux. Note that since the cohomology lattice $H^2(K3,\Z)$ is even, we cannot have $\kappa$ odd for $X=K3$. We can choose $f_2(P)$ appropriately such that $\frac{\kappa}{2}\equiv 0\bmod N$. Moreover, there is only one basic class $\lambda=0$ and the SW invariant is $\SW(0)=1$. Also, since $X_0=1$ and $K^2=0$ for $K3$, we get 
\begin{eqsp}
\universal_{\alpha}^{K3}=\exp\left[\frac{(-1)^{N+1}}{4^{1/N}}\left(pN\Lambda_0+S^2\right)\right]e^{\frac{\pi i (N-1)(Kc_1)}{2}} ~.  
\end{eqsp}
In particular, without observables, we have 
\begin{eqsp}
    \universal_{\alpha}^{K3,\mr{no\,obs}}=e^{\frac{\pi i (N-1)(Kc_1)}{2}}~, 
\end{eqsp}
which is a sign since $K$ is a characteristic vector so that $Kc_1=c_1^2\bmod 2$ and $c_1^2$ is even since $H^2(K3,\Z)$ is an even lattice. 
We thus conclude that
\begin{eqsp}\label{ZD K3}
    Z^{K3,\text{no obs}}_{\mr{D}}=\begin{cases}
% \frac{1}{N}\varphi_{\kappa,N}&\text{if}~\kappa\in 2\Z+1,
%        \\
0&\text{if}~\kappa\in 2\Z,~\frac{\kappa}{2}\not\equiv 0\bmod N,
      \\
      e^{\frac{\pi i (N-1)(Kc_1)}{2}}&\text{if}~\kappa\in 2\Z,~\frac{\kappa}{2}\equiv 0\bmod N. 
    \end{cases}
\end{eqsp}
This is compatible with the known answer for $N=2$~\cite{Aspman:2022sfj,Aspman:2023ate,OGrady1992DonaldsonsPF}.
Since $K3$ is hyperkähler, the twisted and untwisted supersymmetric partition functions can be identified, and~\eqref{ZD K3} is the full untwisted partition function.

\subsection{QCD beyond \texorpdfstring{$\SU(3)$}{SU(3)}}
\label{sec:beyond SU(3)}

\sss{Relation between $z$ and $m$.}
In the following section, we will provide strong evidence for conjecture~\ref{conj:DZSW0=Ua} at rank $N=\rho=3$. An important component of the conjecture for all $N$ is the identification~\eqref{z in terms of m}
\bea\label{z in terms of m, 5.6}
z=-\frac{\Lambda_0^2}{4^{\frac1N}m^2}~,
\eea
between the instanton counting parameter $z$ and the mass $m$. To justify this identification, we use our dictionary, in particular $\tu_{1;s}^{(0)}=-m^2 T_s$~\eqref{Ts dictionary}. With the change of variables $z=z(t)$~\eqref{change of variables}, we have $z=t+\CO(t^2)$ for all values of $\rho$ and $s$. In the decoupling limit $m\to\infty$, we have $z\to0$ and thus $t\to 0$, therefore $t=z+\CO(z^2)$. 
Assuming~\eqref{Vs Ws Xs Qs Rs Ts}
\bea\label{Ts section 5.6}
T_s(t) = \rho t (1+ \tfrac{1}{2}(1-\tfrac{s}{\rho})(2-\tfrac{s}{\rho})t)~,
\eea
we can compute the limit of $\tu_{1;s}^{(0)}$ as $m\to \infty$, and compare with the definition~\eqref{N=1 sing definition} of the $\CN=1$ vacua of $\SU(N)$ QCD: they flow to the $\CN=1$ vacua of the pure $\SU(N)$ theory upon decoupling all hypermultiplets. The $a=1$ coordinate $\tu_{1;0}^{(0)}$ of the $n=0$ singularity for the pure $s=0$ theory is determined in~\eqref{u1;0}; it is $\tu_{1;0}^{(0)}=2^{-\frac 2N}N\Lambda_0^2$. Using these asymptotics, we compute the $z$-series
\bea
\frac{\tu_{1;s}^{(0)}}{\tu_{1;0}^{(0)}}=-\frac{4^{\frac1N}m^2}{\Lambda_0^2}z+\CO(z^2)~,
\eea
such that the leading term is 1 if and only if~\eqref{z in terms of m, 5.6}. This justifies the identification for all ranks $N$.

We should note that this disagrees with the previous $\SU(2)$ result~\cite{Furrer:2026byd} by a minus sign. The discrepancy originates from a choice of convention in~\cite{Furrer:2026byd}, where the $n=0$ vacuum is $-\Lambda_0^2$ rather than $+\Lambda_0^2$. Thus, the vacuum where all the series were matched is the different choice of vacuum, $n=1$ in our notation, rather than $n=0$. The convention $\tu=\pm\Lambda_0^2$  used in most of the previous literature corresponds to the monopole ($+$) and dyon ($-$) singularities, and interchanging those corresponds for $\SU(2)$ to a sign flip $z\mapsto -z$. While this convention is rather inconsequential for $N=2$, our uniform discussion for all $N$ strongly favours the convention where $\tu_{1;0}^{(0)}$ is real and positive for all $N$, and in that case we have the minus sign in~\eqref{z in terms of m, 5.6} for all $N$.

\sss{Factorisation beyond $\SU(3)$.}
This argument is grounded in the assumption that~\eqref{Ts section 5.6} is correct for all $s$ and all ranks $\rho$. To provide evidence for it beyond $\rho=2$ and $3$, we aim to find the $\CN=1$ vacua of the $\SU(N)$ theories for $N\geq 4$, and hope to find agreement with the proposal. 

To do so, we use the factorisation technique to parametrise the $\CN=1$ points. Recall from~\eqref{curve factorised_sec3} that the Seiberg--Witten curve factorises at the $\CN=1$ vacua, $Y^2=H(X)^2F(X)$. We simply demand factorisation by writing 
\bea
H(X)&=\sum_{n=0}^{N-1}h_nX^n~, \\
F(X)&=\sum_{n=0}^2f_nX^n~, 
\eea
where $h_{N-1}=f_2=1$, and solve the system of equations obtained by setting to zero all coefficients in the polynomial $Y^2-H(X)^2F(X)$. The solution space is then parametrised by one undetermined coefficient, for instance $h_{N-2}$. In this way, one can obtain universal functions $T_s^{(a)}$ for all $\rho$, $a=1,\dots, \rho-1$ and $s=1,\dots, 2\rho-1$ such that
\bea\label{ua in terms of Ta}
\tu_{a;s}=-m^{a+1}T_s^{(a)}~.
\eea
By direct calculation, we find universal formulas
\bea
T_s^{(1)}(t)&= \rho t (1+ \tfrac{1}{2}(1-\tfrac{s}{\rho})(2-\tfrac{s}{\rho})t)~, \\
T_s^{(2)}(t)&=-st^2(1+\tfrac23(1-\tfrac s\rho)(2-\tfrac s\rho)t)~, \\
&\;\;\vdots
\eea
The first one of course agrees with $T_s$ predicted by~\eqref{Vs Ws Xs Qs Rs Ts} in Conjecture~\ref{conj_GK}.
For instance, for $\SU(4)$ with $s=1,\dots, 4$, we find the map $h_{2}=-\frac s4mt$ from the undetermined coefficient to the parameter $t$, but for $\SU(5)$ this change of variables becomes more complicated. For the coordinates $a\geq 3$, the series $T_s^{(a)}$ take a more complicated pattern.

One reason for the complication could be the choice of coordinates on the Coulomb branch. We write $P(X)=\det(X-\phi)=X^N-\sum_a u_a X^{N-a-1}$, such that the $u_a$ are coordinates in an elementary-symmetric basis. These coordinates are themselves combinations of the Casimir invariants
\bea
v_k=\frac 1k{\rm Tr}\, \phi^k=\frac1k\sum_{m=1}^N\phi_m^k~,
\eea
where $\langle\phi\rangle=\diag(\phi_1,\dots,\phi_N)$ (keeping in mind that $\sum_m\phi_m=0$).
The two sets of Coulomb branch coordinates are related by a change of basis. To see this, let $e_k$ denote the elementary symmetric polynomials in the eigenvalues of $\phi$, with $e_0=1$ and $e_1=0$. The Newton identities read
\bea
e_k=\frac1k\sum_{j=2}^k(-1)^{j-1}jv_je_{k-j}~.
\eea
In our convention, $u_a=(-1)^a e_{a+1}$, and thus the change of variables from the trace Casimirs $v_k$ to the $u_a$ is generated recursively. For instance, we have
\bea
u_1&=v_2~, \\
u_2&=v_3~, \\
u_3&=v_4-\tfrac12v_2^2~, \\
u_4&=v_5-v_2v_3~, \\
\eea
and so forth. Clearly, for $N\geq 4$ these two bases differ.
One can then replace the parametrisation~\eqref{ua in terms of Ta} by $\tv_{a;s}=-m^{a+1}\widetilde T_s^{(a)}$ and solve for the series $\widetilde T_s^{(a)}$ instead. This turns out to simplify some coefficients for $\SU(4)$ significantly; however, we have not been able to identify a simple closed-form result for $\widetilde T_s^{(3)}$. Another complication appears to come from the cases $N_f\geq N$, where baryonic branches appear and the characteristic polynomial $P(X)$ of the matrix $\phi$ receives a correction in the SW curve~\eqref{SU(N) curve} by the mass polynomial $T(X)$. Some subtleties for these cases indeed appear for $N=2$ and $N=3$.
We hope to address these issues in future work.

\sss{$\SU(N)$ QCD partition function on K3 for all $N$.}
Finally, let us give a non-trivial example of a QCD partition function. For $K3$, we have $\universal_\alpha=X_s^2$, where $X_s$ is given in~\eqref{Vs Ws Xs Qs Rs Ts}. Conjecture~\ref{conj:DZSW0=Ua} then gives $\decoup Z_{\SW,0}=\frac1N X_s^2$, where $\decoup=(\frac{m}{\Lambda_{N_f}})^{\frac{N_f}{N}(N^2+1)}$. Turning off $\bar f_2(P)=0$, we have $\kappa=2(N^2-1)$, such that $\frac12\vd M=Nc_2-(N^2-1)\equiv 1 \mod N$.
Conjecture~\ref{conj:DZSWi=GU} then gives
\bea\label{K3 inst}
Z_{\SW}^{\rm i}=\left.\frac{1}{\decoup}X_s^2\right|_{\{z^{1\bmod N}\}}~.
\eea
This is the exact answer for the instanton component to the $K3$ partition function for $\SU(N)$ with $s=N_f=1,\dots, 2N-1$ flavors. 

To determine explicitly the leading term for large $m$, we determine $X_s^2=1-\frac{\rho^2-1}{\rho}sz+\CO(z^2)$ using the change of variables, where the linear term is leading in~\eqref{K3 inst}. Combining with the decoupling factor, this gives
\bea
Z_{K3}^{\rm i}=\frac{(N^2-1)N_f}{4^{\frac1N}N}\left(\frac{\Lambda_{N_f}}{m}\right)^{2+N_fN}+\CO\left(m^{-(2N+2+N_f(N-1))}\right)~.
\eea
We note again that for $s\geq 1$ this is never the full partition function, since there are also contributions from the reducible branch~\cite{FMM:2026}. We hope to address this contribution in future work.

\section{Algebraic functions from SU(3) couplings}
\label{sec:SU(3) calculation}
In this final section, we compute the physical couplings necessary to formulate topological correlators from the Seiberg--Witten curve for $\SU(3)$ QCD with $N_f=0,\dots, 5$ massive hypermultiplets. This calculation serves as the main check of our Conjecture~\ref{conj:DZSW0=Ua}.
\subsection{Singularities}
\label{sec:singularities}
We start with the $\SU(N)$ Seiberg--Witten curve~\eqref{SU(N) curve}, which for $N=3$ with $0\leq N_f\leq 5$ hypermultiplets in the fundamental representation with equal mass $m$  take the form
\bea\label{SU(3) curve}
Y^2&=P_{N_f}(X)~,
\eea
as in~\eqref{SU(N) curve}. 
Let us write these out explicitly:\footnote{A common notation for the two Casimirs is also $u\coloneqq u_1$ and $v\coloneqq u_2$.} 
\allowdisplaybreaks
\begin{subequations}
\begin{align}
N_f=0\colon \quad Y^2&=(X^3-u_1X-u_2)^2-\Lambda_0^6~, \\
N_f=1\colon \quad Y^2&=(X^3-u_1X-u_2)^2-\Lambda_1^5(X+m)~, \\
N_f=2\colon \quad Y^2&=(X^3-u_1X-u_2)^2-\Lambda_2^4(X+m)^2~, \\
N_f=3\colon \quad Y^2&=(X^3-u_1X-u_2+\tfrac14\Lambda_3^3)^2-\Lambda_3^3(X+m)^3~, \\
N_f=4\colon \quad Y^2&=(X^3-u_1X-u_2+\tfrac14\Lambda_4^2(X+4m))^2-\Lambda_4^2(X+m)^4~, \\
N_f=5\colon \quad Y^2&=(X^3-u_1X-u_2+\tfrac14\Lambda_5(X^2+5mX+10m^2))^2-\Lambda_5(X+m)^5~.
\end{align}
\end{subequations} 
%The decoupling limit is $\Lambda_1\to 0$ and $m\to \infty$ with $m\Lambda_1^5=\Lambda_0^6$.
%
The decoupling limit for generic masses is $m_{N_f}\to\infty$, $\Lambda_{N_f}\to 0$ with 
\bea\label{decoupling limit separate}
\Lambda_{N_f-1}^{6-N_f+1}=m_{N_f} \Lambda_{N_f}^{6-N_f}~.
\eea
For the equal-mass decoupling, we have 
\bea\label{decoupling limit}
m^{N_f} \Lambda_{N_f}^{6-N_f}=\Lambda_0^6~.
\eea
The singular locus for these curves is given by the zero locus of the discriminant $\Delta_{N_f}$~\eqref{SU(N) discriminant}. 
One finds that $\Delta_{N_f}$ is a polynomial in $u_1$ and $u_2$ of degree $6+N_f$ in $u_1$ and $4+N_f$ in $u_2$, and it has total mass dimension $12+N_f$.

\begin{figure}[ht]\centering
	\includegraphics[scale=0.7]{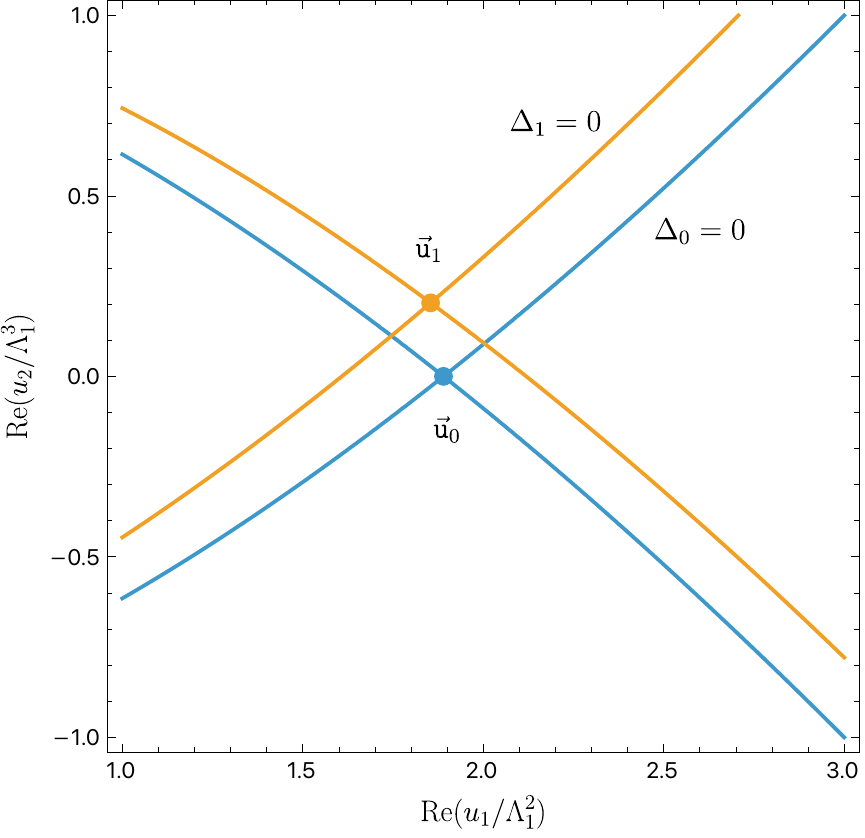}
	\caption{Plot of the singular lines $\Delta_{N_f}=0$ for the $\SU(3)$ theory with $N_f=0$ and $1$, on a real slice through the complex two-dimensional Coulomb branch parametrised by $(u_1,u_2)$. The singular lines intersect at the $\CN=1$ vacua $\vec\tu_{N_f}$. We have chosen the mass $m/\Lambda_1^2=2$, where for $m\to\infty$ the singular locus $\Delta_1=0$ approaches the pure locus $\Delta_0=0$.}
\label{fig:singularlines}
\end{figure}

\sss{The $\CN=1$ vacua.}
Above, we argued that the rank $N$ Segre numbers are given by the contribution from a subset of codimension $r=N-1$ singular points for $\SU(N)$ QCD, with equal and large masses, namely those which become the $\CN=1$ vacua of pure $\SU(N)$ in the decoupling limit. We use this definition~\eqref{N=1 sing definition} in this section as well, and want to compute the moduli $\vec \tu_{N_f}^{(n)}=(\tu_{1;N_f}^{(n)},\tu_{2;N_f}^{(n)})$. See figure~\ref{fig:singularlines} for a picture of the large mass deformation of the pure $\SU(3)$ singular locus.\footnote{See also~\cite[Fig. 4]{Klemm:1995wp} for picture of the global singular locus of pure $\SU(3)$.}
Due to the $\mathbb Z_3$ symmetry, which we will prove in detail below, it is enough to focus on the representative $n=0$, which we denote by $(\tu_{1;N_f},\tu_{2;N_f})$ or $(\tu_1,\tu_2)$,  and the other two vacua $n=1,2$ follow by phase rotations.
As pointed out, the equations~\eqref{sing points pure SU(N)} give a coupled system for all the Casimirs. To obtain the $u_a$ individually, we take a different approach. First, we notice that for $\SU(3)$ and $N_f\geq 2$ the discriminants factor into a simple piece $p_{N_f}$ of degree $N_f$ and a remainder $\widetilde \Delta_{N_f}$:
\bea\label{tilde Delta}
\Delta_{N_f}=p_{N_f}^{N_f} \widetilde \Delta_{N_f}~,
\eea
where the simple piece is explicitly given by
\bea
p_{N_f}=4m^3-4mu_1+4u_2-(N_f-1)(N_f-2)m^{N_f-3}\Lambda_{N_f}^{6-N_f}~.
\eea
This is because for $N_f\geq2$, the polynomial $C(X)$ is divisible by $(X+m)^2$ iff $p_{N_f}=0$, in which case $P_{N_f}$ has a repeated root and $\Delta_{N_f}$ vanishes. We discard these vacua from the calculation since they are quark singularities and originate from the collision of the masses.\footnote{To circumvent this issue, one could start with the curve for generic masses, for which $\Delta_{N_f}$ does not factor automatically. However, the discriminants grow very quickly with $N_f$ generic masses.}

We factor out $p_{N_f}$ since we are interested in the intersecting singularities, and the discriminant of $\Delta_{N_f}$ w.r.t. $u_a$ vanishes whenever it has a double root. The discriminants $\Disc_{u_a}\widetilde\Delta_{N_f}$ are then non-trivial, and factor into a product of several terms. We conjecture that the coordinates $(\tu_1,\tu_2)$ of the $\CN=1$ points are given by the zero locus of the \emph{quadratic} factors of these discriminants:
\bea
\Disc_{u_a}\widetilde \Delta_{N_f}=d_{a}^{(N_f)}(u_b)^2\times \dots~.
\eea
To make this more precise, we define for any univariate polynomial $f$ the greatest common divisors $G_k(f)\coloneqq \gcd(f,f',\dots, f^{(k)})$ with its derivatives. One can show that the quadratic factor of $f$ is given by $G_1(f)G_3(f)/G_2(f)^2$. We therefore define the two univariate polynomials
\bea
d_{a}^{(N_f)}=\frac{G_1(\Disc_{u_a}\widetilde \Delta_{N_f})G_3(\Disc_{u_a}\widetilde \Delta_{N_f})}{G_2(\Disc_{u_a}\widetilde \Delta_{N_f})^2}~,
\eea
such that $d_{1}^{(N_f)}$ is a polynomial in $u_2$ and $d_{2}^{(N_f)}$ is a polynomial in $u_1$.
We list the polynomials for $N_f=0,\dots, 5$:\footnote{For $N_f=0$, there is a further degeneracy which prevents us from using it to calculate $u_2$. However, for the $N_f=0$ cases we have already obtained all the singular moduli in~\eqref{u N=1 SU(N)}. }
\allowdisplaybreaks
%\begin{equation}
{\small\begin{align*}
d_1^{(0)}&=1~,\qquad d_2^{(0)}=4u_1^3-27\Lambda_0^6~,\\
d_1^{(1)}&=314928m^6u_2^3+2519424m^3u_2^4-34992m^4u_2^2\Lambda_1^5-3499200mu_2^3\Lambda_1^5-19683m^5\Lambda_1^{10} \\ &-162000m^2u_2\Lambda_1^{10}+64\left(78732u_2^5+3125\Lambda_1^{15}\right)~,\\
d_2^{(1)}&=432m^4u_1^3-1152m^2u_1^4+768u_1^5-2916m^5\Lambda_1^5+8100m^3u_1\Lambda_1^5-6000mu_1^2\Lambda_1^5+3125\Lambda_1^{10}~,\\
d_1^{(2)}&=39366m^3u_2^3-19683u_2^4+4374m^2u_2^2\Lambda_2^4-19683m^4\Lambda_2^8+10368mu_2\Lambda_2^8-1024\Lambda_2^{12}~,\\
d_2^{(2)}&=729m^4\Lambda_2^4+4\left(3u_1^2+\Lambda_2^4\right)^2-108m^2\left(u_1^3+3u_1\Lambda_2^4\right)~,\\
d_1^{(3)}&=16u_2^3-27m^3\Lambda_3^6~,\qquad d_2^{(3)}=-4u_1^3+27m^3\Lambda_3^3~, \tag{\theequation}\\
d_1^{(4)} &=19683u_2^3-21870mu_2^2\Lambda_4^2-78732m^5\Lambda_4^4+4455m^2u_2\Lambda_4^4+80m^3\Lambda_4^6-4u_2\Lambda_4^6~,\\
d_2^{(4)}&=1296u_1^3-8748m^4\Lambda_4^2-72u_1^2\Lambda_4^2+6m^2\Lambda_4^4+u_1\left(-1944m^2\Lambda_4^2+\Lambda_4^4\right)~,\\
d_1^{(5)}&=20155392u_2^3+80m^2u_2\Lambda_5^2\left(1162188m^2+2455m\Lambda_5-6\Lambda_5^2\right) +256u_2^2\Lambda_5(-306180m^2\\ &-405m\Lambda_5+\Lambda_5^2)-m^4\Lambda_5^2(157464000m^3+35685200m^2\Lambda_5 +91056m\Lambda_5^2-225\Lambda_5^3)~,\\
d_2^{(5)}&=3888u_1^3-16u_1^2\Lambda_5\left(135m+\Lambda_5\right)-60m^2u_1\Lambda_5\left(243m+2\Lambda_5\right)-9m^4\Lambda_5\left(2916m+25\Lambda_5\right)~.
\end{align*}}
%\end{equation} 
By direct calculation, $d_{a}^{(N_f)}$ have degree $\deg d_{a}^{(N_f)}=3,5,4,3,3,3$ for $N_f=1,\dots, 5$. In the decoupling limit~\eqref{decoupling limit}, the leading terms for $m\to \infty$ are $d_1^{(N_f)}\sim u_2^3$ and $d_2^{(N_f)}\sim 4u_1^3-27\Lambda_0^6$, confirming that the large mass limit singles out the three $\CN=1$ vacua~\eqref{u N=1 SU(N)} 
\bea
\vec \tu_{0}^{(n)}=\left(3/2^{\frac23}\Lambda_0^2\varepsilon_3^n,0\right)
\eea
of the pure $N_f=0$ $\SU(3)$ theory, with $n=0,1,2$. To obtain the three $\CN=1$ vacua of the theory with $N_f$ flavors, we require that in the decoupling limit
\bea
\lim_{m\to\infty}\vec\tu_{N_f}^{(n)}=\vec\tu_{0}^{(n)}
\eea
for $n=0,1,2$, which defines the labelling at the same time. In this way, we obtain three singularities $\tu_{N_f}^{(n)}$ from the equations 
\bea\label{SU(3) N=1 vacua def}
\CN=1 \text{ vacua}\colon \qquad d_{a}^{(N_f)}(u_b)=0~.
\eea
Due to a $\mathbb Z_3$ symmetry, it is enough to focus on one of the three $\CN=1$ points, which we also denote by $\tu_{N_f}\coloneqq \tu_{N_f}^{(0)}$. 

\sss{The algebraic functions $T_s^{(1)}$ and $T_s^{(2)}$.} 
The first step to establish the dictionary between topological QCD and Segre numbers is to express the relevant Coulomb branch couplings at the singular points in terms of algebraic functions. To start, we  express the $\CN=1$ vacua in terms of universal functions, for which we introduce the dimensionless quantity
\bea\label{z in terms of m SU(3)}
z=-\frac{\Lambda_0^2}{\sqrt[3]{4}m^2}~,
\eea
and change variables
\bea
z=t(1+(1-\tfrac s3)t)^{1-\frac s3}~,
\eea
where $s=N_f$.
It is a straightforward calculation to verify that in the decoupling limit~\eqref{decoupling limit},
to all orders for $N_f=1,\dots, 5$, we can express the solutions to~\eqref{SU(3) N=1 vacua def} as
\bea\label{u v T P}
\tu_{1,N_f}&=-m^2 T_{s}^{(1)}~, \\
\tu_{2,N_f}&=-m^3 T_{s}^{(2)}~, 
\eea
where
\bea
T_s^{(1)}(t)&=3t(1+\tfrac12(1-\tfrac s3)(2-\tfrac s3)t)~, \\
T_{s}^{(2)}(t)&=-st^2(1+\tfrac23(1-\tfrac s3)(2-\tfrac s3)t)~.
\eea
The function $T_s=T_s^{(1)}$ agrees precisely with~\eqref{Vs Ws Xs Qs Rs Ts} of Conjecture~\ref{conj_GK}, while $T_{s}^{(2)}$ for the cubic Casimir is new. 
For $N_f=5$, we implicitly include a shift,
\bea\label{Nf=5 shift}
\tu_{2,5}=-m^3T_5^{(2)}-\tfrac14\Lambda_5 \tu_{1,5}~,
\eea
with $T_5^{(2)}$ as above. We comment below on this shift.

\sss{$\mathbb Z_3$ symmetry.}
After having identified $\vec\tu_{N_f}= \vec\tu_{N_f}^{(0)}$ with the universal series $T_s^{(1)}$ and $T_s^{(2)}$ for all cases, we can obtain the other two $\CN=1$ vacua $\vec\tu_{N_f}^{(1)}$ and $\vec\tu_{N_f}^{(2)}$ from a $\mathbb Z_3$ symmetry of the curve. To see this, we eliminate $m$ in $d_{a}^{(N_f)}(u_b)$ using~\eqref{z in terms of m SU(3)} and insert the double scaling relation~\eqref{decoupling limit} together with the relations~\eqref{u v T P}. By direct calculation, we find up to overall constants and a monomial in $z$ (which is immaterial for the solutions to $d_{a}^{(N_f)}=0$) that $d_{a}^{(N_f)}(u_b)$ admits a Taylor series in $z^3$. This means that the equation defining the $\CN=1$ vacua has a symmetry $z\mapsto \varepsilon_3z$, where $\varepsilon_3=e^{2\pi i/3}$. From $\tu_{a,N_f}^{(0)}$ this immediately gives us the other two vacua:
\bea
\tu_{1,N_f}^{(n)}=-m^2 T_{N_f}^{(1)}(\varepsilon_3^n z)~,  \\
\tu_{2,N_f}^{(n)}=+m^3 T_{N_f}^{(2)}(\varepsilon_3^n z)~. 
\eea
This $\mathbb Z_3$ symmetry is transferred to all other Coulomb branch functions. It also provides a proof of the scaling arguments of Section~\ref{sec:SW geometry}, see in particular~\eqref{Z_N action on z}--\eqref{Z_N action on T}. Indeed, the functions $T_a$ defined by $u_a=m^{a+1}T_a$ are simply $-T_{s}^{(a)}$. 

\sss{Factorisation of the curve.}
As discussed around~\eqref{curve factorised_sec3}, at the $\CN=1$ points $(u_1,u_2)=(\tu_1,\tu_2)$, the Seiberg--Witten curves~\eqref{SU(3) curve} factorises into two quadratic polynomials as 
\bea\label{curve factorisation}
y^2=H_{N_f}^2(X)F_{N_f}(X)~.
\eea
We label the roots of those polynomials by
\bea
H_{N_f}(X)&=(X-\ts_{N_f}^+)(X-\ts_{N_f}^-)~, \\
F_{N_f}(X)&=(X-\tr_{N_f}^+)(X-\tr_{N_f}^-)~.
\eea
This factorisation is very useful since several CB couplings can be obtained directly from the curve in this factorised form. In~\eqref{dimensionless curve N=1}, we have obtained a simplified curve in the scaling limit, but, as one can check, this curve does not itself provide the factorisation~\eqref{curve factorisation}. 

We obtain the roots $\tr_{N_f}^\pm$ and $\ts_{N_f}^\pm$ by inserting~\eqref{u v T P} into the curve and finding appropriate rescalings of the coordinates $X$ and $Y$ for which it factorises. For $s=N_f=1,\dots, 5$, we find
\bea
\tr_s^\pm =-m\big(
\tfrac{s}{3}t
\pm 2i\,t^{\frac12}
\big(1+(1-\tfrac{s}{3})t\big)^{\frac12}
\big),
\eea
while\footnote{For $s=5$ we include the shift~\eqref{Nf=5 shift}, which makes it more difficult to bring it into a closed-form expression for all $s$}
\allowdisplaybreaks
\begin{subequations}
\begin{align}
\ts_1^\pm&=m\big(\tfrac16t\mp i\big(t(1+\tfrac{5}{12}t)\big)^{\frac12}\big)~, \\
\ts_2^\pm&=m\big(\tfrac13t\mp i\big(t(1+\tfrac13t)\big)^{\frac12}\big)~,  \\
\ts_3^\pm&=m\big(\tfrac12t\mp i\big(t(1+\tfrac34t)\big)^{\frac12}\big)~, \\
\ts_4^\pm&=m\big(\tfrac23t\mp i(1+\tfrac23t)(1-\tfrac13t)^{-\frac12}t^{\frac12}\big)~,  \\
\ts_5^\pm&=m(1-\tfrac23t)^{-2}\left(t(\tfrac56-\tfrac{10}{9}t-\tfrac{7}{54}t^2)\mp i(1+\tfrac13t)^2(1-\tfrac{11}{12}t)^{\frac12}t^{\frac12}\right)~.
\end{align}
\end{subequations} 
For $m\to\infty$, they give the branch points
\bea\label{s^pm}
\ts_{N_f}^{\pm}=\pm 2^{-\frac13}\Lambda_0+\dots,
\eea
hence the $\pm$ labels.

\subsection{Period matrix}
\label{sec:period matrix}
\begin{figure}[ht]\centering
	\includegraphics[scale=1]{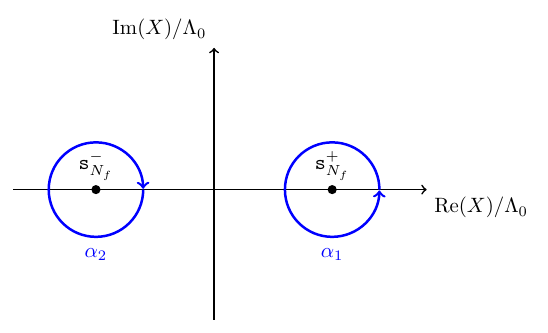}
	\caption{Contours defining dual periods}
\label{fig:contour}
\end{figure}

To determine the period matrix at the strong coupling singularities, recall the matrix $A$~\eqref{A_ab} from Section~\ref{sec:SW geometry}.
Focusing on $\SU(3)$, the two differentials~\eqref{omega_j differentials} are $\omega_1=(2\pi i)^{-1}X\mathrm dX/Y$ and $\omega_2=(2\pi i)^{-1}\mathrm dX/Y$, such that
\bea
A=\frac{1}{2\pi i}\begin{pmatrix}
\displaystyle\int_{\alpha_1}\frac{X\mathrm dX}{Y} &
\displaystyle\int_{\alpha_1}\frac{\mathrm dX}{Y} \\
\displaystyle\int_{\alpha_2}\frac{X\mathrm dX}{Y} & 
\displaystyle\int_{\alpha_2}\frac{\mathrm dX}{Y} \\
\end{pmatrix}~.
\eea
We aim to compute this matrix at the dedicated $\CN=1$ point, after an appropriate duality transformation. We make use of the fact that at the point $(u_1,u_2)=(\tu_1,\tu_2)$, the SW curve factorises as~\eqref{curve factorisation}. Then the period integrals can be evaluated using a residue prescription, as in~\eqref{tA ab}. For this, we need to prescribe the contours $\alpha_1$ and $\alpha_2$. We define $\alpha_1$ as a small counter-clockwise contour around $\ts_{N_f}^+$, while $\alpha_2$ is a small clockwise contour around $\ts_{N_f}^-$ (see Fig~\ref{fig:contour}).
The relative orientation of the two contours is fixed by our choice of canonical symplectic basis for the cycles of the SW curve. 
This prescription allows us to compute the period matrix $A_{ab}$ at strong coupling, which we denote by 
\bea
\tA_{ab}=\left. A_{ab}\right|_{\vec{\tu}_{N_f}}~.
\eea
Since $\alpha_i$ are small contours around $\ts_{N_f}^\pm$, we compute
\bea
\frac{1}{2\pi i}\int_{\alpha_{1,2}}\frac{X^j\mathrm dX}{Y}&=\mathop{\mathrm{Res}}\limits_{X=\ts^\pm_{N_f}}\frac{X^j}{H_{N_f}(X)F_{N_f}(X)^{\frac12}}=\frac{(\ts^\pm_{N_f})^j}{H_{N_f}'(\ts^\pm_{N_f})F_{N_f}(\ts^\pm_{N_f})^{\frac12}}~,
\eea
where $H_{N_f}'(\ts^\pm_{N_f})=\ts^\pm_{N_f}-\ts^\mp_{N_f}$.
Keeping the orientation in mind, this gives explicitly,
\bea\label{tA}
\tA=
\begin{pmatrix}
\displaystyle
\frac{\ts_{N_f}^+}
{(\ts_{N_f}^+-\ts_{N_f}^-)
\sqrt{(\ts_{N_f}^+-\tr_{N_f}^-)
(\ts_{N_f}^+-\tr_{N_f}^+)}}
&
\displaystyle
\frac{1}
{(\ts_{N_f}^+-\ts_{N_f}^-)
\sqrt{(\tr_{N_f}^--\ts_{N_f}^+)
(\tr_{N_f}^+-\ts_{N_f}^+)}}
\\[1.2em]
\displaystyle
\frac{\ts_{N_f}^-}
{(\ts_{N_f}^+-\ts_{N_f}^-)
\sqrt{(\ts_{N_f}^--\tr_{N_f}^-)
(\ts_{N_f}^--\tr_{N_f}^+)}}
&
\displaystyle
\frac{1}
{(\ts_{N_f}^+-\ts_{N_f}^-)
\sqrt{(\tr_{N_f}^--\ts_{N_f}^-)
(\tr_{N_f}^+-\ts_{N_f}^-)}}
\end{pmatrix}~.
\eea
For the calculation of topological correlators, we furthermore require the inverse periods, $J\coloneqq A^{-1}$, and their value at strong coupling:
\bea\label{tJ}
\tJ=\tA^{-1}~.
\eea
Inverting~\eqref{tA} gives explicitly,
\bea
\tJ^{(N_f)}=
\begin{pmatrix}
\sqrt{(\tr_{N_f}^--\ts_{N_f}^+)
(\tr_{N_f}^+-\ts_{N_f}^+)}
&
-\sqrt{(\tr_{N_f}^--\ts_{N_f}^-)
(\tr_{N_f}^+-\ts_{N_f}^-)}
\\[1.2em]
-\ts_{N_f}^-
\sqrt{(\tr_{N_f}^--\ts_{N_f}^+)
(\tr_{N_f}^+-\ts_{N_f}^+)}
&
\ts_{N_f}^+
\sqrt{(\tr_{N_f}^--\ts_{N_f}^-)
(\tr_{N_f}^+-\ts_{N_f}^-)}
\end{pmatrix}~.
\eea
From this, we extract the top row and its determinant:
\bea\label{x_Nf calculation}
\tx_{a,N_f}\coloneqq\left.\tJ^{(N_f)}_{1a}\right|_{\vec{\tu}_{N_f}}~, \quad  
\tx_{N_f}\coloneqq\left.\det_{ab} \tJ_{ab}^{(N_f)}\right|_{\vec{\tu}_{N_f}}~,
\eea
where we read off:
\bea\label{xNf result}
\tx_{1,N_f}&=+\sqrt{(\tr_{N_f}^--\ts_{N_f}^+)
(\tr_{N_f}^+-\ts_{N_f}^+)}~, \\
\tx_{2,N_f}&=-\sqrt{(\tr_{N_f}^--\ts_{N_f}^-)
(\tr_{N_f}^+-\ts_{N_f}^-)}~, \\
\tx_{N_f}&=\sqrt{(\tr_{N_f}^--\ts_{N_f}^-)
(\tr_{N_f}^+-\ts_{N_f}^-)
(\tr_{N_f}^--\ts_{N_f}^+)
(\tr_{N_f}^+-\ts_{N_f}^+)}
(\ts_{N_f}^+-\ts_{N_f}^-)~.
\eea
Using these, we determine the series $S_{a,s}$ of the dictionary for $a=1,2$:
\bea
-\frac 1m \tx_{a,N_f}=S_{a,s}~,
\eea
We find perfect agreement with the universal series $S_{a,s}$ proposed in Conjecture~\ref{conj_GK} and listed in Appendix~\ref{app:algebraic functions}, for $s=N_f=1,2,3,5$. However, for $s=N_f=4$ we propose different functions $S_{a,4}$, and we discuss the discrepancy in Appendix~\ref{app:corrections}. 

One can also easily check the action of charge conjugation as complex conjugation~\eqref{conj A}. Indeed, with $\overline{ \tr_s^\pm}=-\tr_s^\mp$ we have  $\overline{\ts_s^\pm}=-\ts_s^\mp$, and $\bar \tA_{ab} = (-1)^b \tA_{(N-a)b}$ follows. This confirms~\eqref{conj A} by direct calculation.

\subsection{Off-diagonal coupling}
To compute the off-diagonal coupling $\tq_{12}=\tq_{12,N_f}$ at the cusp, we utilise the map from the genus 2 hyperelliptic curve, in which the Seiberg--Witten curves~\eqref{SU(3) curve} are presented, to the so-called Rosenhain form~\cite{Rosenhain:1851}
\bea
\label{curveRosenhain}
Y^2 = X(X-1)(X-\lambda_1)(X-\lambda_2)(X-\lambda_3).
\eea
The map corresponds to a reduction of the degree 6 hyperelliptic curve to degree 5.
This form exploits the fact that a genus-2 curve only depends on three constants $\lambda_1$, $\lambda_2$ and $\lambda_3$ called the Rosenhain invariants. The Rosenhain invariants can be expressed in terms of the even theta constants as
\be \label{thetaconstants}
\lambda_1 = \frac{\Theta_1^2\,\Theta_3^2}{\Theta_2^2\,\Theta_4^2}~,
\hspace{10pt} \lambda_2 =
\frac{\Theta_3^2\,\Theta_8^2}{\Theta_4^2\,\Theta_{10}^2}~,
\hspace{10pt}\lambda_3 =
\frac{\Theta_1^2\,\Theta_8^2}{\Theta_2^2\,\Theta_{10}^2}~,
\ee
with $\Theta_j$ as defined in~\eqref{theta_1..10} in Appendix~\ref{app:modular forms}. The Rosenhain form was used extensively in previous work on pure ($N_f=0$) $\SU(3)$ SYM~\cite{Aspman:2020lmf}, where the cross-ratios of the degree 6 SW curve were painstakingly matched with the Rosenhain invariants $\lambda_i$.

Rather than repeating this analysis for $\SU(3)$ QCD, we can tune the curve to the singular point, in which it takes the factorised form~\eqref{curve factorisation}. The roots of the sextic are then $\ts_{N_f}^\pm$ and $\tr_{N_f}^\pm$, where $\ts_{N_f}^\pm$ have multiplicity two. Since the curve is at a degenerate configuration, all cross-ratios~\eqref{crossratio} are trivial, except a single non-trivial cross-ratio
\bea\label{cross_Nf}
\cross_{N_f}=\frac{(\tr_{N_f}^+-\ts_{N_f}^+)(\tr_{N_f}^--\ts_{N_f}^-)}{(\tr_{N_f}^+-\ts_{N_f}^-)(\tr_{N_f}^--\ts_{N_f}^+)}~.
\eea
Using~\eqref{thetaconstants}, we can map the three cross-ratios to three quotients of theta constants. We match the non-trivial cross-ratio $\cross_{N_f}$ to the theta constants which are non-trivial for $q_{11}, q_{22}\to 0$, where as before $q_{ab}=e^{2\pi i\tau_{ab}}$, after a modular $S$-transformation. Using the modular $S$-transformation~\eqref{S transformation Theta}, which flips the characteristic of the Theta functions, we find
\bea
\lambda_1(\Omega)&=1+\CO(q_{11}^{\frac12},q_{22}^{\frac12})~, \\
\lambda_2(\Omega)=\lambda_3(\Omega)&=\left(\frac{q_{12}^{\frac12}+1}{q_{12}^{\frac12}-1}\right)^2+\CO(q_{11}^{\frac12},q_{22}^{\frac12})~,
\eea
for $q_{11}, q_{22}\to 0$. Here $\Omega=\begin{pmatrix}
    \tau_{11}&\tau_{12}\\\tau_{21}&\tau_{22}
\end{pmatrix}$. We thus compare the non-trivial cross-ratio to the non-trivial Rosenhain invariants:
\bea
\CC_{N_f}=\lim_{q_{11},q_{22}\to 0}\lambda_{2,3}(\Omega) = \left(\frac{\tq_{12}^{\frac12}+1}{\tq_{12}^{\frac12}-1}\right)^2~.
\eea
The equation has two branches of solutions for $\tq_{12}^{\frac12}$ which do not give the same $\tq_{12}$.
To fix this ambiguity, we use the infinite mass limit to pure $\SU(3)$ SYM, for which~\cite{Klemm:1995wp,Marino:1998bm}
\bea\label{q12 large m}
\lim_{m\to\infty}\tau_{12}=\frac{i}{\pi}\log2~,
\eea
at the $\CN=1$ vacuum $n=0$, 
such that $\lim_{m\to\infty}\tq_{12}= \frac14$. This gives the correct solution
\bea
\tq_{12,N_f}^{\frac12}=\frac{\cross_{N_f}^{\frac12}-1}{\cross_{N_f}^{\frac12}+1}~,
\eea
which, using~\eqref{cross_Nf}, we may also write as
\bea\label{q12}
\tq_{12,N_f}^{\frac12}=\frac{\sqrt{(\tr_{N_f}^{+}-\ts_{N_f}^{-})(\tr_{N_f}^{-}-\ts_{N_f}^{+})}-\sqrt{(\tr_{N_f}^{-}-\ts_{N_f}^{-})(\tr_{N_f}^{+}-\ts_{N_f}^{+}})}{\sqrt{(\tr_{N_f}^{+}-\ts_{N_f}^{-})(\tr_{N_f}^{-}-\ts_{N_f}^{+})}+\sqrt{(\tr_{N_f}^{-}-\ts_{N_f}^{-})(\tr_{N_f}^{+}-\ts_{N_f}^{+}})}~.
\eea
From this, we determine the universal series $Z_{12,s}$ from the dictionary for all $s=N_f=1,\dots, 5$:
\bea
\tq_{12,N_f}=Z_{12,s}^{-1}~, 
\eea
and find perfect agreement for all universal functions $Z_{12,s}$ determined in~\cite{Gottsche:2020ass}, which we list also in Appendix~\ref{app:algebraic functions}.

\sss{Generalisation to $N>3$.}
This calculation of the off-diagonal coupling is the only one in this section that is specialised to $\SU(3)$, since the Rosenhain form is only available for genus $N-1=2$. However, the expression~\eqref{q12} also suggests a natural generalisation to $\SU(N)$. At an $\CN=1$ point, the curve factorises as $Y^2=H(X)^2F(X)$~\eqref{curve factorised_sec3}, with $N-1$ double roots $\ts_a$ and two simple roots $\tr^\pm$, as in~\eqref{H and F}. Since this maximally degenerate curve is rational, the finite off-diagonal couplings $\tau_{ab}$, $a\neq b$, are expected to be determined by cross-ratios involving the preimages of the corresponding roots $\ts_a$ and $\ts_b$. This suggests that expressions for $\tq_{ab}$ analogous to~\eqref{q12} should exist for general $\SU(N)$.
A consistency check comes from the off-diagonal couplings at the $\CN=1$ singularities for pure $\SU(4)$ SYM, which will give $\tq_{ab}$ for $m\to \infty$ and generalise our input~\eqref{q12 large m} to higher rank. Ref.~\cite{Edelstein:2000jc} determined the off-diagonal $a<b$ couplings for pure $\SU(N)$ (see also~\cite{DHoker:2020qlp}):
\bea
\tau_{ab}=\frac{1}{\pi i}\log \frac{\tan \frac{\pi b}{2N}-\tan \frac{\pi a}{2N}}{\tan \frac{\pi b}{2N}+\tan \frac{\pi a}{2N}}~.
\eea
From this, we determine $\tq_{ab;0}=e^{2\pi i\tau_{ab}}$:
\bea\label{tq jk SU(N)}
\tq_{ab;0}=\left(\frac{\sin \frac{(a-b)\pi}{2n}}{\sin \frac{(a+b)\pi}{2n}}\right)^2~,
\eea
which are real and positive. We have immediately
\bea
\tq_{(N-a)(N-b);0}=\tq_{ab;0}~,
\eea
which is a consequence again of charge conjugation symmetry, see~\eqref{tau CC}.
For $N=3$, this gives $\tq_{jk;0}=\frac14$ as before~\eqref{q12 large m}, while for $N=4$ we have $\tq_{12;0}=\tq_{23;0}=3-2\sqrt2$, and $\tq_{13;0}=\frac12$, and so forth.

\subsection{Discriminant derivatives and Igusa cusp form}
From the physical discriminant, we require several couplings. To compute the coupling
\bea\label{tmNf def}
\tm_{N_f}=\left.\frac{\Delta_{N_f}}{a^1 a^2}\right|_{\vec{\tu}_{N_f}}~,
\eea
we consider the change of variables $\vec u=\vec u(\vec a)$, where $\vec a=(a^1,a^2)^\tra$ and $\vec u=(u_1,u_2)^\tra$.
The discriminant admits a Taylor series in $\vec u$ at the singularity $\vec \tu=(\tu_1,\tu_2)^\tra$,
\bea\label{Delta SU(3) series}
\Delta_{N_f}(\vec u)=\Delta_{N_f}(\vec \tu)+(\nabla_{\vec u}\Delta_{N_f})^\tra(\vec u-\vec \tu)+\tfrac12(\vec u-\vec \tu)^\tra\tH_\Delta(\vec u-\vec \tu)+\dots~.
\eea
with Hessian $(\tH_\Delta)_{ab}=\frac{\partial^2\Delta_{N_f}}{\partial u_a\partial u_b}|_{\tu}$. 
Since $\Delta_{N_f}(\vec \tu)=\nabla_{\vec  u}\Delta_{N_f}(\vec \tu)=0$ by assumption of $\vec \tu$ being an $\CN=1$ point\footnote{see comment below~\eqref{sing points pure SU(N)}}, we can substitute $\vec u-\vec \tu=\tJ \vec a+\dots$ into this equation, with $\tJ$ the change of variables~\eqref{tJ} from $\vec a$ to $\vec u$, which gives to second order
\bea
\Delta_{N_f}(\vec u)=\tfrac12a^\tra(\tJ^\tra\tH_\Delta \tJ)a+\dots~.
\eea
The coupling~\eqref{tmNf def} is then twice the off-diagonal of this symmetric matrix,
\bea\label{m_Nf calculation}
\tm_{N_f}%=\frac{\Delta_{N_f}}{a^1 a^2}
=(\tJ^\tra\tH_\Delta \tJ)_{12}~.
\eea
With this, we can compute the $\chih$ coupling to $\SU(3)$ correlators:
\bea
X_{s} = 3\alpha_{N_f}^{12} \beta_{N_f}^{-8} \left( \frac{\Lambda_0}{\Lambda_{N_f}} \right)^{10}\Lambda_{N_f}^{-2} \tx_{N_f}^6 \tm_{N_f}^{-1}~.
\eea
Inserting $\tm_{N_f}$~\eqref{m_Nf calculation},  explicit calculation of the Hessian $\tH_{\Delta}$ and insertion of $\vec\tu$~\eqref{u v T P} and
$\tx_{N_f}$~\eqref{xNf result}, gives perfect agreement with $X_s$ with~\eqref{Vs Ws Xs Qs Rs Ts} for $s=N_f=1,\dots, 5$. The gravitational couplings necessarily take the form~\eqref{alpha beta SU(3)}, where we determine for all $N_f=1,\dots, 5$:
\bea\label{SU(3) grav couplings}
\alpha_{N_f}&=2^{\frac23}\Lambda_{N_f}^{-\frac32}~, \\
\beta_{N_f}&=2^{\frac23}e^{\frac{\pi i}{8}}\Lambda_{N_f}^{-\frac38(4+N_f)}~.
\eea
To compute the $c_1^2$ coupling, we moreover require the quantity
\bea
\tn_{N_f}\coloneqq\left.\frac{\Delta_{N_f}}{q_{11}q_{22}}\right|_{\vec{\tu}_{N_f}}~,
\eea
which we compute as follows. 
Write the first order degeneration in the $q_{ij}$ series as 
\bea
\vec u=\vec \tu~+\tU_{11}q_{11}+\tU_{22}q_{22}+\CO(q_{11}^2,q_{11}q_{22},q_{22}^2)~,
 \eea
where $\vec u=(u_1,u_2)^\tra$ and $\vec\tu=(\tu_1,\tu_2)$ as before, while $\vec \tU_{11}\coloneqq \frac{\partial \vec u}{\partial q_{jj}}\big|_{q_{11}=q_{22}=0}$. The $q_{12}$ coefficient vanishes because $q_{12}$ does not vanish at the cusp, as is clear from above, and the series has to become $\vec u=\vec\tu$ at the cusp.
We insert into~\eqref{Delta SU(3) series}, which gives
\bea\label{Delta q11q22}
\Delta_{N_f}=q_{11}q_{22} \tU_{11}^\tra \tH_{\Delta}\tU_{22}+\dots,
\eea
which means that at the singularity, $\tn_{N_f} = \tU_{11}^\tra \tH_{\Delta}\tU_{22}$. 
To compute this quantity, we make use of an identity which is a consequence of Theorem~\ref{thm:discriminant} in Appendix~\ref{subsec:hyperelliptic-discriminants}: 
\bea\label{delta det = chi10}
-\tfrac{1}{64}\Lambda_{N_f}^{18-3N_f}\Delta_{N_f} (\det J)^{-10}=\chi_{10}~.
\eea
Here $J_{ab}= \frac{\partial u^a}{\partial a^b}$ as before, and $\chi_{10}$ is Igusa's cusp form. The theorem determines the relative factor between the three functions, but is formulated in a convention different from ours. To determine the relative factor, we fix the power of $\Lambda_{N_f}$ by dimensional analysis and the decoupling limit, and the numerical constant can be fixed by a weak-coupling analysis in pure $\SU(3)$, as in~\cite{Klemm:1995wp,Aspman:2020lmf}.
Since $\Delta_{N_f}$ is the discriminant of the curve, while $\chi_{10}$ is a cusp form,  both vanish at all the singular points. The Jacobian determinant $\det J$ forms the bridge between the physical and mathematical discriminant. This identity is a higher genus generalisation of identities more familiar from rank one theories, such as~\cite{Aspman:2021vhs,Manschot:2021qqe}.

The cusp form $\chi_{10}$ is defined as the product of all genus 2 theta constants $\Theta_j$~\eqref{theta_1..10} of even characteristic,
\bea
\chi_{10}=2^{-12}\prod_{j=1}^{10}\Theta_j^2~.
\eea
It is a genus-two Siegel modular form of weight $10$ for the symplectic group ${\rm Sp}(4,\mathbb Z)$, that is, it transform as
\bea
\chi_{10}(\gamma\Omega)=\det (C\Omega+D)^{10}\chi_{10}(\Omega)~,
\eea
for all $\gamma=\begin{psmallmatrix} A&B\\ C& D\end{psmallmatrix}\in {\rm Sp}(4,\mathbb Z)$ and all elements $\Omega=\begin{pmatrix}
    \tau_{11}&\tau_{12}\\\tau_{21}&\tau_{22}
\end{pmatrix}\in \mathbb{H}_2$ of the Siegel upper-half plane and $\gamma$ acts on $\Omega$ as $\gamma\Omega= (A\Omega+B)(C\Omega+D)^{-1}$.
At the cusp, the $q_{ab}$ expansion reads
\bea\label{chi10 expansion}
\chi_{10}(\Omega)=q_{11}q_{22}\big(q_{12}^{\frac12}-q_{12}^{-\frac12}\big)^2+\CO(q_{11}^2,q_{22}^2)~,
\eea
where $q_{ab}=e^{2\pi i \tau_{ab}}$. The $q_{ab}$ series is identical after the genus two $S$-transformation. Inserting in~\eqref{delta det = chi10} and using~\eqref{Delta q11q22}, the $q_{11}q_{22}$ factors cancel on both sides, and the first-order degeneration allows us to solve for $\tn_{N_f}$:
\bea\label{tn in terms of tx}
\tn_{N_f}=-64\Lambda_{N_f}^{3N_f-18}\tx_{N_f}^{10}\big(\tq_{12}^{\frac12}-\tq_{12}^{-\frac12}\big)^2~,
\eea
where by $\tq_{12}$ we denote the coupling $q_{12}=e^{2\pi i\tau_{12}}$ at the $\CN=1$ point.
Inserting $\tx_{N_f}$~\eqref{xNf result} and $\tq_{12,N_f}^{\frac12}$~\eqref{q12} gives

\bea
\tn_{N_f}^{\frac18}=e^{\frac{5\pi i}{8}}2^{\frac54}\Lambda_{N_f}^{\frac38(N_f-6)}
\frac{(\tr_{N_f}^+-\ts_{N_f}^+)^{\frac34}(\tr_{N_f}^+-\ts_{N_f}^-)^{\frac34}(\tr_{N_f}^--\ts_{N_f}^+)^{\frac34}(\tr_{N_f}^--\ts_{N_f}^-)^{\frac34}(\ts_{N_f}^--\ts_{N_f}^+)}{(\tr_{N_f}^--\tr_{N_f}^+)^{\frac14}}~,
\eea
where we also fix the phase required to define the 8th root of~$\tn_{N_f}$.
With this quantity, we can compute the coupling to $c_1^2$. For the comparison with the dictionary, we however require first the couplings to the background fluxes, for which we use the contact term of the surface observable.

\subsection{Contact term and background fluxes}
\sss{Contact term for surface observable.}
The contact terms for $\SU(2N)$ QCD with
$N_f$ massive hypermultiplets has been studied in~\cite{Edelstein:2000wg,Edelstein:2000aj,Takasaki:1998vm}. For the surface observable $\tilde I^1(x)
$~\eqref{surface observable} based on the quadratic Casimir $u_1$, the contact term  $G_{N_f}= T_{1,1}$  is determined in~\cite[(4.10)]{Edelstein:2000wg} in terms of a higher genus theta quotient multiplied by the periods $\frac{\partial u^1}{\partial a^a}$, corrected by a rational function in all the Casimirs $u_a$. The theta quotient becomes a constant at strong coupling, while the periods evaluate to the dual periods $\tx_{a,{N_f}}$. The Casimirs become the location of the singularities $\tu_{a,N_f}$. 
Their formula does not study the case with even rank; nevertheless, we conjecture for all $\SU(N)$ theories with $N_f$ massive flavors:
\bea\label{GNf from curve}
\tG_{N_f}=\frac{1}{2(2N-N_f)\Lambda_{N_f}^2}\left(
\tu_{1,N_f}-\tfrac12(N-N_f)\sum_{a=1}^{N-1} \tx_{a,{N_f}}^2\right)+\delta_{N_f}~,
\eea
which includes possible shifts $\delta_{N_f}$ familiar from rank one~\cite{Moore:1997pc,LoNeSha,mm1998,Mari_o_1999,Labastida:1998sk,Manschot:2021qqe}. 
For $N=3$, we propose that the shifts $\delta_{N_f}=0$ vanish for $N_f\leq 3$, while 
\bea
\delta_4&=-\frac{1}{48}~, \\
%\delta_5&=-\frac{1}{96}+\frac{5(3+t)}{144z^3}~,
\delta_5&=-\frac{1}{96}-\frac{5}{12}\frac{m}{\Lambda_5}+\frac{5i}{36}\tH_5~.
\eea
For $N_f=5$, we find a curious shift by the other coupling $\tH_5$ of the surface to the background fluxes.\footnote{This is likely an artifact from the parametrisation of the curve. Recall that there are other subtleties for the $N_f=5$ theory that make it break ranks, such as~\eqref{Nf=5 shift}, and the function $R_5$ obtained from $\tH_5$, as will be discussed below and in~\eqref{R5 discrepancy}.}
Another curiosity is that for $N_f=3$, the contact term $\tG_3$ does not depend on the periods $\tx_{a,3}$ at all. This is due to the transition at $N_f=3$ from theories without to those with baryonic branches, see section~\ref{sec:SW geometry}.

For  $N=3$, inserting the above results for $\tu_{1,N_f}$, $ \tx_{a,{N_f}}$, and the shifts, we find agreement with the dictionary for $s=N_f=1,\dots, 5$:
\bea\label{Qs = GNf}
Q_s=-\left(\frac{\Lambda_{N_f}}{m}\right)^2 \tG_{N_f}~,
\eea
with $Q_s$ as in~\eqref{Vs Ws Xs Qs Rs Ts}.

\sss{Contact term and background fluxes.}
To derive the couplings to the background fluxes, we use the expression~\eqref{eq:C11C12=theta} from the blowup relation, which expresses the contact term $G_{N_f}$ as:
\bea\label{GNf log theta}
G_{N_f}=\frac{1}{8\pi^2\Lambda_{N_f}^2}\frac{\partial u_1}{\partial a^a}\frac{\partial u_1}{\partial a^b}\Theta^{(a,b)}_{\log}[\vec\Delta,0]\big(\tfrac{N_f}{2}\vec v\mid\tau\big)~,
\eea
where the sum over $a,b\in\{1,2\}$ is understood. Moreover, the logarithmic derivatives are defined in~\eqref{logarithmic theta derivative}, and $\vec\Delta=(-\frac12,-\frac12)$.
Here,  $\vec v$ has components $v_a=\tau_{ai}$, which are independent of the flavor index $i=1,\dots, N_f$ since these are derivatives with respect to the mass $m_i$ and are equal after a subsequent equal mass limit. Their exponentials are the couplings~\eqref{tDa}
\bea
\tD_{a,N_f}\coloneqq \prod_{i=1}^{N_f}q_{ai}=q_{ai}^{N_f}=e^{2\pi i N_f v_a}
\eea
we wish to determine. For this, we consider the $S$-transformation of~\eqref{GNf log theta}, which becomes
\bea
\tG_{N_f}=\frac{1}{8\pi^2\Lambda_{N_f}^2}\tx_{a,{N_f}}\tx_{b,{N_f}}\Theta^{(a,b)}_{\log,{\rm D}}[\vec\Delta,0]\big(\tfrac{N_f}{2}\vec \tv\mid\tau \big)~.
\eea
We calculate the constant terms of the logarithmic derivatives after the $S$-transformation,
\bea
\Theta^{(1,1)}_{\log,{\rm D}}[\vec\Delta,0](\vec z|\tau_{\rm D})&=-4\pi^2\frac{p_1(p_2+(1+p_2^2)\tq_{12}^{\frac12}+p_2\tq_{12})}{(p_1+p_2+(p_1p_2+1)\tq_{12}^{\frac12})^2}~, \\
\Theta^{(1,2)}_{\log,{\rm D}}[\vec\Delta,0](\vec z|\tau_{\rm D})&=-4\pi^2\frac{p_1p_2(\tq_{12}-1)}{(p_1+p_2+(p_1p_2+1)\tq_{12}^{\frac12})^2}~, \\
\Theta^{(2,2)}_{\log,{\rm D}}[\vec\Delta,0](\vec z|\tau_{\rm D})&=-4\pi^2\frac{p_2(p_1+(1+p_1^2)\tq_{12}^{\frac12}+p_1\tq_{12})}{(p_1+p_2+(p_1p_2+1)\tq_{12}^{\frac12})^2}~,
\eea
where we denote $\vec z=(z_1,z_2)$ with $p_a=e^{2\pi i z_a}$  for readability. Inserting now $p_a=\tD_{a,{N_f}}^{\frac{N_f}{2}}$, this gives
\begin{equation}\small\begin{aligned}\label{Gnf in terms of Da}
\tG_{N_f}=-\frac{\tx_{a,{N_f}}\tx_{b,{N_f}}}{2\Lambda_{N_f}^2}\frac{\delta_{ab}\tD_{a,{N_f}}^{\frac{N_f}{2}}\left(\tD_{\bar a,{N_f}}^{\frac{N_f}{2}}+(1+\tD_{\bar a,{N_f}}^{N_f})\tq_{12}^{\frac12}+\tD_{\bar a,{N_f}}^{\frac{N_f}{2}}\tq_{12}\right)+(1-\delta_{ab})\tD_{1,{N_f}}^{\frac{N_f}{2}}\tD_{2,{N_f}}^{\frac{N_f}{2}}\left(\tq_{12}-1\right)}{\left(\tD_{1,{N_f}}^{\frac{N_f}{2}}+\tD_{2,{N_f}}^{\frac{N_f}{2}}+\left(1+\tD_{1,{N_f}}^{\frac{N_f}{2}}\tD_{2,{N_f}}^{\frac{N_f}{2}}\right)\tq_{12}^{\frac12}\right)^2}~,
\end{aligned}\small\end{equation} where $\bar a=3-a$, and the sum over $a,b=1,2$ is understood.
Inserting moreover $\tq_{12}^{\frac12}$ from~\eqref{q12} and  
$\tx_{a,N_f}$ from~\eqref{xNf result}, we can solve for all $s=N_f=1,\dots, 5$ the equation $Q_s=-\left(\frac{\Lambda_{N_f}}{m}\right)^2 \tG_{N_f}$~\eqref{Qs = GNf} for $\tD_{a,N_f}$ with $a=1,2$.
We find that this equation does \emph{not} have a unique solution for the pair $\{\tD_{1,N_f},\tD_{2,N_f}\}$, which is in particular because there is only one equation to solve for two unknowns. However, if we demand that as large mass series they are complex conjugates,
\bea
\tD_{2,N_f}=\overline{\tD_{1,N_f}}~,
\eea
then there is only one unknown, e.g. $\tD_{1,N_f}$, and we can solve for it. It turns out that~\eqref{Qs = GNf} has two solutions which are related by a change of signs of their coefficients. We fix this sign by requiring that
\bea
\tD_{1,s}=1+\sqrt 3z^{\frac12}+\CO(z^2)~.
\eea
Then one can show that the equation has a unique pair of complex conjugate solutions $\{\tD_{1,N_f},\tD_{2,N_f}\}$. We distinguish between the conjugates by fixing the second coefficients:\footnote{Curiously, for $N_f=3$ the solutions have real coefficients and $\tD_{1,3}=\tD_{2,3}$, which is also where many algebraic functions on the mathematical side simplify. }
\bea
\tD_{1,N_f}=1+\sqrt3z^{\frac12}+\left(\tfrac32+\tfrac{N_f-3}{2\sqrt3}i\right)z+\CO(z^{\frac32})~, \\
\tD_{2,N_f}=1+\sqrt3z^{\frac12}+\left(\tfrac32-\tfrac{N_f-3}{2\sqrt3}i\right)z+\CO(z^{\frac32})~.
\eea
For $s=N_f=1,\dots, 5$, we find perfect agreement for the dictionary 
\bea\label{tD=Y}
 \tD_{a,N_f}=Y_{a,s}~,
\eea
with the same functions $Y_{a,s}$ proposed in Conjecture~\ref{conj_GK}, and listed in Appendix~\ref{app:algebraic functions}. We should note that~\eqref{Gnf in terms of Da} provides an explicit algebraic equation for $Y_{a,s}$ through~\eqref{tD=Y}, and can be used to \emph{determine} the defining equation of $Y_{a,s}$ as listed in Appendix~\ref{app:algebraic functions}. This also provides a method to determine these defining equations for ranks larger than 3, where the strategy of~\cite[Remark 4.1]{Gottsche:2020ass} likely becomes cumbersome.

\sss{Couplings of background fluxes to surface.}
To derive the couplings $\tH_{N_f}$ of the background fluxes to the surface $x$, we use again the result from the blowup function~\eqref{eq:C11C12=theta}, which expresses it as a logarithmic theta derivative,
\bea
H_{N_f}= \frac{1}{\pi N_f\Lambda_{N_f}}\frac{\partial u_1}{\partial a^a}\Theta^{(a)}_{\log}[\vec\Delta,0]\left(\tfrac{N_f}{2}\vec v\mid\tau\right)~.
\eea
where the sum over $a=1,2$ is understood. After the $S$-transformation, this becomes
\bea
\tH_{N_f}=\frac{1}{\pi N_f\Lambda_{N_f}}\tx_{a,{N_f}}\Theta^{(a)}_{\log,{\rm D}}[\vec\Delta,0]\left(\tfrac{N_f}{2}\vec v\mid\tau\right)~. 
\eea
For this, we compute similarly as above the constant terms after the $S$-transformation:
\bea
\Theta^{(1)}_{\log,{\rm D}}[\vec\Delta,0](\vec z|\tau_{\rm D})&=\pi i\frac{p_1-p_2+(p_1p_2-1)\tq_{12}^{\frac12}}{p_1+p_2+(p_1p_2+1)\tq_{12}^{\frac12}}~, \\
\Theta^{(2)}_{\log,{\rm D}}[\vec\Delta,0](\vec z|\tau_{\rm D})&=\pi i\frac{-p_1+p_2+(p_1p_2-1)\tq_{12}^{\frac12}}{p_1+p_2+(p_1p_2+1)\tq_{12}^{\frac12}}~. \\
\eea
Inserting  $p_a=\tD_{a,{N_f}}^{\frac{N_f}{2}}$, this gives
\bea
\tH_{N_f}=\frac{i}{N_f\Lambda_{N_f}}\sum_{a=1}^2\tx_{a,1}\frac{(-1)^a(\tD_{2,{N_f}}^{\frac{N_f}{2}}-\tD_{1,{N_f}}^{\frac{N_f}{2}})+(\tD_{1,{N_f}}^{\frac{N_f}{2}}\tD_{2,{N_f}}^{\frac{N_f}{2}}-1)\tq_{12}^{\frac12}}{\tD_{1,{N_f}}^{\frac{N_f}{2}}+\tD_{2,{N_f}}^{\frac{N_f}{2}}+(\tD_{1,{N_f}}^{\frac{N_f}{2}}\tD_{2,{N_f}}^{\frac{N_f}{2}}+1)\tq_{12}^{\frac12}}~. 
\eea
With $ \tD_{a,N_f}=Y_{a,s}$~\eqref{tD=Y}, and the other quantities from above, we find for $s=N_f=1,\dots, 4$:
\bea
\frac{\Lambda_{N_f}}{im}\tH_{N_f}=R_{s}
\eea
with $R_s=t$,
as required by the dictionary~\eqref{dictionary}. For $N_f=5$, we find a correction $R_5=t-\tfrac15z^3\delta_{N_f,5}$, which corrects the identity $5R_5+2S_5+S_{1,5}+S_{2,5}=0$~\eqref{constraints}. We discuss this discrepancy further in Appendix~\ref{app:corrections}.

\subsection{Couplings to background fluxes}
Finally, we consider the couplings $C_{ij}$~\eqref{Cij def} to the background fluxes. To determine their values at strong coupling, we consider again the blowup equation, which we determine for the background fluxes $k_i'=\{\frac12,\frac12,\dots, \frac12\}$ and $k_i'=\{\frac12,\dots,\frac12,-\frac12\}$. Since in the equal mass limit there are only two independent couplings $C_{ii}$ and $C_{i\neq j}$, these two configurations are sufficient to solve for them. The equations read in the equal mass limit:
\bea\label{plus minus Cij}
A^{-1}B C_{11}^{\frac{N_f}{4}}C_{12}^{\frac{N_f}{4}(N_f-1)}&=\Theta[\vec\Delta,0|]\big(\tfrac{N_f}{2}\vec v|\tau\big)~, \\
A^{-1}B C_{11}^{\frac{N_f}{4}}C_{12}^{\frac{N_f-4}{4}(N_f-1)}&=\Theta[\vec\Delta,0|]\big(\tfrac{N_f-2}{2}\vec v|\tau\big)~, \\
\eea
At the strong coupling singularity, the coupling $C_{11}$ becomes $\tC_{N_f}$, while $C_{12}$ becomes $\hat \tC_{N_f}$. To determine $\hat \tC_{N_f}$, we simply divide the two equations and solve for it. To determine $\tC_{N_f}$, we eliminate $\hat \tC_{N_f}$ from the system, and use the relation~\eqref{delta det = chi10} between $A$ and $B$ using Igusa's cusp form $\chi_{10}$ to eliminate the dependence on the discriminant in favour of the functional determinant $\det J$, which becomes $\tx_{N_f}$~\eqref{x_Nf calculation} at the singularity. For this, we use~\eqref{gravitational couplings} with the gravitational couplings~\eqref{SU(3) grav couplings}. We furthermore require the leading term in the expansion of the theta functions at the cusp:
\bea
\Theta_{\log,{\rm D}}[\vec\Delta,0](\vec z|\tau_{\rm D})&=\frac{p_1+p_2+(1+p_1p_2)q_{12}^{\frac12}}{p_1^{\frac12}p_2^{\frac12}q_{12}^{\frac14}}q_{11}^{\frac18}q_{22}^{\frac18}~,
\eea
where again $p_a=e^{2\pi i z_a}$. We use this equation with $p_a=\tD_{a,N_f}^{\frac{N_f}{2}}$ for the first equation in~\eqref{plus minus Cij}, and $p_a=\tD_{a,N_f}^{\frac{N_f-2}{2}}$ for the second equation. Inserting everything gives for $N_f=1,\dots, 5$, where the superscripts $N_f$ and $N_f-1$ are exponents:\footnote{Recall that $\hat \tC_{1}$ does not exist}
\allowdisplaybreaks
\begin{subequations}
\begin{align}
\tC_{N_f}^{N_f}&=\left(\tD_{1,N_f}^{\frac{N_f}{2}}+\tD_{2,N_f}^{\frac{N_f}{2}}+\left(1+\left(\tD_{1,N_f}\tD_{2,N_f}\right)^{\frac{N_f}{2}}\right)\tq_{12,N_f}^{\frac12}\right)^{4-N_f} \notag\\ 
&\quad\;  \times \frac{\left(\tD_{1,N_f}\tD_{2,N_f}^{\frac{N_f}{2}}+\tD_{1,N_f}^{\frac{N_f}{2}}\tD_{2,N_f}+\left(\tD_{1,N_f}\tD_{2,N_f}+\left(\tD_{1,N_f}\tD_{2,N_f}\right)^{\frac{N_f}{2}}\right)\tq_{12,N_f}^{\frac12}\right)^{N_f}\Lambda_{N_f}^{9}}{8\tq_{12,N_f}^{\frac12}(1-\tq_{12,N_f})\left(\tx_{N_f}\tD_{1,N_f}^{\frac{N_f}{2}}\tD_{2,N_f}^{\frac{N_f}{2}}\right)^3}~, \\
\hat\tC_{N_f}^{N_f-1}&=\frac{(\tD_{1,N_f}\tD_{2,N_f})^{\frac12}\left(\tD_{1,N_f}^{\frac{N_f}{2}}+\tD_{2,N_f}^{\frac{N_f}{2}}+\left(1+(\tD_{1,N_f}\tD_{2,N_f})^{\frac{N_f}{2}}\right)\tq_{12,N_f}^{\frac12}\right)}{\tD_{1,N_f}\tD_{2,N_f}^{\frac{N_f}{2}}+\tD_{1,N_f}^{\frac{N_f}{2}}\tD_{2,N_f}+\left((\tD_{1,N_f}\tD_{2,N_f})^{\frac{N_f}{2}}+\tD_{1,N_f}\tD_{2,N_f}\right)\tq_{12,N_f}^{\frac12}}~.
\end{align}
\end{subequations} 
Using~\eqref{tD=Y}, we find for all $s=N_f=1,\dots, 5$:
\bea
\Big(\tfrac{\Lambda_0}{\Lambda_{N_f}}\Big)^{\frac{9}{N_f}}\tC_{N_f}&=W_s~,\\
\Big(\tfrac{\Lambda_0}{\Lambda_{N_f}}\Big)^{-\frac{18}{N_f}}\hat\tC_{N_f}^2 \tC_{N_f}^{-2}&=V_s~,
\eea
where the series $V_s$ and $W_s$ agree precisely with the ones predicted~\eqref{Vs Ws Xs Qs Rs Ts}.

\sss{Remaining algebraic functions.}
The remaining algebraic functions, $S_s$, $Y_s$, $Z_{11,s}$, $Z_{22,s}$ and $Z_s$, are fixed by consistency of the dictionary~\eqref{dictionary}:
\allowdisplaybreaks
\begin{subequations}
\begin{align}
S_s&=\frac{iN_f}{2}\left(\frac{\Lambda_{N_f}}{m}\right)\tH_{N_f}+\frac{1}{2m}(\tx_{1,N_f}+\tx_{2,N_f})~, \\
Y_s&=\left(\frac{\Lambda_0}{\Lambda_{N_f}}\right)^{-\frac{9}{N_f}}\tC_{N_f}^{-1}\hat\tC_{N_f}^{1-N_f}\tD_{1,N_f}^{-\frac12}\tD_{2,N_f}^{-\frac12}~, \\
Z_{11,s} &=\tD_{1,N_f}^{-\frac{s}{N_f}} \tq_{12,N_f}^{\frac{1}{2}}~, \\
Z_{22,s} &=\tD_{2,N_f}^{-\frac{N_f}{2}} \tq_{12,N_f}^{\frac{1}{2}}~, \\
Z_s &= \tfrac13 \alpha_{N_f}^{-1} \beta_{N_f} \tx_{N_f}^{-\frac{1}{2}} \mathtt{n}_{N_f}^{\frac{1}{8}} \hat{\tC}_{N_f}^{\frac{N_f(N_f-1)}{4}} \tC_{N_f}^{\frac{N_f}{4}}\tD_{1,N_f}^{\frac{N_f}{4}}\tD_{2,N_f}^{\frac{N_f}{4}}\tq_{12,N_f}^{-\frac{1}{4}}~.
\end{align}
\end{subequations} 
We find perfect agreement again with the universal functions proposed by~\cite{Gottsche:2020ass}, except for the correction of $S_4$ proposed in Appendix~\ref{app:corrections}.

%\sss{Asymptotics of universal series.}
With our normalisations, we determine the asymptotics for all $s=1,\dots, 5$:
\allowdisplaybreaks
\begin{subequations}
\begin{align} \label{asymptotics}
Q_s&=\tfrac12z+\CO(z^2)~, \notag\\
R_s&=z+\CO(z^2)~, \notag\\
S_{s}&=+\sqrt{3}z^{\frac12}-\tfrac{s}{2}z+\CO(z^{\frac32})~,\notag \\
S_{a,s}&=-\sqrt{3}z^{\frac12}\pm\tfrac{i s}{2\sqrt3}z+\CO(z^{\frac32})~, \notag\\
T_s&=z+\CO(z^2)~, \notag\\
V_s&=1+3z+\CO(z^2)~, \notag\\
W_s&=1-z+\CO(z^2)~, \notag\\
X_s&=1 - \tfrac{4 s}{3} z+\CO(z^2)~, \notag\\
Y_s&=1-\sqrt3z^{\frac12}+\tfrac{6-s}{2}z+\CO(z^{\frac32})~, \notag\\
Y_{a,s}&=1+\sqrt3z^{\frac12}+\left(\tfrac32\pm\tfrac{3-s}{2\sqrt3}i\right)z+\CO(z^{\frac32})~, \notag\\
Z_s&=2+\sqrt3 sz^{\frac12}+\CO(z)~,\notag \\
Z_{12,s}&=4+\tfrac13(6-s)sz+\CO(z^2)~, \notag\\
Z_{aa,s}&=\tfrac12-\tfrac{\sqrt3}{4}sz^{\frac12}+\tfrac{s}{24}\left(5s-3\pm i\sqrt3\,(s-3)\right)z+\CO(z^{\frac32})~,
\end{align}
\end{subequations} 
where $+$ corresponds to $a=1$ and $-$ corresponds to $a=2$.

\appendix
\section{Modular forms and hyperelliptic curves}
\label{app:modular forms}
In this appendix, we introduce Siegel modular forms and their relation to hyperelliptic curves through the Rosenhain form and a discriminant identity. For further reading see~\cite{Bruinier08,freitag1983,pitale2019siegel}.

\subsection{Siegel modular forms}\label{sec:SiModForms}
Ordinary modular forms are constructed by the action of an
$\slz$ Möbius transformation on the upper half-plane
$\mathbb H$. Siegel modular forms
generalise this notion by introducing an action of $\mr{Sp}(2g,\mathbb Z)$
on the so-called Siegel upper half-plane $\mathbb{H}_g$, which works
for any \emph{genus} $g\in\mathbb N$. 

\sss{Siegel modular group.}
The Siegel upper half-plane
\be \label{Siegel upperhalf}
\mathbb H_g =\{\Omega\in\text{Mat}(g;\mathbb C)\,|\, \Omega^\tra=\Omega, \, \ip\,  \Omega >0\}
\ee
consists of complex symmetric $g\times g$ matrices whose component-wise imaginary part is positive definite. This generalises the ordinary upper half-plane $\mathbb H=\mathbb{H}_1$. For example, for $g=2$  this means that 
\begin{equation}
\Omega=\begin{pmatrix}\tau_{11}&\tau_{12}\\ \tau_{12}&\tau_{22}\end{pmatrix},\quad \ip \tau_{11}>0, \quad \ip\tau_{11}\ip \tau_{22}-(\ip \tau_{12})^2>0.
\end{equation}
Define the Siegel modular group of genus $g$ as
\be\label{sp2gzdef}
\Sp(2g,\mathbb Z) = \{M\in \text{Mat}(2g;\mathbb Z)\,|\, M^\tra J M =J\} \quad\text{with } J = \left(\begin{smallmatrix} 0& \mathbbm 1_g\\ -\mathbbm 1_g&0\end{smallmatrix}\right).
\ee
The group $\Sp(4,\BZ)$ can be generated by the
elements $J$ and  $T=\left(\begin{smallmatrix} \mathbbm{1}_g & s \\0 &
    \mathbbm{1}_g \end{smallmatrix}\right)$ with $s=s^\tra$~\cite{Bruinier08}.
Let $\Gamma_N$ denote the congruence subgroup of level $N$ for the symplectic modular group of degree $g$, so
\bea\label{Gamma N}
\Gamma_g(N)=\left\{\gamma\in \Sp(2g,\mathbb Z)\mid \gamma\equiv \mathbbm 1_{2g}\mod N\right\}~.
\eea
An element $\gamma=\left(\begin{smallmatrix} A&B\\C&D\end{smallmatrix}\right)\in \Sp(2g,\mathbb Z)$ acts on the Siegel upper half-plane by 
\be\label{symplectictransf}
\Omega\longmapsto\gamma(\Omega)=(A\Omega+B)(C\Omega+D)^{-1}.
\ee
A (classical) Siegel modular form of weight $k$ and genus $g$ for a subgroup $\Gamma\subset \Sp(2g,\mathbb Z)$ is a holomorphic function $f\colon\mathbb{H}_g\to\mathbb C$ satisfying
\be f(\gamma(\Omega))=\det(C\Omega+D)^kf(\Omega) \qquad \forall \gamma =\begin{pmatrix}A&B\\C&D\end{pmatrix}\in \Gamma, \ee
where for $g=1$ holomorphicity at $i\infty$ is required in addition.

\sss{Theta functions.}
\label{app:thetaconstants}
Theta series provide an explicit class of classical Siegel modular forms. For $a$, $b \in \BQ^g$ and $\bz\in\mathbb{C}^g$, $\Omega\in \mathbb{H}_g$, define the genus $g$ Riemann theta function by 
\be\label{definitionstheta}
\Theta\begin{bmatrix}a\\b\end{bmatrix}(\bz|\Omega) = \sum_{k\in \BZ^g}\exp\left(\pi i(k+a)^\tra\Omega(k+a)+2\pi i(k+a)^\tra\, (\bz+b)  \right).
\ee
They have the following transformation property under $\gamma =\begin{psmallmatrix}A&B\\C&D\end{psmallmatrix}\in\mathrm{Sp}(2g,\Z)$:
\begin{eqsp}
    \Theta\begin{bmatrix}Da-Cb+\frac{1}{2}\mathrm{diag}[CD^{\mathsf{T}}]\\-Ba+Ab+\frac{1}{2}\mathrm{diag}[AB^{\mathsf{T}}]\end{bmatrix}\left([(C\Omega+D)^{-1}]^{\mathsf{T}}\bz\bigg{|}(A\Omega+B)(C\Omega+D)^{-1}\right)\\=\kappa(a,b,\gamma)\sqrt{\mr{det}(C\Omega+D)}e^{\pi i\bz^{\mathsf{T}}[(C\Omega+D)^{-1}C]\bz}\Theta\begin{bmatrix}a\\b\end{bmatrix}(\bz|\Omega) ~,
\end{eqsp}
where $\kappa(a,b,\gamma)$ is a complex number independent of $\bz,\Omega$. For $\gamma=\begin{psmallmatrix}
    0&\mathbbm 1_g\\-\mathbbm 1_g&0
\end{psmallmatrix}\in \mathrm{Sp}(2g,\Z)$, we get 
\begin{eqsp}\label{S transformation Theta}
    \Theta\begin{bmatrix}b\\a\end{bmatrix}(-(\Omega^{\mathsf{T}})^{-1}\bz|-\Omega^{-1})=\kappa(a,b,\gamma)\sqrt{\det(-\Omega)}e^{\pi i\bz^{\mathsf{T}}\Omega^{-1}\bz}\Theta\begin{bmatrix}a\\b\end{bmatrix}(\bz|\Omega)~. 
\end{eqsp}
\par 

\sss{Genus one.}
For genus $g=1$, the theta series become the usual elliptic Jacobi theta functions $\vartheta_j$. With $q=e^{2\pi i\tau}$, they are defined as
\be\label{jt_elliptic}
\begin{split}
&\vartheta_1(\tau,z)=i \sum_{r\in
  \mathbb{Z}+\frac12}(-1)^{r-\frac12}q^{r^2/2}e^{2\pi i
  rz}~, \\
&\vartheta_2(\tau,z)= \sum_{r\in
  \mathbb{Z}+\frac12}q^{r^2/2}e^{2\pi i
  rz}~,\\
&\vartheta_3(\tau,z)= \sum_{n\in
  \mathbb{Z}}q^{n^2/2}e^{2\pi i
  n z}~,\\
&\vartheta_4(\tau,z)= \sum_{n\in 
  \mathbb{Z}} (-1)^nq^{n^2/2}e^{2\pi i
  n z}~.
\end{split}
\ee
We also use the notation $\jt_j(\tau)=\jt_j(\tau,0)$.
The Dedekind eta function is defined as the infinite product
\bea\label{dedekind eta}
\eta(\tau)=q^{\frac{1}{24}}\prod_{n=1}^\infty(1-q^n)~,
\eea
and it relates to the theta functions as
\bea\label{jt234=eta}
\jt_2\jt_3\jt_4=2\eta^3~.
\eea

\sss{Genus two.}
We are especially interested in the case $g=2$ and when the entries of the column vectors $a,b$ take values in the set $\{0,\frac{1}{2} \}$. The corresponding theta functions are usually referred to as the theta characteristics. We call $\gamma = \left[\begin{smallmatrix}a\\b\end{smallmatrix}\right]$ an even (odd) characteristic if $4a^\tra b$ is even (odd). In the case of genus two there are  ten even  theta constants~\cite{Previato:2013},
\begin{equation}\label{theta_1..10}
    \begin{alignedat}{4}
    &\Theta_{1} = \Theta\begin{bmatrix}0&0\\0&0\end{bmatrix},\hspace{5pt} \Theta_{2} = \Theta\begin{bmatrix}0&0\\\frac{1}{2}&\frac{1}{2}\end{bmatrix}, \hspace{5pt} \Theta_{3} = \Theta\begin{bmatrix}0&0\\\frac{1}{2}&0\end{bmatrix}, \hspace{5pt} \Theta_{4} = \Theta\begin{bmatrix}0&0\\0&\frac{1}{2}\end{bmatrix}, \hspace{5pt} \Theta_{5} = \Theta\begin{bmatrix}\frac{1}{2}&0\\0&0\end{bmatrix}, \\
&\Theta_{6} = \Theta\begin{bmatrix}\frac{1}{2}&0\\0&\frac{1}{2}\end{bmatrix},\hspace{5pt}\Theta_{7} = \Theta\begin{bmatrix}0&\frac{1}{2}\\0&0\end{bmatrix},\hspace{5pt}\Theta_{8} = \Theta\begin{bmatrix}\frac{1}{2}&\frac{1}{2}\\0&0\end{bmatrix},\hspace{5pt}\Theta_{9} = \Theta\begin{bmatrix}0&\frac{1}{2}\\\frac{1}{2}&0\end{bmatrix},\hspace{5pt}\Theta_{10} = \Theta\begin{bmatrix}\frac{1}{2}&\frac{1}{2}\\\frac{1}{2}&\frac{1}{2}\end{bmatrix}.
 \end{alignedat}
\end{equation}
All even theta constants can be related through algebraic identities
to four \emph{fundamental} ones,  $\Theta_1$, $\Theta_2$,
$\Theta_3$, $\Theta_4$~\cite{Previato:2013}. They are weight $\frac 12$ Siegel modular forms for a subgroup of $\text{Sp}(4,\mathbb Z)$. Their transformation properties can be found in~\cite{freitag1983}. 

A particularly interesting genus-2 Siegel modular form is Igusa's cusp form $\chi_{10}$~\cite{igusa1967}. It is defined as the product of all genus 2 theta constants $\Theta_j$~\eqref{theta_1..10} of even characteristic,
\bea\label{chi10 def}
\chi_{10}=2^{-12}\prod_{j=1}^{10}\Theta_j^2~.
\eea
It is a genus-two Siegel modular form of weight $10$ for the full symplectic group ${\rm Sp}(4,\mathbb Z)$,

\subsection{Hyperelliptic curves}
\label{SWRHform}
A generic genus $g$ hyperelliptic curve takes the form
\bea\label{genus g curve}
Y^2=\prod_{j=1}^{2g}(X-r_j)~.
\eea
where $r_j$ are the roots. 
We particularly focus on the case of genus 
$g=2$. 

\sss{Rosenhain form for genus $2$.}
A canonical form of a genus 2 curve is the so-called Rosenhain form~\cite{Rosenhain:1851}
\bea
\label{curveRosenhain_app}
Y^2 = X(X-1)(X-\lambda_1)(X-\lambda_2)(X-\lambda_3).
\eea
It is a degree $2g-1=5$ equation, rather than the generic hyperelliptic curve~\eqref{genus g curve} of degree $2g=6$. Every genus
two hyperelliptic curve can be brought to the Rosenhain form. The three roots $\lambda_i$  of the polynomial are also referred to as
\emph{Rosenhain invariants}. These invariants can be understood as being complementary to the Igusa invariants~\cite{igusa1967,igusa1962}.
The Rosenhain invariants can be expressed in terms of even theta constants as
\be \label{thetaconstants_app}
\lambda_1 = \frac{\Theta_1^2\,\Theta_3^2}{\Theta_2^2\,\Theta_4^2},
\hspace{10pt} \lambda_2 =
\frac{\Theta_3^2\,\Theta_8^2}{\Theta_4^2\,\Theta_{10}^2},
\hspace{10pt}\lambda_3 =
\frac{\Theta_1^2\,\Theta_8^2}{\Theta_2^2\,\Theta_{10}^2}.
\ee
with $\Theta_j$ as in~\eqref{theta_1..10}, defines by~\eqref{definitionstheta}.

To relate the Rosenhain curve~\eqref{curveRosenhain_app} to an ordinary sextic~\eqref{genus g curve}, note that a degree 5 polynomial can be obtained from degree 6 by a linear fractional transformation, which maps three of the roots $r_j$ to $\infty$, $0$ and 1. Linear fractional maps leave
cross-ratios invariant, which is a convenient way to relate the two curves. If we define the cross-ratio of four
points $z_i\in \mathbb{CP}^1$ as
\be
\label{crossratio}
C(z_1,z_2,z_3,z_j)=\frac{(z_1-z_3)(z_2-z_j)}{(z_1-z_j)(z_2-z_3)},
\ee
the curves can be related as  $C(\{\infty,0,1,\lambda_j\})=\lambda_j$ for $j=1,2,3$.

\subsection{Discriminants and Siegel modular forms}
\label{subsec:hyperelliptic-discriminants}
We furthermore need an identity relating the discriminant of a hyperelliptic curve to the period matrix through a Siegel modular form. Such identities are due to P. Lockhart~\cite{Lockhart1994}, based on the  hyperelliptic Thomae correspondence~\cite{thomae1870beitrag,Fay1973}.\footnote{(see also de Jong~\cite{Jong2007})}

\sss{Discriminant identity.} Let $y^2=f(x)$ be hyperelliptic curve, where $f$ is monic, separable, and of degree $2g+1$ (an example of which is the Rosenhain form~\eqref{curveRosenhain_app} for $g=2$). Let $\Delta=\Disc(f)$ be the discriminant. We choose the basis of holomorphic differentials
\bea
    \omega_j=\frac{x^{j-1}dx}{2y}~,\qquad j=1,\ldots,g~,
\eea
and let $\{\alpha_1,\ldots,\alpha_g,\beta_1,\ldots,\beta_g\}$ be a symplectic basis of $H_1(X,\mathbb Z)$, with intersections
\bea
    \alpha_i\cdot\alpha_j=0~,\qquad \beta_i\cdot\beta_j=0~,\qquad \alpha_i\cdot\beta_j=\delta_{ij}~.
\eea
We form the period matrices
\bea
    A_{ij}=\int_{\alpha_j}\omega_i~,\qquad B_{ij}=\int_{\beta_j}\omega_i~,
\eea
such that the Riemann period matrix is $   \Omega=A^{-1}B\in\mathbb H_g$.

\begin{theorem}[\cite{Lockhart1994}]\label{thm:discriminant}
Define the binomial coefficients
\bea
    r=\binom{2g+1}{g+1}~,\qquad n=\binom{2g}{g+1}~.
\eea
There exists a Siegel modular form $\varphi_g$ of weight $4r$ on $\Gamma_g(2)$, constructed from the hyperelliptic even theta constants, such that
\bea\label{theorem identity}
    \pi^{-4gr}(\det A)^{4r}\Delta^n=\varphi_g(\Omega)~.
\eea
\end{theorem}

The modular form $\varphi_g$ is constructed as follows. For a characteristic $\eta=(a,b)\in(\frac12\mathbb Z/\mathbb Z)^{2g}$, the corresponding theta constant is obtained from~\eqref{definitionstheta} by setting $\bz=0$, $\Theta\left[\eta\right](0|\Omega)$.
After choosing an ordering of the $2g+1$ finite branch points together with the point at infinity, the hyperelliptic Thomae correspondence associates to each subset $T$ of $g+1$ branch points, modulo complementation, an even theta characteristic $\eta_T=(a_T,b_T)$ (see also~\cite{thomae1870beitrag,Fay1973}). Equivalently, after distinguishing the point at infinity, one may take $T$ to run over the subsets of the finite branch points of cardinality $g+1$. Then
\bea
\varphi_g(\Omega)=\prod_T\Theta\left[\eta_T\right](0|\Omega)^8~.
\eea
The eighth powers remove the sign ambiguities in the theta transformation law, and the resulting product is a Siegel modular form on $\Gamma_g(2)$ of weight $4r$.

\sss{Examples for $g=1,2,3$.}
For $g=1$, the relevant even theta constants are the three classical theta constants $\jt_2$, $\jt_3$ and $\jt_4$~\eqref{jt_elliptic}. Therefore, $\varphi_1=\jt_2^8\jt_3^8\jt_3^8$, which we may simplify using Dedekind's eta function~\eqref{jt234=eta} to give $\varphi_1=2^8\eta^{24}$. Since $r=3$ and $n=1$, up to a normalisation factor $c_1$, we have 
\bea
\Delta\, A^{12}=c_1\, \eta(\tau)^{24}~.
\eea
For $g=2$, the construction selects the ten even theta characteristics. The product over those is the fourth power of Igusa's cusp form $\chi_{10}$~\eqref{chi10 def}, up to prefactors. This gives
$\varphi_2=2^{48}\chi_{10}^4$. With $r=10$ and $n=4$, we can take the fourth root of the identity~\eqref{theorem identity}, after which it becomes 
\bea
    \Delta(\det A)^{10}=c_2\, \chi_{10}(\Omega)~,
\eea
with another constant $c_2$.

For $g=3$, the construction gives the product over the $35$ hyperelliptic even theta characteristics associated to four-element subsets of the seven finite branch points:
\bea
    \varphi_3(\Omega)=\prod_{1\leq i<j<k<\ell\leq 7}\Theta[\eta_{ijk\ell}](0,\Omega)^8~.
\eea
This product has weight $140$, and~\eqref{theorem identity} becomes
\bea
\Delta^{15}(\det A)^{140}=c_3\, \varphi_3(\Omega)~.
\eea
Unlike the genus-two case, this is not the product over all even theta constants: a genus-three curve has $36$ even theta characteristics, while on the hyperelliptic locus one even theta constant vanishes. The form $\varphi_3$ is the product of the remaining $35$ non-vanishing hyperelliptic even theta constants to the eighth power. It becomes clear that we can not reduce the identity as $\Delta\times\dots$ as is the case for $g=1,2$, since this gives fractional powers of the period matrix determinant as well as of $\varphi_3$.

\section{The algebraic functions for rank 3}
\label{app:algebraic functions}
In this appendix, we list all the universal functions for rank $\rho=3$ Segre numbers, obtained in~\cite[Section 4.3]{Gottsche:2020ass}. We follow the definitions and notation closely.
\subsection{The algebraic functions}
\sss{Definitions.}
The algebraic functions $S$, $Y$ and $Z$ for rank $\rho$ can be defined in more than one way. Let us explain it for the example of the function $S$. Ref~\cite{Gottsche:2020ass} write the algebraic functions $S_{j,s}$ for subsets $J\subset [\rho-1]\coloneqq \{1,\dots, \rho-1\}$. From these, define
\bea
S_{J,s}=S_s+\sum_{j\in J}S_{j,s}~.
\eea
For example for $\rho=3$, this gives the four functions
\bea\label{4 functions rho=3}
S_{\varnothing,s}&=S_s~, \\
S_{\{1\},s}&=S_s+S_{1,s}~, \\
S_{\{2\},s}&=S_s+S_{2,s}~, \\
S_{\{1,2\},s}&=S_s+S_{1,s}+S_{2,s}~, \\
\eea
which are solutions to a quartic equation for each rank $s$. Changing variables from $z$ to $t$ using~\eqref{changeofvar}, we write the four functions~\eqref{4 functions rho=3} as series in $z$, which have half-integer exponents with complex coefficients. The solutions for $\rho=3$ come in pairs of two, which are related by complex conjugation and alternating signs:
\bea\label{conjugation sign}
\upsilon S_{\{1,2\},s}&=S_{\varnothing,s}~, \\
\varkappa\, S_{\{2\},s}&=S_{\{1\},s}~.
\eea
Here, $\upsilon\colon z^{\frac12}\mapsto -z^{\frac12}$, and $\varkappa\colon i\mapsto -i$, as in~\eqref{upsilon varkappa}.

The generating function of rank $\rho$ Segre numbers depends only on the functions $S_s$, and $S_{j,s}$ for $j=1,\dots, \rho-1$. For $\rho=3$, we construct them from the solutions $S_{\varnothing,s}$ and $S_{j,s}$ as:
\bea
S_s&=S_{\varnothing,s}~, \\
S_{j,s}&=S_{\{j\},s}-S_{\varnothing,s}~.
\eea
Similarly, the functions $Y$ and $Z$ carry indices with different meanings. Define again
\bea
Y_{J,s}\coloneqq Y_s\prod_{j\in J}Y_{j,s}~, \quad Z_{J,s}\coloneqq Z_s\prod_{i\le j\in J}Z_{ij,s}~,
\eea
for all subsets $J\subset [\rho-1]$. For $\rho=3$, this gives the functions
\begin{equation}
\begin{alignedat}{4}
Y_{\varnothing,s}&=Y_s~, \qquad &Z_{\varnothing,s}&=Z_s~,  \\
Y_{\{1\},s}&=Y_sY_{1,s}~, \qquad  &Z_{\{1\},s}&=Z_sZ_{11,s}~, \\
Y_{\{2\},s}&=Y_sY_{2,s}~,  \qquad  &Z_{\{2\},s}&=Z_sZ_{22,s}~,  \\
Y_{\{1,2\},s}&=Y_sY_{1,s}Y_{2,s}~, \qquad &Z_{\{1,2\},s}&=Z_sZ_{11,s}Z_{12,s}Z_{22,s}~. \\
\end{alignedat}
\end{equation}
The solutions are again chosen such that~\eqref{conjugation sign} holds, with $S$ replaced by the functions $Y$ and $Z$.
The generating function depends on $Y_s$, $Y_{j,s}$, $Z_s$ and $Z_{ij,s}$, which we extract as
\begin{equation}
\begin{alignedat}{4}
Y_s&=Y_{\varnothing,s}~, \qquad &Z_s&=Z_{\varnothing,s}~, \\
Y_{j,s}&=Y_{\{j\},s}Y_{\varnothing,s}^{-1}~, \qquad &Z_{jj,s}&=Z_{\{j\},s}Z_s^{-1}~, \\
&  \qquad &Z_{12,s}&=Z_{\{1,2\},s}\left(Z_{\varnothing,s}Z_{11,s}Z_{22,s}\right)^{-1}~. 
\end{alignedat}
\end{equation}

For $s=1,2,4,5$, Ref.~\cite{Gottsche:2020ass} finds explicit power series
$s_{1},\ldots, s_{4}$, $y_{1},\ldots, y_{4}$, $z_{1},\ldots, z_{4}$, with $s_2=0$,
such that\footnote{We change the branch of the square root for the definition of $S_{\varnothing,s}$ compared to~\cite{Gottsche:2020ass}, which interchanges $S_{\varnothing,s}$ with $S_{\{1,2\},s}$. This choice is compatible with the universal asymptotics~\eqref{asymptotics}.}
\bea\label{S Y Z in terms of s y z}
S_{\varnothing,s}&=\tfrac{1}{2}\big(s_{1}+s_{2}+3^{\frac{1}{2}}t^{\frac{1}{2}}\big(\smfr{1}{3t}(s_{3}+s_{4})\big)^{\frac{1}{2}}\big)~,\
S_{\{1\},s}=\tfrac{1}{2}\big(s_{1}-s_{2}+3^{\frac{1}{2}}i t\big(-\smfr{1}{3t^2}(s_{3}-s_{4})\big)^{\frac{1}{2}}\big)~,\\
Y_{\varnothing,s}&=\tfrac{1}{2}\big(y_{1}+y_{2}-3^{\frac{1}{2}}t^{\frac{1}{2}}\big(\smfr{1}{3t}(y_{3}+y_{4})\big)^{\frac{1}{2}}\big)~,\
Y_{\{1\},s}=\tfrac{1}{2}\big(y_{1}-y_{2}+3^{\frac{1}{2}}i t\big(-\smfr{1}{3t^2}(y_{3}-y_{4})\big)^{\frac{1}{2}}\big)~,\\
Z_{\varnothing,s}&=\tfrac{1}{2}\big(z_{1}+z_{2}+3^{\frac{1}{2}}t^{\frac{1}{2}}\big(\smfr{1}{3t}(z_{3}+z_{4})\big)^{\frac{1}{2}}\big)~,\
Z_{\{1\},s}=\tfrac{1}{2}\big(z_{1}-z_{2}-3^{\frac{1}{2}}i t\big(-\smfr{1}{3t^2}(z_{3}-z_{4})\big)^{\frac{1}{2}}\big)~.
\eea
These four functions satisfy a quartic equation with coefficients being polynomials in the $s_i$, $y_i$ or $z_i$.

\sss{Quartic equations for $\rho=3$.}
In~\cite{Gottsche:2020ass}, the functions $S$, $Y$ and $Z$ for $\rho=3$ are conjecturally written as roots of a quartic polynomial, with coefficients rational in $t$. These quartic polynomials have different presentations depending on the rank $s$.

\sss{$\boldsymbol{s=1}$.} 
For $s=1$, the change of variables is $z = t(1+\tfrac{2}{3}t)^{\frac{2}{3}}$. Conjecturally, the power series
$S_{\varnothing,1}$, $S_{\{1,2\},1}$, $S_{\{1\},1}$, $S_{\{2\},1}$ are the four solutions of 
\bea
x^4 + 2tx^3 - (3t+ t^2)x^2 - (3t^2 + 2t^3)x -(t^3 + \smfr{2}{3}t^4)=0~.
\eea
Similarly for $Y_{\varnothing,1}$, $Y_{\{1,2\},1}$, $Y_{\{1\},1}$, $Y_{\{2\},1}$: 
\bea
x^4-(4+\smfr{17}{3}t)(1+\smfr{2}{3}t)^{\frac{1}{2}}x^3+(6 + 18t + 16t^2 + \smfr{31}{9}t^3)x^2-(4+\smfr{17}{3}t)(1+\smfr{2}{3}t)^{\frac{7}{2}}x+(1+\smfr{2}{3}t)^6=0~,
\eea
and for $Z_{\varnothing,1}$, $Z_{\{1,2\},1}$, $Z_{\{1\},1}$, $Z_{\{2\},1}$: 
\bea
x^4-6\frac{1+\smfr{10}{9}t}{1+\smfr{2}{3}t}x^3+\frac{(13+\smfr{58}{3}t+\smfr{55}{9}t^2)(1+\smfr{10}{9}t)}{(1+\smfr{2}{3}t)^3}x^2+\frac{(4+\smfr{5}{3}t)(1+\smfr{10}{9}t)^2}{(1+\smfr{2}{3}t)^3}(-3x+1)=0~.
\eea
Using the notation introduced above, these can be written as
\bea
s_1&=-t~, \quad s_3=6t(1+\smfr{5}{6}t)~, \quad s_4=6t(1+\smfr{2}{3}t)^{\frac{1}{2}}(1+\smfr{10}{9}t)^{\frac{1}{2}}~,\\
y_1&=(2+\smfr{17}{6}t)(1+\smfr{2}{3}t)^{\frac{1}{2}}~,\quad
y_2=\smfr{3}{2}t(1+\smfr{10}{9}t)^{\frac{1}{2}}~,\\
y_3&=6t+\smfr{25}{2}t^2+\smfr{20}{3}t^3~,\quad
y_4=(6t+\smfr{17}{2}t^2)(1+\smfr{2}{3}t)^{\frac{1}{2}}(1+\smfr{10}{9}t)^{\frac{1}{2}}~, \\
z_1&=\frac{3+\smfr{10}{3}t}{1+\smfr{2}{3}t}~, \quad
z_2=\frac{(1+\smfr{5}{3}t)(1+\smfr{10}{9}t)^{\frac{1}{2}}}{(1+\smfr{2}{3}t)^{\frac{3}{2}}}~,\quad
z_3=\frac{6t+\smfr{35}{3}t^2+\smfr{50}{9}t^3}{(1+\smfr{2}{3}t)^3}~,\quad 
z_4=\frac{6t(1+\smfr{10}{9}t)^{\frac{3}{2}}}{(1+\smfr{2}{3}t)^{\frac{5}{2}}}.
\eea

\sss{$\boldsymbol{s=2}$.} For $s=2$, the change of variables is $z = t(1+\tfrac{1}{3}t)^{\frac{1}{3}}$. Conjecturally, the power series
$S_{\varnothing,2}$, $S_{\{1,2\},2}$, $S_{\{1\},2}$, $S_{\{2\},2}$ are the four solutions of 
\bea
x^4 + 4tx^3 + (-3t+ 4t^2)x^2 - 6t^2x -4t^3 - \smfr{4}{3}t^4=0~.
\eea
Similarly for $Y_{\varnothing,2}$, $Y_{\{1,2\},2}$, $Y_{\{1\},2}$, $Y_{\{2\},2}$:
\bea
x^4-(4+\smfr{7}{3}t)(1+\smfr{4}{3}t)^{\frac{1}{2}}x^3+(6+8t)(1+\smfr{1}{3}t)^2 x^2-(4+\smfr{7}{3}t)(1+\smfr{4}{3}t)^{\frac{1}{2}}(1+\smfr{1}{3}t)^3x+(1+\smfr{1}{3}t)^6=0~,
\eea
and for $Z_{\varnothing,2}$, $Z_{\{1,2\},2}$, $Z_{\{1\},2}$, $Z_{\{2\},2}$:
\bea
x^4-\frac{(6+5t)(1+\smfr{4}{3}t)(1+\smfr{4}{9}t)}{(1+\smfr{1}{3}t)^3}x^3+\frac{13(1+\smfr{4}{3}t)^2(1+\smfr{4}{9}t)}{(1+\smfr{1}{3}t)^3}x^2+\frac{4(1+\smfr{4}{3}t)^2(1+\smfr{4}{9}t)^2}{(1+\smfr{1}{3}t)^4}(-3x+1)=0~.
\eea
Using the notation~\eqref{S Y Z in terms of s y z}, these can be written as
\bea
s_1&=-2t~, \quad s_3=6t+4t^2~, \quad s_4=6t(1+\smfr{4}{3}t)^{\frac{1}{2}}(1+\smfr{4}{9}t)^{\frac{1}{2}}~,\\
y_1&=(2+\smfr{7}{6}t)(1+\smfr{4}{3}t)^{\frac{1}{2}}~,\quad
y_2=\smfr{3}{2}t(1+\smfr{4}{9}t)^{\frac{1}{2}},\\
y_3&=6t+\smfr{17}{2}t^2+\smfr{8}{3}t^3~,\quad
y_4=(6t+\smfr{7}{2}t^2)(1+\smfr{4}{3}t)^{\frac{1}{2}}(1+\smfr{4}{9}t)^{\frac{1}{2}}~,\\
z_1&=\frac{3(1+\smfr{5}{6}t)(1+\smfr{4}{9}t)(1+\smfr{4}{3}t)}{(1+\smfr{1}{3}t)^3}~, \, 
z_2=\frac{(1+\smfr{25}{6}t+\smfr{16}{9}t^2)(1+\smfr{4}{3}t)^{\frac{1}{2}}(1+\smfr{4}{9}t)^{\frac{1}{2}}}{(1+\smfr{1}{3}t)^3}~,\\
z_3&=\frac{(24t + \smfr{101}{2}t^2 + \smfr{98}{3}t^3 + \smfr{20}{3}t^4)(1+\smfr{4}{3}t)(1+\smfr{4}{9}t)}{(1+\smfr{1}{3}t)^6}~, \, 
z_4=\frac{(24t+  \smfr{59}{2}t^2 + \smfr{26}{3}t^3)(1+\smfr{4}{3}t)^{\frac{3}{2}}(1+\smfr{4}{9}t)^{\frac{3}{2}}}{(1+\smfr{1}{3}t)^6}.
\eea

\sss{$\boldsymbol{s=3}$.}
For $z = t$, the series are
\bea
S_{\varnothing,3}&=-\smfr{3}{2}t-3^{\frac{1}{2}}t^\frac{1}{2}(1+\smfr{3}{4}t)^{\frac{1}{2}}~,\quad 
S_{\{1\},3}=-\tfrac{1}{2}(3+3^{\frac{1}{2}}i)t~,\\
Y_{\varnothing,3}&=1+\smfr{3}{2}t-3^{\frac{1}{2}}t^\frac{1}{2}(1+\smfr{3}{4}t)^{\frac{1}{2}}~,\quad 
Y_{\{1\},3}=1~, \\
Z_{\varnothing,3}&=\frac{2(1+t)(1+\frac{3}{4}t)^{\frac{1}{2}}}{(1+3t)(1+\frac{3}{4}t)^{\frac{1}{2}}-\frac{3}{2}3^{\frac{1}{2}}t^{\frac{1}{2}}(1+t)}~,\quad
Z_{\{1\},3}=1+t~.
\eea

\sss{$\boldsymbol{s=4}$.}
For $s=4$, the change of variables is $z = t(1-\tfrac{1}{3}t)^{-\frac{1}{3}}$. Conjecturally, the power series
$S_{\varnothing,4}$, $S_{\{1,2\},4}$, $S_{\{1\},4}$, $S_{\{2\},4}$ are the four solutions of 
\bea\label{S s=4}
x^4 + 2tx^3 - (3t + t^2)x^2 - (3t^2 + 2t^3)x - (t^3 + \smfr{2}{3}t^4)=0~.
\eea
Similarly for $Y_{\varnothing,4}$, $Y_{\{1,2\},4}$, $Y_{\{1\},4}$, $Y_{\{2\},4}$:
\bea
x^4-(4-\smfr{1}{3}t)(1+\smfr{2}{3}t)^{\frac{1}{2}}x^3+(6+4t)(1-\smfr{1}{3}t)^2x^2-(4-\smfr{1}{3}t)(1-\smfr{1}{3}t)^3(1+\smfr{2}{3}t)^{\frac{1}{2}}x+(1-\smfr{1}{3}t)^6=0~.
\eea
and for $Z_{\varnothing,4}$, $Z_{\{1,2\},4}$, $Z_{\{1\},4}$, $Z_{\{2\},4}$:
\bea
x^4&-\frac{(6 + 14t - \smfr{7}{3}t^2)(1-\smfr{2}{9}t)(1+\smfr{2}{3}t)^3}{(1-\smfr{1}{3}t)^6}x^3+\frac{(13 +36t + \smfr{23}{3}t^2 - \smfr{298}{27}t^3)(1-\smfr{2}{9}t)(1+\smfr{2}{3}t)^4}{(1-\smfr{1}{3}t)^8}x^2\\&+\frac{4(1-\smfr{2}{9}t)^2(1+\smfr{2}{3}t)^8}{(1-\smfr{1}{3}t)^{10}}(-3x+1)=0~.
\eea

\sss{$\boldsymbol{s=5}$.}
The change of variables is $z = t(1-\tfrac{2}{3}t)^{-\frac{2}{3}}$. Conjecturally, the power series
$S_{\varnothing,5}$, $S_{\{1,2\},5}$, $S_{\{1\},5}$, $S_{\{2\},5}$ are the four solutions of 
\bea
x^4 &+\frac{2t(5 - \smfr{20}{3}t + \smfr{11}{9}t^2)}{(1-\smfr{2}{3}t)^2}x^3+  \frac{-3t + 45t^2 - \smfr{305}{3}t^3 + \smfr{740}{9}t^4 - \smfr{220}{9}t^5 + \smfr{193}{81}t^6}{(1-\smfr{2}{3}t)^4}x^2\\&-\frac{3t^2(5 - \smfr{20}{3}t + \smfr{11}{9}t^2)(1 - \smfr{14}{3}t + t^2)}{(1-\smfr{2}{3}t)^3}x-\frac{t^3(5 - \smfr{20}{3}t + \smfr{11}{9}t^2)^2}{(1-\smfr{2}{3}t)^3}=0~.
\eea
Similarly, for  $Y_{\varnothing,5}$, $Y_{\{1,2\},5}$, $Y_{\{1\},5}$, $Y_{\{2\},5}$: 
\bea
x^4-(4+\smfr{1}{3}t)(1-\smfr{2}{3}t)^{\frac{1}{2}}x^3+(6-6t-\smfr{1}{9}t^3)x^2-(4+\smfr{1}{3}t)(1-\smfr{2}{3}t)^{\frac{7}{2}}x+(1-\smfr{2}{3}t)^6=0~,
\eea
and for $Z_{\varnothing,5}$, $Z_{\{1,2\},5}$, $Z_{\{1\},5}$, $Z_{\{2\},5}$:
\bea
x^4&-\frac{(6 + 22t - \smfr{7}{3}t^2)(1-\smfr{2}{9}t)(1+\smfr{1}{3}t)^4}{(1-\smfr{2}{3}t)^7}x^3\\&+\frac{(13 + \smfr{14}{3}t - \smfr{851}{9}t^2 + \smfr{1943}{27}t^3 + \smfr{5455}{81}t^4 - \smfr{19009}{243}t^5 +\smfr{11623}{729}t^6 - \smfr{202}{2187}t^7 - \smfr{689}{6561}t^8)(1-\smfr{2}{9}t)(1+\smfr{1}{3}t)^6}{(1-\smfr{2}{3}t)^{15}}x^2\\&+\frac{4(1-\smfr{2}{9}t)^2(1+\smfr{1}{3}t)^{12}(1-\smfr{11}{12}t)}{(1-\smfr{2}{3}t)^{15}}(-3x+1)=0~.
\eea

\subsection{Corrections}
\label{app:corrections}
By the explicit calculation of physical couplings for $\SU(3)$ using the Seiberg--Witten geometry, as completed in Section~\ref{sec:SU(3) calculation}, we find remarkable agreement with the conjectured universal series. Indeed, for each flavor $s=N_f=1,\dots, 5$, this amounts to calculating $16$ universal series ($j,k=1,2$)
\bea
Q_s,~ R_s,~ S_s,~ S_{j,s},~ T_s,~ V_s,~ W_s,~ X_s,~ Y_s,~ Y_{j,s},~ Z_s,~  Z_{jk,s},~
\eea
from 14 physical couplings ($a=1,2$, we include also $\tu_{2;N_f}$ which is not required for the dictionary)
\bea
 \tC_{N_f},~ \hat \tC_{N_f},~ \tD_{a;N_f},~ \tx_{N_f},~ \tG_{N_f},~ \tH_{N_f},~ \tm_{N_f},~ \tn_{N_f},~ \tq_{12;N_f},~  \tu_{a;N_f},~  \tx_{a;N_f}~. 
\eea
In particular, we have checked $5\times 16=80$ identities. We merely find \emph{two} discrepancies between the physical calculation and the proposed universal series, which we discuss in this Appendix. 

The first discrepancy concerns the $S$-functions for $\rho=3$ and $s=4$, where we suggest that the defining quartic~\eqref{S s=4} should be replaced by
\bea\label{S correction}
x^4+8tx^3-\frac{3t(1-8t+\tfrac{20}{9}t^2)x^2}{1-\tfrac13t}-\frac{12t^2(1-\tfrac{24}{9}t+\tfrac{4}{9}t^2)x}{1-\tfrac13t}-16t^3(1-\tfrac13t)=0~,
\eea
giving new series $\widetilde S_{a,4}$ and $\widetilde S_4$.
This solution has
\bea
s_1&=-4t~, \\
s_2&=0~, \\
s_3&=6t(1-\tfrac49t^2)(1-\tfrac13t)^{-1}~, \\
s_4&=6t(1-\tfrac29t)^{\frac12}(1+\tfrac23t)^{\frac32}(1-\tfrac13t)^{-1}~,
\eea
in the notation of~\eqref{S Y Z in terms of s y z}.
With this correction, we have the uniform asymptotics~\eqref{asymptotics} for $s=1,\dots, 5$, which is not valid with the $S$-functions proposed in~\eqref{S s=4}.

The second correction concerns the function $R_5$. We propose a correction $\widetilde R_5$ of a single coefficient in $R_5$, as 
\bea\label{R5 discrepancy}
\widetilde R_5=t-\tfrac15z^3~.
\eea

Evidence for the corrections comes from the constraints~\eqref{constraints} of the dictionary for $s=0,\dots, 5$: 
\begin{subequations}\label{constraints app}
\begin{align}
V_s^{\frac{s-1}{2}} W_s^sY_s Y_{1,s}^{\frac12}Y_{2,s}^{\frac12}&=1~,  \\
Y_{1,s}^{\frac s2}Z_{11,s}Z_{12,s}^{\frac12}&=1~, \\
Y_{2,s}^{\frac s2}Z_{22,s}Z_{12,s}^{\frac12}&=1~, \\
sR_s+2S_{s}+S_{1,s}+S_{2,s}&=0~, \label{last constraint}
\end{align}
\end{subequations}
We find that all equations hold for the series conjectured in~\cite{Gottsche:2020ass}, except~\eqref{last constraint} for $s=4$ and $s=5$. For $s=4$, with the functions $S_{a,s}$ of~\eqref{S s=4}, the lhs of~\eqref{last constraint} is $3t$ rather than $0$. For $s=5$, the lhs is $z^3$ rather than 0. 
With the corrections $\widetilde S_{a,4}$~\eqref{S correction} and $\widetilde R_5$~\eqref{R5 discrepancy} as suggested above, all four equations~\eqref{constraints app} hold to higher order. 

A similar issue occurs as well for $\rho=2$, where the constraint for $s=3$ reads $3R_3+2S_{3}+S_{1,3}+S_{2,3}=-z^2$, and a correction of $\widetilde R_3=t+\tfrac13z^2$ has been suggested in~\cite{Furrer:2026byd}.

\section{Lie algebras and root systems}\label{app:lie_alg}
In this appendix, we collect some basic material on Lie algebras used in section~\ref{sec:sum_flux}. The reader is referred to~\cite{humphreys1972introduction} for proofs of the claimed results.
\par Let $\mf{g}$ be a finite dimensional Lie algebra. An element $x \in \mf{g}$ is called \textit{nilpotent} or \textit{ad-nilpotent} if $\operatorname{ad}_x: \mf{g} \rightarrow \mf{g}$ defined by
\begin{equation}
\operatorname{ad}_x(y)=[x, y] \text {, }
\end{equation}
is nilpotent. If ad$(x)$ is diagonalisable, then $x$ is called \textit{semisimple}. A subalgebra of $\mf{g}$ consisting of semisimple elements is called a \textit{toral subalgebra}. One can prove that a toral subalgebra of a complex Lie algebra is abelian\footnote{This result works for Lie algebras over algebraically closed fields. Thus, we need to consider complex Lie algebras.}. A maximal toral subalgebra $\mf{t}_{\mf{g}}$ of $\mf{g}$ is called the \textit{Cartan subalgebra} of $\mf{g}$. Considering the complexification $\mf{g}^{\mathbb{C}}$, since $\mf{t}_{\mf{g}}^{\mathbb{C}}$ is abelian, the elements of $\mf{t}^{\mathbb{C}}_{\mf{g}}$ can be simultaneously diagonalized. More precisely, the set of operators $\left\{\mr{ad}_x: x \in \mathfrak{t}^{\mathbb{C}}_{\mf{g}}\right\}$ are simultaneously diagonalizable. This means that the complexification $\mf{g}^{\mathbb{C}}$ can be decomposed as a direct sum
\begin{equation}
\mf{g}^{\mathbb{C}}=\bigoplus_{\alpha \in \mf{t}_{\mf{g}}^*} \mf{g}^{\mathbb{C}}_\alpha~,
\end{equation}
where $\mf{g}_\alpha\coloneqq \left\{y \in \mf{g} : \mr{ad}_x(y)=\alpha(x) y \quad \forall~ x \in \mf{t}_{\mf{g}}\right\}$. Let $\Phi\coloneqq \left\{\alpha \in \mf{t}_{\mf{g}}^*:\alpha\neq 0, \mf{g}_\alpha \neq 0\right\}$.
Then $\Phi$ is called the \textit{set of roots} of $\mf{g}$. Note that $\mf{g}_0=C_{\mf{g}}(\mf{t}_{\mf{g}})$, the centralizer of $\mf{t}_{\mf{g}}$ in $\mf{g}$. One can show that $C_{\mf{g}}(\mf{h})=\mf{h}$. Thus we have the decomposition
\begin{equation}
\mf{g}=\mf{h} \oplus \bigoplus_{\alpha \in \Phi} \mf{g}_\alpha .
\end{equation}
One can show that the restriction of the Killing form $\kappa$ to $\mf{h}$ is non-degenerate. Thus given $\phi \in \mf{h}^*$, there exists a unique $t_\phi \in \mf{h}$ such that $\phi(x)=\kappa\left(t_\phi, x\right)$ for all $x \in \mf{h}$. In particular, $\Phi \subset \mf{h}^*$ corresponds to $\left\{t_\alpha: \alpha \in \Phi\right\} \subset \mf{h}$.
\begin{theorem}
The following is true:
\begin{enumerate}
\item $\kappa\left(\mf{g}_\alpha, \mf{g}_\beta\right)=0$ if $\alpha, \beta \in \mf{h}^*$ with $\alpha+\beta \neq 0$. In particular $\kappa\left(\mf{h}, \mf{g}_\alpha\right)=0$ for all $\alpha \in \Phi$. \item  $\Phi$ spans $\mf{h}^*$.
\item If $\alpha \in \Phi$ then $-\alpha \in \Phi$.
\item If $\alpha \in \Phi$, then for $x \in \mf{g}_\alpha, y \in \mf{g}_{-\alpha}$, we have $[x, y]=\kappa(x, y) t_\alpha$. In particular $\left[\mf{g}_\alpha, \mf{g}_{-\alpha}\right]$ is one-dimensional with basis $t_\alpha$.
\item $\alpha\left(t_\alpha\right)=\kappa\left(t_\alpha, t_\alpha\right) \neq 0$ for $\alpha \in \Phi$.
\item If $\alpha \in \Phi$ and $x_\alpha \in \mf{g}_\alpha \backslash\{0\}$, then there exists $y_\alpha \in \mf{g}_{-\alpha}$ such that the subalgebra spanned by $x_\alpha, y_\alpha$ and $z_\alpha\coloneqq \left[x_\alpha, y_\alpha\right]$ is isomorphic to $\mf{su}(2)$ under the map
\begin{equation}
x_\alpha \mapsto\left(\begin{array}{ll}
0 & 1 \\
0 & 0
\end{array}\right), \quad y_\alpha \mapsto\left(\begin{array}{ll}
0 & 0 \\
1 & 0
\end{array}\right), \quad z_\alpha \mapsto\left(\begin{array}{cc}
1 & 0 \\
0 & -1
\end{array}\right) \text {. }
\end{equation}
Moreover
\begin{equation}
z_\alpha=\frac{2 t_\alpha}{\kappa\left(t_\alpha, t_\alpha\right)}, \quad z_\alpha=-z_{-\alpha} .
\end{equation}
\item $\operatorname{dim} \mf{g}_\alpha=1$ for $\alpha \in \Phi$.
\end{enumerate}
\end{theorem}
Since $\Phi$ spans $\mf{h}^*$, we can choose a basis $\left\{\alpha_1, \ldots, \alpha_r\right\} \subset \Phi$ of $\mf{h}^*$.
For $\beta \in \Phi$, we can write
\begin{equation}
\beta=\sum_{i=1}^r c_i \alpha_i~.
\end{equation}
Then one can prove that $c_i \in \mathbb{Q}$. Let $E_{\mathbb{Q}}$ be the $\mathbb{Q}$-span of all the roots. Then $\operatorname{dim}_{\mathbb{Q}} E_{\mathbb{Q}}=\operatorname{dim}_{\mathbb{C}} \mf{h}^*=\tex{rk}(\mf{g})$. Introduce a bilinear form on $\mf{h}^*$ as
\begin{equation}
(\alpha, \beta)=\kappa\left(t_\alpha, t_\beta\right)
\end{equation}
Then one can show that
\begin{equation}
(\alpha, \beta) \in \mathbb{Q} \text { and } \frac{2(\alpha, \beta)}{(\beta, \beta)} \in \mathbb{Z} \text {, for } \alpha, \beta \in \Phi \text {. }
\end{equation}
Moreover, this form is positive-definite on $E_{\mathbb{Q}}$. Let $E=\mathbb{R} \otimes_{\mathbb{Q}} E_{\mathbb{Q}}$.
Then $\Phi$ spans $E$ and $0 \notin \Phi$. Moreover if $\alpha \in \Phi$ then the only multiple of $\alpha$ in $\Phi$ is $-\alpha$. For $\alpha, \beta \in \Phi$, we have
\begin{equation}
\beta-\frac{2(\beta, \alpha)}{(\alpha, \alpha)} \alpha \in \Phi \text {. }
\end{equation}
A subset $\Delta \subset \Phi$ is called a \textit{base} if $\Delta$ is a basis of $E$ and for each root $\beta \in \Phi$, when we write
\begin{equation}
\beta=\sum_{\alpha \in \Delta} k_\alpha \alpha
\end{equation}
then $k_\alpha \in \mathbb{Z}$ and either all $k_\alpha$ are non-negative or non-positive. The crucial result is that there exists a basis of $\Phi$. The elements of $\Delta$ are called \textit{simple roots}. If all $k_\alpha \geq 0 \quad\left(k_\alpha \leq 0\right)$ then we call $\beta$ \textit{positive (negative)}. We write $\beta \succ 0 \quad(\beta \prec 0)$. The collection of positive (negative) roots is denoted by $\Phi^{+}\left(\Phi^{-}\right)$. Clearly $\Phi^{-}=-\Phi^{+}$. We put a partial order on $E$ by defining
\begin{equation}
\beta \prec \alpha \Leftrightarrow \alpha-\beta>0 \text { or } \beta=\alpha \text {. }
\end{equation}
If $\Delta$ is a base of $\Phi$ then $(\alpha, \beta) \leq 0$ for $\alpha \neq \beta$ in $\Delta$ then $\alpha-\beta \notin \Phi$. If $\alpha$ is a positive root but not simple, then $\alpha-\beta$ is a root for some $\beta \in \Delta$. In this case, $\alpha-\beta$ is necessarily a positive root. Let $\left\{\alpha_1, \ldots, \alpha_r\right\}=\Delta$ be a set of simple roots. Then the matrix $\left[\left\langle\alpha_i, \alpha_j\right\rangle\right]$ is called the Cartan matrix, where
\begin{equation}
\left\langle\alpha_i, \alpha_j\right\rangle\coloneqq 2 \frac{\left(\alpha_i, \alpha_j\right)}{\left(\alpha_j, \alpha_j\right)} \in \mathbb{Z} \text {. }
\end{equation}
Let $\Lambda_{\tex{w}} \subset E$ be defined as
\begin{equation}
\Lambda_{\tex{w}}\coloneqq \{\lambda \in E:\langle\lambda, \alpha\rangle \in \mathbb{Z} \quad \forall \alpha \in \Phi\} .
\end{equation}
$\Lambda_{\tex{w}}$ is called the set of \textit{weights} of $\mf{h}$. Clearly $\Lambda_{\tex{w}}$ is an additive subgroup of $E$ since $\langle\lambda, \alpha\rangle$ is linear in $\lambda$. One can show that $\lambda \in \Lambda_{\tex{w}}$ if and only if $\langle\lambda, \alpha\rangle \in \mathbb{Z}$ for all $\alpha \in \Delta$. Clearly $\Phi \subset \Lambda_{\tex{w}}$. Let $\Lambda_{\tex{r}}$ be the subgroup of $\Lambda_{\tex{w}}$ generated by $\Phi$. We call it the \textit{root lattice}. It is a lattice in $E$ since it is the $\mathbb{Z}$-span of an $\mathbb{R}$-basis of $E$. Fix a base $\Delta \subset \Phi$. Then $\lambda \in \Lambda_w$ is called \textit{dominant} if $\langle\lambda, \alpha\rangle \geq 0$ for all $\alpha \in \Delta$ and \textit{strongly dominant} if $\langle\lambda, \alpha\rangle>0$ for all $\alpha \in \Delta$. Let $\Lambda_{\tex{w}}^{+} \subset \Lambda_{\tex{w}}$ be the set of all dominant weights. A $\lambda \in \Lambda^{+}$ is called \textit{minimal} if $\mu \in \Lambda^{+}, \mu \prec \lambda$ implies $\mu=\lambda$. One can show that the set of minimal roots is in 1-1 correspondence with the coset $\Lambda_{\tex{w}} / \Lambda_{\tex{r}}$. In other words, every coset of $\Lambda_{\tex{w}} / \Lambda_{\tex{r}}$ contains a unique minimal weight.

\section{Symbol list} 
\begin{longtable}{|p{2.3cm}|p{10cm}|} 
\hline
Symbol & Meaning \\
\hline 
$a^a$ & Low-energy scalar fields \\
$a_{\rm D}^a$ & Scalars dual to $a^a$ \\
$\bm{a}$ & Vector $(a^1,\dots,a^{N-1})$ of scalar fields \\
$\bm{a}_{\rm D}$ & Vector $(a_{\rm D}^1,\dots,a_{\rm D}^{N-1})$ of dual scalar fields \\
$\alpha$ & $K$-theory class in $K(S)$ \\
$\alpha_a$ & Homology cycles on SW curve \\
$\alpha_M$ & Tautological class on moduli space $M$ \\
$\alpha_{N_f},\alpha_{N_f,0}$ & Constant part of effective gravitational coupling $A$ \\
$A$ & Effective gravitational coupling \\
$B$ & Effective gravitational coupling \\
$B(\cdot,\cdot)$ & Intersection form of 4-manifold $X$ \\
$\beta_a$ & Homology cycles on SW curve \\
$\vec{\beta}$ & Shift of integer cohomology lattice in the IR \\
$\beta_{N_f},\beta_{N_f,0}$ & Constant part of effective gravitational coupling $B$ \\
$b_i$ & Betti numbers of 4-manifold $X$ \\
$b_2^{\pm}$ & Dimension of the space of (anti-)self-dual 2-forms on $X$ \\
$c(\alpha_M)$ & Total Chern class of $\alpha_M$ \\
$c_1$, $c_2$ & Chern classes of sheaves on $S$ \\
$c_1(\alpha)$, $c_2(\alpha)$ & Chern classes of $K$-theory class $\alpha\in K(S)$ \\
$c_1(E),c_2(E)$& Chern classes of the rank $N$ vector bundle $E$ \\
$\ch(E)$ & Chern character of $E$\\
$\mathsf C$ & Charge conjugation operator\\
$C^{\rm phys}$& Subgroup of the center of $\widetilde G^{\rm phys}$\\
$C^{\rm phys}_{\rm max}$& Maximal admissible subgroup of the center of $\widetilde G^{\rm phys}$\\
$C^{\rm tw}$& Subgroup of the center of $\widetilde G^{\rm tw}$\\
$\chi$ & Topological Euler characteristic of $X$ \\
$\chi_{10}$ & Igusa cusp form \\
$\chih$ & Holomorphic Euler characteristic, or $\frac14(\chi+\sigma)$ \\
$C_{ij}$ , $\tC_{N_f}$, $\hat \tC_{N_f}$ & Exponentiated quadratic coupling to background fluxes \\
$C'$ & Exceptional divisor of the blowup \\
$\mathcal{D}$ & Vanishing locus of the discriminant \\
$\mathcal{D}^{(\ell)}$ & Codimension $\ell$ strata of $\mathcal{D}$\\
$\slashed D_j$ & Dirac operator for the $j$-th hypermultiplet \\
$\decoup,\decoup_{N_f}$ & Decoupling factor \\
$\Delta_{N_f},\Delta$ & Discriminant of SW curve \\
$\delta_{4},\delta_5$ & Shift in the universal function $\tG_{N_f}$ for $N_f=4,5$ \\
$E$ & Rank $N$ vector bundle associated to the gauge bundle \\
$\tD_{a,N_f}$ & The quantity $q_{ai}^{N_f}$ at the $\CN=1$ singularity \\
$e^{i\phi_n}$ & Global phase of SW contribution at $n$-th $\CN=1$ vacua\\
$\varepsilon_N$ & The phase $e^{2\pi i/N}$ \\
%$E_k$ & Eisenstein series \\
$\eta$ & Dedekind $\eta$-function \\
$\eta^{s}_{pqr},\eta^{\mr{f};j}_{pqr},\eta^{g}_{pqr}$ & Integral lift of various cocycles \\
$f_2(P),f_2(E)$ &'t Hooft flux for gauge bundle
\\
$\ov{f}_2(P),\ov{f}_2(E)$ &Integral lift of 't Hooft flux 
\\
$f_a$ & Complex parameters for surface observable \\
$\CF$ & Prepotential \\
$\CF_{IJK\dots}$ & Successive derivatives of prepotential with respect to $a^I$\\
${\rm Fr}(X)$ & Frame bundle of $X$\\
$g$ & Genus of hyperelliptic curve \\
$G_{N_f}$, $\tG_{N_f}$ & Contact term for surface observables \\
$G^{\rm phys}$& Structure group of the physical theory \\
$\widetilde G^{\rm phys}$& Simply connected cover of $G^{\rm phys}$ \\
$G_\rho$ & Galois group of field extension $\mathbb{Q}(i,\varepsilon_{2\rho})$ \\
$G^{\rm tw}$& Structure group of the twisted theory \\
$\widetilde G^{\rm tw}$& Simply connected cover of $G^{\rm tw}$ \\
$\Gamma$ & Lattice $L^{N-1}$ of IR photon partition function \\
$\Gamma(W_j^{\pm}\otimes E)$ & Space of sections of the chiral $\spinc$ bundle coupled to $E$ \\
$\gamma$ & Electromagnetic charge of BPS state or element of ${\rm Sp}(2g,\Z)$ \\
$\segregen_{\segre}$ & Generating function of virtual Segre numbers \\
$\segregen_{\universal}$ & Projection of universal Segre series $\universal_\alpha$ \\
$G_{N_f},\tG_{N_f}$ & Contact term for surface observable \\
$h^\vee(G)$& Dual Coxeter number of Lie group $G$\\
$\mathbb{H}_g$ & Genus $g$ Siegel upper half space \\
$H_{N_f,j},\tH_{N_f}$ & Contact term for surface observable\\
$\iota$ & Inclusion map from $\Z_{n}$ to $\CZ$\\
$\tilde{I}(S),\tilde{I}^a(S)$ & Surface observables for surface class $S$ \\
%$\CI_\mu^X$ & $u$-plane integrand on $X$ \\
$J$ & Period point \\
$k,k(P)$ & Instanton number of principal bundle\\
$\bfk_{\pm}$ & Projection of a class $\bk$ to (anti)selfdual class \\
$\bfk_j$ & Background fluxes \\
$k_j'$ & Background fluxes along exceptional divisor \\
$(\bfk),\llbracket \boldsymbol{k}^2 \rrbracket $ & Combinations of $\bfk_j$ \\
$\kappa$ & The combination $(N^2-1)\chih+(N-1)\ov{f}_2(P)^2$ \\
$\varkappa$ & The complex conjugation map \\
$K$ & Canonical class of $X$ \\
$K_n$ & Normalization of SW contribution at $n$-th $\CN=1$ vacua \\
$K_\rho$ & Field extension of $\mathbb{Q}$ by $i,\varepsilon_{2\rho}$ \\
$K_X$ & Canonical bundle of $X$ \\
$K(S)$ & Grothendieck group of coherent sheaves on $S$ \\
$L$ & Middle cohomology lattice $H^2(X,\mathbb Z)$ modulo torsion \\
 $\CL$ & Zero mode Lagrangian \\
$\CL_j$ & Line bundle coupled to $j$-th hypermultiplet \\
$\Lambda_{N_f}$ & Scale of SQCD with $N_f$ flavors \\
$\lambda$ & Seiberg--Witten differential \\
$\lambda_1,\lambda_2,\lambda_3$ & Rosenhain invariants \\
$\lambda^a$ & Shifted lattice vectors of the IR lattice sum \\
$\lambda^a_{\Z}$ & Integer cohomology part of $\lambda^a$ \\
$\Lambda_{\rm r}$ & Root lattice of $\SU(N)$ \\
$\Lambda_{\rm w}$ & Weight lattice of $\SU(N)$ \\
$m_1,\dots,m_{N_f}$ & Bare masses of $N_f$ hypermultiplets \\
$m$ & Bare equal mass of the $N_f$ hypermultiplets \\
$\bfm $ & Mass vector $(m_1,\dots, m_{N_f})$ \\
$\vec{m}_a$ & Minimal weights of $\SU(N)$ \\
$\CM_{\rm Coulomb} $ & Regularized Coulomb branch \\
$\tm_{N_f}$ & The quantity $\Delta_{N_f}/(a^1\cdots a^{N-1})$ at the $\CN=1$ singularity \\
$\bfmu(b)$ & Cohomology class of gerbe $b$ \\
$\bfmu(\mf{b})$ & Representative of $\mu(b)$ in $H^2(X,\Lambda_{\rm w}/\Lambda_{\rm r})$ \\
$\ov{\bfmu}(\mf{b})$ & Lift of $\bfmu(\mf{b})$ to $H^2(X,\Lambda_{\rm w})$ \\
$\mud$ & Donaldson map \\
$M$ & Moduli space of semistable sheaves on $S$ with fixed $(\rho, c_1,c_2)$ \\
$\CM_{v(E)}^{\CQ,N_f}$ & Moduli space of solutions to non-abelian monopole equations \\
$\CM^{\rm i}_{v(E)}$ & Instanton moduli space with $v(E)=(\rk(E),c_1(E),c_2(E))$ \\
$\CM_{v(E)}^{\rm r}$ & Reducible component of above moduli space \\
$M^{[\rm vir]}$ & Virtual fundamental class\\
$\bm{n}_e$ & Electric charge of a BPS state \\
$N_f$ & Number of hypermultiplets \\
$\bm{n}_m$ & Magnetic charge of a BPS state \\
$\tn_{N_f}$ & The quantity $\Delta_{N_f}/(q_{11}\cdots q_{(N-1)(N-1)})$ at the $\CN=1$ vacua \\
$\nabla^{\rm tw}$ & Connection on twisted bundle $P^{\rm tw}$\\
%$\nu$ & $u$-plane measure factor \\
$\CO$ & Observable in TQFT \\
$\langle \CO\rangle$ & Correlation function of $\CO$ on $X$ \\
$\langle \CO\rangle^{\rm i}$ & Contribution of instanton component to $\langle\CO\rangle$ \\
$\Omega$ & A point in the Siegel upper half space\\
$\omega_a$ & Holomorphic differentials on SW curve \\
$p,p^a,p'$ & Fugacity for the point class \\
$P$ & Principal $\SU(N)/\Z_N$-bundle \\
$P^{\rm gauge}$ & Gauge bundle \\
$P^{\rm f}$ & flavor $\mr{U(1)}^{N_f}$-bundle \\
$P^{\mr{f};i}$ & $i$-th flavor U(1)-bundle  \\
$P^{\rm R}$ & R-symmetry bundle \\
$P^{\rm tw}$ & Principal $G^{\rm tw}$-bundle for twisted theory\\
% $\PM_{N_f}$ & Matone's polynomial \\
%$\wp$ & \ws{} $\wp$-function \\
%$\Phi$ & $u$-plane integral \\
$\phi$ & Adjoint vector multiplet scalar \\
$\phi_m$ & Component of adjoint vector multiplet scalar \\
$\Phi^m_{\,n}$ & $\CN=1$ chiral multiplet in adjoint\\
$\varphi^{\rm tw}$ & Twisting homomorphism\\
$\varphi_{g}$ & Genus $g$ Siegel modular form\\
$\Pi_{am}$ & Period matrix of a genus $g$ curve \\
$\Psi,\widetilde{\Psi}$ & Lattice sum in the IR \\
$Q(\cdot)$ & Quadratic form associated to intersection form on $X$ \\
$q_{IJ},\tq_{ab,N_f}$ & Exponentiated coupling matrix\\
$\widetilde{q}_{aa}$ & Ratio of $a^a$ and $q_{aa}$\\
$\widetilde{q}_{ab}$ & Inverse of $q_{ab}$\\
$Q_m^i$ & $\CN=1$ chiral multiplet in fundamental\\
$\widetilde{Q}^m_i$ & $\CN=1$ chiral multiplet in anti-fundamental\\
$Q_s$ & Universal function \\
$r$ & Rank of the gauge group
\\
$\rk(E)$ & Rank $N$ of the vector bundle $E$
\\
$r_n$ & The reduction modulo $n$ map from $\Z$ to $\Z_n$\\
$\rho$ & Rank of semistable sheaves on $S$ \\
$\vec{\rho}$ & Weyl vector of $\SU(N)$ \\
$\rho_s$ & The homomorphism from $\Z_2$ to $\CZ$\\
$\rho^{\rm gauge}$ & The homomorphism from $\Z_N$ to $\CZ$\\
$R_s$ & Universal function \\
$R[[z]]$ & Formal power series in $z$ over a ring $R$ \\
$\CR$ & Blowup factor \\
$s$ & Rank of $\alpha\in K(S)$ \\
$\sigma$ & Signature $b_2^+-b_2^-$ of $X$ \\
$S$ & Smooth projective surface (sec.~\ref{sec:top_cor_seg_num}) and surface class (sec.~\ref{sec:IR theory}) \\
$S_+$ & Self-dual part of Pontryagin dual of surface $S$ \\
$S^\pm$ & Chiral spinor bundles
\\
$\segre_\alpha$ & Virtual Segre numbers of $S$ \\
$S_s$, $S_{j,s}$ & Universal function \\
$\SW$ & Seiberg--Witten invariant, differential geometry convention \\
$\mathcal{SW}$ & Seiberg--Witten invariant, algebraic geometry convention \\
$t$ & Change of variables \\
$T_{a,b}$ & Contact term for surface observables \\
$\Theta_\Gamma$ & Lattice theta function associated to lattice $\Gamma$ \\
$\Theta_{1},\Theta_{2},\Theta_{3},\Theta_{4}$ & Fundamental even theta constants \\
$\Theta[\vec\gamma,\vec\delta]$ & Riemann theta function with characteristic vector $\vec\gamma,\vec\delta$\\
$\Theta_{\log}^{(a_1,\dots,a_k)}$ & $k$-th order logarithmic derivative of $\Theta$\\
$t_{pq},t_{pq}^{\mathrm{R}},t_{pq}^{\mathrm{f};i},t_{pq}^g$& Transition functions for various bundles
\\
$(\widetilde{t}^+_{pq},\widetilde{t}_{pq}^{-})$& Local lift of transition functions of frame bundle
\\
$\widetilde{t}_{pq}^{\mathrm{R}},\widetilde{t}_{pq}^{\mathrm{f};i},\widetilde{t}_{pq}^g$& Local lifts of transition functions for various bundles
\\
$\tau_{ab},\tau_{IJ}$ & Low-energy effective coupling matrix on the CB \\
% $\ov{\tau}_{ab},\ov{\tau}_{IJ}$ & Complex conjugate of coupling matrix \\
$\tau,\tau_{\rm uv}$ & UV coupling \\
$\jt_j$ & Jacobi theta function \\
$T_a,T^{(a)}_s$ & Universal function related to $\tu_{s;a}^{(0)}$ \\
$U$ & Point observable on the Coulomb branch\\
$u_a$ & Coulomb branch coordinates \\
$U_{pqr\dots}$ & Overlap $U_p\cap U_q\cap U_r\cap\cdots$ of open sets of a good cover \\
$\upsilon$ & Monodromy operator \\
$\vec{\tu}_{N_f}^{(n)}$ & Coordinate of $n$-th $\CN=1$ vacua of SQCD with $N_f$ flavors\\
$\tu_{a;N_f}^{(n)}$ & Components of $\vec{\tu}_{N_f}^{(n)}$\\
$\universal_\alpha$ & Product of universal functions \\
$\universal_\alpha^{\text{no obs}}$ & Universal Segre series without observable insertion \\
$\vd M$ & Virtual dimension of $M$\\
$v_j$ & Linear coupling to background fluxes \\
$V^j,V^j_k$ & Index bundle for adjoint Dirac operator of $\slashed D_j^\dagger$ \\
$V_s$ & Universal function \\
$w_2(X)$ & Second Stiefel--Whitney class of $X$\\
$W^{m}_{\,n}$ & $\CN=1$ chiral field strength multiplet\\
$W_s$ & Universal function \\
$W_j^{\pm}$ & Chiral $\spinc$ bundle for $j$-th hypermultiplet \\
$W^j,W_k^j$ & Index bundle for Dirac operator of $\slashed D_j$ \\
$x,x'$ & Surface class \\
$\tx_{a,N_f}$  & The quantity $\frac{\partial u_1}{\partial a^a}$ at singularity
\\
$\tx_{N_f}$  & Determinant of period $\frac{\partial u_a}{\partial a^b}$ at singularity \\
$X$ & Smooth, oriented, compact, simply-connected four-manifold \\
$\widetilde X$ & Blowup of $X$ at a point \\
$X_s$ & Universal function \\
$y_{IJ}$ & Imaginary part of coupling matrix $\tau_{IJ}$\\
$Y_s$, $Y_{j,s}$ & Universal function \\
$z$ & Fugacity for Segre numbers \\
$\vec z$ & Elliptic variable of Riemann theta function\\
$\CZ$ & The group $\Z_{\mr{lcm}(2,N)}$ \\
$Z_{\rm Coulomb}(\bm{m})$ & Contribution from Coulomb branch to partition function \\
$Z_{\rm D},Z_{\rm D}(\bm{m})$ & Topological partition function of $\CN=2$ $\SU(N)$ SQCD \\
$Z^{\rm i}_{\rm D},Z^{\rm i}_{\rm D}(\bm{m})$ & Contribution of instanton component to partition function \\
$Z^{\rm r}(\bm{m})$ & Contribution of reducible component to partition function \\
$Z_s$, $Z_{jk,s}$ & Universal function \\
$Z_\SW(\bm{m}),Z_\SW$ & Seiberg--Witten contribution to partition function \\
$Z_{\SW}^{(\ell)}$ & Seiberg--Witten contribution from substrata $\mathcal{D}^{(\ell)}$ \\
$Z_{\SW,n}$ & Seiberg--Witten contribution from $n$-th $\CN=1$ vacua \\
$Z^{\rm i}_\SW(\bm{m}),Z^{\rm i}_\SW$ & Instanton component of SW contribution \\
$Z_{\SW}^{\rm coll}$ & SW contribution of colliding quark singularities \\
$Z_{\SW}^{\CN=1}$ & SW contribution of $\CN=1$ vacua \\
$Z_{\SW}^{\rm SCFT}$ & SW contribution of multicritical SCFT points \\
$Z_{\gamma}$ &Central charge of BPS state with charge $\gamma$ \\
$Z_u$ & Contribution from $u$-plane to partition function for rank 1 \\
$\Z_N^{(1)}$ & One-form symmetry group of pure $\SU(N)$ theory\\
$\Z_2^{\mathsf C}$ & $\Z_2$ charge conjugation symmetry group\\
$\Z^{(r)}_{4N}$ & Broken $\mr{U(1)}_r$-symmetry on the Coulomb branch\\
$\zeta^{\pm}_{pqr},\zeta^{\rm R}_{pqr}$ & Cocycle representing various cohomology classes \\
$\zeta^{\mr{f};i}_{pqr},\zeta^{g}_{pqr}$ & Cocycle representing various cohomology classes \\ 
\hline 
\end{longtable}

\renewcommand{\baselinestretch}{1.2} 
\bibliography{bib} 
\bibliographystyle{ytamsalpha} 
\end{document}